\documentclass[a4paper,11pt,twoside,openright]{book}

\usepackage[T1]{fontenc}
\usepackage[utf8]{inputenc}
\usepackage[english]{babel}

\usepackage{amsmath, amssymb, amsfonts, latexsym}
\usepackage{commath}
\usepackage{bm}

\usepackage{newtxtext,newtxmath}

\usepackage{graphicx, wrapfig, float, subcaption, longtable, caption}
\usepackage{tikz}
\usetikzlibrary{arrows,positioning,shapes.geometric}
\usepackage{tikz-3dplot}
\usepackage{pdfpages}
\usepackage{pgfplots}
\pgfplotsset{width=7cm,compat=1.10}
\usepackage{circuitikz}

\usepackage{pst-optic, pst-optexp, pstricks-add, pst-eucl, pst-intersect}

\usepackage{booktabs, enumerate, quoting, siunitx, rotating}

\usepackage{fancyhdr}
\renewcommand{\chaptermark}[1]{\markboth{\thechapter.\ #1}{}}

\usepackage{geometry}
\usepackage[colorlinks=false,hidelinks]{hyperref}
\usepackage[version=3]{mhchem}
\usepackage{pdflscape}
\usepackage{cite}
\usepackage{environ}

\begin{document}

\thispagestyle{empty}
\cleardoublepage
\begin{titlepage}
\thispagestyle{empty}
	\begin{center}
    {\Large \textbf{RF Helicon Plasma Thruster for an Atmosphere-Breathing Electric Propulsion System (ABEP)}\par}
    \vspace{3cm}
	{A thesis accepted by the Faculty of Aerospace Engineering and Geodesy\\ of the University of Stuttgart in fulfilment of the requirements\\ for the degree of Doctor of Engineering Sciences (Dr.-Ing.)\par}
	\vspace{3cm}
	{By\par
	\Large \textbf{Francesco Romano}\par}
	{born in\\ Sassari\par}
    \vspace{3cm}
    \end{center}

    {Main referee: Priv. Doz. Dr.-Ing. Georg Herdrich\par
    Co-referee: Prof. Dr.-Ing. Stefanos Fasoulas, Prof. Daniele Pavarin \par}
    {Day of defense: 30.07.2021\par}
\begin{center}
	\vspace{2cm}
    {Institute of Space Systems (IRS)\\ University of Stuttgart\par
    \textbf{2021}}
	\end{center}
\end{titlepage}

{\markboth{}{} 
\thispagestyle{empty}
\cleardoublepage
{\markboth{}{} 
\thispagestyle{empty}

\begin{flushright}
\textit{Dedicato alla mia famiglia}
\end{flushright}

\vspace{80pt}
\begin{flushright}
\textit{The first principle is that you must not fool yourself -\\
and you are the easiest person to fool.} \\
Richard Feynman, \textit{"Surely You're Joking, Mr. Feynman!"}\\
\end{flushright}

\clearpage 
{\markboth{}{} 
\thispagestyle{empty} 

\frontmatter

\tableofcontents
\pagestyle{empty}
\chapter*{\centering Abstract}
\addcontentsline{toc}{chapter}{Abstract}
This dissertation deals with the development of Atmosphere-Breathing Electric Propulsion (ABEP) technology, that can enable propellant-less continuous orbiting in very low Earth orbits (VLEO). It uses an intake in front of the spacecraft  to collect the residual atmosphere and deliver it to an electric thruster as propellant, finally utilizing the cause of aerodynamic drag as source of thrust. 
A literature review is presented to give the ABEP state-of-the-art of the technology and the most relevant performance parameters are highlighted. The application of ABEP in VLEO is investigated by applying analytical equations based on atmospheric models and intake efficiencies based on the outcome of this work, and available state-of-the-art thruster efficiencies. Such analysis derives the collectible propellant flow, the aerodynamic drag, and the power required to fully compensate the drag.  The case of GOCE using an ABEP system is presented, as well as its application in very low Mars orbit (VLMO).

The intake and the thruster are investigated and designed within this dissertation. 

Three ABEP intakes designs are hereby presented, based on gas-surface-interaction properties. Two are based on fully diffuse reflections, delivering collection efficiencies $\eta_c<0.5$ and one based on fully specular reflections of $\eta_c<0.95$. Their sensitivity to misalignment with the flow is analysed as well highlighting the specular design of being more robust compared to the diffuse one by maintaining relatively high $\eta_c$ even for large angles. 

The ABEP thruster is based on contactless technology: there is no component in direct contact with the plasma, and a quasi-neutral plasma jet is produced. This enables operation with multiple propellant species (also aggressive such as atomic oxygen in VLEO) and densities, and does not require a neutraliser. The thruster is based helicon plasma discharges to provide higher efficiency compared to inductive ones.

A numerical investigation to determine the basic thruster design parameters, such as input frequency, antenna, plasma density and required magnetic field is performed using the HELIC code. The analysis lead to the application of the birdcage antenna, that is simulated, assembled and experimentally verified. It operates at resonance and it is impedance-matched, leading to high electrical efficiency. The thruster is integrated and tested on \ce{Ar}, \ce{N2}, \ce{O2}. The successful testing highlights a low power consumption $P_f<\SI{60}{\watt}$, and the capability of ignition and operation for each propellant and for multiple mass flows that are within the ABEP expected range. Furthermore, a B-dot probe is designed and assembled aiming to detect helicon waves within the plasma plume.

Finally, the greatest challenge to continuously overcome the aerodynamic drag at low orbits is investigated within this dissertation, proposing the ABEP solution. This lead to three efficient intake designs based on verified particle simulations, and a new contactless helicon-based plasma thruster. The latter is verified, built, and tested, delivering high electrical efficiency, $>99\%$ in the vacuum case, capable to operate on atmospheric propellant, and requiring very low input powers, $P_f\sim\SI{60}{\watt}$. Therefore, both intake and thruster are promising technologies for a near-future real space application of the ABEP system.

Furthermore, outlooks and improvements for ABEP development and the required diagnostics are presented including potential spacecraft ABEP-based configurations for the long-term development of the ABEP system.

\mainmatter
\pagestyle{fancy}
\chapter{Introduction}
\label{ch:introduction}
The current fast grow of the satellite market together with that of the private space sector~\cite{book_OUR} requires a more affordable and reliable way to reach space and to stay there. As the re-usability of rockets became a reality with SpaceX, the cost of launches decreased and has been decreasing further while at the same time driving other companies of doing the same. This enables to launch more and more satellites into space. Recently, the focus is getting into lower altitude orbiting satellites around the Earth for a large variety of reasons. Higher resolution imaging with less complex optics, larger signal-to-noise ratio for radar applications, improved geospatial position accuracy and communication link-budgets, reduced radiation doses that might enable the use of less expansive electronics onboard the satellite represent  the most frequently discussed potential benefits. Moreover, orbiting at low altitudes reduces the risk of collision with space debris and at the same time ensures that once the target satellite lifetime is reached, self-disposal is enabled by the drag produced by the residual atmosphere itself. Indeed, at very low Earth orbits (VLEO), $h<\SI{450}{\kilo\meter}$, the satellite lifetime is limited due to the aerodynamic drag from a couple of months to some days depending on both the altitude and the satellite's configuration, see Fig.~\ref{fig:lifetime}.

\begin{figure}[h]
\centering
\includegraphics[width=6cm]{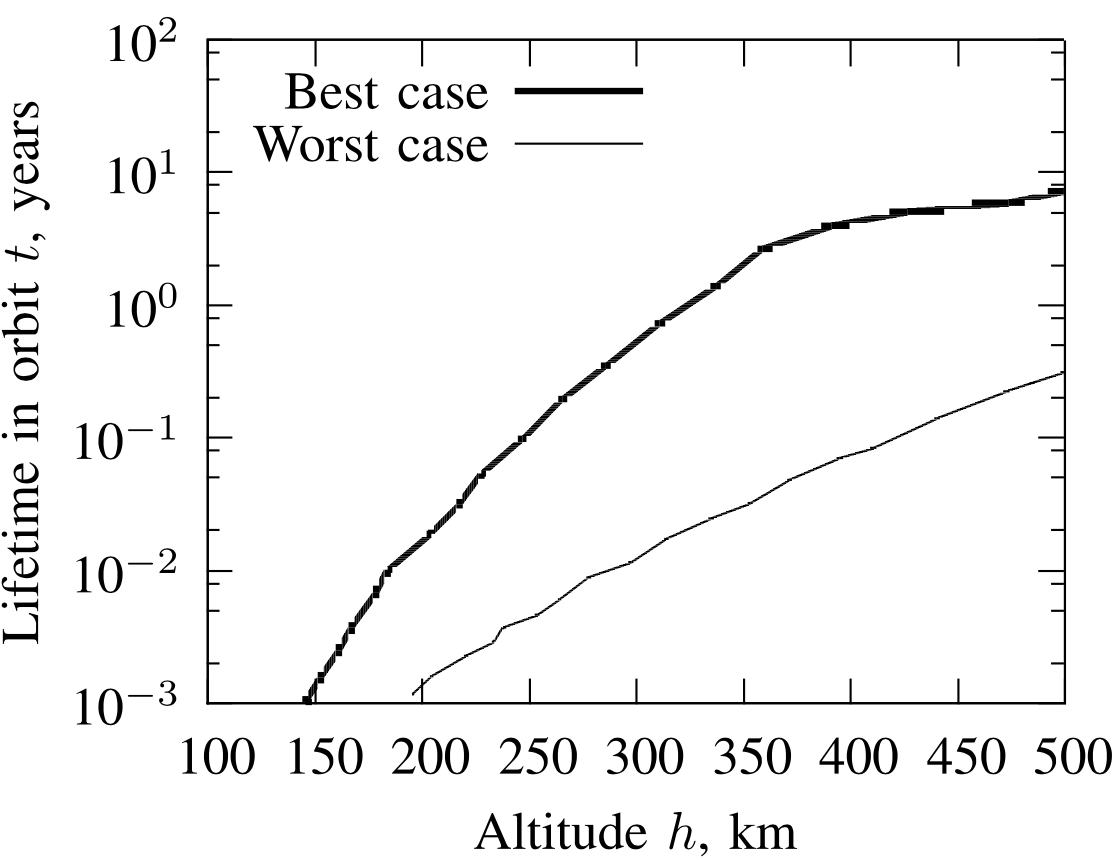}
\caption{Satellite's Lifetime vs Altitude. Best Case: Ballistic Coefficient of $Ba=\SI{200}{\kilo\gram\per{\meter^2}}$ at Min. Solar Activity. Worst Case: $Ba=\SI{20}{\kilo\gram\per{\meter^2}}$ at Max. Solar Activity~\cite{6945885}.}
\label{fig:lifetime}
\end{figure}

A propulsion system to compensate for the drag would mitigate such limited lifetime, however a large amount of propellant must be carried on board, increasing both mass and size of the spacecraft (SC). Atmosphere-Breathing Electric Propulsion (ABEP) is a concept that utilizes the same cause of the drag as a source for the thrust. The system would collect the atmospheric particles by means of an intake and use them as propellant for an electric thruster. Such systems can be a part of a satellite orbiting any celestial body with atmosphere given that, theoretically, enough electrical power is available, but also that the requirements in terms of mass, volume, and cost are met. 

A preliminary approach to highlight the basic dependencies,  is based on the equilibrium between aerodynamic drag $D$ and thrust $T$, that is $T=D$ which is directly linked to the atmospheric environment and satellite design, the frontal area $A_f$ and the drag coefficient $C_D$, but especially on the two main parameters that define the ABEP system efficiency:
\begin{itemize}
\item Intake collection efficiency $\Rightarrow$ $\eta_c=\frac{\dot{N}_{thr}}{\dot{N}_{in}}$
\item Total thruster efficiency of an electric thruster $\Rightarrow$ $\eta_T=\frac{P_{jet}}{P_{in}}$
\end{itemize} 
The intake collection efficiency $\eta_c$ is the ratio between the amount of particles collected and delivered to the thruster by the intake $\dot{N}_{thr}$, and the amount of those encountered by the intake $\dot{N}_{in}$. The thruster efficiency $\eta_T$ is the ratio between the total amount of input electric energy $P_{in}$, and that which is converted into kinetic energy, the jet power  $P_{jet}=\frac{1}{2}\dot{m}_{thr}c^2_e$, where $c_e$ is the exhaust velocity~\cite{goebel2008fundamentals}. Therefore, to have an efficient ABEP-system, the single efficiencies of thruster and intake, $\eta_T$ and $\eta_c$ must be maximized. 

The aerodynamic drag force can be calculated as $D=\frac{1}{2}\rho(h)v^2(h) A_f  C_D$. Considering the SC aligned with the flow, and so the intake, the incoming particle mass flow is calculated as $\dot{m}_{in}(h)=\rho(h)v(h)A_{in}$, where $\rho(h)$ is the atmospheric density and $v(h)$ is the orbital velocity, both depending on altitude $h$, and $A_{in}$ is the intake collecting area, with the condition $A_{in}\leq A_f$. Finally, the relation between the jet power $P_{jet}=\frac{1}{2}\dot{m}_{thr}(h)c_e^2(h)$ and input power $P_{in}$ is $P_{in}=P_{jet}/\eta_T$. Combining such equations, leads to the electrical power required by the thruster $P_{in}$ for an ABEP system to fully compensate the drag, $D=T$. The respective dependencies are highlighted in Eq.~\ref{eq:reqpowera} and are summarized in the following bullet list:
\begin{itemize}
\item SC design: drag coefficient $C_D$ and frontal area $A_f$;
\item Orbit and environment: velocity and density of the incoming flow $v(h)$, $\rho(h)$;
\item ABEP system: intake area and efficiency, thruster efficiency $A_{in}$,~$\eta_c$,~$\eta_T$.
\end{itemize} 

\begin{equation}
P_{in}= \frac{1}{8} \overbrace{(C_D A_f)^2}^\text{SC Design} \underbrace{v^3(h)\rho(h)}_\text{Orbit, Environment} \overbrace{\biggl(\frac{1}{\underbrace{A_{in} \eta_c}_\text{Intake /}} \frac{1}{\underbrace{\eta_T}_\text{Thruster}}\biggr) }^\text{ABEP System}
\label{eq:reqpowera}
\end{equation}

This dissertation deals with the development of an ABEP system, with particular focus on the electric thruster, which aims to ionize and accelerate the collected atmospheric particles to produce thrust in a contactless manner by using radio frequency (RF) waves and electromagnetic (EM) fields. This is to be achieved with a combination of an optimized RF antenna, from both the electrical and plasma points of view, and by an applied static magnetic field, that is also aimed to trigger the formation of helicon waves within the plasma discharge. The contactless nature of the thruster is crucial to operate with the non-uniform atmospheric environment (pressures $p_i$ and densities $\rho_i$ for the different species $i$) and aggressive propellant species while, at the same time, reducing thruster's complexity by producing a quasi-neutral plasma plume that makes the implementation of a neutraliser not necessary. The main potential advantages are, in summary:	

\begin{itemize}
\item Contactless $\Rightarrow$ wide operation regime ($p_i$,~$\rho_i$), use of aggressive propellants;
\item Optimized RF antenna $\Rightarrow$ high electrical efficiency, plasma ionization, and acceleration;
\item Applied magnetic field $\Rightarrow$ plasma acceleration, Helicon waves for high efficiency;
\item Quasi-neutral plasma plume $\Rightarrow$ neutraliser not required.
\end{itemize}

\chapter{Structure of the Dissertation}
This dissertation is structured as following:

Chapter~1 and 2 provide introduction to the work presented hereby, highlighting the main ABEP-related parameters of interest as well as an overview of the dissertation's structure itself. 

Chapter~3 describes the ABEP working principle, and provides the literature review of ABEP-systems state-of-the-art as well as the respective system analysis for a VLEO mission using ABEP. The chapter concludes with an hypothetical application for an Earth as well as for a Mars observation mission. 

Chapter~4 describes the ABEP intake working principle, the balance model for the analytical performance estimation, and concludes with three intake designs based on Direct simulation Monte Carlo.

Chapter~5 provides with the basic plasma physics required for the ABEP thruster design, its working principle, the pre-design phase based on top level requirements, the radio-frequency technology fundamentals, the thruster's circuit design, and, finally, the detailed birdcage antenna principle of operation, design, and experimental verification. 

Chapter~6 describes the facility set-up used for the thruster test-campaign in each subsystem including an impedance model for the thruster as well. 

Chapter~7 describes the principle of operation, design, and calibration of a 3-axes magnetic-inductive B-dot probe. 

Chapter~8 presents the thruster's test campaign using Argon, Nitrogen, and Oxygen as input propellant.

Chapter~9 contains the discussion and conclusion, while Chapter~10 presents the outlook for future work. 

\chapter{Atmosphere-Breathing Electric Propulsion}
This chapter presents the concept of Atmosphere-Breathing Electric Propulsion (ABEP), its definition and the main parameters of concern. Orbiting at very low altitudes around any celestial body with atmosphere can open a new range of opportunities for many kind of space missions and applications, where "very low" is defined as the altitudes at which the aerodynamic force plays a role onto spacecraft design~\cite{vleobenefit}.  Mission's lifetime at low altitudes is finally limited due to the aerodynamic drag, as it causes a rapid orbit decay leading to missions shorter than 30 days depending on the altitude, see Fig.~\ref{fig:lifetime}. To extend it, the SC needs to minimize the resulting aerodynamic drag force. Passive methods are a tailored SC aerodynamic geometrical design, and the implementation of materials with specific particle-interaction properties to reduce energy exchange between incoming atmospheric particles and the SC itself~\cite{MOE2005793}. Active methods include an efficient propulsion system that actively compensates the aerodynamic drag, the use of steerable/fixed surfaces with specific particle-interaction material properties to actively control lift and drag combined with the attitude determination and control subsystem (ADCS)~\cite{CRISP202185}, conventional propulsion~\cite{GOCE}, an ABEP system, or the hybridization of aforementioned methods. This can enable the change of orbital parameters by combining drag and lift forces with propulsion~\cite{aerodynamiccontrol}. Finally, if a conventional propulsion system is implemented, the amount of propellant that can be carried on-board and used to counteract for the drag will, most probably, define the mission's lifetime~\cite{GOCE}.\begin{itemize}
	\item Passive methods: minimize aerodynamic drag:\begin{itemize}
		\item drag-reducing SC tailored geometry/design;
		\item advanced materials with specific particle-interaction properties.
	\end{itemize}
	\item Active methods: counteract aerodynamic drag:
	\begin{itemize}
		\item combination of ADCS and fixed/steerable surfaces that have specific particle-interaction material properties;
		\item conventional propulsion for drag compensation and/or manoeuvring;
		\item ABEP-based drag compensation and/or manoeuvring;
		\item hybridization of fixed/steerable surfaces with conventional or ABEP-based propulsion.
	\end{itemize}
\end{itemize}

New and more advanced missions in low Earth orbits are, for example, continuous measurements of low atmospheric properties, gravitational and magnetic field measurements~\cite{GOCE}, more detailed agriculture, water-security, climate-change related monitoring, higher resolution data for humanitarian and crisis managements, and improved imaging~\cite{vleobenefit}. An ABEP system is not limited to small satellites only, but can be also conceptually applied to much larger SC that dive into the atmosphere to gather propellant, store it if necessary, and later use it to produce thrust~\cite{demetriades1959novel,demetriades1962plasma,demetriades1962preliminary,demetriades1963propulsive,ABEP0002,ABEP0003}. Finally, it could ideally nullify the on-board propellant requirement, provide drag compensation, and produce additional $\Delta v$ for orbit raising, lowering, and controlled de-orbiting. 

\section{ABEP Concept}
The concept of an Atmosphere-Breathing Electric Propulsion System (ABEP) is to capture the residual atmosphere of the planet and to use it as propellant for the electric thruster to compensate for the aerodynamic drag, see Fig.~\ref{fig:ABEP}. The main functions of an ABEP system are:
\begin{itemize}
	\item Collecting the residual atmospheric particles to be used as propellant;
	\item ionizing and accelerating the collected propellant to produce thrust.
\end{itemize}
\begin{figure}[H]
	\centering
	\includegraphics[width=14cm]{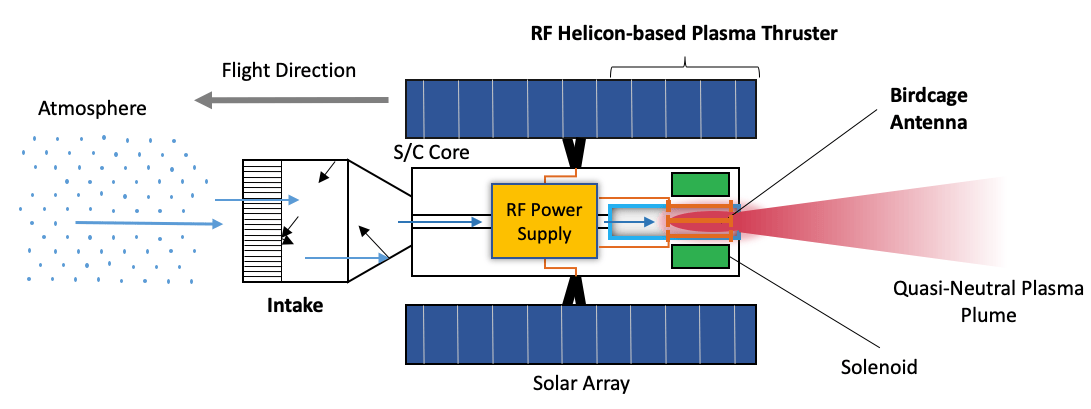}
	\caption{Atmosphere-Breathing Electric Propulsion (ABEP) Spacecraft using the RF Helicon-based Plasma Thruster.}
	\label{fig:ABEP}
	\vspace{-10pt}
\end{figure}
An ABEP system can be theoretically implemented on SCs orbiting at very low altitudes around any celestial body with atmosphere, and can continuously ingest the atmosphere in front of it and use it to produce thrust (and compensate for the aerodynamic drag). The intake collects the atmospheric particles in front of the SC and delivers them to the thruster. The intake design is based on the free molecular flow (FMF) condition which is highly dependent on the interaction between particles and surfaces. The reference performance parameter of an ABEP intake is the intake collection efficiency $\eta_c$. It is defined as the ratio between the collected particle flow and the incoming particle flow. In Earth's atmosphere, an ABEP system operates in the very low Earth orbit (VLEO) $h<\SI{450}{\kilo\meter}$~\cite{vleobenefit}, in which the most dominant components are \ce{N2} and atomic oxygen \ce{O}, also named AO. Due to the aggressive nature of AO, its use as propellant for a conventional thruster/EP system such as gridded ion thruster (GIT) or Hall-effect thruster (HET) lead to rapid performance degradation over time due to erosion of the grids (GIT) or of the discharge channel (HET)~\cite{presitael1,presitael2}. The second main issue of conventional EP operating on atmospheric propellant, is the need of a neutraliser: such devices either need a supplemental propellant tank of some noble gas such as \ce{Xe}, or a new design to allow long time operation on atmospheric propellant~\cite{SITAEL2015,SITAEL2016,SITAEL2017,SITAEL2019a,SITAEL2019b}. Moreover, the atmosphere of a planet is a non-homogeneous environment, and the system must cope with variable propellant density and composition. This requires an ABEP system to be throttleable, ignitable at low pressure, and flexible in terms of propellant as to cope with variations in its composition and density over time. Therefore, the chosen approach is of an RF plasma thruster that has no component in direct contact with the plasma, namely contactless. The reference parameter of an electric thruster is the thruster efficiency $\eta_T$, defined as the ratio between the jet power and the input electrical power.  In summary, when dealing with conventional EP systems, the major challenges of an ABEP system are the following:
\begin{itemize}
	\item Reactive propellants deteriorate thruster's components that are in direct contact with it (e.g. electrodes, accelerating grids, discharge channels, etc.);
	\item Density and composition of the ingested propellant are variable along the orbit;
	\item Need to operate a neutraliser on atmospheric propellant.
\end{itemize}

	\section{Literature Review}
	A literature review of the most relevant ABEP studies is briefly presented in the following. Early air-breathing concept studies date back to 1959, see~\cite{demetriades1959novel,demetriades1962plasma,demetriades1962preliminary,demetriades1963propulsive,ABEP0002,ABEP0003}. In this section, only the most advanced studies are briefly presented with the corresponding main mission requirements and system performances.\\

\subsubsection{Air-Breathing Cylindrical Hall-Effect Thruster (ABCHT)}
The ABCHT concept has been proposed by Diamant in 2009~\cite{dia1,dia2}. 
ABCHT  is a 2-stage thruster made of an electron cyclotron resonance (ECR) ionisation stage, and a cylindrical HET. Tests have been performed with \ce{Xe} as propellant. The spacecraft concept has a frontal area of $A_f=\SI{0.5}{\square\meter}$ and a $C_D=2.2$. The estimated intake efficiency, of a not given design, is $\eta_c=0.35$ and a respective intake area of $A_{in}=\SI{0.25}{\square\meter}$. Using a thruster power of $P=\SI{0.95}{\kilo\watt}$, the spacecraft can theoretically perform the mission at an altitude of $h=\SI{220}{\kilo\meter}$ with full drag compensation. The thruster, delivering $\eta_T=25\%$, is provided with a pressure of $p_{in}=\SI{0.01}{\pascal}$ to ignite, requiring the intake to perform a compression of 500:1. Based on the data presented in the study, a mass flow to the thruster is estimated to be $\dot{m}_{thr}=\SI{0.108}{\milli\gram\per{\second}}$, requiring $T=\SI{5}{\milli\newton}$ resulting in $T/P=\SI{5.3}{\milli\newton\per{\kilo\watt}}$ with $I_{sp}=\SI{4720}{\second}$. In the study, it is pointed out that the high-temperature cathode exposed to oxidizing substances will restrict the thruster's lifetime. A further issue is the minimum pressure for cathode operation, that requires another compression mechanism. The additional stage for ECR is implemented to increase the thruster's efficiency required by the operation conditions. Finally, the use of \ce{Ar} or \ce{Xe} is proposed to operate a microwave cathode for neutralization, and it is calculated that a tank of \SI{8}{\kilo\gram} of \ce{Xe} is required for a 5-year mission.


 \subsubsection{University of Michigan: RF Plasma Thruster and Helicon Hall-Effect Thruster}
Shabshelowitz~\cite{shabshelowitz2013study} at the University of Michigan investigated both HET and RF technology for a thruster applied to an ABEP system and published the results in 2013. The assumed spacecraft's mass is $m=\SI{325}{\kilo\gram}$, the orbit is circular at $h=\SI{200}{\kilo\meter}$ around the Earth for 3 years. The frontal area is $A_f=\SI{0.39}{\square\meter}$, and the length $L=\SI{2.1}{\meter}$. The spacecraft has a cylindrical shape and is covered with solar cells. The inlet area is $A_{in}=\SI{0.25}{\square\meter}$ and the intake efficiency is arbitrarily assumed to $\eta_c=90\%$. The RF plasma thruster (RPT), see Fig.~\ref{fig:RPT}, is made by an helical antenna and six coils that generate the magnetic field required to trigger helicon waves within the plasma discharge. The RPT has been tested with \ce{Ar} and some thrust has been produced. By operating the RPT on \ce{N2} and air, it did not produce any measurable thrust.\begin{figure}[h]
 	\centering
 	\includegraphics[width=11cm]{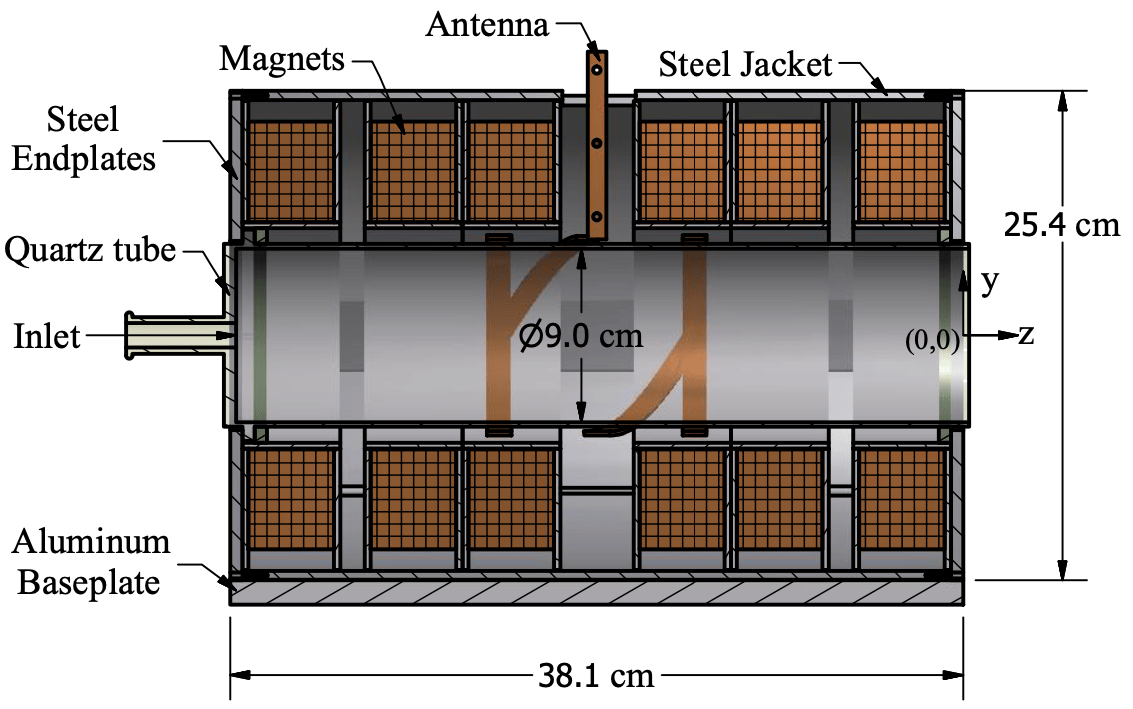}
 	\caption{RPT Concept~\cite{shabshelowitz2013study}.}
 	\label{fig:RPT}
 \end{figure} 
The helicon Hall-effect thruster (HHT), see Fig.~\ref{fig:HHT}, is a conventional HET with an helicon-wave based ionization stage. This can theoretically improve the thrust significantly, finally delivering $T/P < \SI{21}{\milli\newton\per{\kilo\watt}}$ on \ce{N2}. 
Finally, it is concluded that can a better understanding on helicon waves is required, and that the low performances can be increased by optimizing the thruster, and by improving several components of the propulsion system itself~\cite{shabshelowitz2013study}.\begin{figure}[h]
\centering
 	\includegraphics[width=7.5cm]{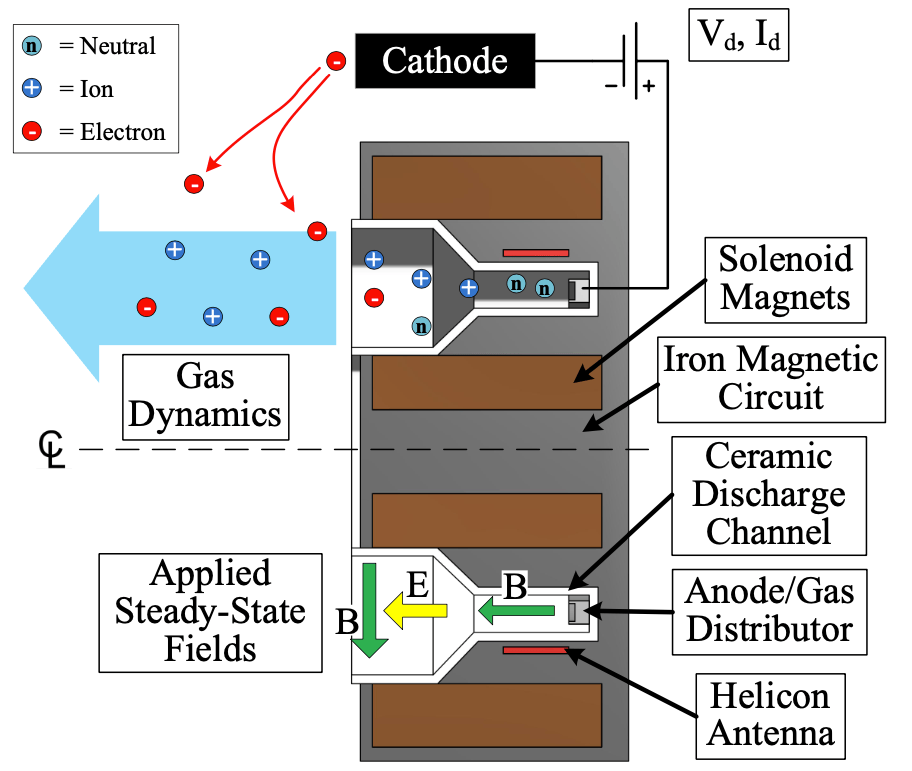}
 	\caption{HHT Concept~\cite{shabshelowitz2013study}.}
 	\label{fig:HHT}
 \end{figure} 
 
\subsubsection{Air-Breathing Ion Engine (ABIE)}
The studies from Nishiyama and JAXA date back to 2003~\cite{JAXA,JAXA2,JAXA3,JAXA4,JAXA5}, the concept is of an Air-Breathing Ion Engine (ABIE) as shown in Fig.~\ref{fig:ABIE}. Atmospheric propellant is ionized by an ECR-based thruster with grids for plasma acceleration. The microwave antenna ionizes the plasma within a region of magnetic field such that the ECR condition is satisfied.\begin{figure}[h]
 	\centering
 	\includegraphics[width=10.5cm]{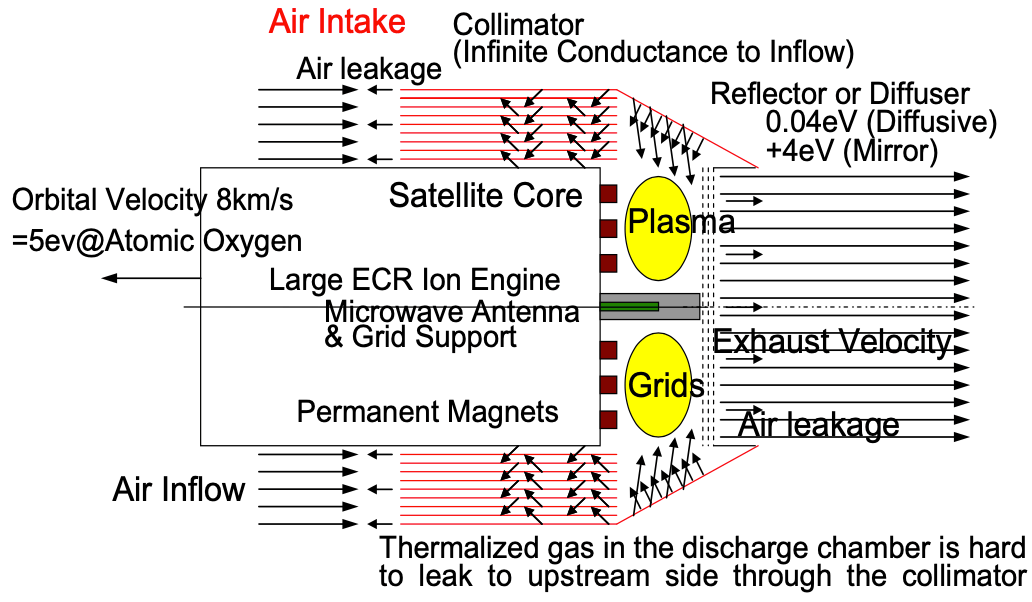}
 	\caption{JAXA ABIE Concept~\cite{JAXA3}.}
 	\label{fig:ABIE}
 \end{figure}
 Different applied voltages to the grids ensures that ions are extracted and accelerated while an external cathode neutralizes the plume. The concept is an all-in-one device, the outer ring region of the spacecraft is a ring-shaped intake with a honeycomb structure of small ducts in the front that works as trap for the collected particles. The intake is designed to provide at least $p_{in}=\SI{0.5}{\pascal}$ of pressure at the ionization region.
 The spacecraft concept is designed for an altitude of $h=150-\SI{200}{\kilo\meter}$.  The front area is assumed to be $A_f=\SI{1.5}{\square\meter}$ and $C_D=2$. The power required for full drag compensation at $h=\SI{170}{\kilo\meter}$ is  $P=4-\SI{5.59}{\kilo\watt}$ and $T/P\sim \SI{56}{\milli\newton\per{\kilo\watt}}$. Improvement of the ion source and of the intake is proposed by increasing their resistance to corrosion caused by AO. This can increase lifetime and thrust density. The mission lifetime at $h=\SI{170}{\kilo\meter}$ shall be of at least 2 years. With the same $A_f$ and $A_{in}=\SI{0.48}{\square\meter}$ is available of providing $\eta_c=46\%$. Finally, the thruster has been operated on basis of laser detonation generated flow and provided a thrust of $T=\SI{0.12}{\milli\newton}$ running at $P=\SI{54}{\watt}$, details on the neutralizer operation have not been found by the author.
 
\subsubsection{BUSEK Mars-Atmosphere-Breathing Hall-Effect Thruster}
The BUSEK company~\cite{BUSEK} developed an ABEP concept applied to a small SC orbiting Mars: the Martian Atmosphere-Breathing Hall-Effect Thruster (MABHET). Solar arrays aligned with the flow provide electrical power. The concept is depicted in Fig.~\ref{fig:BUSEK}. The HET has been tested with a propellant mixture that reproduces Mars' atmosphere: $95.7\%$ \ce{CO2}, $2.7\%$ \ce{N2} and $1.6\%$ \ce{Ar}. The measured thrust to power peak ratio is of $T/P<\SI{33}{\milli\newton\per{\kilo\watt}}$. The intake area is $A_{in}=\SI{0.15}{\square\meter}$ while the frontal area is $A_f=\SI{0.30}{\square\meter}$. The length of the intake is $L_{in}=\SI{3.7}{\meter}$ with $d_{in}=\SI{0.6}{\meter}$ diameter. The achieved collection efficiency in simulations is $\eta_c=35\%$ allowing a compression of up to 100:1. Finally, a power of $P=\SI{1.2}{\kilo\watt}$ is estimated to be required to provide full drag compensation in an altitude range between $h=120-\SI{180}{\kilo\meter}$. The MABHET shall operate better in Mars atmosphere compared to Earth's, because of lower atmospheric density and temperature, but also the accommodation coefficient, how atmospheric particles and spacecraft surfaces interact, would be different.\begin{figure}[h]
	\centering
	\includegraphics[width=7.5cm]{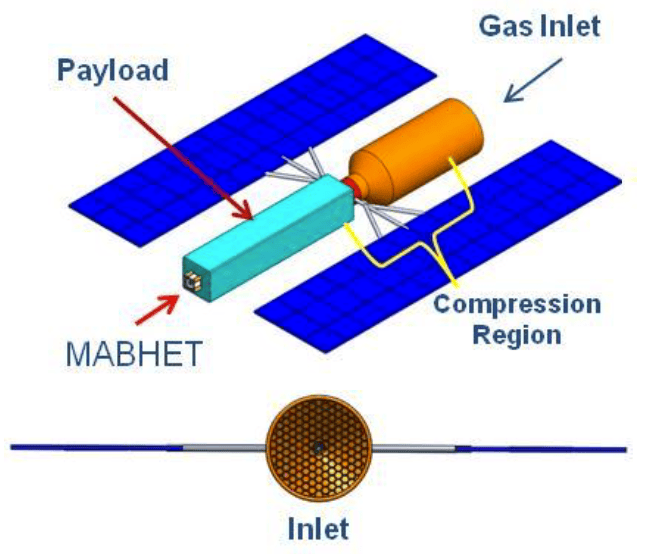}
	\caption{BUSEK MABHET Concept~\cite{BUSEK}.}
	\label{fig:BUSEK}
\end{figure} 
 
\subsubsection{RAM-EP, ESA and SITAEL}
ESA~\cite{di2007ram} proposed a technology demonstration mission featuring ABEP. The spacecraft weights \SI{1000}{\kilo\gram} equipped with four RIT-10 operating on atmospheric propellant. Intake and thrusters are physically separated. The orbit is a circular Sun-Synchronous Orbit (SSO) at an altitude of $h=180-\SI{250}{\kilo\meter}$ for a $3-8$ years mission. Front area is $A_f=\SI{1}{\meter^2}$ with a $C_D=2$, the maximum available power for the propulsion system is $P=\SI{1}{\kilo\watt}$, enabling a theoretical thrust level of $T=2-\SI{20}{\milli\newton}$. The intake has a front array of diffuser baffles to reduce back flow and, according to simulations, can provide up to $p=\SI{E-3}{\pascal}$ at the thruster. The total solar array (SA) area is of $A_{SA}=\SI{19.74}{\square\meter}$ capable to generate an EoL power of $P=\SI{2.9}{\kilo\watt}$ combined with a \SI{612}{\watt\hour} \ce{Li}-Ion battery to cope with power peaks. The work has been further developed later by testing RIT and HET~\cite{presitael1,presitael2} at Alta Spa (now SITAEL) with atmospheric propellant. The tests have highlighted the lifetime limitation of both thruster technologies when operating on atmospheric propellant and, for the same reason, the need to upgrade the cathode (neutralizer).
The experiments have been performed with the HET PPS 1350 and the RIT-10-EBB with pure \ce{N2} and mixtures of \ce{O2} and \ce{N2} as propellat, both at a power level of \SI{450}{\watt}. After a 10 hours test, the RIT showed no signs of erosion operating with \ce{N2}, whilst when operating on \ce{O2}, erosion of the accelerator grid was visible. For both thrusters, the performance was lower than with \ce{Xe} while operating on atmospheric propellant because of the thruster not being optimized for a different propellant, and because of the overall lower molar mass. The work has been reinitiated again in 2014 with the new SITAEL RAM-EP concept, a device that is composed by both intake and thruster as one unit~\cite{SITAEL2015,SITAEL2016}. The thruster is a modified HET with a pre-ionization stage right after the intake with a specific magnetic field configuration. The RAM-EP system is designed for altitudes $h<\SI{200}{\kilo\meter}$ and a lifetime of more than 4 years. The intake is built with a front cylindrical honeycomb structure of small ducts in a split-ring configuration, and a conical convergent section at the back, which condenses the ingested gas in front of the thruster, see Fig.~\ref{fig:SITAEL_Concept}.  The intake ducts have a length of $L_{duct}=\SI{0.3}{\meter}$ with an intake area of $A_{in}=\SI{0.126}{\square\meter}$ providing a simulated intake efficiency of $\eta_c=0.28-0.32$.  An HET has been used as particle flow generator: it creates a beam of high velocity plasma to the RAM-EP device, so that the functionality of both intake and thruster could be assessed, see Fig.~\ref{fig:SITAEL_Built}. Ignition is achieved with pure \ce{Xe}, followed by a transition to the desired $\ce{N2}-\ce{O2}$ mixture. The neutralizer is continuously operated with \SI{2}{\milli\gram\per{\second}} of \ce{Xe}. The generated thrust was $T=6\SI{\pm 1}{\milli\newton}$ compared to the drag to be counteracted of $D=26\SI{\pm1}{\milli\newton}$~\cite{SITAEL2017}. A new roadmap for RAM-EP development started in 2019, with updated mission requirements: an altitude range of $h=120-\SI{250}{\kilo\meter}$, a spacecraft length of $L=1.27-\SI{1.52}{\meter}$ a width of $W=1.79-\SI{3.82}{\meter}$, a solar array area between $A_{SA}=2-\SI{5.8}{\square\meter}$ to fit into a VEGA launcher, and power required between $P=480-\SI{1570}{\watt}$ with an expected drag of $D=4.9-\SI{22.4}{\milli\newton}$ and a required thrust of $T=6.9-\SI{25.8}{\milli\newton}$~\cite{SITAEL2019a,SITAEL2019b}.
\begin{figure}[h]
	\centering
	\includegraphics[width=10cm]{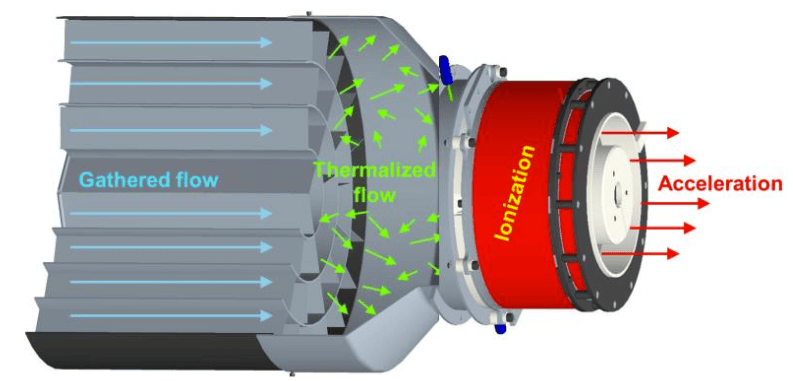}
	\caption{SITAEL RAM-EP Concept~\cite{SITAEL2017}.}
	\label{fig:SITAEL_Concept}
\end{figure} 
\begin{figure}[h]
	\centering
	\includegraphics[width=13cm]{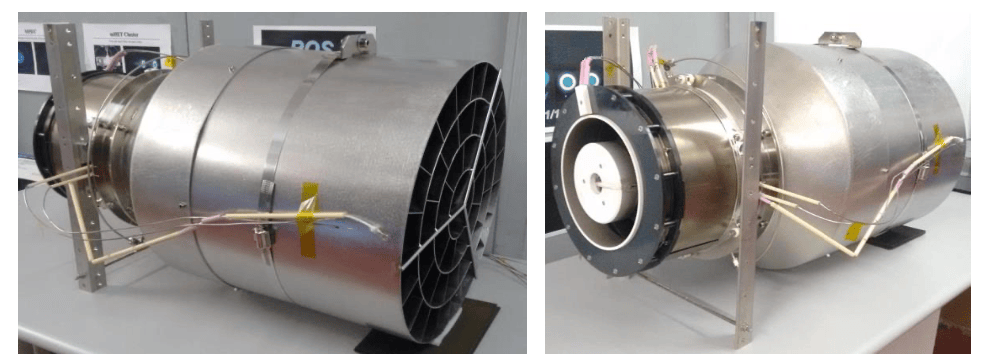}
	\caption{SITAEL RAM-EP Concept Assembled~\cite{SITAEL2017}.}
	\label{fig:SITAEL_Built}
\end{figure} 

\subsubsection{ABEP Studies at TsAGI and RIAME}
Notable are also the studies performed in Russia considering ABEP from the Central Aerohydrodynamic Institute (TsAGI) and the Research Institute of Applied Mechanics and Electrodynamics (RIAME) of Moscow Aviation Institute (MAI) considering the following: $A_f=\SI{1}{\square\meter}$, $A_{in}=\SI{0.1}{\square\meter}$, $C_D=2$, $h=185-\SI{222}{\kilo\meter}$, $\eta_T=0.4-0.6$ for an ion thruster on $\dot{m}_{thr}=0.075-\SI{0.11}{\milli\gram\per{\second}}$ of atmospheric propellant, $\eta_c=0.333-0.336$ requiring $I_{sp}=2500-\SI{4500}{\second}$ with $P<\SI{800}{\watt}$~\cite{TSAGI1,TSAGI2,filatyev2019control,TSaGI2018a,TSAGI2018ab,erofeev2017air,kanev2015electro}.

\newpage

The main mission, spacecraft, and ABEP systems parameters of this brief literature review are summed up within Tab.~\ref{tab:lite}.
\begin{table}[h]
\caption{ABEP Literature Review Summary}
\label{tab:lite}
\centering
\resizebox{\textwidth}{!}{
\begin{tabular}{llcllll}
	\toprule
	& ABCHT & Michigan & ABIE & BUSEK & RAM-EP & TsAGI/RIAME\\
 	& \cite{dia1,dia2} & \cite{shabshelowitz2013study} & ~\cite{JAXA,JAXA2,JAXA3,JAXA4,JAXA5} & \cite{BUSEK,BUSEK2} & \cite{SITAEL2015,SITAEL2016,SITAEL2017, SITAEL2019a, SITAEL2019b} & MAI~\cite{TSAGI1,TSAGI2,filatyev2019control,TSaGI2018a,TSAGI2018ab,erofeev2017air,kanev2015electro}\\
 	\midrule
	\textbf{Thruster} & ECR+HET & RPT / HHT & ECR$+$Grid & HET & HET & RIT\\
	\midrule
	Required$/$Assumed &&&&&&\\
	\midrule
	$T, \SI{}{\milli\newton}$ & 5 &$8.8$ & $(40-78.4)^{*}$  & $5-100$ & $33.23$ & $50-100$\\
	$T/P, \SI{}{\milli\newton\per{\kilo\watt}}$ &$5.3^*$ & $29-59$ & $10-14$ & $30$& NA & $(100-200)^*$\\
	$P$, \SI{}{\kilo\watt} & 0.95 & 0.3 & $4-5.6$ & $0.1-2.5$ & NA& $<0.5$\\
	$I_{sp}$, \SI{}{\second} & $4721^*$ & 1750& $3000$ & $1200-2000$& $2000-4000$ & $3543-3573$\\
	$\eta_T$, - & 0.25 & 0.25  & NA & $>0.2$ & NA & $0.4-0.6$ \\
	\midrule
	Experiment &&&&&&\\
	\midrule
	$T, \SI{}{\milli\newton}$ & NA & NA / $40$  & $0.12$  & $46.5^{*}$ & $6$ & $1.7-7.3$ \\
	$T/P, \SI{}{\milli\newton\per{\kilo\watt}}$ &NA & NA / $<21 $ & $2.4^{*}$ & $31$& $1.74^{*}$& $9.13^{*}$\\
	$P$, \SI{}{\kilo\watt} & $<0.096$ & 1 & $0.054$ & $1.5$ & $3.44^{+}$ & $<0.8$\\
	$I_{sp}$, \SI{}{\second} & NA & NA / $121^{*}$& NA & NA& $(406-506)^{*}$ & NA\\
	$\eta_T$, - & NA & NA  & NA & $<0.27^{+}$ & NA & NA \\
	\midrule
	\textbf{Intake}  &&&&&&\\
	\midrule
	$A_{in}$, \SI{}{\square\meter}& 0.25 & 0.25 & 0.48 & 0.15 & $0.126$ & $1$\\
	$p_{in}$, \SI{}{\pascal}& 0.01 & NA & ${\SI{1.5E-3}{}}^{+}$ & NA & NA & $\SI{1E-3}{}$\\
	$\eta_c$, - & 0.35 & 0.9 & $<0.46^{-}$ & $(0.2-0.4)^{-}$ & $(0.28-0.32)^{-}$ & $0.33-0.34$\\
	$\dot{m}_{thr}$, \SI{}{\milli\gram\per{\second}} & $0.108^*$ & 0.51 & NA & NA & $(1.2-1.5)^{*}$ & $0.015-0.068$\\
	& & & & & $+\sim 2.6(\ce{Xe})^{+}$ &\\
	\midrule
	\textbf{Satellite} &&&&&&\\
	\midrule
	$m$, \SI{}{\kilo\gram} & NA & 325 & $152-208$ & 180 & $250-1000$ & $100$ \\
	$A_f$, \SI{}{\square\meter}& 0.5 & 0.39 & 1.5 & 0.3 & $<0.126$ & 1 \\
	$C_D$ & 2.2 & $2-4$ & 2 & $2.2-4$ & 2 & 2\\
	$h$, \SI{}{\kilo\meter}& $>220$ & 200 & $150-200$ & $120-180$ & $120-250$ & $185-222$\\
	years & 5 & $>3$& $>2$& NA & $2-8$ & NA\\   
	\bottomrule
	\footnotesize{$^+$ Experiment,} & \footnotesize{$^{++}$ Required,} &   \footnotesize{$^*$ Derived,} &\footnotesize{$^{-}$ Simulation}& & &\\ 
	\end{tabular}%
	}
	\end{table}


 \subsubsection{Literature Conclusions}
 The studies on ABEP in literature show trends in the targeted mission's lifetime, required $T/P$ ratio, and early estimation of the satellite's geometry. Most of the studies assess modification of proven conventional EP such as HET due to the high technology readiness level (TRL) and the high thrust-to-power ratio when running on noble propellants. However, the operation with atmospheric propellant generally leads to lower performances and requires the thruster unit of the following:
 \begin{itemize}
 \item operate on atmospheric propellant and provide high $T/P>\SI{30}{\milli\newton\per{\kilo\watt}}$;
 \item operate with variable mass flow and composition over time due to the atmospheric environment;
 \item ignition and operation at low pressures.
 \end{itemize}
Most importantly, the reactive nature of atomic oxygen AO that is present in VLEO (but also in VLMO) poses a threat when implementing conventional EP for three main reasons:
 \begin{itemize}
 \item acceleration grids erosion $\Rightarrow$ thruster performance degradation over time;
 \item discharge channel erosion $\Rightarrow$ thruster performance degradation over time;
 \item neutralizer erosion when operating on atmospheric propellant $\Rightarrow$ performance degradation over time.
 \end{itemize}
 
 These issues can be potentially solved at once by employing a contact-less solution for both ionization and acceleration that produces a quasi-neutral plasma plume. Therefore, this places a contact-less plasma thruster based on helicon-discharge in a very advantageous position for the use in an ABEP system due to:
 \begin{itemize}
 \item contact-less operation;
 \item compatibility with a large variety (theoretically any kind) of propellants~\cite{takahashi2019helicon};
 \item compatible with variable mass flow and propellant composition~\cite{takahashi2019helicon};
 \item ignition and operation possible at low pressures~\cite{takahashi2019helicon};
 \item helicon-wave based discharges provides higher plasma density vs power compared to inductive ones~\cite{chen2015};
 \item quasi-neutral plasma plume does not require a neutralizer.
 \end{itemize}
 
 The issue of AO erosion does also impact the intake of the ABEP system, as well as all the SC surfaces that face the flow. Finally, the appropriate choice of materials must be made to ensure the respective properties are maintained throughout the whole mission duration.

	\section{ABEP Design Justification}
	
The development of the RF contactless plasma thruster for an Atmosphere-Breathing Electric Propulsion System (ABEP), which is presented in this dissertation, began with a 3-years scholarship funded by the Landesgraduiertef\"{o}rderung of the Baden-W\"{u}rttemberg region and then merged onto the DISCOVERER project for the development of the intake and the thruster. DISCOVERER is a project that received funding from the European Union's Horizon 2020 research and innovation programme under grant agreement No.~737183. Within this project, the Institute of Space Systems (IRS) of the University of Stuttgart is responsible for the development of an intake and an RF contactless plasma thruster for an ABEP system. The project started in January 2017 with a total duration of $5.25$ years. The consortium is lead by the University of Manchester, United Kingdom. The ABEP system development approach can be divided into the intake and the thruster development. 

\subsection{Intake}
The intake of the ABEP system, is the device that must efficiently collect the atmospheric particles at very low altitude orbits (VLO) and drives them to the electric thruster, or to a storage tank. Such devices can be active~\cite{li2015design} or passive~\cite{romanoiepc,romanoacta3,JAXA,JAXA2,JAXA3,JAXA4,JAXA5,filatyev2019control,di2007ram,BUSEK,shabshelowitz2013study,6945885,SITAEL2015,SINGH201515,SITAEL2016,SITAEL2017,erofeev2017air}. The intake faces the incoming flow to maximize the collection of atmospheric particles. In the ABEP altitude range and for small spacecrafts, the flow is a free molecular flow (FMF), meaning that gas-gas interactions (collisions) can be typically neglected. The intake design is hereby designed as a passive device based on the FMF regime laws, and its performance depends on the gas surface interaction (GSI) properties of the materials of which it is composed, on the intake geometry, and on the alignment with the incoming flow. The challenge is to provide the thruster with enough propellant flow to compensate the drag, while, at the same time, minimizing its contribution to the drag. There are two extremes GSI properties of materials that are of interest for the intake design: 
\begin{itemize}
\item diffuse reflecting material: all the particles energy is absorbed upon impact and the direction of reflection is based on a half-space Maxwellian velocity vector distribution corresponding to the surface temperature;
\item specular reflecting material: the particles maintain their kinetic energy after impact and the direction of reflection is symmetric.
\end{itemize}
Based on the GSI model, an intake can be designed as follows~\cite{romanoacta3}:
\begin{itemize}
\item \textbf{diffuse} reflection $\rightarrow$ the intake operates as "molecular trap": a small honeycomb structure of small ducts in the front of the intake allows the atmospheric particles to get inside the intake while, at the same time, blocking their backflow, see Fig.~\ref{fig:intakeD};
\item \textbf{specular} reflections $\rightarrow$ the intake operates as "parabolic mirror" based on optics-like design, see Fig.~\ref{fig:intakeS}: the honeycomb structure of small ducts is foreseen to reduce backflow as well;
\item \textbf{hybrid/combination} of more or less specular and diffuse-based materials $\rightarrow$ multi- or single stage intake that combines the respective GSI properties aiming to intake performance maximization.
\end{itemize} 
%
%
\begin{figure}[h]
	\centering
	\includegraphics[width=15cm]{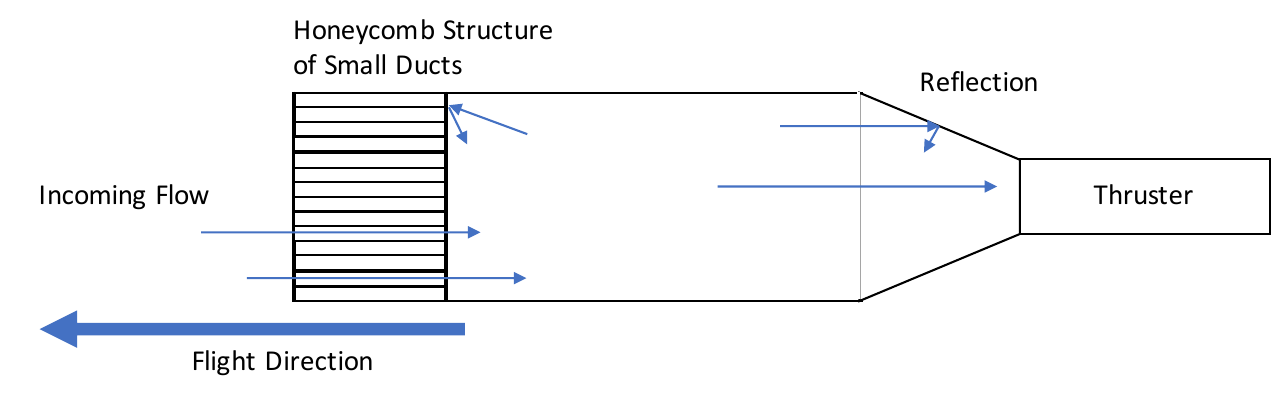}
	\caption{ABEP Intake Based on Diffuse-Reflecting Surfaces.}
	\label{fig:intakeD}
\end{figure}
\begin{figure}[h]
	\centering
	\includegraphics[width=15cm]{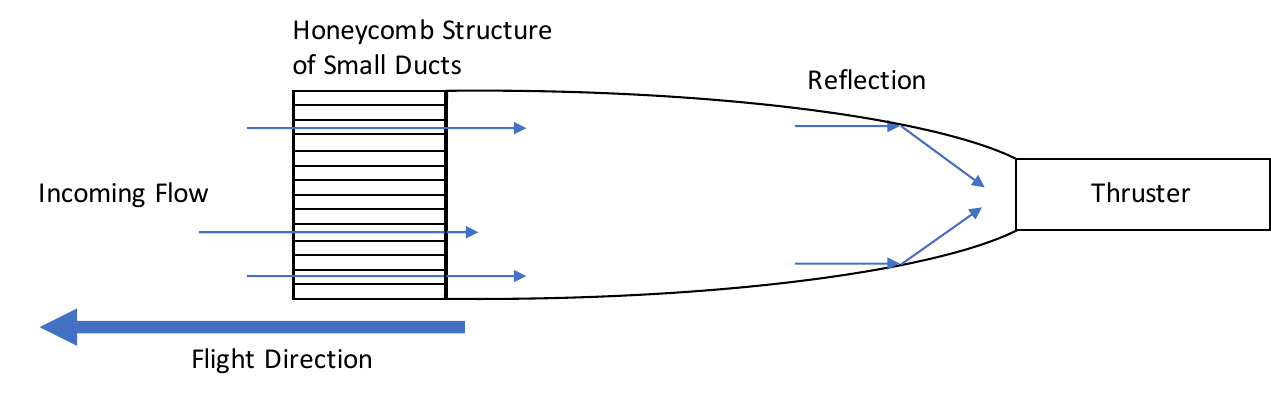}
	\caption{ABEP Intake Based on Specular-Reflecting Surfaces.}
	\label{fig:intakeS}
\end{figure}

\subsection{Thruster}
The thruster of the ABEP system is the device that transforms the atmospheric particles collected by the intake into the plasma state, and accelerates them to produce thrust. The choice is of developing an electric contactless RF plasma thruster, as it offers advantages for the operation with atmospheric propellant:
\begin{itemize}
\item \textbf{contactless design} removes any component in direct contact with the plasma, \textbf{no} performance \textbf{degradation} over time due to operation with aggressive species such as AO in VLEO (erosion);
\item \textbf{quasi-neutral plasma exhaust}: a \textbf{neutraliser} is not needed; 
\item contactless devices are able to \textbf{cope with any propellant} and with respective variations of its density and composition (natural non-uniformity of VLO environments). 
\end{itemize}

However, such thrusters have limited flight heritage, as well as a maximum efficiency, as of today, of $20\%$~\cite{taka2021}. The development of the RF helicon-based plasma thruster (IPT) began from the heritage at IRS on inductively coupled plasma sources (ICP)/inductively heated plasma generators (IPG)~\cite{dropmann}. The thruster concept is of an RF-fed antenna surrounding a discharge channel in which the propellant is injected. The antenna launches electromagnetic waves within the discharge channel that excite the propellant particles, and ionizes them finally achieving the plasma state. In combination with an applied static magnetic field, the plasma is accelerated through the exhaust to produce thrust, see Fig.~\ref{fig:IPTbaby}. The RF discharge is based on helicon waves, as it yields higher plasma density to power ratios compared to the less efficient inductive and capacitive discharges respectively~\cite{chen220}. Such high efficiency is necessary to minimize power consumption and to maximize ionization efficiency, therefore enabling the electromagnetic acceleration necessary to produce the highest exhaust velocities. The thruster's preliminary conceptual design is shown in Fig.~\ref{fig:IPTbaby}, in which an antenna (of general configuration, in this case a solenoid) is fed by an RF power supply and creates the electromagnetic fields to ionize the gas that is injected by the injector into the discharge channel. A solenoid is fed by a DC power supply, alternatively this can be a system of permanent magnets, that creates the static magnetic field needed for enabling the formation of helicon waves within the plasma, and for the creation of a magnetic nozzle effect for (quasi-neutral) plasma acceleration.
\begin{figure}[h]
	\centering
	\includegraphics[width=9cm]{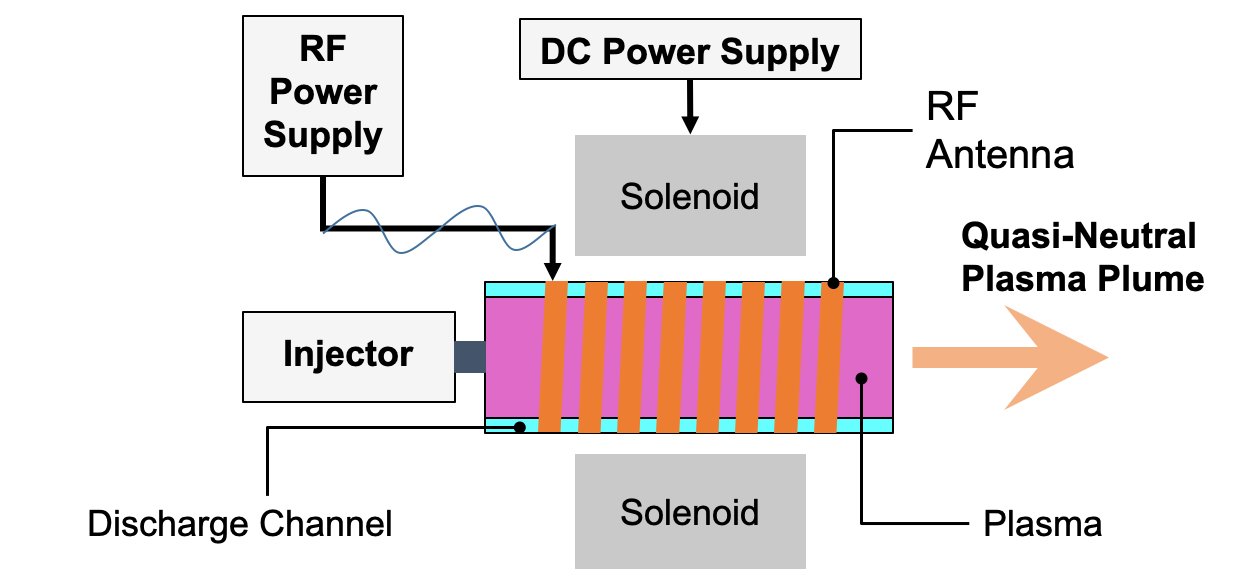}
	\caption{Thruster's Preliminary Conceptual Design.}
	\label{fig:IPTbaby}
\end{figure}

	\section{System Analysis}
	\label{ch:SA}
	This chapter presents the system analysis for an ABEP-based spacecraft and mission. It provides the respective dependencies of the system design based on the mission parameters. Standard required inputs for Earth-based satellites comprise size, payload, orbit, and available power. Further inputs are depending on the mission type, such as required coverage, revisit and eclipse time, field of view of the payload, link budget, disturbance torques to be compensated, battery and solar array sizing, and, finally, cost estimation. 
In case of an ABEP-based mission those inputs must be integrated with detailed information on the atmospheric density $\rho(h)$ and composition $n_i(h)$, as they depend on:
	\begin{itemize}
	\item altitude $h$ and location: latitude (Lat.) and longitude (Long.);
	\item solar and geomagnetic activity, $F10.7_{avg}$, $Ap$.
	\end{itemize}
 Such values are strongly linked to the design and the performance of the ABEP system, which are mainly driven by:
 \begin{itemize}
  \item frontal and intake area $A_f$,~$A_{in}$;
  \item collection/intake and thruster efficiency $\eta_c$,~$\eta_T$;.
\end{itemize}
Based on those values, the most relevant ABEP parameters can be calculated, such as
\begin{itemize}
\item aerodynamic drag $D(h)$;
\item the mass flow that can be collected and reach the thruster $\dot{m}_{thr}(h)$;
\item the composition of the flow reaching the thruster $n_i(h)$.
\end{itemize}
Consequently, given the ABEP intake and thruster performances, the ABEP-required electric power $P_{ABEP}$, and the required exhaust velocity $c_e$, can be preliminarily estimated for a given thrust to drag ratio $T/D$. 
The system analysis logic is presented in Fig.~\ref{fig:systemanal}. Within this work, the total drag $D$ and its respective translation in required thrust $T$ and $c_e$ and $P_{ABEP}$, the power required only by the ABEP system, are considered for deriving the ABEP operation envelope based on given VLEO altitude ranges and averaged values for both solar activities and densities over orbit. 

\tikzstyle{block} = [draw, fill=white, rectangle, 
    minimum height=3em, minimum width=6em]
\tikzstyle{sum} = [draw, fill=white, circle, node distance=1cm]
\tikzstyle{input} = [coordinate]
\tikzstyle{output} = [coordinate]
\tikzstyle{pinstyle} = [pin edge={to-,thin,black}]
\begin{figure}[h]
\centering
\resizebox{0.95\textwidth}{!}{
\begin{tikzpicture}[node distance=1cm,auto,>=latex']
    \node [input, name=input] {};
    \node [block, right of=input, node distance=4cm, align=left] (controller) {Atmospheric Model \\ NRLMSISE-00};
     \draw [draw,->] (input) -- node {
     \begin{tabular}{l}
    Orbit (Lat.~Long.)\\
    Altitude~$h$\\
    $F10.7$, $Ap$\\
    \end{tabular}
    } (controller);
    \node [block, right of=controller, node distance=6cm] (system) {ABEP System Analysis};
    \node [block, above of=system, node distance=2cm] (measurements) {$\eta_c$,~$\eta_T$,~$A_f$,~$A_{in}$,~$C_D$};
    \draw [->] (controller) edge node {$\rho(h)$, $n_i(h)$} (system);
    \draw [<->] (measurements) -- node {$D(h)$,~$\dot{m}_{thr}(h)$,~$T/D$} (system);
    \draw [->] (controller) |- node [above] {\hspace{4.5cm}$\rho(h)$, $n_i(h)$} (measurements);
    \node [coordinate] (end) [right of=system, node distance=5cm]{};
    \draw [->] (system) edge node {$P_{ABEP}(h)$,~$c_e(h)$} (end);
    \coordinate[below=of system]   (aux1);
    \draw[->] (system) -- (aux1) -| (input.south west);
\end{tikzpicture}}
\caption{ABEP System Specific Mission Inputs and Dependencies}
\label{fig:systemanal}
\end{figure}
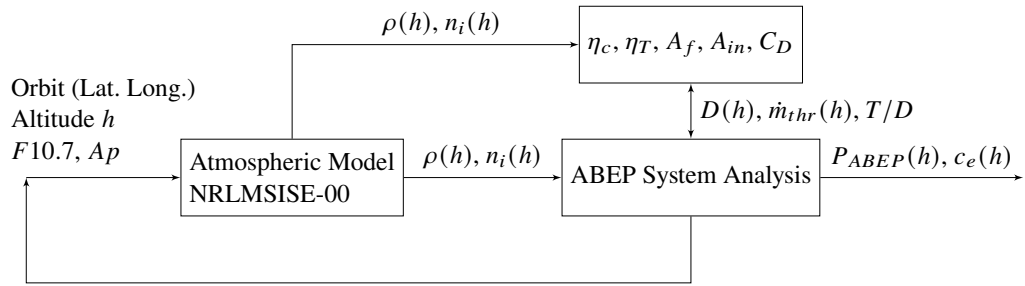

Finally, an ABEP system can be theoretically applied to any celestial body with atmosphere, given that enough electric power is provided. 
This chapter focuses in the VLEO region of Earth's atmosphere. As an addition for further studies, the exemplary cases for Mars, Venus, Titan, gas giants, and the Sun are also briefly analysed within this dissertation.

\section{VLEO Environment}
The reference atmospheric model NRLMSISE-00~\cite{picone} is used for the system analysis. Many atmospheric models exist and behave differently depending on the main parameters of interest, and the considered altitude range. However, according to~\cite{VALLADO2014141}, the inherent inaccuracy of the available atmospheric models is of $10-15\%$ or more. Therefore, a single model that is best for all applications does not exist. The NRLMSISE-00 is an empirical global model of Earth's atmosphere under different conditions of solar and geomagnetic activities. Inputs are date, geographical coordinates, and solar activity parameters: the solar radio flux at a wavelength $\lambda=\SI{2800}{\mega\hertz}$ named $F10.7$, and the geomagnetic planetary index $Ap$, which indicates the intensity of the solar activity averaged around the globe. The standard of minimum, average, and high solar activity are applied as defined in~\cite{ISO14222}.
 Very Low Earth Orbit (VLEO) is defined as the altitude range between the Karman line at~$h=\SI{100}{\kilo\meter}$ up to $h=\SI{450}{\kilo\meter}$~\cite{Crisp2020} in which the aerodynamic forces cannot be neglected in the mission's design. The NRLMSISE-00 model shows that the most present elements in VLEO are atomic oxygen (AO) and \ce{N2}, with AO more dominant at higher altitudes as shown in Fig.~\ref{fig:rho_h_comp}. The solar activity, which cycles every $\sim11$~years, influences the atmospheric properties, resulting in changes of the density vs.~altitude profiles, see Fig.~\ref{fig:rho_h_solact}, as the solar wind compresses and releases the atmosphere, the variation is larger the higher the altitude in the VLEO range. 
\begin{figure}[h]
	\centering
	\begin{subfigure}[b]{0.47\linewidth}
		\includegraphics[width=\linewidth]{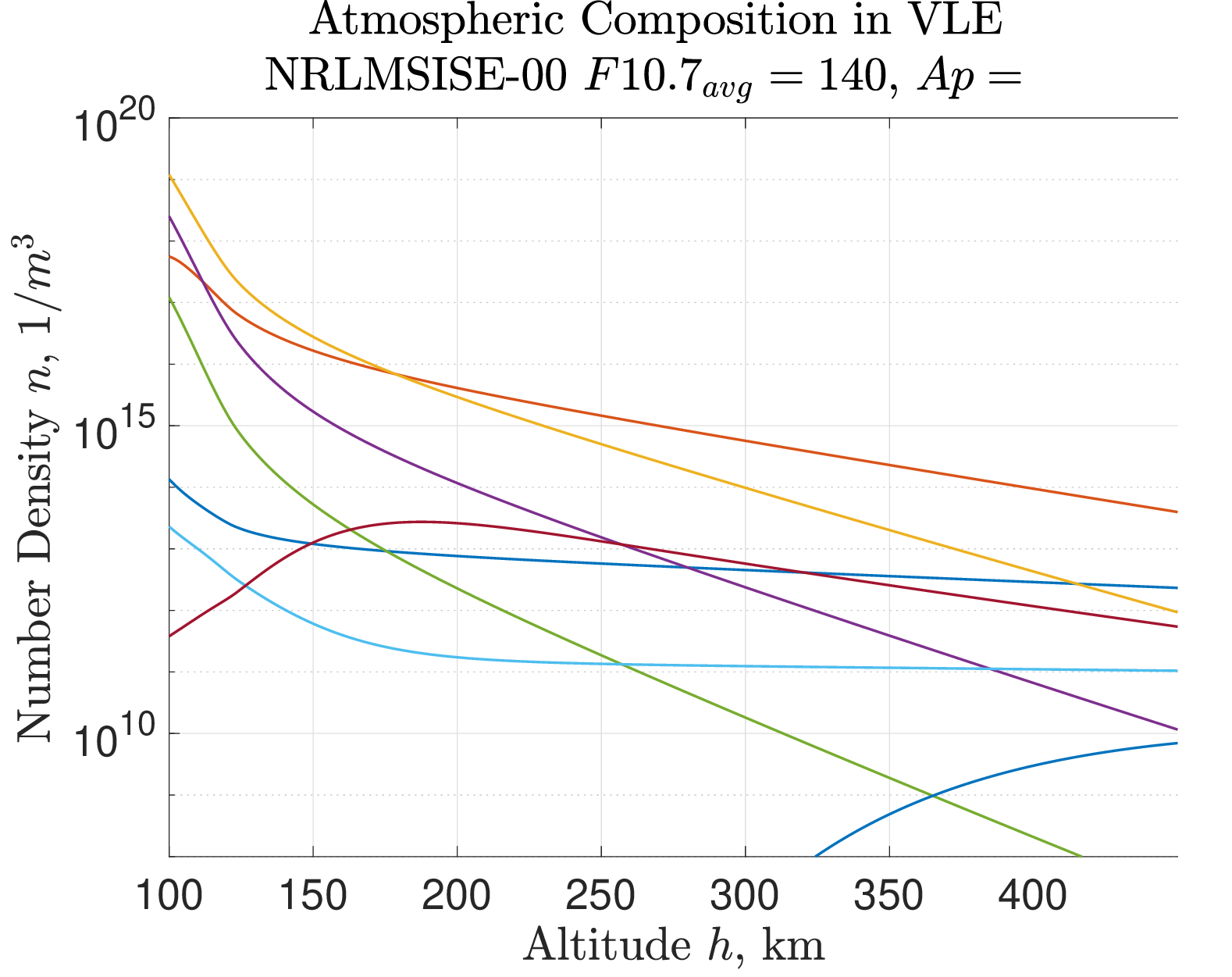}
		\caption{Composition vs Altitude.}
		\label{fig:rho_h_comp}
	\end{subfigure}
	\begin{subfigure}[b]{0.45\linewidth}
		\includegraphics[width=\linewidth]{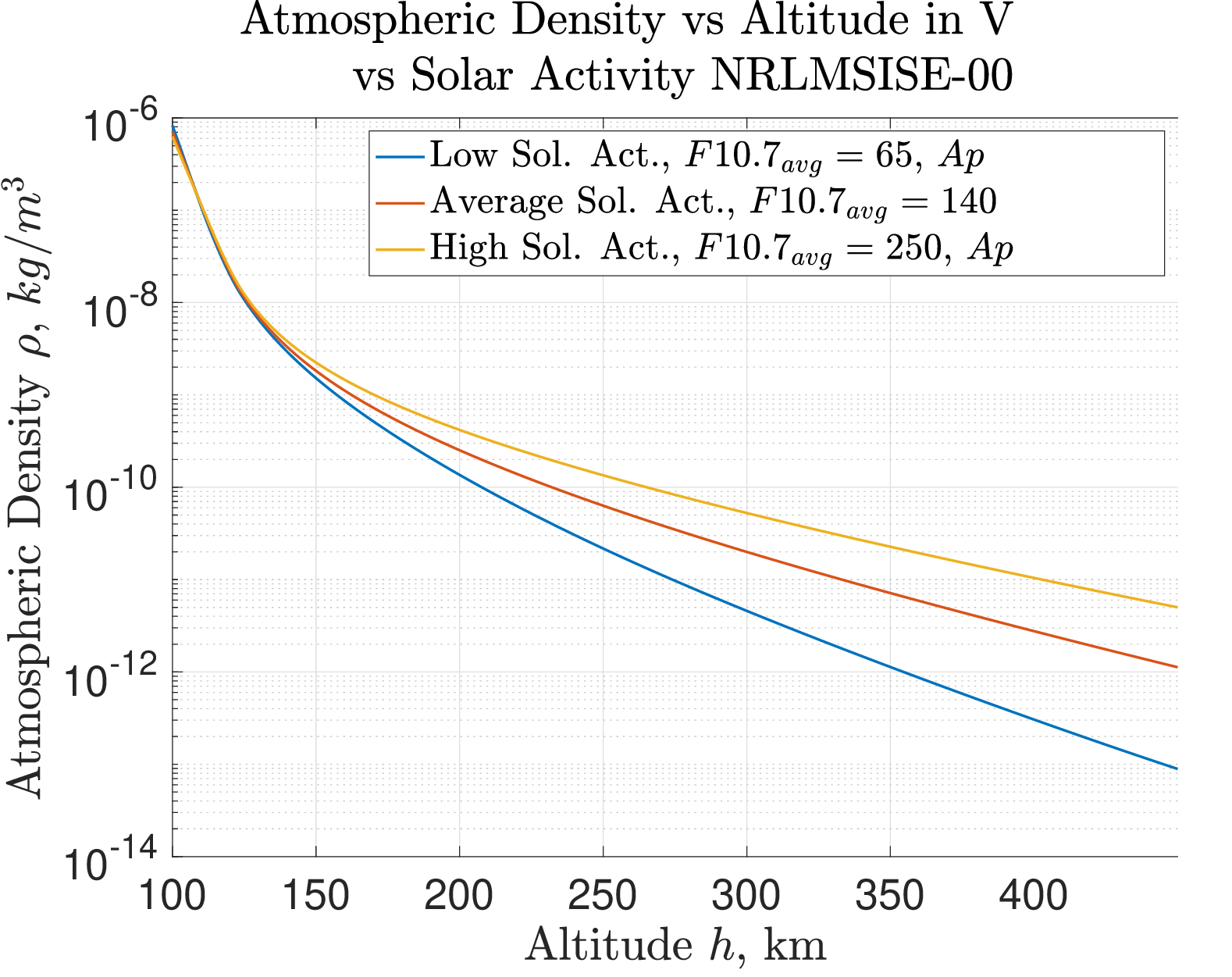}
		\caption{Density vs Altitude and Solar Activity.}
		\label{fig:rho_h_solact}
	\end{subfigure}
	\caption{VLEO Environment, NRLMSISE-00 at Lat.$=\SI{0}{\degree}$, Long.$=\SI{0}{\degree}$.}
\end{figure}
Density variation also arises on shorter time scales when orbiting either through the illuminated or eclipsed side of the Earth, for example in one orbital period, as well as over variations in latitude and longitude due to the non-uniformity of Earth's atmosphere, see Fig.~\ref{fig:rho_h_latitude} and Fig.~\ref{fig:rho_h_longitude}. 
This means that variations of atmospheric density and composition are intrinsic of VLEO operation and, thus, not only directly influence the aerodynamic drag, but also the ABEP system operation and performance. Therefore, the thruster is required to operate with a certain flexibility in terms of propellant composition and density to cope with such variation and preserve its performance. Finally, the altitude range for ABEP operation is reduced to $h=100-\SI{250}{\kilo\meter}$ as at higher altitudes the resulting atmospheric density is too small for thruster operation and the use of conventional EP might be, to this point of time, more convenient~\cite{di2007ram,romanoacta}.
\begin{figure}[h]
		\centering
		\begin{subfigure}[b]{0.47\linewidth}
		\includegraphics[width=\linewidth]{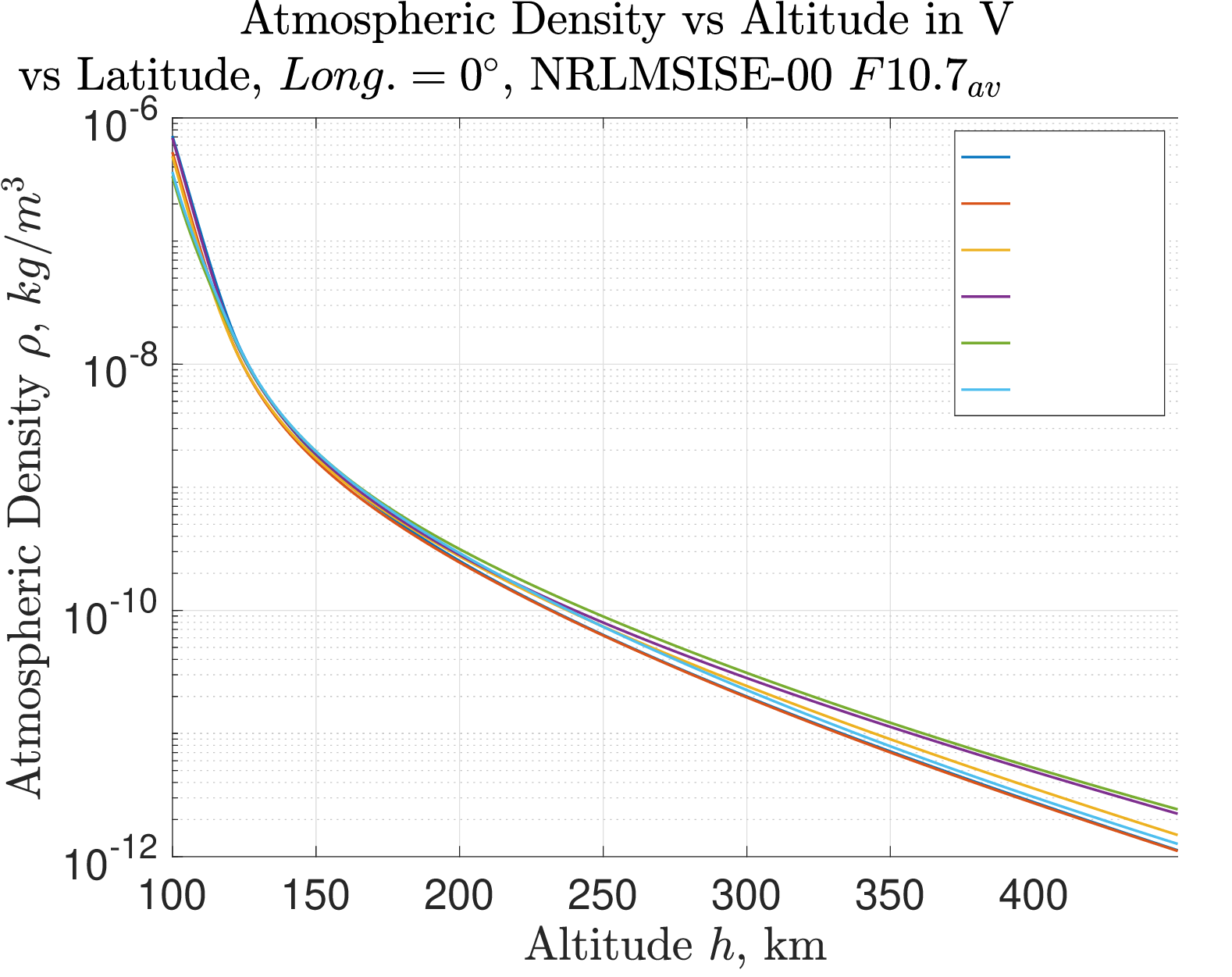}
		\caption{Density vs Altitude and Latitude.}
		\label{fig:rho_h_latitude}
		\end{subfigure}
		\begin{subfigure}[b]{0.47\linewidth}
		\includegraphics[width=\linewidth]{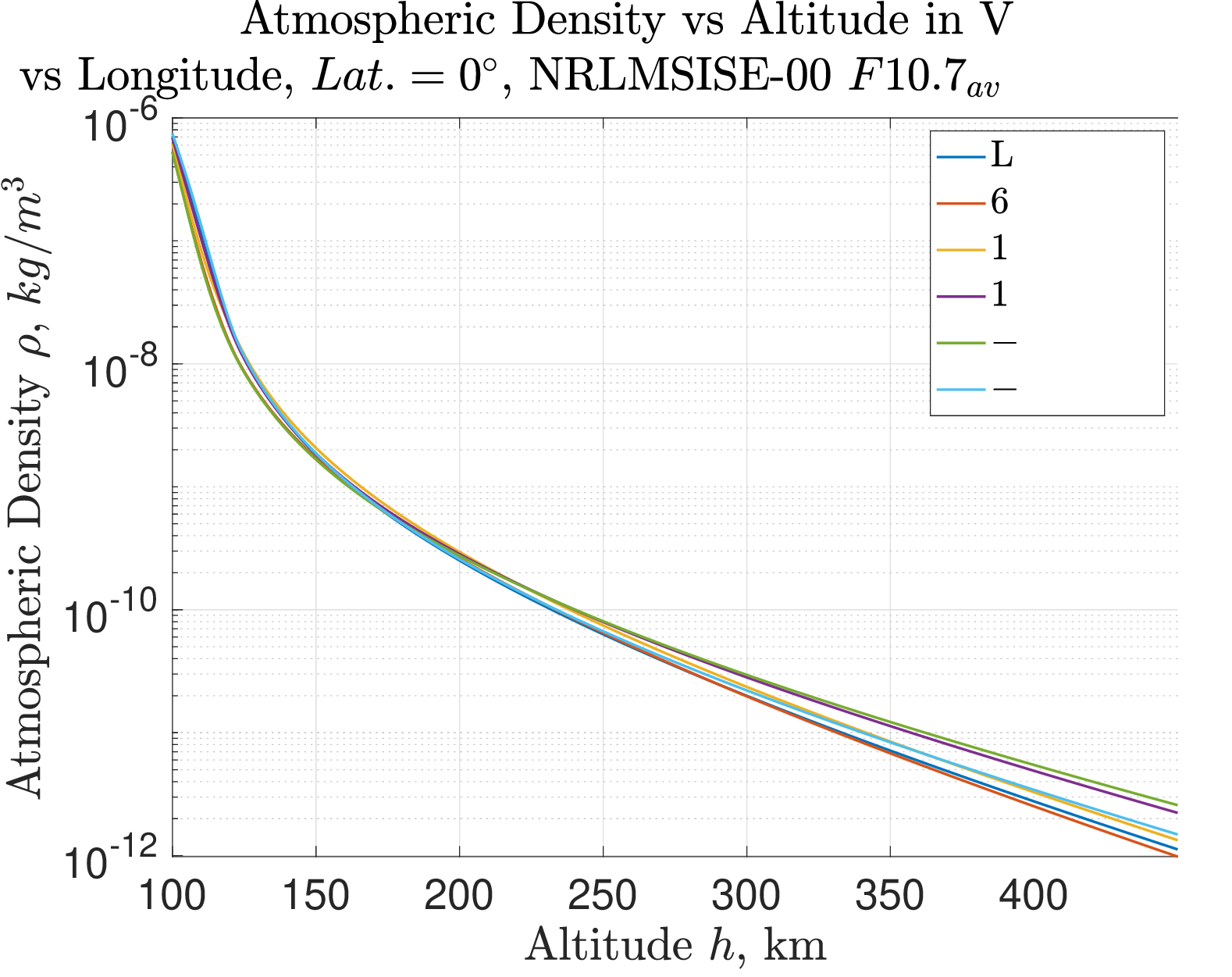}
		\caption{Density vs Altitude and Longitude.}
		\label{fig:rho_h_longitude}
		\end{subfigure}
			\caption{VLEO Environment, NRLMSISE-00, Latitude and Longitude Dependencies.}
\end{figure}

\section{Drag, Mass Flow, Exhaust Velocity Estimation}
The aerodynamic drag on the spacecraft can be calculated by using Eq.~\ref{eq:drag}.
\begin{equation}
D(h)=\frac{1}{2}\rho(h)v^2(h)C_D A_{f}
\label{eq:drag}
\end{equation}
The density $\rho(h)$ and the orbital velocity (relative to the flow) $v(h)$ are considered altitude dependent only, see Eq.~\ref{eq:vrel}.
\begin{equation}
v(h)=\biggl( 1-\frac{\Omega_E R_E}{v_{O}(h)}\biggr) \cos{(i)} v_{O}(h)
\label{eq:vrel}
\end{equation}
Where $v_{O}(h)=\sqrt{GM/(r_E+h)}$, with $GM=\SI{3.986E14}{{\meter}^3{\second}^{-2}}$, $r_E=\SI{6378E3}{\meter}$ the average Earth radius, $\Omega_E=\SI{7.28921150e-5}{\radian\per{\second}}$ the Earth's rotational velocity with the hypothesis that the atmosphere rotates rigidly with the Earth~\cite{cook1968contraction}, and $i$ the orbit's inclination set to $\SI{0}{\degree}$. Indeed, $v(h)$ might not be constant over time also due to the presence of very strong high altitudes winds as observed by GOCE that orbited at $h<\SI{250}{\kilo\meter}$~\cite{GOCE}. Finally, the frontal area of the spacecraft is $A_f$, therefore neglecting surfaces that are aligned with the flow, and the drag coefficient is $C_D$. In literature, for low altitude orbiting small satellites $C_D=2.2$~\cite{shen,romanoacta}. The in-flight measurements of GOCE, highlighted a measured $C_D<3.7$~\cite{GOCECD}, much larger than the usual $C_D=2.2$. Based on this, and to be more conservative, $C_D=3.7$ is used. This shall also partially compensate for neglecting spacecraft surfaces other than the frontal section. Moreover, specific geometries, materials, their finishes, and the resulting accommodation coefficient can be tuned to further decrease $C_D$ and, therefore, the aerodynamic drag~\cite{cdreduction,vleobenefit,CRISP202185}. The resulting drag is calculated by assuming $A_f=A_{in}$ for the "EFD", "Diffuse", and "Specular" intake designs that are described in Ch.~\ref{ch:intake}. The drag order of magnitude in VLEO is shown in Fig.~\ref{fig:d_h}. The differences are due to the resulting $A_{in}$ in each intake design, respectively of $A_{in}=\SI{0.005}{\meter^2}$ for the "EFD", $A_{in}=\SI{0.008}{\meter^2}$ for the "Diffuse", and $A_{in}=\SI{0.019}{\meter^2}$ for the "Specular", derived from the intake analysis~\cite{romanoacta3}.

\begin{figure}[h]
		\centering
		\begin{subfigure}[b]{0.47\linewidth}
		\includegraphics[width=\linewidth]{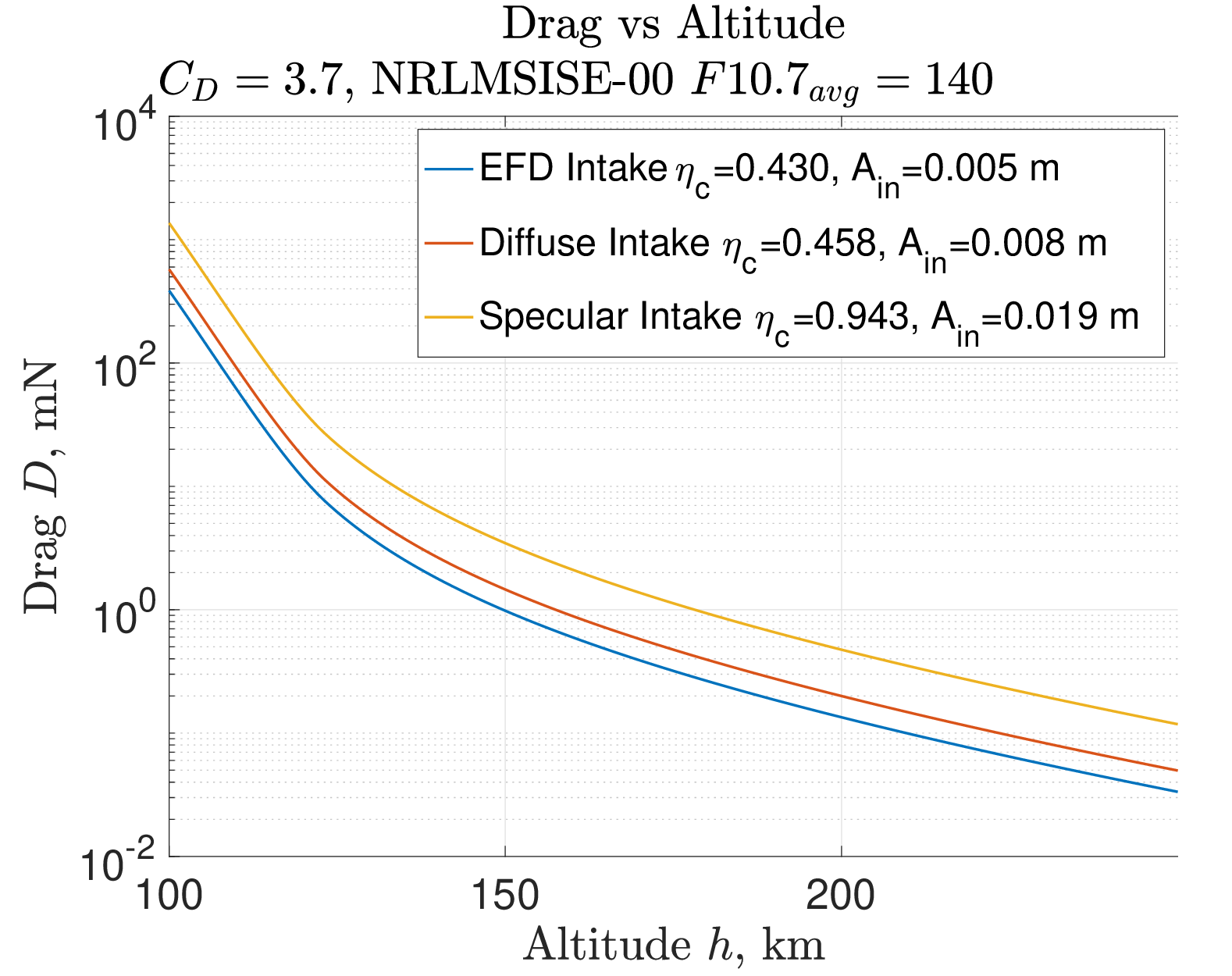}
		\caption{Aerodynamic Drag in VLEO, Eq.~\ref{eq:drag}.}
		\label{fig:d_h}
		\end{subfigure}
		\begin{subfigure}[b]{0.47\linewidth}
		\includegraphics[width=\linewidth]{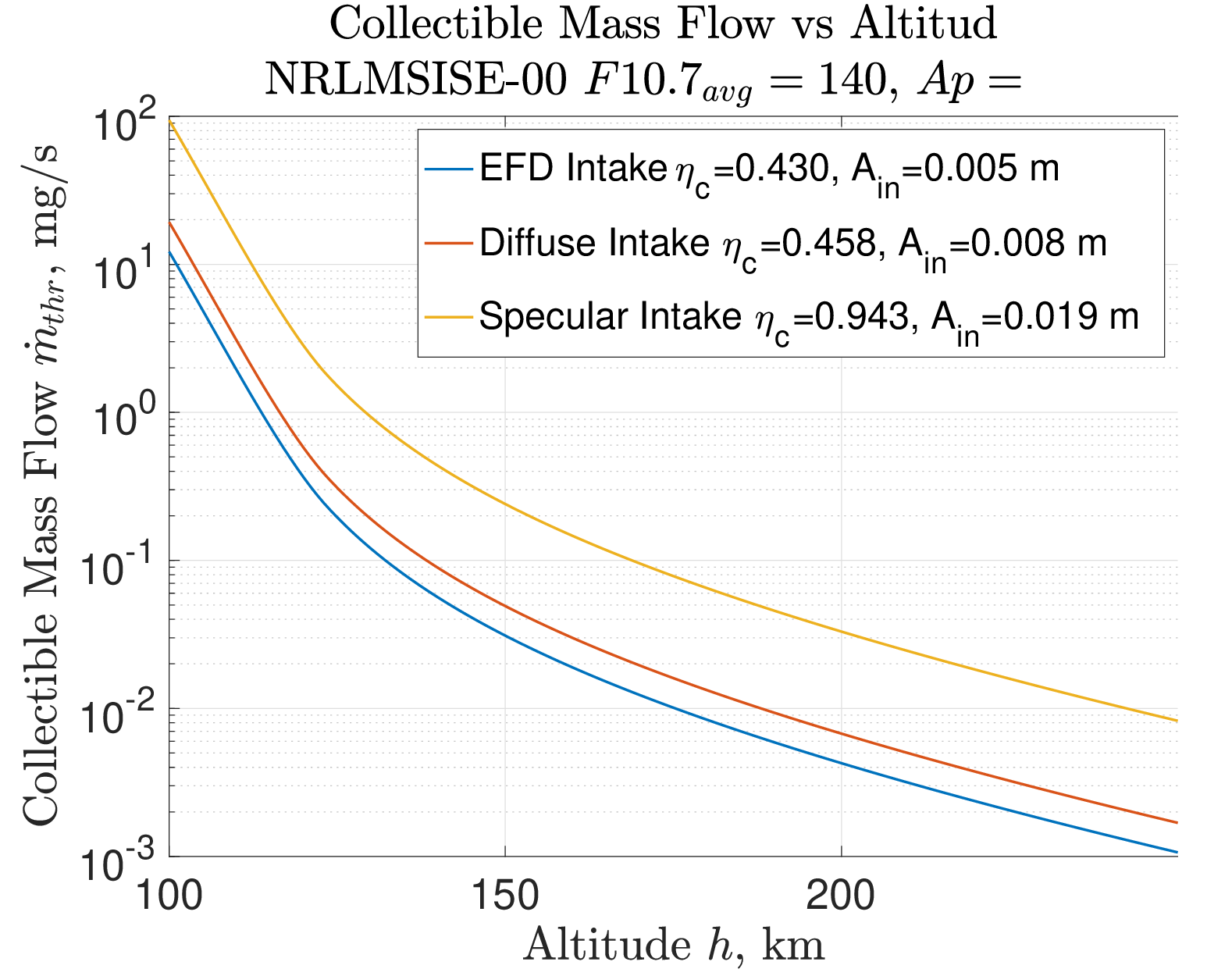}
		\caption{Collectible Mass Flow in VLEO, Eq.~\ref{eq:mdot}.}
		\label{fig:mdot_h}
		\end{subfigure}
		\caption{ABEP in VLEO, NRLMSISE-00.}
\end{figure}

The collectible mass flow $\dot{m}_{thr}$ is shown within Fig.~\ref{fig:mdot_h}, resulting in $\dot{m}_{thr}=\SI{1E2}{}-\SI{1E-3}{\milli\gram\per{\second}}$. This is calculated according to Eq.~\ref{eq:mdot}.
\begin{equation}
\dot{m}_{thr}(h)=\rho(h) A_{in} v(h) \eta_c
\label{eq:mdot}
\end{equation}

 Higher values can be reached, but the intake has to be respectively optimized. Similarly, it depends on thruster requirements and the respective operational condition envelope.


\section{ABEP Power Requirements}
The calculation of the ABEP system power requirement can be extracted by combining thrust and drag equations at the condition $T=D$, and the jet power, Eq.~\ref{eq:jetpower} $P_{jet}$~\cite{goebel2008fundamentals}. This includes the collection (intake) efficiency $\eta_c$, the thruster efficiency $\eta_T$, and the required exhaust velocity $c_e(h)$.
\begin{equation}
P_{jet}(h)=\frac{1}{2}\dot{m}_{thr}(h)c_e^2(h)
\label{eq:jetpower}
\end{equation}
The resulting equation for the ABEP power calculation is Eq.~\ref{eq:reqpower}. 
\begin{equation}
P_{ABEP}=\frac{1}{2}\frac{\dot{m}_{thr}(h)c_e^2(h)}{\eta_T}=\frac{1}{8}\rho(h) \frac{A_f^2}{A_{in}} {v}^3(h) C^2_D \frac{1}{\eta_c \eta_T}
\label{eq:reqpower}
\end{equation} 
This can be used to first approximate the required electric power range for a given ABEP-based spacecraft, $P_{ABEP}=P_{jet}/\eta_T$, based on the respective efficiencies of designed intakes $\eta_c=0.430,~0.458,~0.943$, presented in Ch.~\ref{ch:intake}, and thruster $\eta_T=0.2$, based on the maximum experimentally achieved value for an helicon plasma thruster as of beginning of 2021~\cite{takahashi2019helicon,taka2021}.

Concerning the required exhaust velocity $c_e(h)$, under the assumption of full drag compensation $T(h)=D(h)$, the calculation is performed as in Eq.~\ref{eq:ce}, with the additional condition of $A_f\geq A_{in}$, where $\dot{m}_{thr}(h)=\rho(h) A_{in} v(h) \eta_c$, resulting in $c_e(h)<\SI{35}{\kilo\meter\per{\second}}$, within the values provided by conventional EP devices~\cite{levchenko2020perspectives}. Such a value is almost constant as it follows the behaviour of $v(h)$. 

\begin{equation}
c_e(h)=\frac{T(h)}{\dot{m}_{thr}(h)}=\frac{D(h)}{\dot{m}_{thr}(h)}=\frac{1}{2} \frac{A_f}{A_{in}}\frac{v(h) C_D}{\eta_c}
\label{eq:ce}
\end{equation}

The results of the ABEP required power $P_{ABEP}$ over altitude $h$ are shown in Fig.~\ref{fig:P_h}, in which two horizontal lines highlight \SI{100}{\watt} and \SI{1}{\kilo\watt} power levels. The $P_{ABEP}$ increases by lowering the orbit as $\rho(h)$ and, consequently also $D(h),~\dot{m}_{thr}(h)$, become greater. The range is of $P_{ABEP}=3\times10^{-2}-\SI{6E2}{\kilo\watt}$ for $h=100-\SI{250}{\kilo\meter}$. The calculations based on the ``Diffuse" and ``Specular" intakes require more power due to the combination of larger $A_{in}$ and $\eta_c$. Moreover, by selecting the exemplary altitudes of $h=180,~200,~\SI{250}{\kilo\meter}$, the variation of $P_{ABEP}$ over latitude is shown in Fig.~\ref{fig:P_h_LL_180},~\ref{fig:P_h_LL_200},~\ref{fig:P_h_LL_250}, and over longitude is shown in Fig.~\ref{fig:P_h_LA_180},~\ref{fig:P_h_LA_200},~\ref{fig:P_h_LA_250}. The results exhibits a sinusoidal-like dependency on latitude and longitude of $P_{ABEP}$. This is given to the fact that, for a given altitude, $\rho(Lat.,~Long.)$ depends on latitude and longitude mostly because of the presence or absence of direct solar radiation, illuminated or eclipsed side of the orbit. Finally, this corresponds, in a simplified manner, in a greater drag force in the illuminated side, compared to the eclipsed side for a given altitude~\cite{cook1968contraction}.

\begin{figure}[h]
		\centering
		\includegraphics[width=0.6\linewidth]{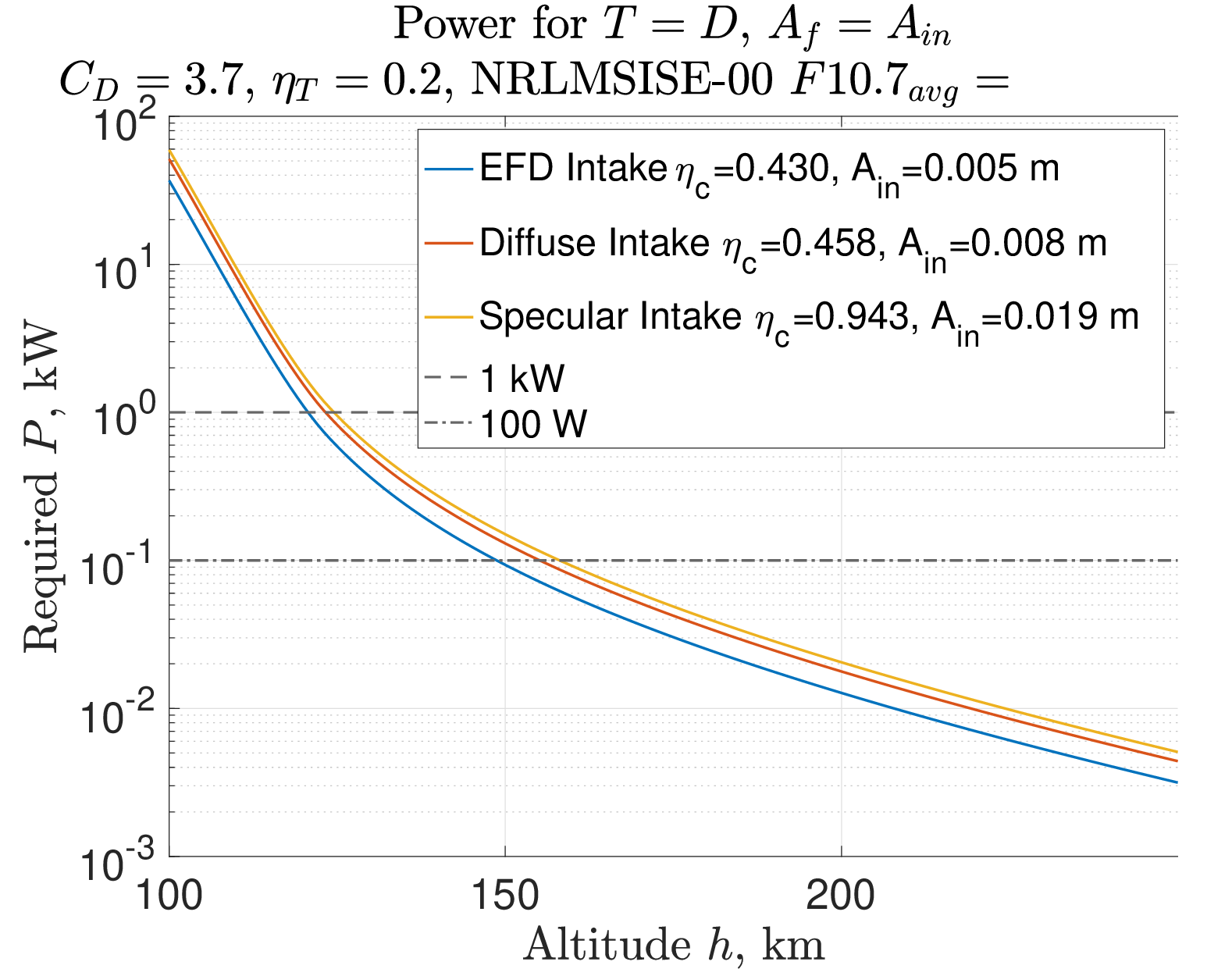}
		\caption{ABEP Requirements in VLEO, NRLMSISE-00.}
		\label{fig:P_h}
\end{figure}

%
%
%

\begin{figure}[h]
		\centering
		\begin{subfigure}[b]{0.47\linewidth}
		\includegraphics[width=\linewidth]{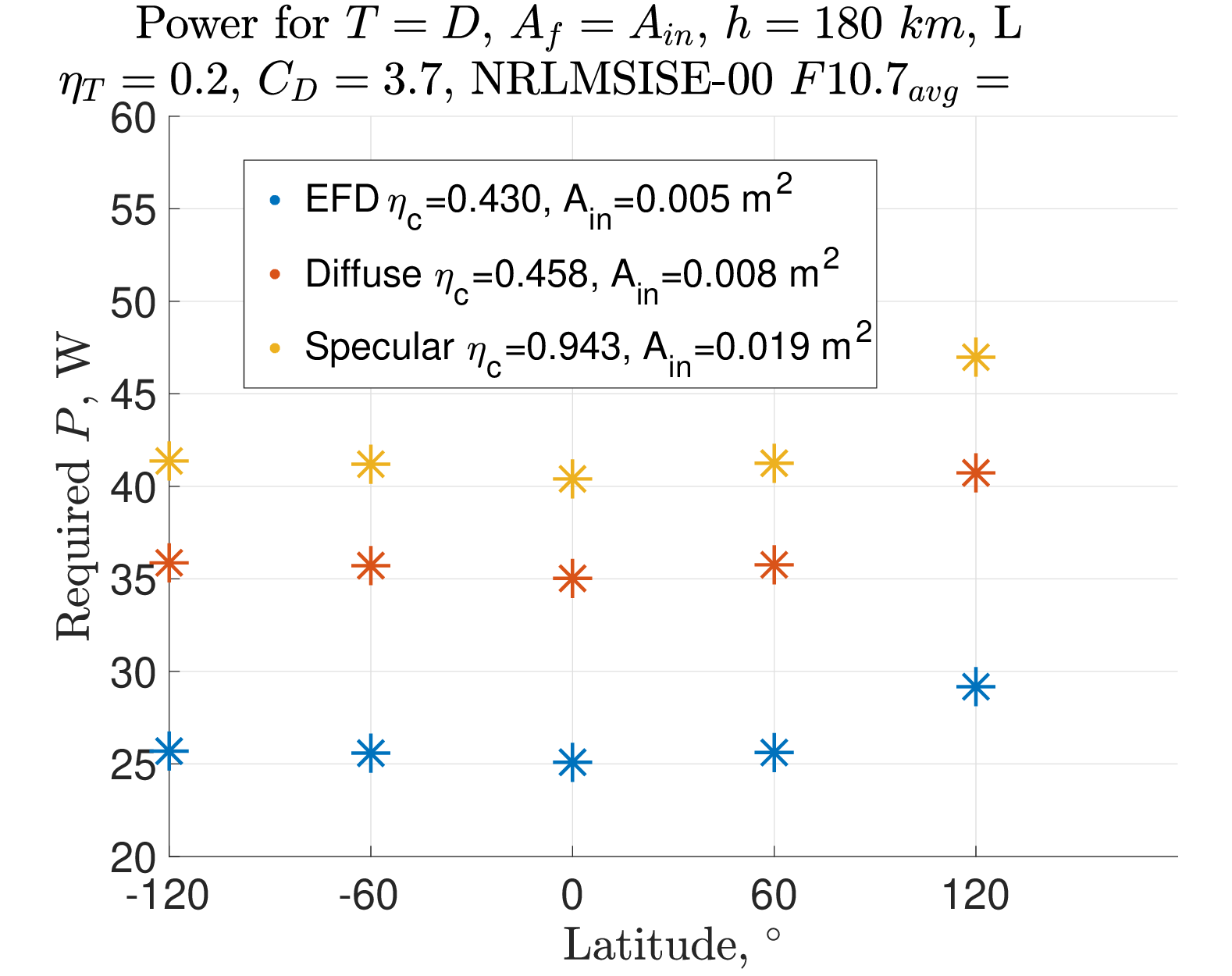}
		\caption{$\SI{180}{\kilo\meter}$}
		\label{fig:P_h_LL_180}
		\end{subfigure}
		\begin{subfigure}[b]{0.47\linewidth}
		\centering
		\includegraphics[width=\linewidth]{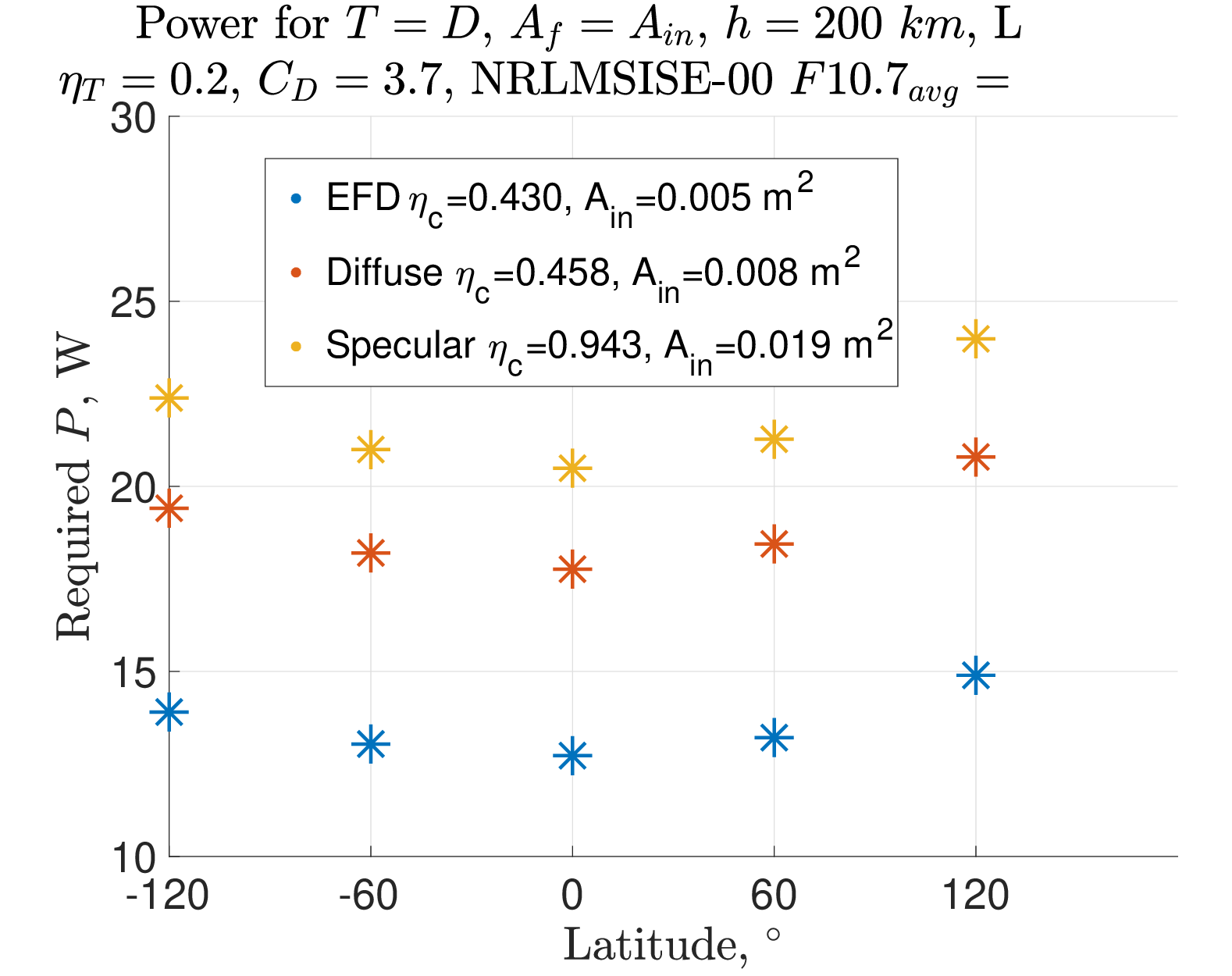}
		\caption{$\SI{200}{\kilo\meter}$}
		\label{fig:P_h_LL_200}
		\end{subfigure}
		\begin{subfigure}[b]{0.5\linewidth}
		\centering
		\includegraphics[width=\linewidth]{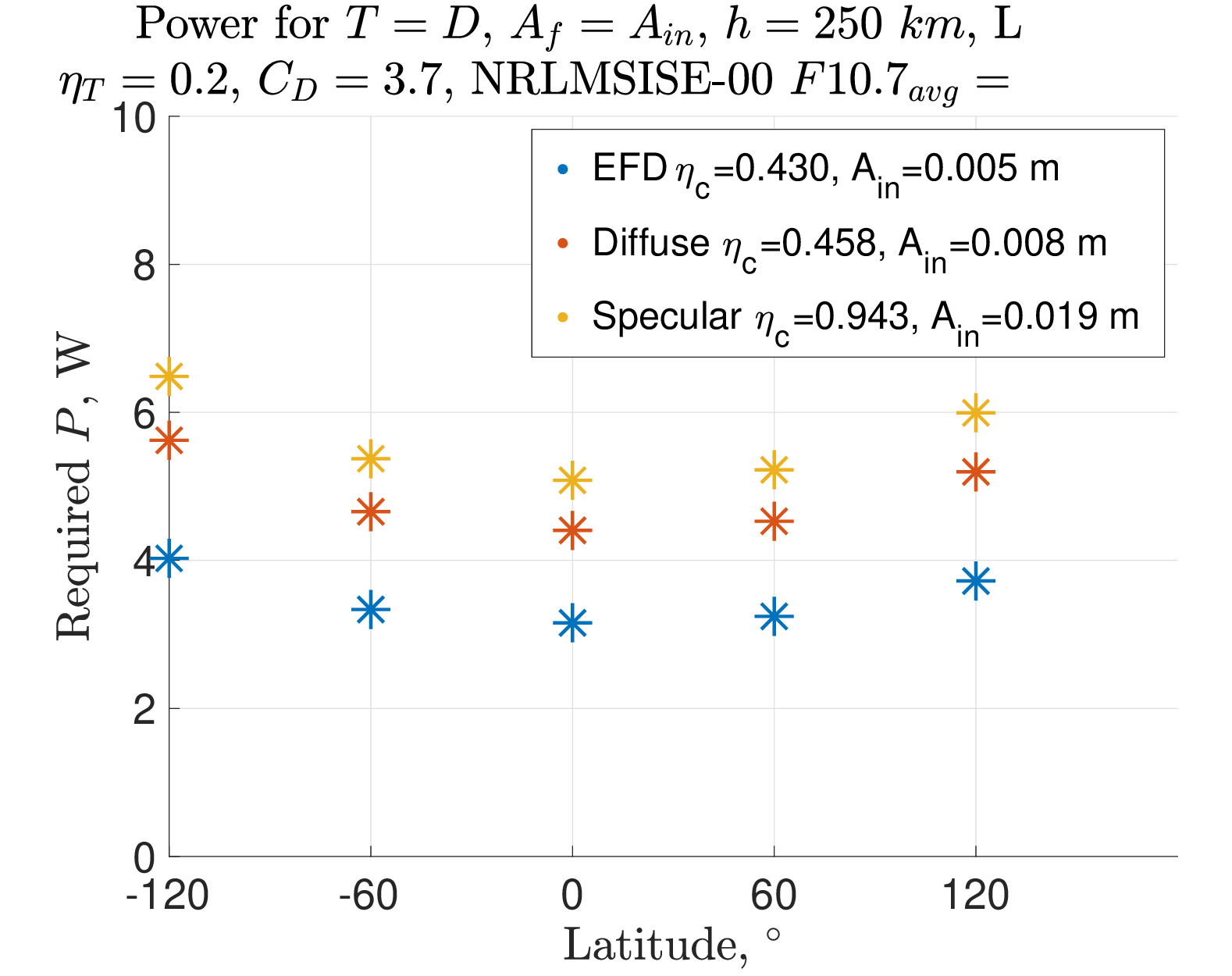}
		\caption{$\SI{250}{\kilo\meter}$.}
		\label{fig:P_h_LL_250}
		\end{subfigure}
\caption{Required Power for $T=D$ in VLEO over Latitude, NRLMSISE-00 Model.}
\end{figure}

\begin{figure}[h]
	\centering
	\begin{subfigure}[b]{0.47\linewidth}
	\includegraphics[width=\linewidth]{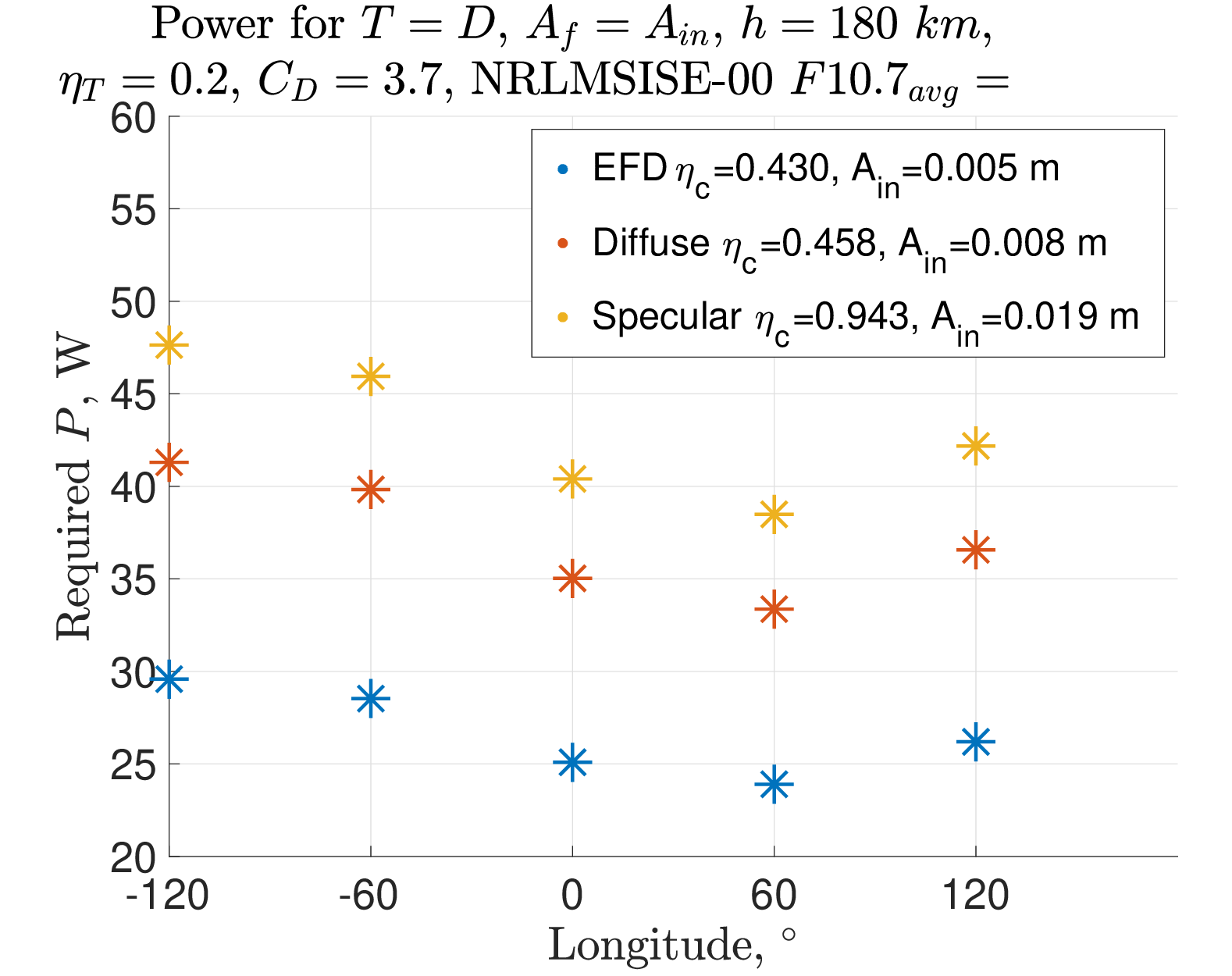}
	\caption{$\SI{180}{\kilo\meter}$}
	\label{fig:P_h_LA_180}
	\end{subfigure}
	\begin{subfigure}[b]{0.47\linewidth}
	\centering
	\includegraphics[width=\linewidth]{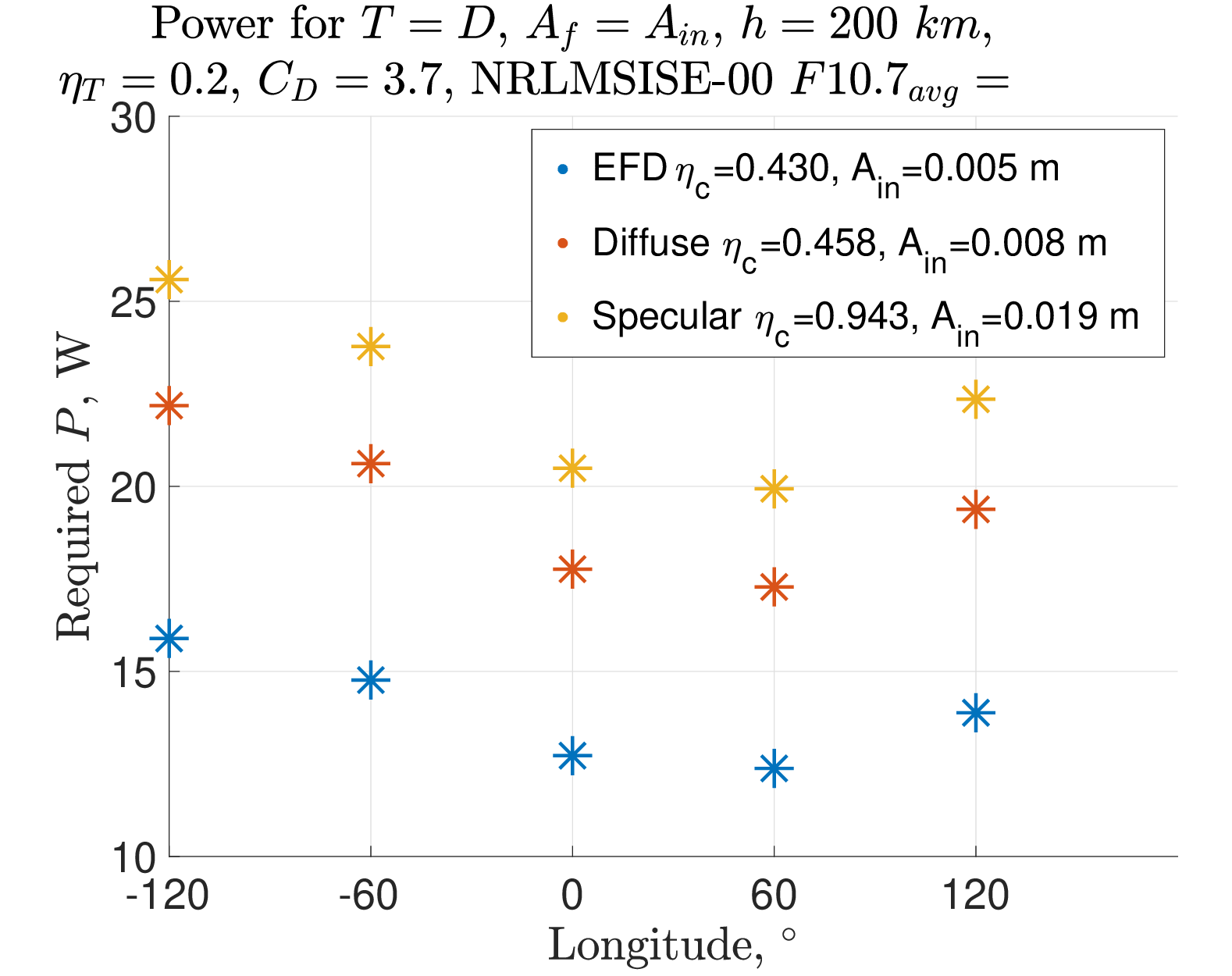}
	\caption{$\SI{200}{\kilo\meter}$}
	\label{fig:P_h_LA_200}
	\end{subfigure}
	\begin{subfigure}[b]{0.5\linewidth}
	\centering
	\includegraphics[width=\linewidth]{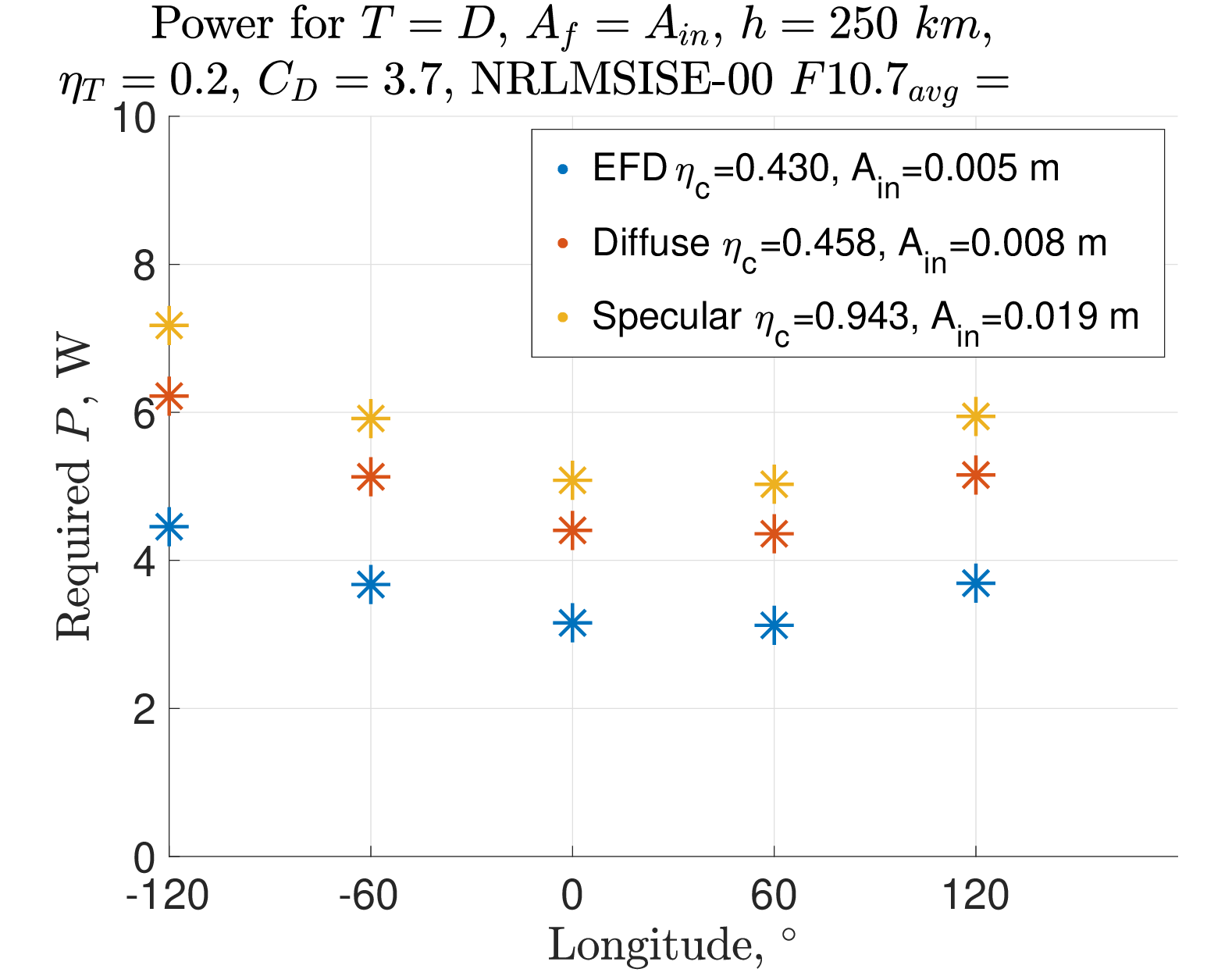}
	\caption{$\SI{250}{\kilo\meter}$}
	\label{fig:P_h_LA_250}
	\end{subfigure}
\caption{Required Power for $T=D$ in VLEO over Longitude, NRLMSISE-00 Model.}
\end{figure}

The results highlight the dependency of required $P_{ABEP}$ on latitude and longitude, at a given altitude, and over altitude. The required $P_{ABEP}$ is larger at low altitudes and it decreases as $\rho(h)$, therefore, the solar activity also influences $P_{ABEP}$, especially at higher altitudes. Moreover, for a given fixed date, the required $P_{ABEP}$ for full drag compensation is dependent on latitude and longitude. By fixing the orbit along the equator, $i=\SI{0}{\degree}$ and extrapolating at different longitudes, $P_{ABEP}$ varies within $20-30\%$ along one orbit. By fixing the orbit along the Greenwich line and extrapolating for different latitudes, instead, $P_{ABEP}$ varies within $12-15\%$ along one orbit. Therefore, the ABEP system must be capable to operate at different power levels with variable propellant flows and compositions while maintaining the overall required performance. 

\subsubsection{Conclusion}
The system analysis shows that in the VLEO environment, the collected propellant species are mostly \ce{N2} and AO  at different ratios and densities depending on location, time, and altitude. The solar activity must be taken into account for the mission design, as the solar wind compresses and releases the atmosphere over time changing the density profile, the variation is larger at higher altitudes. Latitude and longitude variations, therefore also day and night time, also results in changes of the atmospheric density and composition as the atmosphere is not a uniform environment and as illuminated and eclipsed side of the Earth results in different density vs altitude profiles, those then result in $12-30\%$ variation of the required $P_{ABEP}$ over one orbit. The $P_{ABEP}$ is therefore strongly linked to $\rho$ and $n$ that depend on $h,~t,~Lat.,$ and~$Long.$, in which great influence is given by the solar activity. Finally, as the flight of a mission with an ABEP system does not have a date yet, an average solar activity is chosen. From such initial condition, the respective $\dot{m}_{thr}$ over $h$ for the different intake designs is estimated, see Fig.~\ref{fig:mdot_h}. The altitude range is reduced to $h<\SI{250}{\kilo\meter}$, as at higher orbits the collectible mass flow for the thruster becomes too small and the use of conventional EP might be more convenient. By assuming $A_f=A_{in}$, the required powers is $P_{ABEP}<\SI{1}{\kilo\watt}$ at any given configuration at $\SI{130}{\kilo\meter}<h<\SI{250}{\kilo\meter}$, requiring $c_e<\SI{35}{\kilo\meter\per{\second}}$ which is within the conventional EP $c_e(h)$. In particular, for $h<\SI{130}{\kilo\meter}$, $P_{ABEP}$ increases along with $\rho(h)$ up to $P_{ABEP}>\SI{10}{\kilo\watt}$. The calculations are performed by assuming intake efficiencies $\eta_c$ based on~\cite{romanoacta3} and discussed in Ch.~\ref{ch:intake}, and the state-of-the-art experimental helicon plasma thruster efficiency $\eta_T=0.2$~\cite{takahashi2019helicon,taka2021}. Therefore:
\begin{itemize}
\item $\SI{130}{\kilo\meter}<h<\SI{250}{\kilo\meter}$;
\item $A_f=A_{in}$;
\item $P_{ABEP}<\SI{1}{\kilo\watt}$ for $\eta_c=0.430,~0.458,~0.943$, $\eta_T=0.2$;
\item $c_e<\SI{35}{\kilo\meter\per{\second}}$.
\end{itemize}

For higher thruster and intake efficiencies, $P_{ABEP}$ can be further reduced and the altitude range adjusted. In any case, ABEP system performance and requirements are strongly and directly linked to the mission, especially in terms of spacecraft geometry, orbit, lifetime, and must be continuously iterated in a feedback loop to reach the final design. The thruster developed within this dissertation is scalable~\cite{heliconscalable} and can be optimized for given mass flow ranges. Once this is characterized, the range of operating conditions is experimentally found and then traced back to the corresponding intake and orbit configuration.

\section{ABEP Applications}
Hereby example of application of an ABEP system to a GOCE-like spacecraft in Earth orbit is presented. Moreover, an ABEP system, as mentioned earlier, can be applied to any celestial body with atmosphere, given that the required electrical power is provided. Within this section, a brief overview on the application of ABEP to Mars is presented, followed by a very brief outlook on other celestial bodies such as Venus, Titan, gas giants, and the Sun.

\subsection{Requirements for a GOCE-like Spacecraft using ABEP}
GOCE represents one of the most recent missions that orbited in VLEO and, due to its size and power level, represents a candidate mission for an ABEP system. Moreover, previous calculations are based on $A_f$ only, discarding other surfaces, i.e. lateral ones, and do not represent a real spacecraft. Therefore, the frontal area of GOCE, $A_f=\SI{1.1}{\meter^2}$, is taken to perform the calculations, as well as an available $P_{ABEP}<\SI{1.6}{\kilo\watt}$ at EoL~\cite{GOCE}. The drag $D$ is estimated based on both GOCE's $A_f=\SI{1.1}{\meter^2}$ and its measured $C_D=3.7$, see Fig.~\ref{fig:d_h_GOCE}, and the $\dot{m}_{thr}$ based on $A_{in}<A_f$ for the three intakes presented in Ch.~\ref{ch:intake}. Since an extra-investigation is required to ensure the intake direct scalability, the total $A_{in,i}$ is calculated based on clustering. The number of single intakes of $A_{in,int}$ that fits into $A_f$, $N_{int}=A_f/A_{in,int}$, is rounded to the smaller integer to account for the structural area, resulting in $A_{in,i}=N_{int} A_{in,int}$. Concerning the thruster, one unit is considered operating at $\eta_T=0.2$ as mentioned previously.

\begin{figure}[h]
		\centering
		\begin{subfigure}[b]{0.48\linewidth}
		\includegraphics[width=\linewidth]{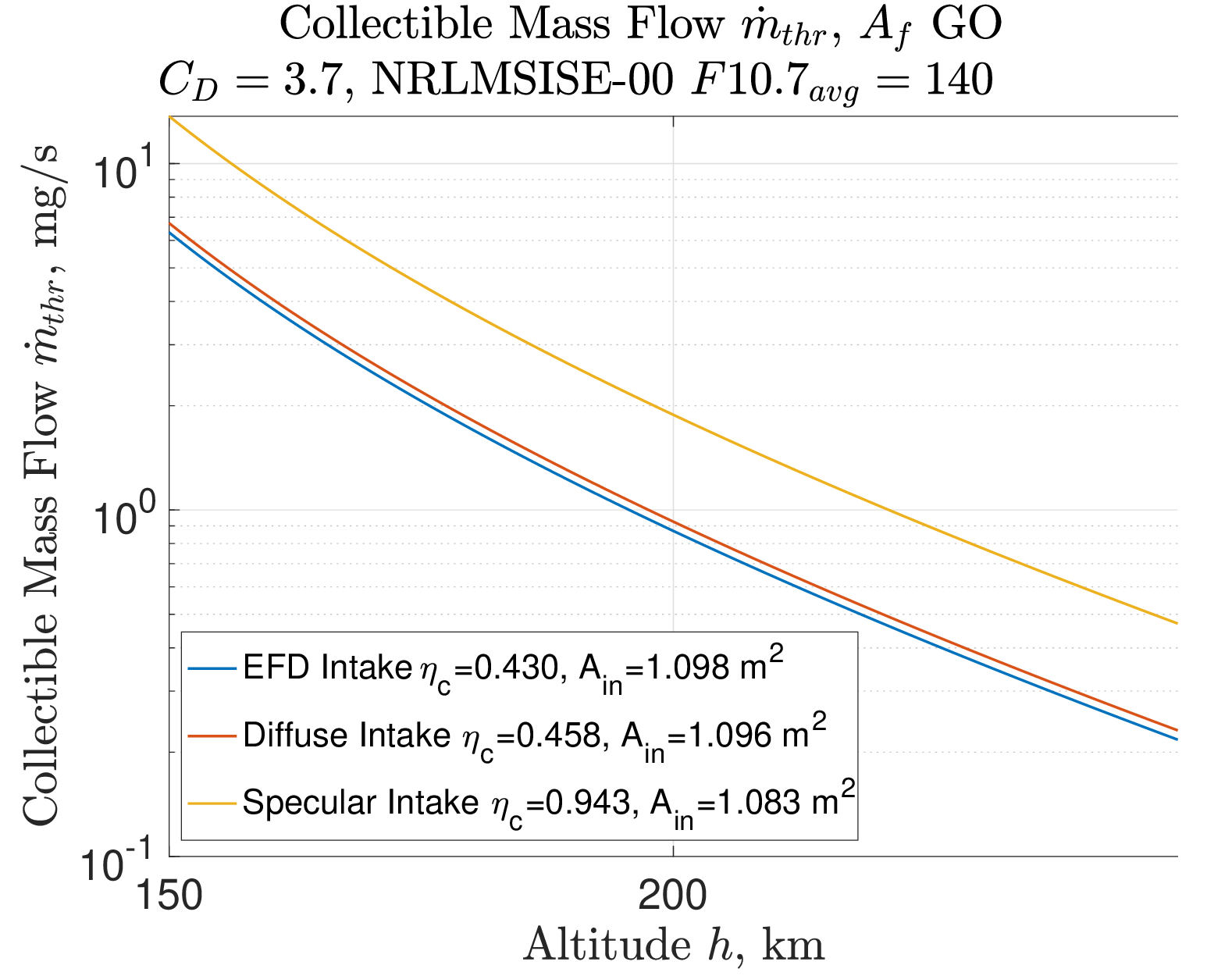}
		\caption{Collectible Mass Flow in VLEO.}
		\label{fig:mdot_h_GOCE}
		\end{subfigure}
		\begin{subfigure}[b]{0.44\linewidth}
		\includegraphics[width=\linewidth]{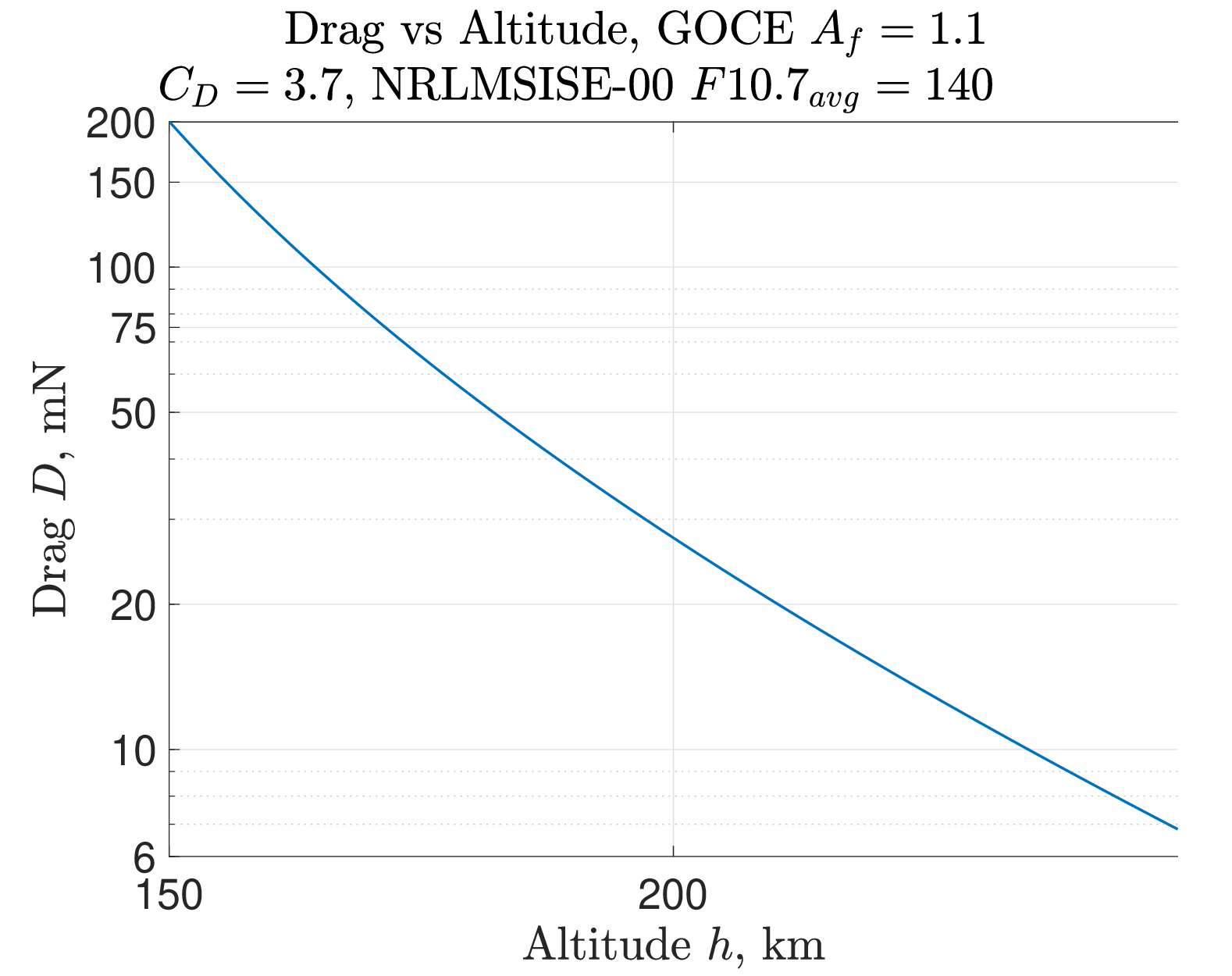}
		\caption{Aerodynamic Drag in VLEO.}
		\label{fig:d_h_GOCE}
		\end{subfigure}
			\caption{ABEP GOCE in VLEO, NRLMSISE-00.}
\end{figure}


The results show that the power requirement increases due to the large $A_f$, requiring $P_{ABEP}>\SI{10}{\kilo\watt}$, see Fig.~\ref{fig:P_h_GOCE}, for $h=150-\SI{160}{\kilo\meter}$, therefore the VLEO altitude range is reduced to $h=150-\SI{250}{\kilo\meter}$. The drag to be compensated within this altitude range is between $D=6-\SI{200}{\milli\newton}$, see Fig.~\ref{fig:d_h_GOCE}. The respective $\dot{m}_{thr}$, see Fig.~\ref{fig:mdot_h_GOCE}, is between $\dot{m}_{thr}=20-\SI{0.2}{\milli\gram\per{\second}}$, and the required $c_e\sim15-\SI{32}{\kilo\meter\per{\second}}$, still within the limits of current available EP technologies. 

From the ABEP point of view, it is interesting to visualize the dependency of $c_e$ on $\eta_c$, see Fig.~\ref{fig:ceetacGOCE}. Here, $c_e$ is shown as the average value over the VLEO altitude range $h=100-\SI{250}{\kilo\meter}$ for $A_f=A_{in}$, and results in $c_e<\SI{50}{\kilo\meter\per{\second}}$ if $\eta_c > 0.2$. Furthermore, the power requirement is of $P_{ABEP}<\SI{1.6}{\kilo\watt}$ for the EFD and diffuse intakes for $h>\SI{215}{\kilo\meter}$, while for the specular intake for $h>\SI{190}{\kilo\meter}$, reaching $P_{ABEP}<\SI{1}{\kilo\watt}$ for $h>\SI{230}{\kilo\meter}$ and $h>\SI{205}{\kilo\meter}$ respectively, see Fig.~\ref{fig:P_h_GOCE}. 
\begin{figure}[h]
		\centering
		\begin{subfigure}[b]{0.48\linewidth}
		\includegraphics[width=\linewidth]{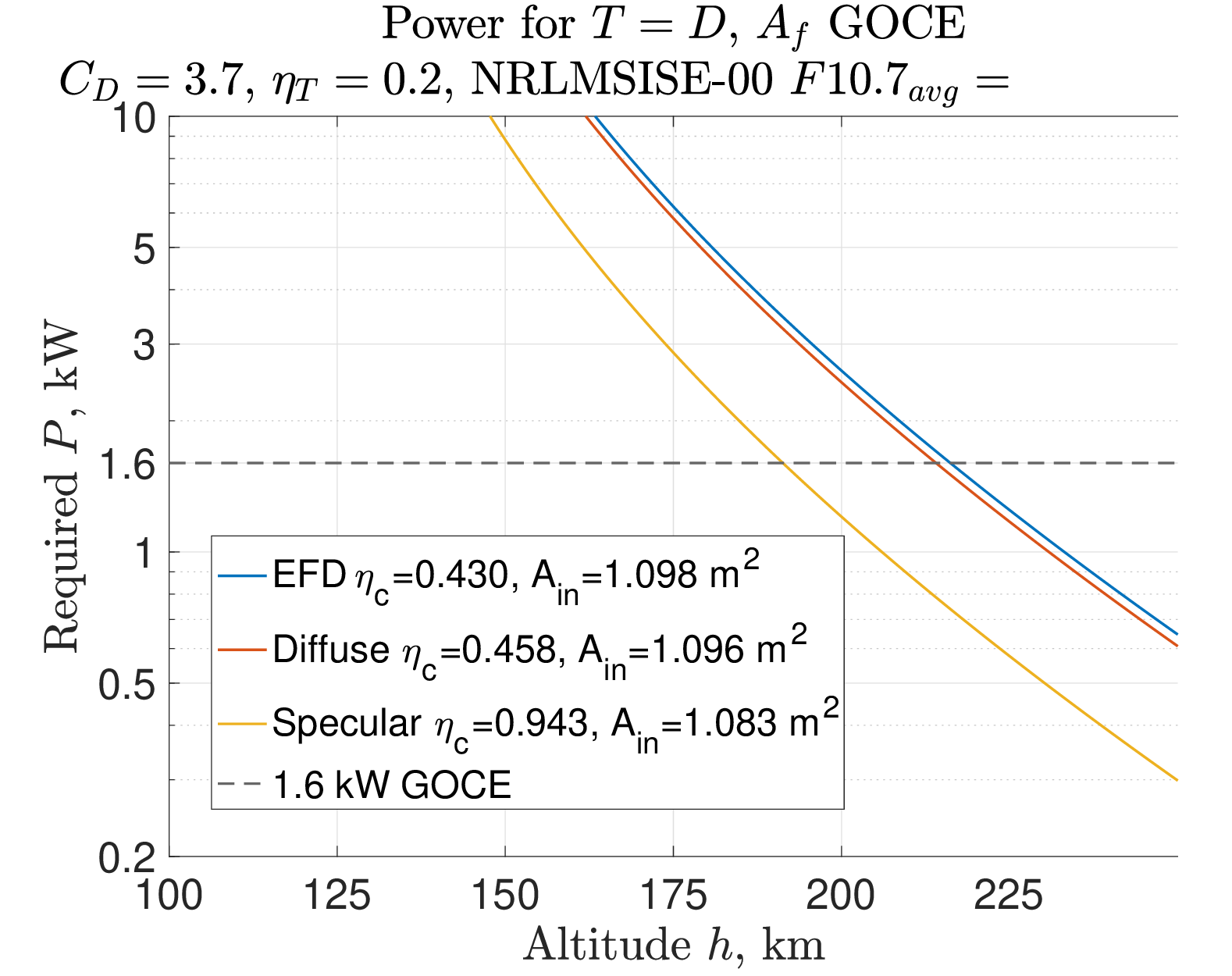}
		\caption{Required $P_{ABEP}$ in VLEO.}
		\label{fig:P_h_GOCE}
		\end{subfigure}
		\begin{subfigure}[b]{0.43\linewidth}
		\includegraphics[width=\linewidth]{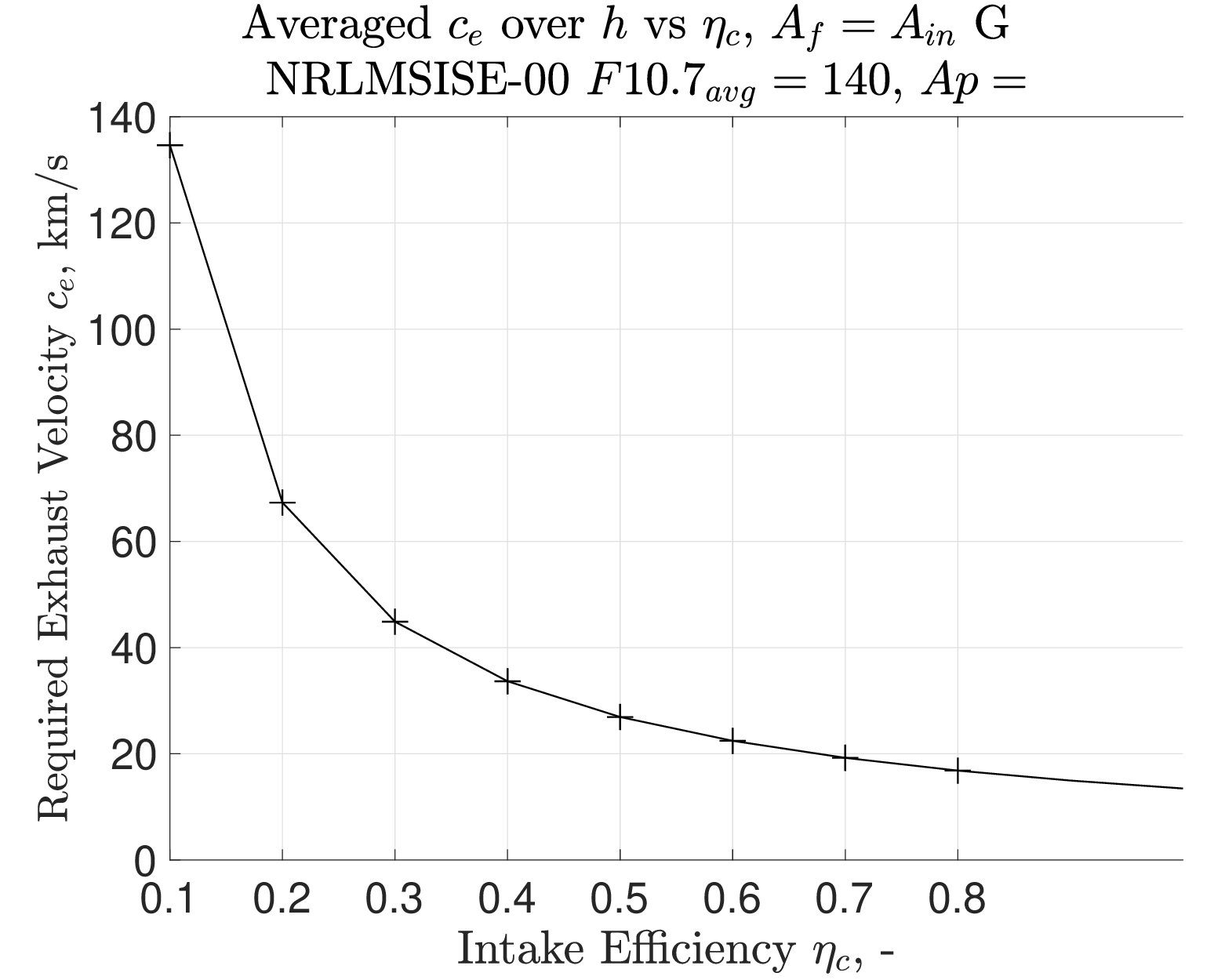}
		\caption{Averaged $c_e$ Required vs $\eta_c$ in VLEO.}
		\label{fig:ceetacGOCE}
		\end{subfigure}
			\caption{ABEP GOCE Requirements in VLEO, NRLMSISE-00.}
\end{figure}

Finally, an ABEP system based on a helicon plasma thruster could be applied to a GOCE-like mission, that has $A_f=\SI{1.1}{\meter^2}$, with both diffuse and specular-based intakes developed here, for circular orbits at altitudes $h>\SI{215}{\kilo\meter}$ requiring $P_{ABEP}<\SI{1.6}{\kilo\watt}$, the EoL power available for GOCE, based on drag calculations using $C_D=3.7$. This corresponds to a thrust to power ratio between $T/P=11-\SI{23}{\milli\newton\per{\kilo\watt}}$ and an $I_{sp}=1500-\SI{3000}{\second}$. The main parameters of the ABEP-equipped, GOCE-like spacecraft in VLEO are presented in Tab.~\ref{tab:ABEP_GOCE_VLEO}.
\begin{table}[h]
\caption{GOCE ABEP in VLEO, $\eta_c=0.430,~0.458,~0.943$,~$\eta_T=0.2$.}
\centering
\begin{tabular}{ccccc}
\toprule
$P_{ABEP}$ & $h$ & $T/P$ & $I_{sp}$  & $C_D$ \\
\SI{}{\kilo\watt} & \SI{}{\kilo\meter} & \SI{}{\milli\newton\per{\kilo\watt}} & \SI{}{\second} &- \\
\midrule
$1.6-0.3$ & $190-250$ & $10-24$ & $1500-3000$ & $3.7$ \\ 
\bottomrule
\end{tabular}
\label{tab:ABEP_GOCE_VLEO}
\end{table}
If less power is to be applied to the ABEP system, the minimum altitude would increase. Instead, if higher thrust efficiencies can be achieved, the minimum altitude of operation, for a given power, would be lowered. 


\subsection{Example of Extended ABEP Mission in Earth Orbit}
As the dissertation's objective is the development of an ABEP system, the chosen mission scenario is technology demonstration. It includes orbit raising, maintaining, and lowering while combining it with nadir pointing for Earth observation. To ensure electrical power to be continuously delivered by the solar arrays, a Sun-synchronous Dawn-Dusk orbit (SSO) is chosen, which also provides constant illumination conditions that are optimal for Earth observation. The spacecraft is oriented such that the intake always faces the incoming flow to maximize the propellant collection. The mission scenario is illustrated in Fig.~\ref{fig:ABEP_Mission}.
\begin{figure}[h]
	\centering
	\includegraphics[width=\textwidth]{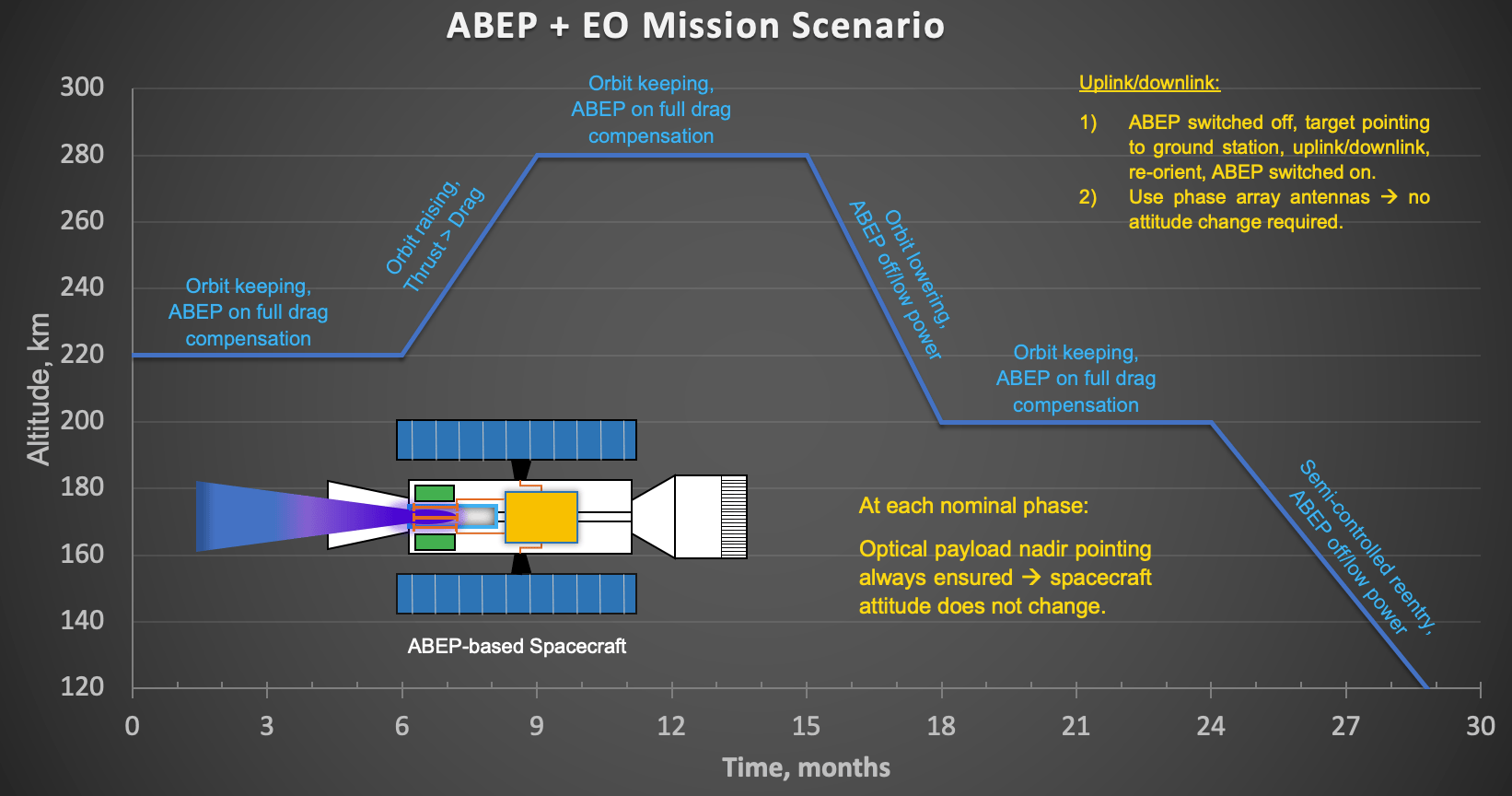}
	\caption{ABEP Technology Demonstration Mission Scenario.}
	\label{fig:ABEP_Mission}
\end{figure}

A first phase of 6 months is of orbit keeping at $h=\SI{220}{\kilo\meter}$ on full drag compensation. At month 6, the power input for the ABEP system is increased such to achieve $T>D$ with an acceleration that is able to bring the spacecraft to an altitude of $h=\SI{280}{\kilo\meter}$ within 3 months. From month 9 to month 12, a circular orbit at $h=\SI{280}{\kilo\meter}$ is maintained with full drag compensation $T=D$. At month 15, the ABEP input power is decreased to allow the orbit to drop to a $h=\SI{200}{\kilo\meter}$ circular orbit within 3 months. The power is then again increased at month 18 to perform orbit keeping at an altitude of $h=\SI{200}{\kilo\meter}$ for 6 months. By switching the ABEP system off, or alternatively maintaining it at minimum power, a semi-controlled descent is set, to reach an altitude of $h=\SI{120}{\kilo\meter}$ within 5 months, while keep naturally descending and burn in the Earth's atmosphere, completing the mission for a total duration of $>24$ months. 


\subsection{ABEP in Very Low Mars Orbit (VLMO)}
The atmosphere of Mars is mostly composed of \ce{CO2}, AO, \ce{N2}, and \ce{CO} as shown in Fig.~\ref{fig:marsatm} based on the Mars Climate Database (MCD) v5.2 atmospheric model~\cite{millour2015mars}. The solar activity, set based on Earth values as required by the MCD v5.2, influences the higher orbits as in the Earth case, see Fig.~\ref{fig:marssol}.
\begin{figure}[H]
	\centering
	\begin{subfigure}[b]{0.47\linewidth}
		\includegraphics[width=\linewidth]{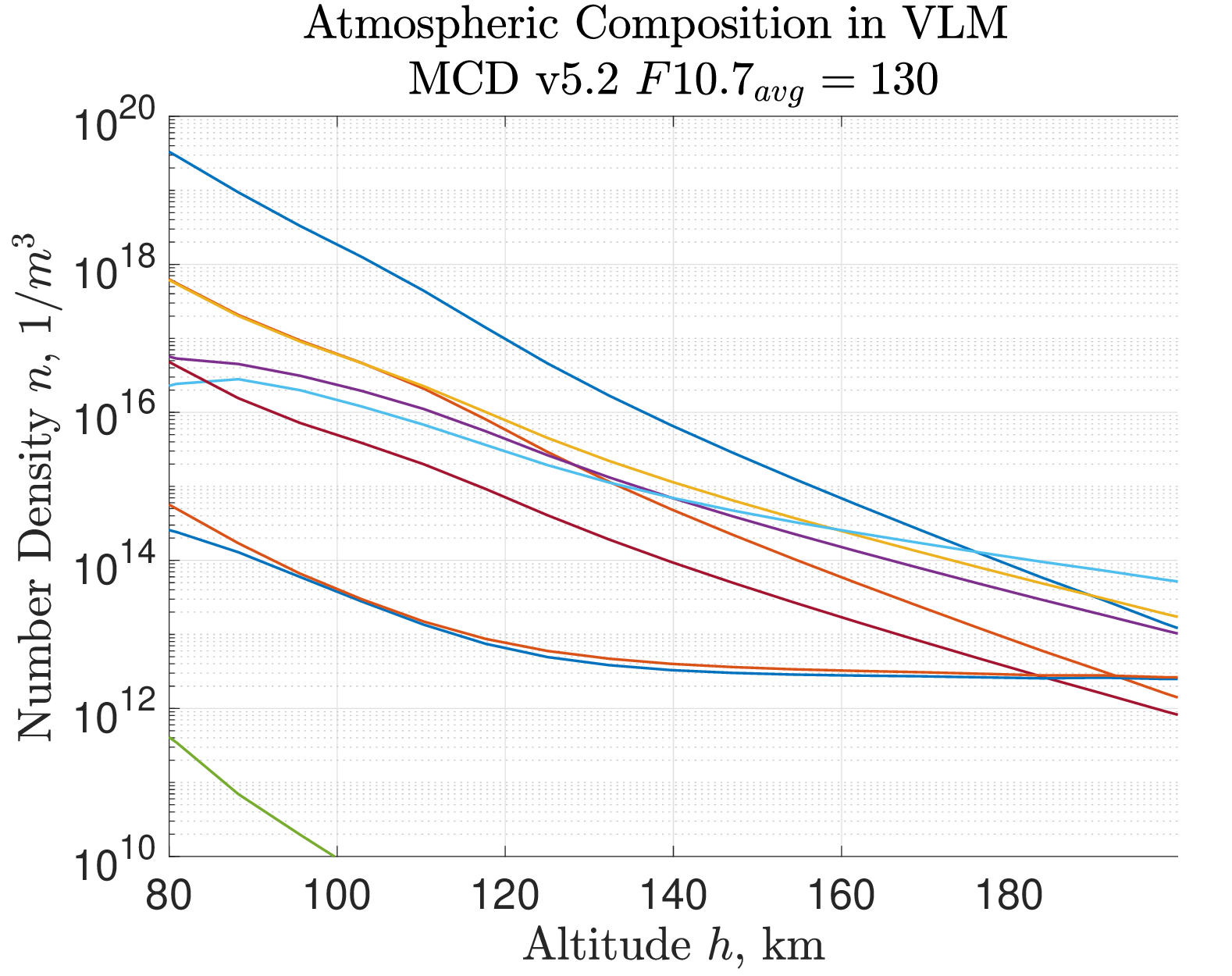}
		\caption{Composition vs Altitude.}
		\label{fig:marsatm}
	\end{subfigure}
	\begin{subfigure}[b]{0.47\linewidth}
		\includegraphics[width=\linewidth]{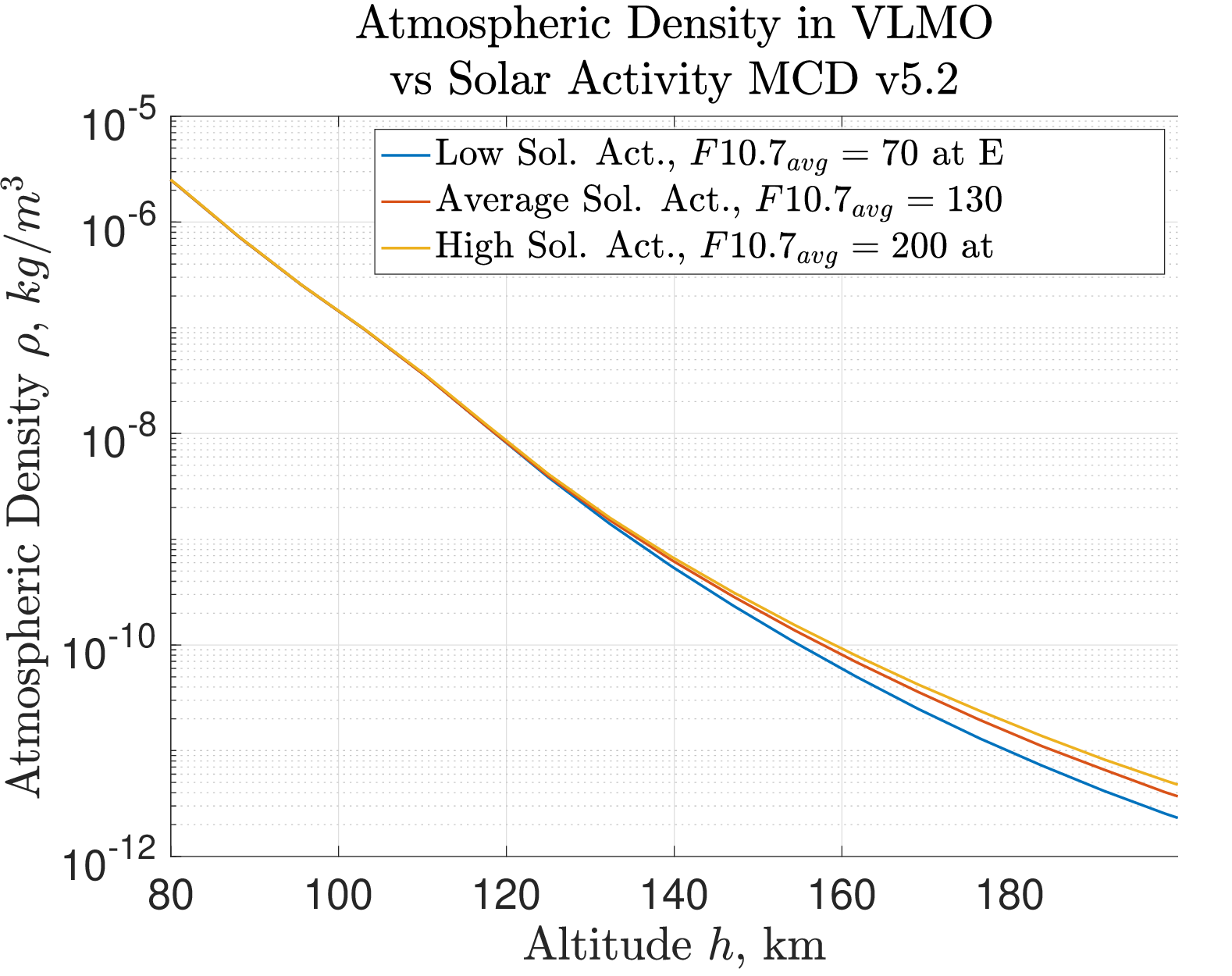}
		\caption{Density vs Altitude and Solar Activity.}
		\label{fig:marssol}
	\end{subfigure}
	\caption{VLMO Environment, MCD v5.2 at Lat.$=\SI{0}{\degree}$, Long.$=\SI{0}{\degree}$.}
\end{figure}
By assuming a drag coefficient $C_D=3$ based on~\cite{BUSEK}, higher than what conventionally assumed $C_D=2$~\cite{marscd}, and the same presented intake performances as well as thruster efficiency considered in the Earth case, the estimations of $\dot{m}_{thr}$, $D$, and $P_{ABEP}$ are presented in Fig.~\ref{fig:mdot_h_Mars},~Fig.~\ref{fig:d_h_Mars}, and Fig.~\ref{fig:P_h_Mars}. The limits of very low Mars orbit (VLMO), are set to $h=\SI{80}{\kilo\meter}$ due to the possibility of dust storms to achieve such orbits, and the maximum to $h=180-\SI{200}{\kilo\meter}$ according to the studies from BUSEK~\cite{BUSEK}. The collectible mass flow range is between $\dot{m}_{thr}=2\times10^{-5}-\SI{2e2}{\milli\gram\per{\second}}$, while the drag to be compensated is $D=2\times10^{-3}-\SI{7e2}{\milli\newton}$.
\begin{figure}[H]
	\centering
	\begin{subfigure}[b]{0.48\linewidth}
		\includegraphics[width=\linewidth]{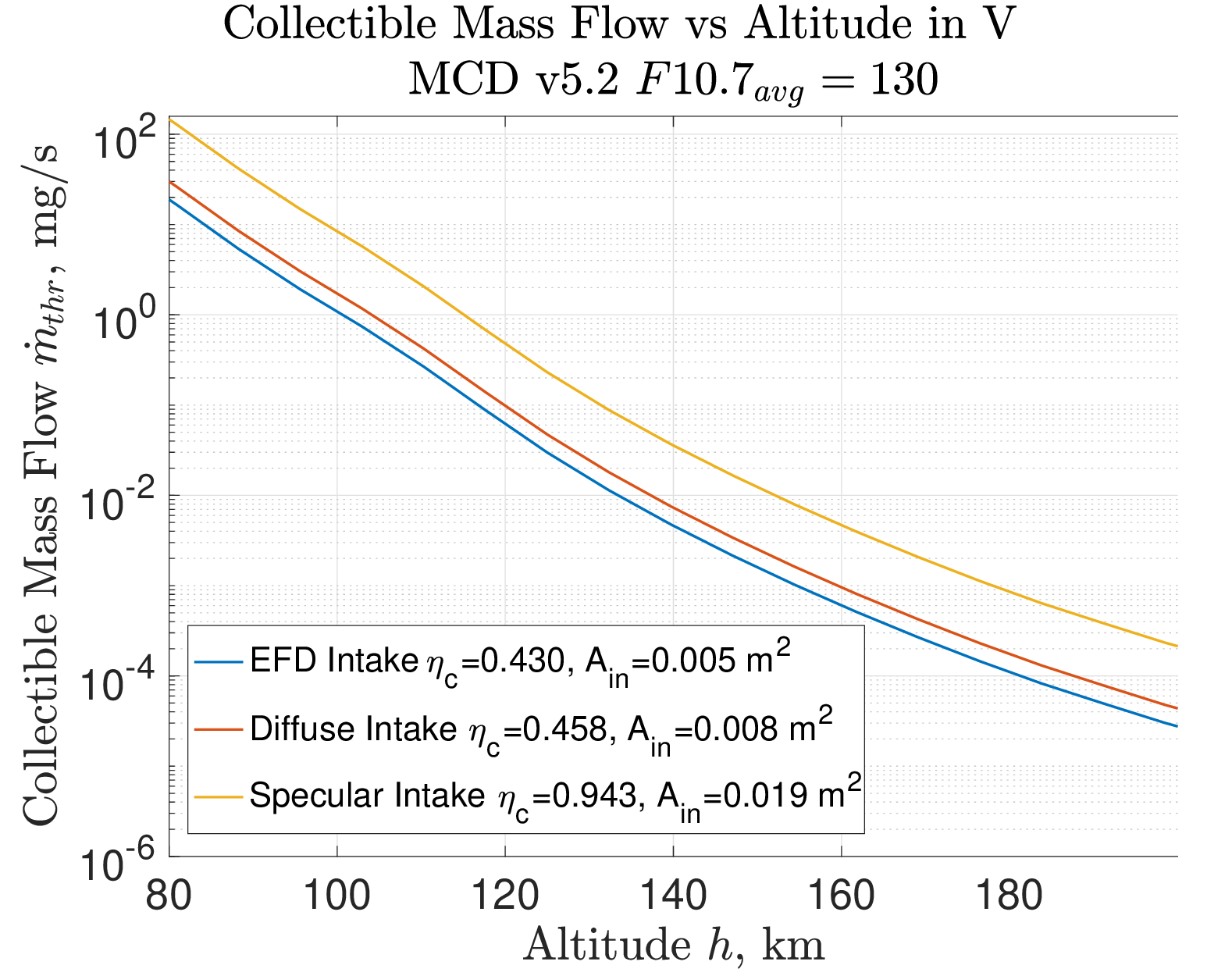}
	\caption{Collectible Mass Flow in VLMO.}
		\label{fig:mdot_h_Mars}
	\end{subfigure}
	\begin{subfigure}[b]{0.44\linewidth}
		\includegraphics[width=\linewidth]{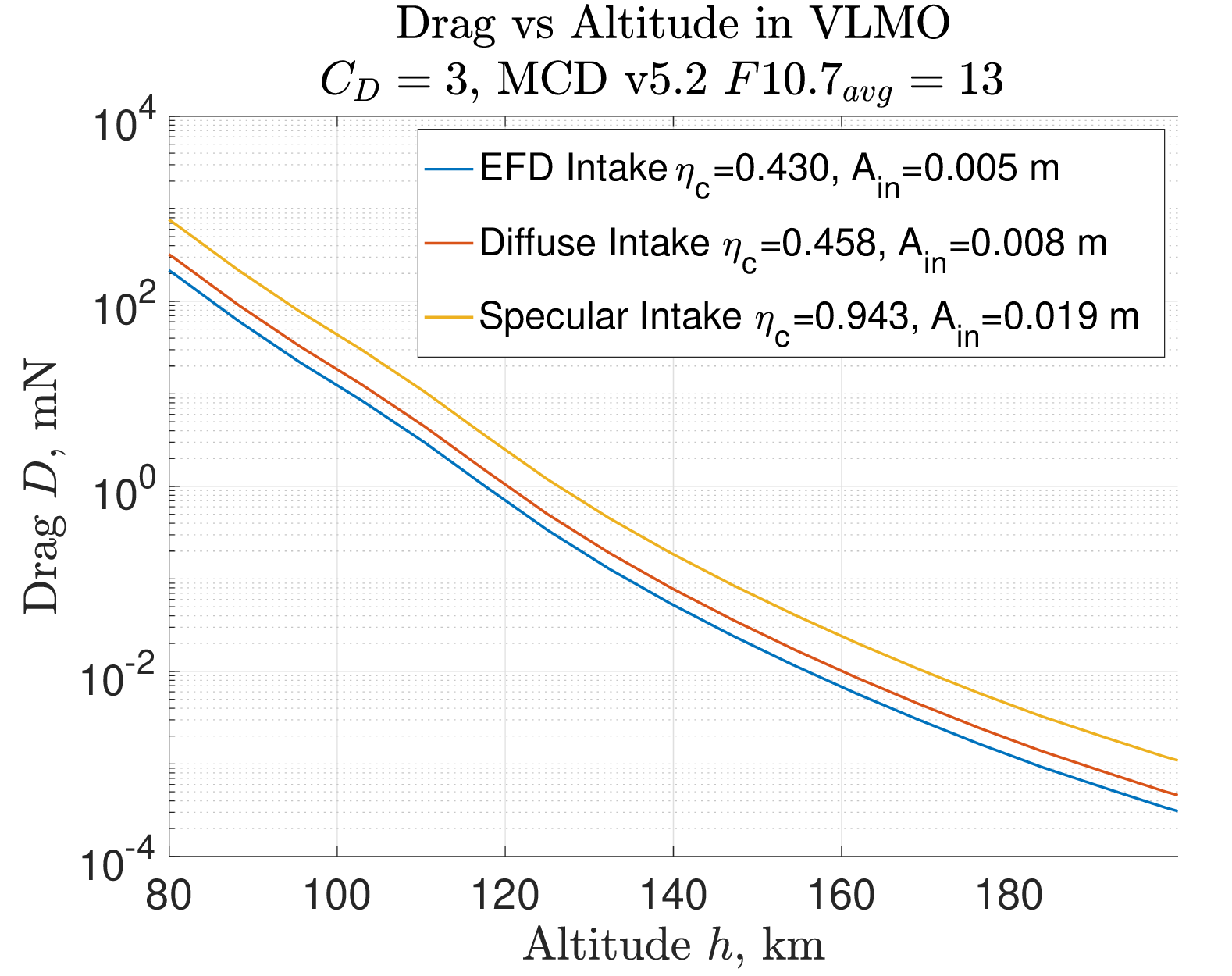}
		\caption{Aerodynamic Drag in VLMO, $A_{in}=A_f$.}
		\label{fig:d_h_Mars}
	\end{subfigure}
	\caption{ABEP in VLMO, MCD v5.2.}
\end{figure}

Finally, the ABEP systems require an exhaust velocity $c_e<\SI{12}{\kilo\meter\per{\second}}$ and a power $P_{ABEP}<\SI{1}{\kilo\watt}$ for any altitude $h>\SI{100}{\kilo\meter}$.
\begin{figure}[h]
	\centering
		\includegraphics[width=0.6\linewidth]{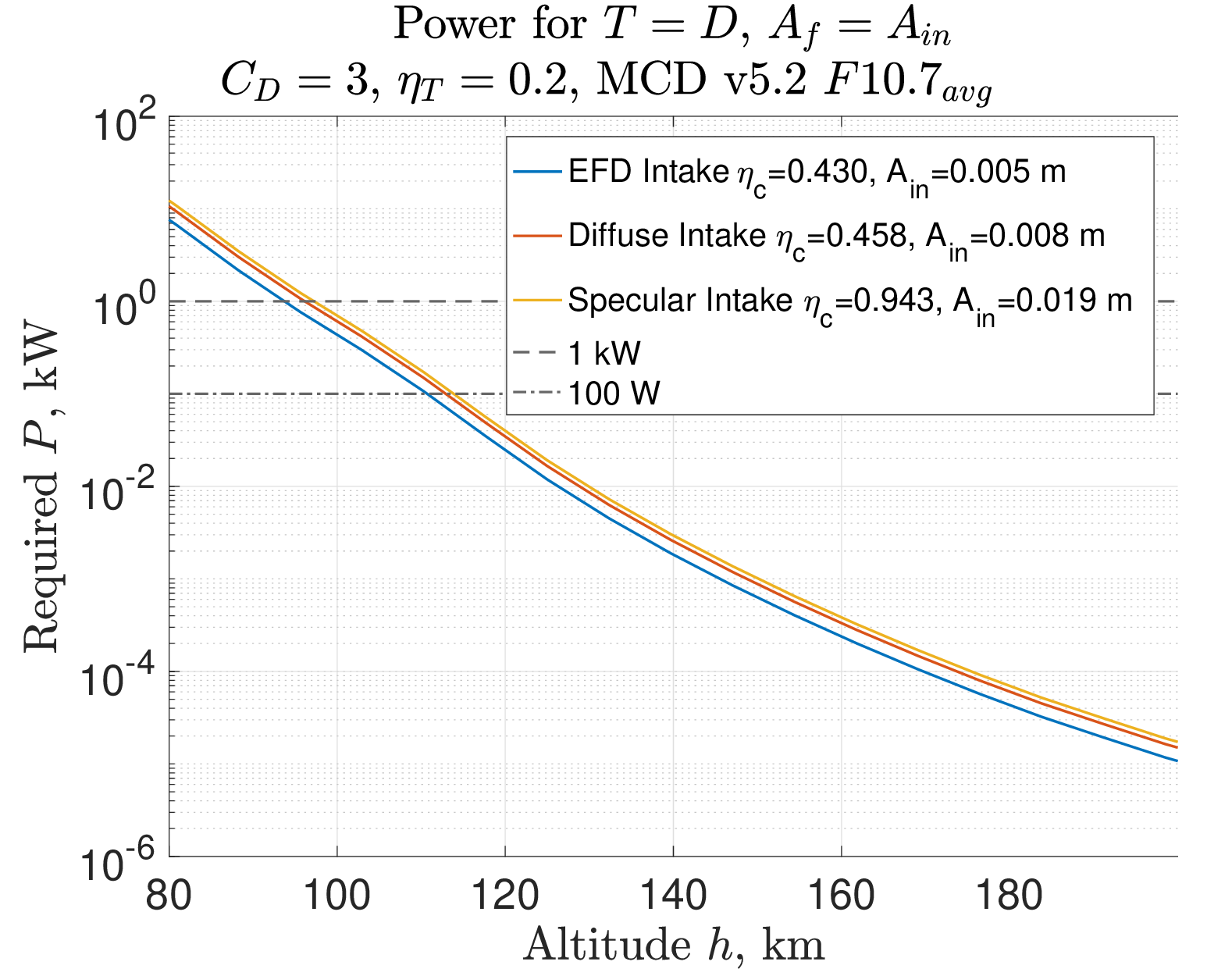}
	\caption{Required $P_{ABEP}$ in VLMO, MCD v5.2.}
		\label{fig:P_h_Mars}
\end{figure}

Similarly, the case of GOCE is analysed for the application in VLMO. The same power available of \SI{1.6}{\kilo\watt} is applied, but it must take into account larger solar arrays due to the increased distance from the Sun. 
Results show a collectible mass flow between~$\dot{m}_{thr}=6\times10^{-1}-\SI{7E4}{\milli\gram\per{\second}}$, see Fig.~\ref{fig:mdot_h_Mars_busek}, and a drag to be compensated between $D=6\times10^{-1}->\SI{2E3}{\milli\newton}$, see Fig.~\ref{fig:d_h_Mars_busek}. The drag is above $D=\SI{2E3}{\milli\newton}$ for $h<\SI{100}{\kilo\meter}$, while compared to the ABEP GOCE for VLEO, the $D=\SI{200}{\milli\newton}$ is achieved at $h_{Mars}=\SI{117}{\kilo\meter}$ compared to the $h_{Earth}=\SI{150}{\kilo\meter}$.
\begin{figure}[h]
	\centering
	\begin{subfigure}[b]{0.48\linewidth}
		\includegraphics[width=\linewidth]{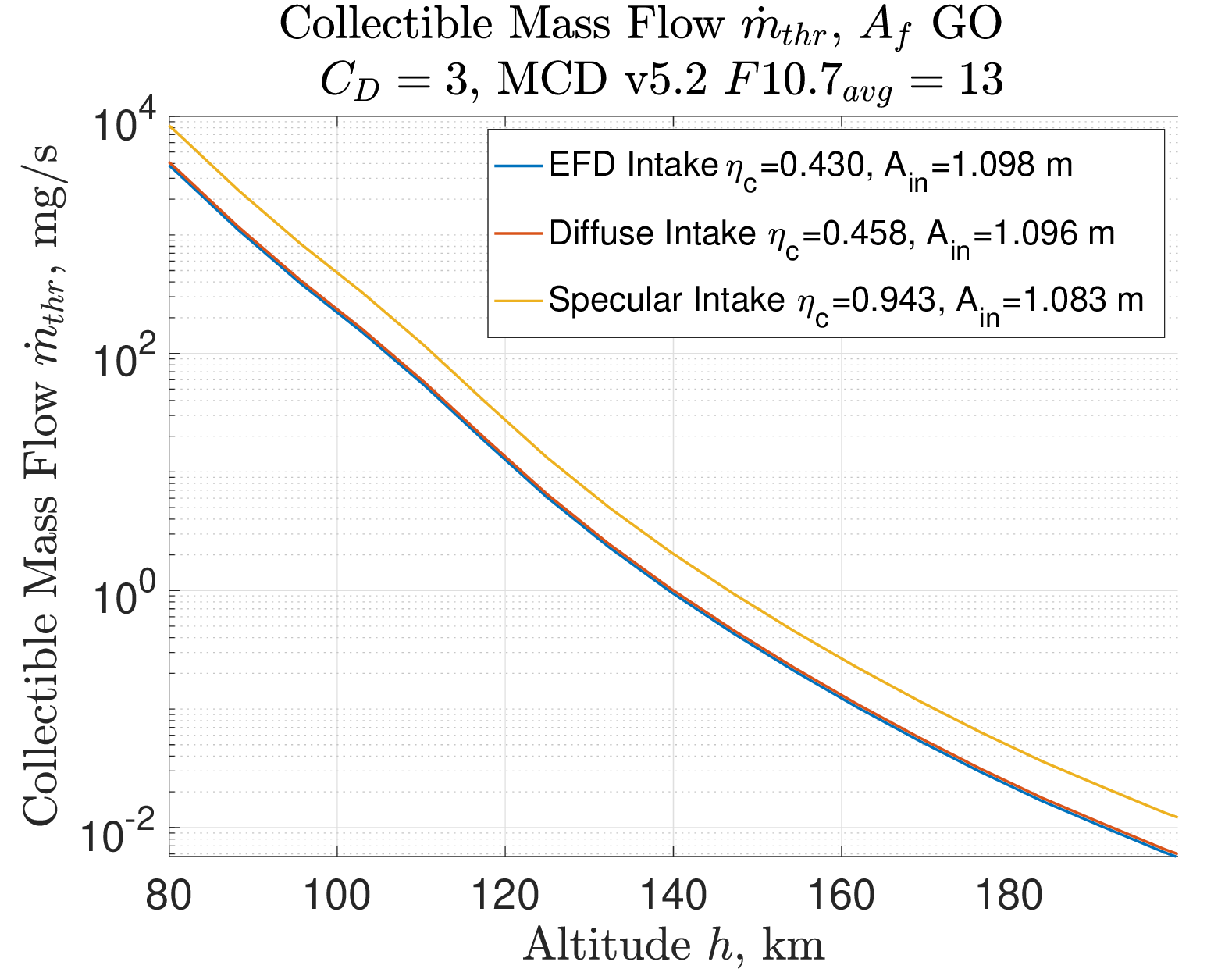}
	\caption{GOCE Collectible Mass Flow in VLMO}
		\label{fig:mdot_h_Mars_busek}
	\end{subfigure}
	\begin{subfigure}[b]{0.46\linewidth}
		\includegraphics[width=\linewidth]{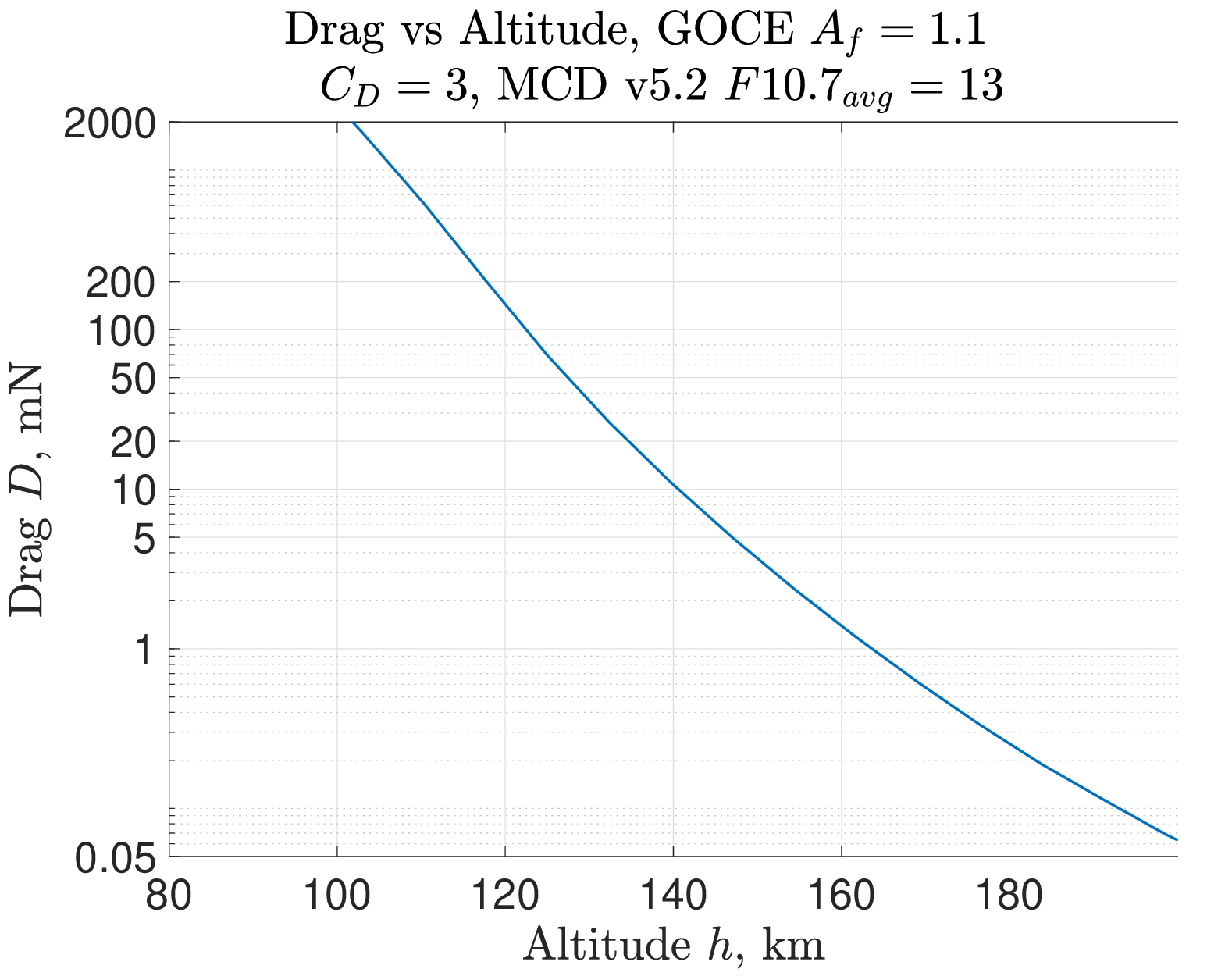}
		\caption{GOCE Aerodynamic Drag in VLMO.}
		\label{fig:d_h_Mars_busek}
	\end{subfigure}
	\caption{ABEP GOCE in VLMO, MCD v5.2.}
\end{figure}

The ABEP systems require $c_e<\SI{12}{\kilo\meter\per{\second}}$ for the available intake performances, see Fig.~\ref{fig:ceetacGOCE_Mars}. Here $c_e$ is shown as the average value over VLMO altitude for $A_f=A_{in}$ and is below $c_e<\SI{50}{\kilo\meter\per{\second}}$ for any $\eta_c > 0.1$. The required power $P_{ABEP}<\SI{1}{\kilo\watt}$ for any altitude $h>\SI{130}{\kilo\meter}$, see Fig.~\ref{fig:P_h_Mars_busek}.
\begin{figure}[h]
	\centering
	\begin{subfigure}[b]{0.47\linewidth}
			\includegraphics[width=\linewidth]{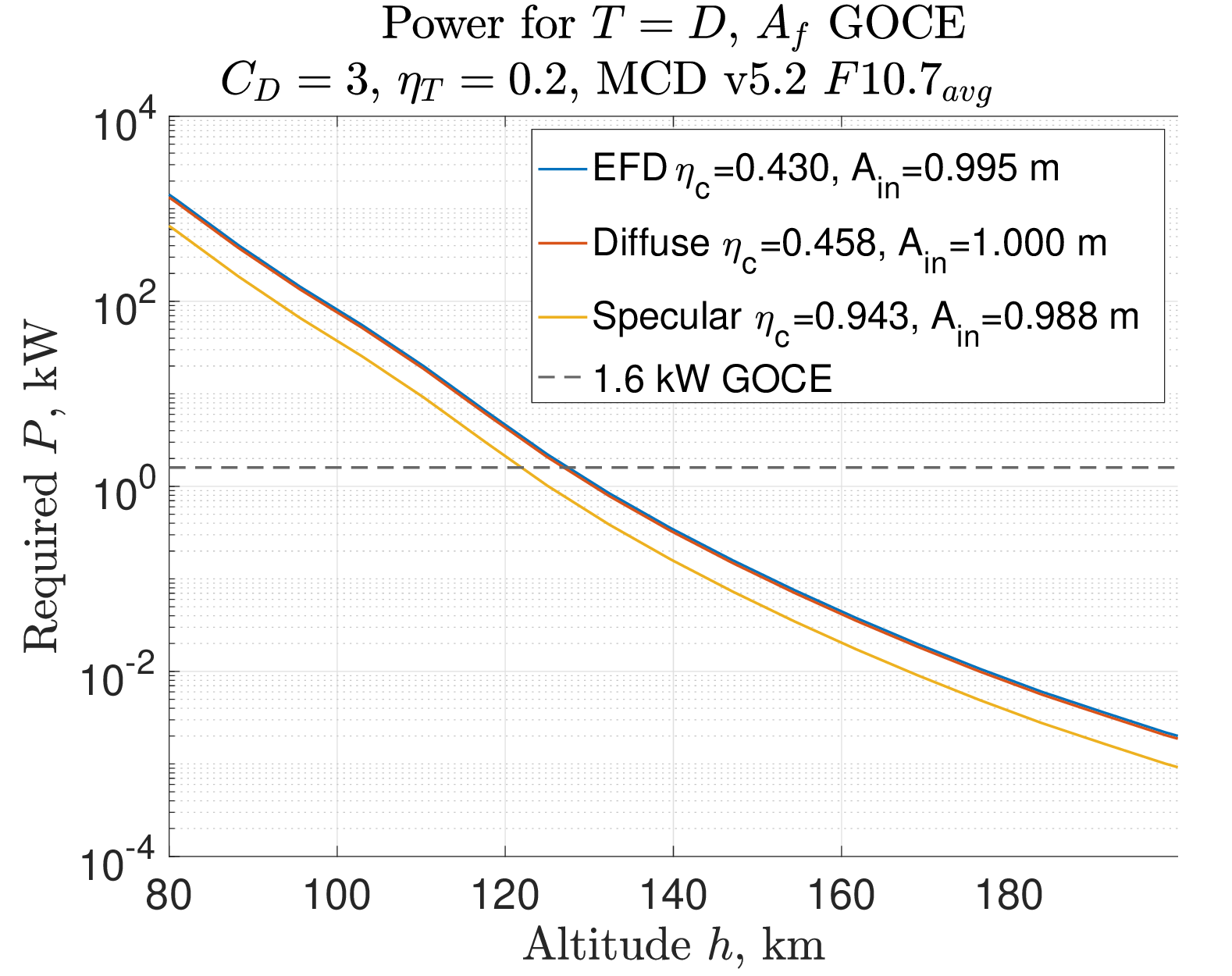}
			\caption{Required GOCE $P_{ABEP}$ in VLMO}
			\label{fig:P_h_Mars_busek}
	\end{subfigure}
		\begin{subfigure}[b]{0.44\linewidth}
			\includegraphics[width=\linewidth]{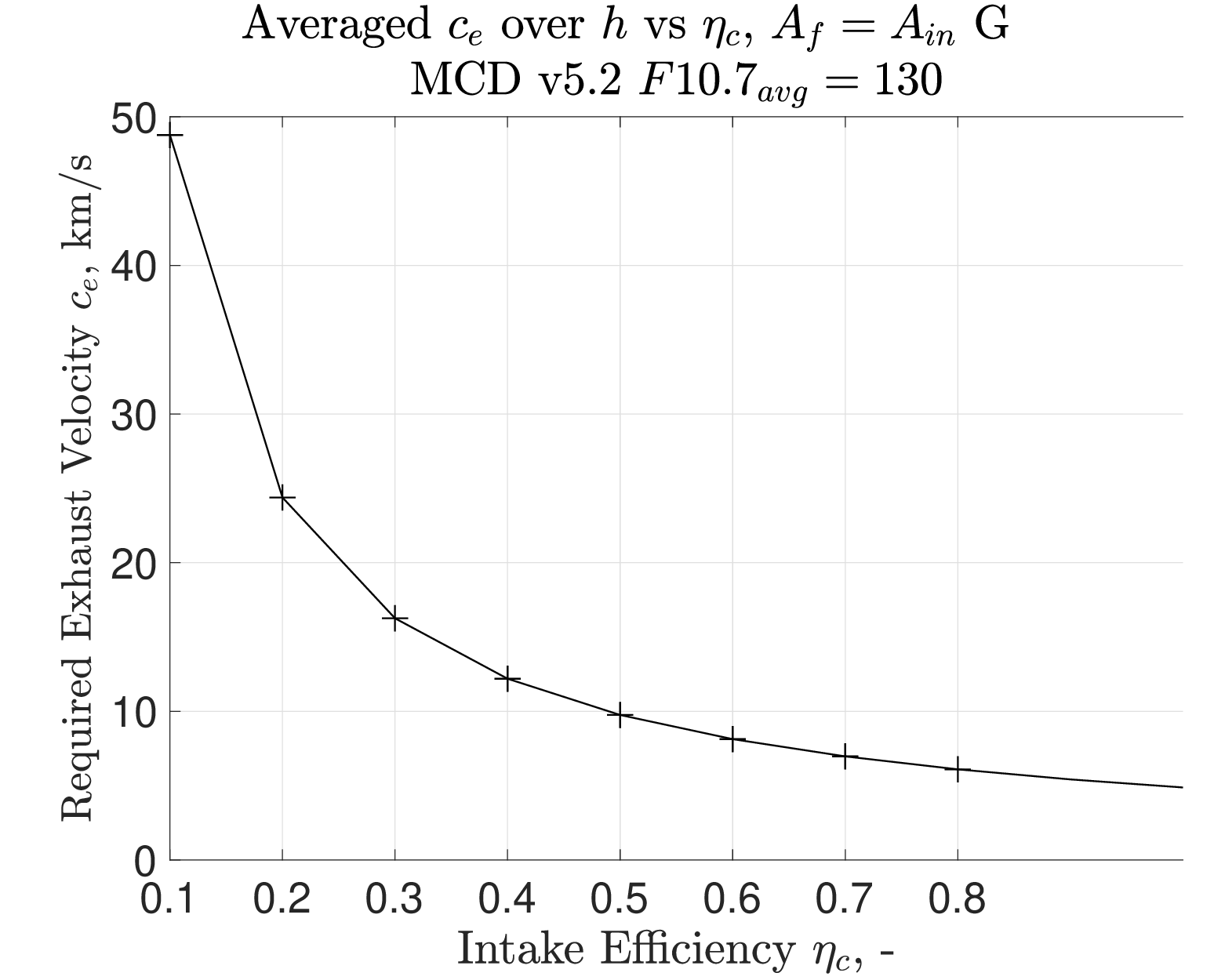}
		\caption{Averaged $c_e$ Required vs $\eta_c$ in VLMO.}
			\label{fig:ceetacGOCE_Mars}
		\end{subfigure}
	\caption{ABEP GOCE Requirements in VLMO, MCD v5.2.}
\end{figure}
Finally, an ABEP system in VLMO might be easier to apply compared to one in VLEO. It requires less $P_{ABEP}$ and $c_e$ due to the reduced orbital velocity. But, as the distance of Mars from the Sun is greater than Earth's, larger solar arrays are required. At $h_{Mars}\sim\SI{160}{\kilo\meter}$, the density, therefore $\dot{m}_{thr}$, $\rho=\SI{9e9}{\kilo\gram\per{\meter^{-3}}}$ is comparable to that in VLEO at $h_{Earth}\sim\SI{250}{\kilo\meter}$ (the ABEP upper $h$ limit) therefore, the range of operation in VLMO is shifted further below, setting a preliminary upper limit for VLMO to $h_{Mars}=\SI{160}{\kilo\meter}$. The analysis yields $T/P=30-\SI{65}{\milli\newton\per{\kilo\watt}}$ and an $I_{sp}=500-\SI{1100}{\second}$. The main parameters of the ABEP-equipped, GOCE-like spacecraft in VLMO are presented in Tab.~\ref{tab:ABEP_GOCE_VLMO}.
\begin{table}[h]
\centering
\caption{GOCE ABEP in VLMO, $\eta_c=0.430,~0.458,~0.943$,~$\eta_T=0.2$.}
	\begin{tabular}{ccccc}
	\toprule
	$P_{ABEP}$ & $h$ & $T/P$ & $I_{sp}$  & $C_D$ \\
	\SI{}{\kilo\watt} & \SI{}{\kilo\meter} & \SI{}{\milli\newton\per{\kilo\watt}} & \SI{}{\second} & - \\
	\midrule
	$1.6-0.02$ & $120-160$ & $30-65$ & $500-1100$ & $3$ \\ 
	\bottomrule
	\end{tabular}
\label{tab:ABEP_GOCE_VLMO}
\end{table}
The use of ABEP in VLMO can be of high interest for multiple autonomous drones that can assist future human activity on Mars by providing communication, planetary observation, as well as climate monitoring, especially dust storms, and generating a high quality geoid of the planet, similarly to what GOCE performed for Earth.
\subsection{Future ABEP Applications}
The use of ABEP can be extended to Venus, with the atmosphere almost fully composed of \ce{CO2} at a relatively high density~\cite{mahieux2012densities}, where solar arrays can be applied as main power source. Instead, given that another power source is provided, i.e. nuclear, further celestial bodies can be explored such as Titan (\ce{N2} and \ce{CH4}), or any gas giant, Jupiter, Saturn, Uranus, Neptune with atmospheres that contain mostly \ce{H2} and \ce{He} at different fractions. Also, low power, ABEP-based drones, i.e. a 3U CubeSat sized, could be released in the lower atmospheres of gas giants by larger mother-ships. Finally, if power, materials, and thermal protection system technologies are developed further, the use of ABEP could be as well imagined in the future for the application in (relatively) low Sun (or star) orbits, in which spacecrafts dive and gather \ce{H2} and \ce{He}, store it in a propellant tank, and use it to travel throughout the solar system. 


\chapter{Intake}
\label{ch:intake}
This chapter describes the design and principle of operation of an intake for an Atmosphere-Breathing Electric Propulsion System.
 The intakes described within this dissertation are designed as passive devices, therefore without moving parts and optimized for a spacecraft orbiting around the Earth. The absence of moving parts in the final designs mitigates any failure mode that might compromise the whole ABEP system operation during the mission. An intake faces the flow to collect the atmospheric particles. For the ABEP altitude range, it is a free molecular flow, meaning that gas-gas interactions can be generally neglected.
 Three finalized designs are presented, two are based on diffuse reflecting materials, and one on specular ones, all optimized for the thruster developed within this dissertation, having a discharge channel diameter of  $d_{out}=\SI{37}{\milli\meter}$. Parallelly a hybrid intake concept is also introduced.
 
\section{The Balance Model (Full Accommodation)}
Diffuse reflecting materials have an energy accommodation coefficient of $\alpha=1$ (full accommodation), see Eq.~\ref{eq:accom}, where $E_i$ is the energy of incoming molecule, $E_r$ is that of the re-emitted molecule, and $E_w$ is that of the re-emitted molecule if it left the surface at wall temperature.
\begin{equation}
\alpha=\frac{E_i - E_r}{E_i-E_w}
\label{eq:accom}
\end{equation}

Full accommodation, $\alpha=1$ means that once a particle hits such material, it transfers all the energy to it assuming the material's temperature. After impact, the direction at which the particle is based on a half-space Maxwellian velocity vector distribution, and its resulting movement is given only by thermal diffusion. 

To develop such intake, a balance model (BM) depending only on geometry, flows, and transmission probabilities is developed based on the following hypotheses: 
\begin{itemize}
\item Free molecular flow, no inter-particle collisions;
\item full accommodation, diffuse reflection at the walls, $\alpha=1$;
\item thermal diffusion movement inside the intake's chamber;
\item ideal gas behaviour.
\end{itemize}
The intake is divided into four main sections:
\begin{itemize}
\item The inlet ducts $\Rightarrow$ honeycomb-like structure of small ducts; 
\item the main duct $\Rightarrow$ particles are free to flow; 
\item the chamber section $\Rightarrow$ particles hit the walls at the back and fully accommodate; 
\item the outlet section  $\Rightarrow$ the thruster's discharge channel.
\end{itemize}
\begin{figure}[h]
	\centering
	\includegraphics[width=9cm]{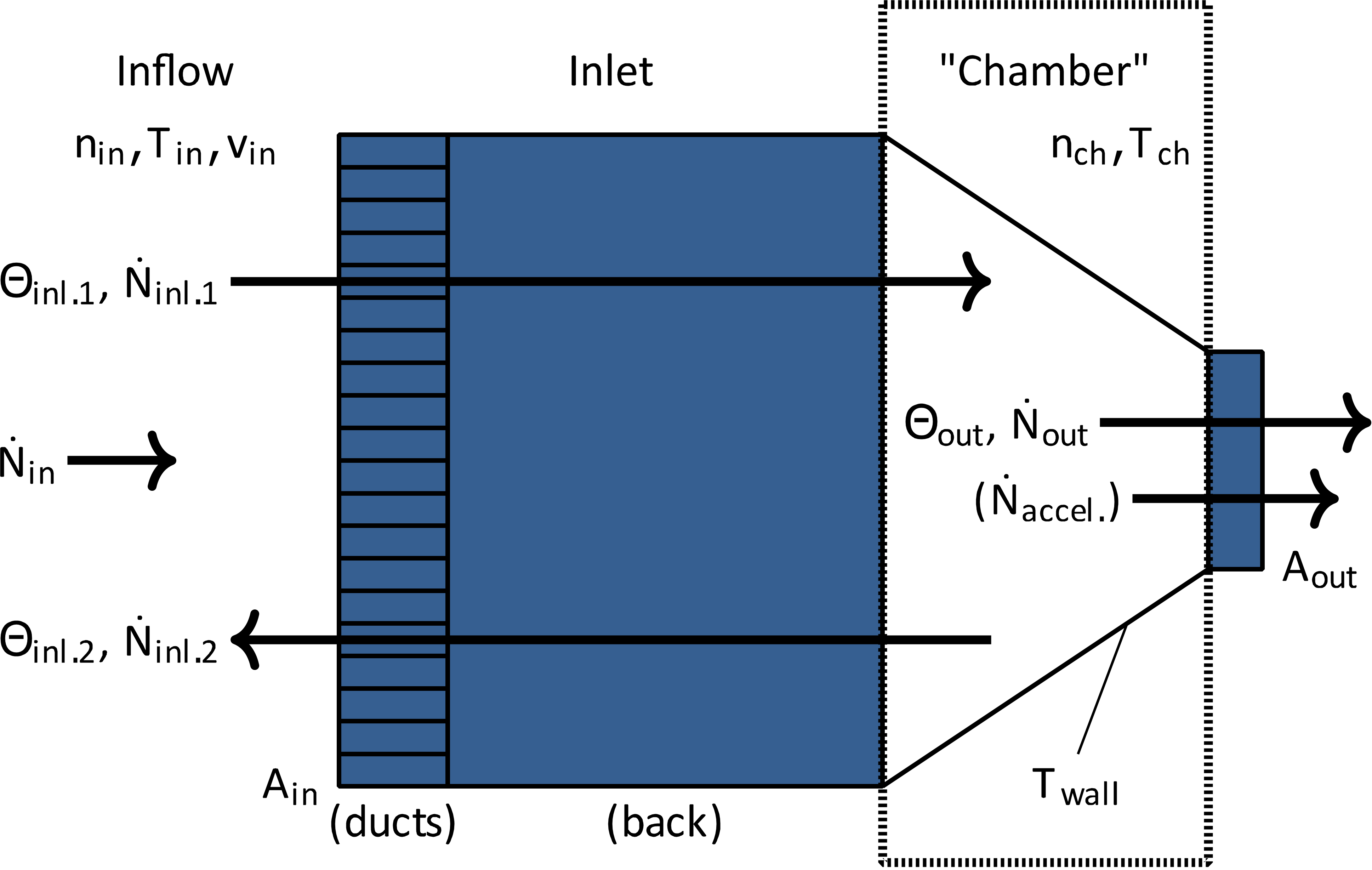}
	\caption{Intake Scheme for the Balance Model (BM)~\cite{tilman}}
		\label{fig:Bala}
\end{figure}

Each section is given a transmission probability $0<\Theta<1$, defined as the probability of a particle that enters a given structure to reach the end of it without being reflected back~\cite{clausing1932stromung}. These can be analytically calculated for the fully diffuse case and serve to estimate each number flow rate $\dot{N}$. 
The inlet duct section is a honeycomb-like structure of small tubes (ducts) that are aligned with the flow. Their task is to let in the incoming particles, with $\Theta_{inl.1}$, and later block them from escaping the intake, as they flow back in a random direction after full accommodation with the chamber walls with $\Theta_{inl.2}$. Finally, the outlet section is and/or leads into the thruster discharge channel (or to an eventual storage tank). Correspondingly, such area is defined by the discharge channel diameter and its respective $\Theta_{out}$. The intake and its sections are shown in Fig.\ref{fig:Bala}. The incoming flow condition is extracted from the NRLMSISE-00 atmospheric model~\cite{picone} for Earth's atmosphere: composition, density, temperature, orbital velocity, and thruster discharge channel diameter are the main parameters of interest needed for the intake design.

\subsection{Required Input Parameters}
The input parameters required for the intake design are:
\begin{itemize}
	\item Upper limit is the available area $A_f>A_{in}$, and the resulting drag (based on $A_f$);
	\item outer section of the intake: generally the thruster discharge channel size $\Rightarrow$ $A_{out}$;
	\item inflow condition: velocity, composition and density $\Rightarrow$ depending on altitude, latitude and longitude, epoch (if in daylight or in eclipse).
\end{itemize}
With such inputs, the initial intake geometry can be defined, and the respective transmission probabilities $\Theta$ estimated. This process is then iterated to maximize/optimize the desired parameter, such as the collection efficiency $\eta_c$, the mass flow reaching the thruster $\dot{m}_{thr}$, or the chamber pressure $p_{ch}$.

\subsection{Balance Model Main Equations (Full Accommodation)}
The following equations are used in the BM to evaluate the intake performance (full accommodation), especially the intake efficiency $\eta_c$ in Eq.~\ref{eq:effc}, and the density ratio $\frac{n_{ch}}{n_{in}}$ in Eq.~\ref{eq:effn}, that are then used to calculated the mass flow to the thruster $\dot{m}_{thr}$. The detailed derivation of the BM is given in Appendix~A. 
\begin{equation}
\dot{N}_{inl.{1}}=\dot{N}_{in}\Theta_{inl.{1}}
\label{eq:BM1}
\end{equation}

\begin{equation}
\eta_{c}=\frac{\dot{N}_{out}}{\dot{N}_{in}}=\frac{\Theta_{inl.1}}{\frac{A_{in}}{A_{out}}\frac{\Theta_{inl.2}}{\Theta_{out}}+1}
\label{eq:effc}
\end{equation}

\begin{equation}
\frac{n_{ch}}{n_{in}}=\frac{m_p v_{in}\Theta_{inl.1}}{\Theta_{inl.2}+\frac{A_{out}}{A_{in}}\Theta_{out}}\sqrt{\frac{2\pi}{m_p k_B T_{ch}}}
\label{eq:effn}
\end{equation}
Based on Eq.~\ref{eq:BM1},~\ref{eq:effc}, and~\ref{eq:effn} the intake performance depending on areas and $\Theta$ can be calculated. The latter can be analytically estimated within a certain range and are divided in those for the particles moving only with thermal diffusion, the Clausing case, and those which have velocities much higher than the thermal velocity. For the Clausing case, $\Theta$ is calculated, namely $\Theta_{inl.2}$ and $\Theta_{out}$, assuming $\alpha=1$, using the analytical expression of Clausing~\cite{clausing1932stromung} which results are shown in Fig.~\ref{fig:Clausing} for the given aspect ratios $L/R$ and for cylindrical structures~\cite{tilman}.

For fast particles, $\Theta_{inl.1}$, see Fig.~\ref{fig:dirindir}, is the sum of three components:
\begin{itemize}
\item $\Theta_{direct}$: particles travelling \textbf{directly} through the structure without wall interactions;
\item $\Theta_{indirect}$: particles travelling \textbf{indirectly} through the structure incurring into one or multiple collision with the walls;
\item $\Theta_{back}$: particle exchanging all their energy with the wall: \textbf{Clausing} case.
\end{itemize}

\begin{figure}[h]
	\centering
	\begin{subfigure}[b]{0.46\linewidth}
		\includegraphics[width=\linewidth]{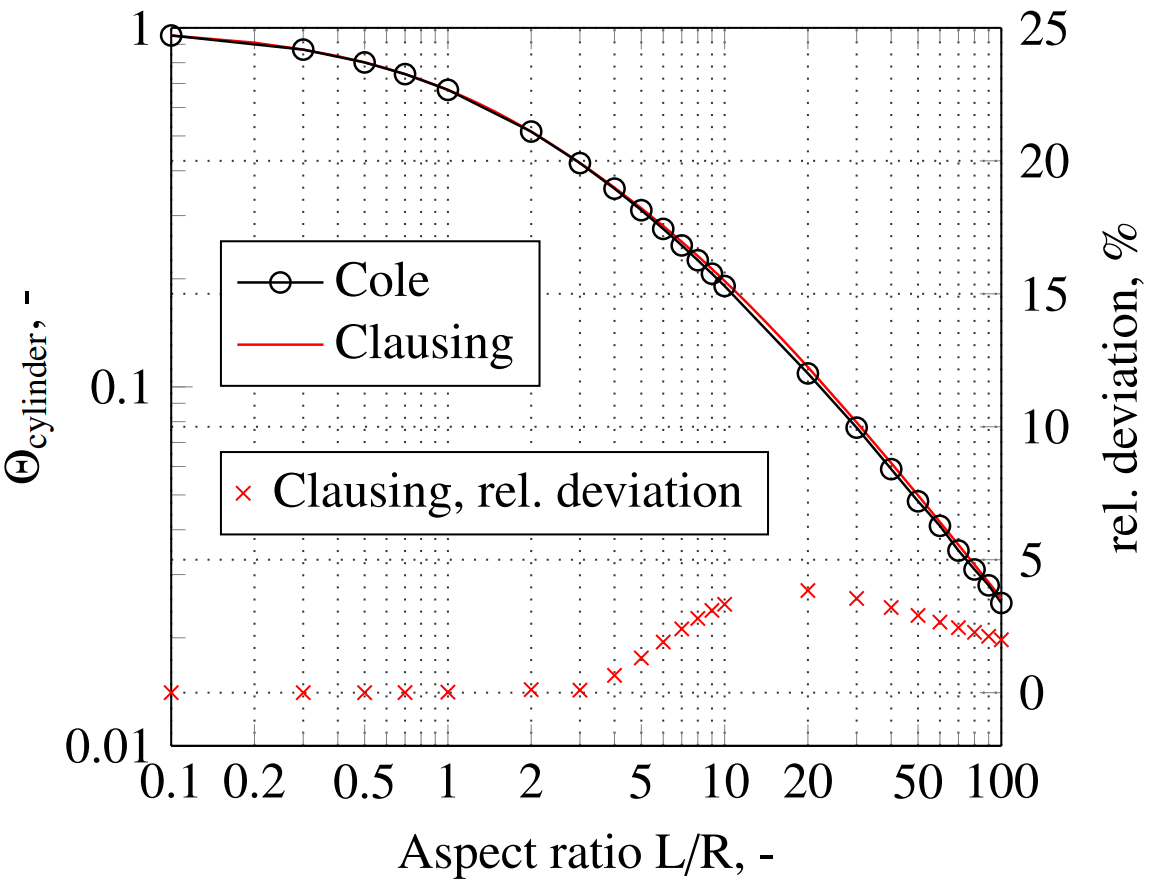}
	\caption{Clausing $\Theta$ analytical solutions~\cite{tilman}.}
		\label{fig:Clausing}
	\end{subfigure}
	\begin{subfigure}[b]{0.45\linewidth}
	\includegraphics[width=\linewidth]{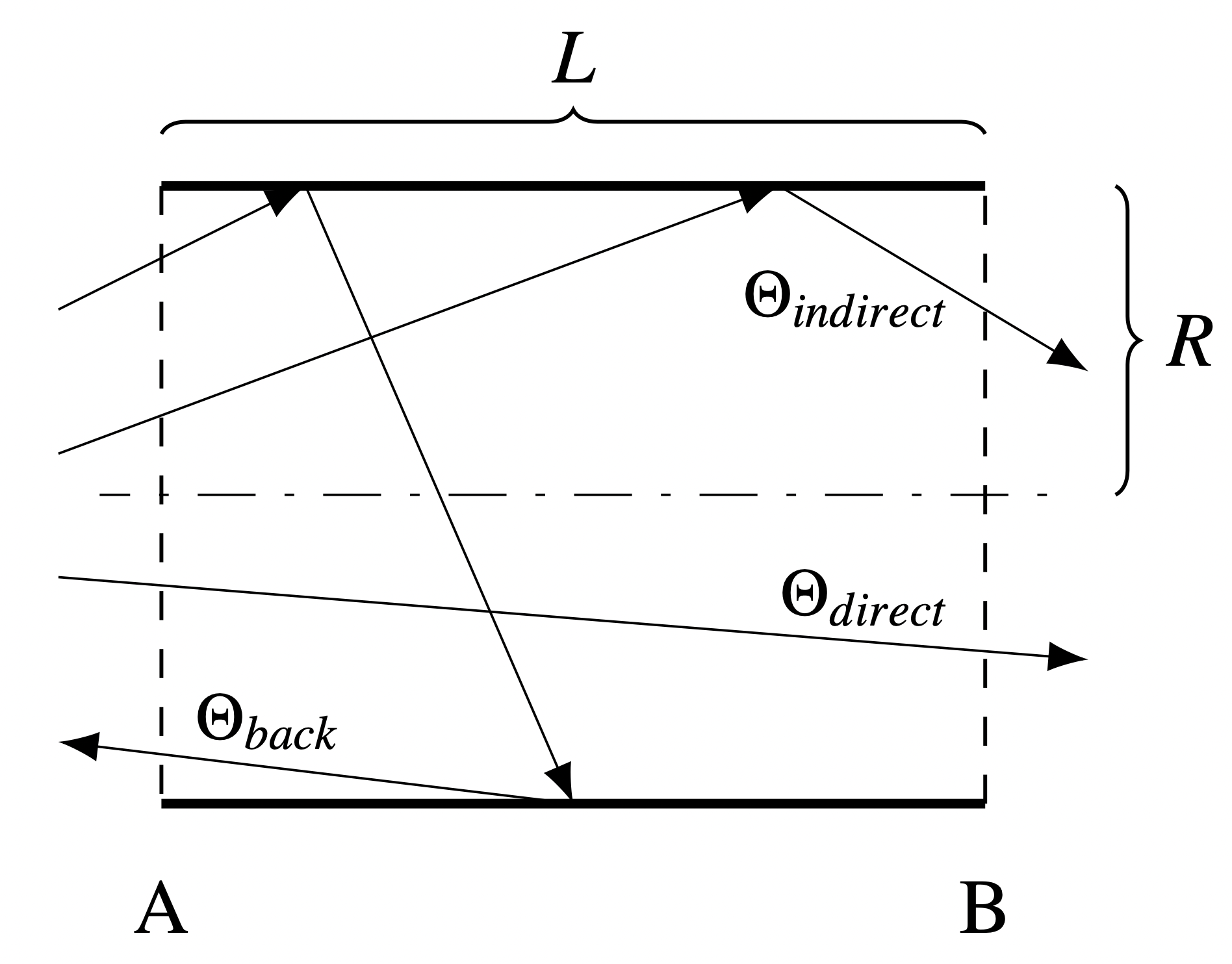}
	\caption{Definition of $\Theta$ through a cylinder~\cite{tilman}.}
	\label{fig:dirindir}
	\end{subfigure}
	\caption{Transmission Probabilities $\Theta$.}
\end{figure}
 
Finally, the overall transmission probability must be a combination of the three, Shown in Fig.~\ref{fig:transcyl} and Fig.~\ref{fig:transhex} for circular and hexagonal ducts, is (in green) $\Theta_{total}=\Theta_{cylinder,hexagon}=\Theta_{direct}+\Theta_{indirect}=\Theta_{inl.1}$~\cite{boldini}, (the latter allows better filling ratio of the inlet front surface), where $\Theta_{fast}=\Theta_{direct}$, $\Theta_{scat.}=\Theta_{indirect}$, and $\Theta_{back}=\Theta_{Clausing}$. 
In Fig.~\ref{fig:transcyl} and Fig.~\ref{fig:transhex}, the horizontal axis is the dimensionless inflow parameter defined as $X=(L/R)/(\sqrt{2}S)$, where
 $S=v_{in}/\sqrt{2k_B T_{in}/m_p}$ is defined as the molecular speed ratio ranging between $7<S<20$ for average solar activity in VLEO, being $S>5$ the hyperthermal case~\cite{MOSTAZAPRIETO201456}, in which the drift caused by random thermal motion can be neglected~\cite{LIVADIOTTI2020}.
The values of $\Theta_{cylinder,hexagon}$ are extracted from Fig.~\ref{fig:transcyl} and Fig.~\ref{fig:transhex}, and by varying $L/R$ and areas, the intake design is optimized. In particular, according to Fig.~\ref{fig:transcyl} and Fig.~\ref{fig:transhex}, the subscripts $1,mid,2$ refer to simulation results of different ranges of $X$ while, for $\Theta_{Clausing}$, the subscripts $1,2,3$ refer to three different simulation inflow conditions, see Tab.~\ref{tab:conditions} in Appendix~A. 
As displayed in Fig.~\ref{fig:transcyl} and Fig.~\ref{fig:transhex} by increasing $X$, $\Theta_{direct}$ decreases, while $\Theta_{indirect}$ increases presenting a maximum at $X=1.5$ for both cases. This leads to an overall decrease of $\Theta_{total}$ with $X$, hence with $L/R$ and inversely proportional to $S$. 	
\begin{figure}[h]
	\centering
	\includegraphics[width=0.9\linewidth]{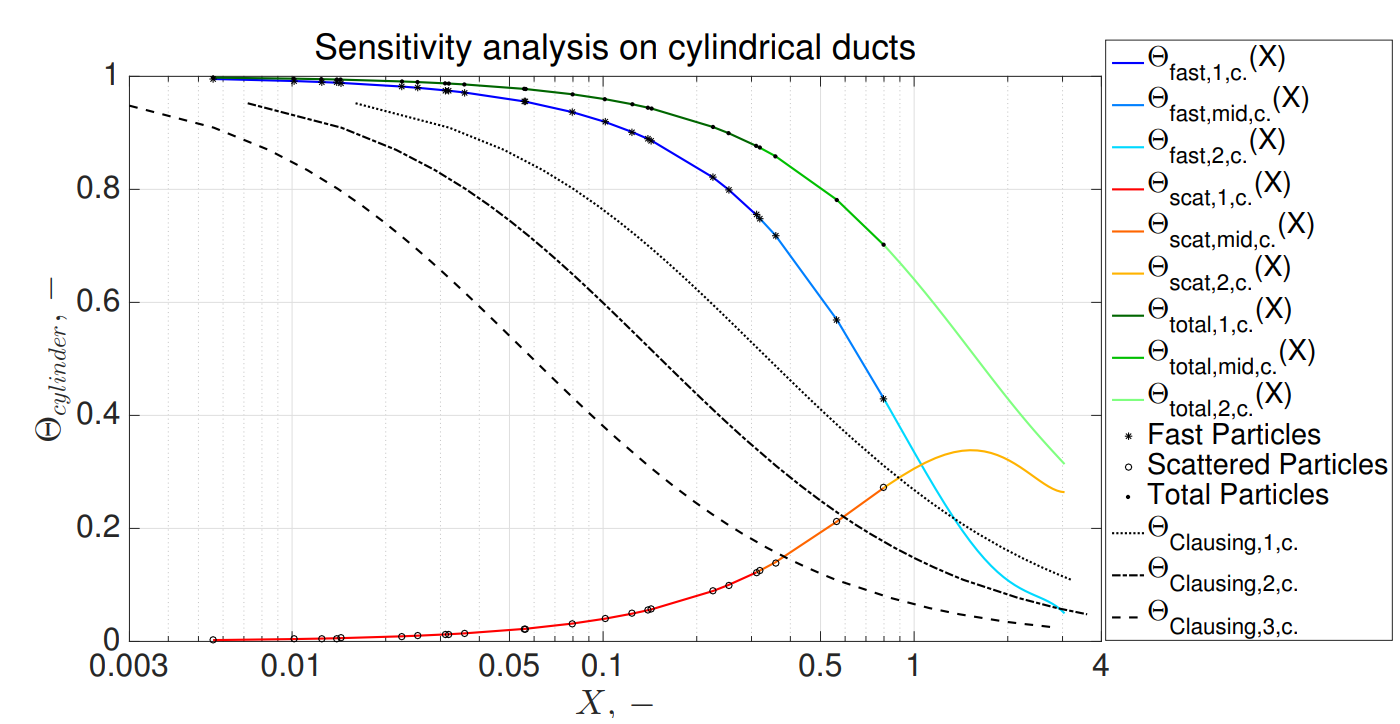}
	\caption{Transmission Probability over $X$ for Cylindrical Ducts~\cite{tilman,boldini}.}
		\label{fig:transcyl}
\end{figure}
\begin{figure}[h]
	\centering
	\includegraphics[width=0.9\linewidth]{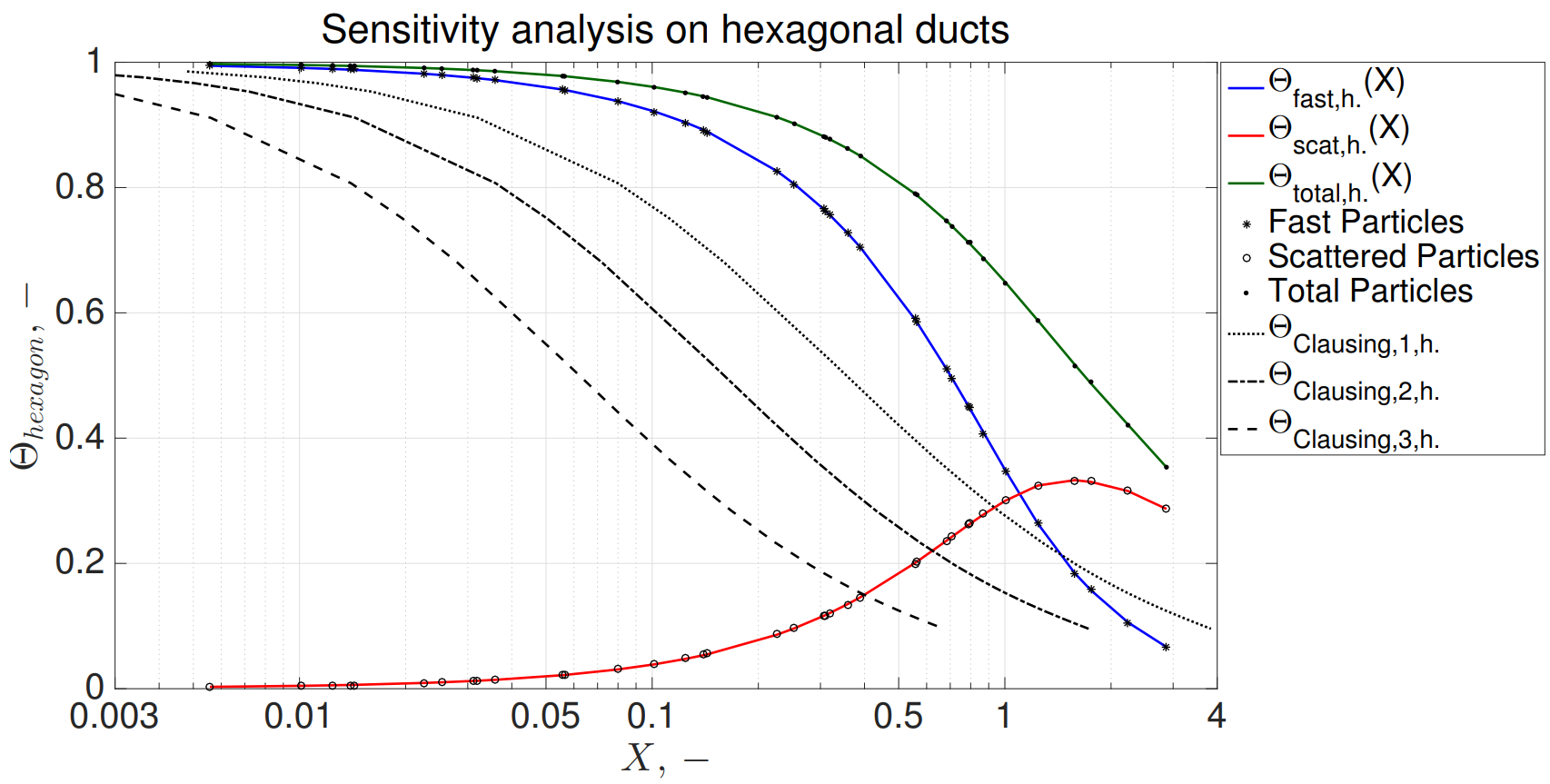}
	\caption{Transmission Probability over $X$ for Hexagonal Ducts~\cite{tilman,boldini}.}
		\label{fig:transhex}
\end{figure}

\subsection{PICLas Numerical Tool and Inflow Conditions}
PICLas is a numerical tool for simulating non-equilibrium gas and plasma flows developed by the Institute of Space Systems (IRS) and the Institute of Aerodynamics and Gasdynamics (IAG) at the University of Stuttgart~\cite{PICLAS}. The code is a three-dimensional, parallelized simulation framework for the coupling of, among others, DSMC and Particle-in-Cell~\cite{Munz2014662,PICLAS}. For the scope of the intake design, the DSMC module with the Maxwell model for surface interactions is used. The proportion of diffuse and specular reflections is regulated, in the code, by the momentum accommodation coefficient (MomentumACC). For the case MomentumACC~$>R_{DSMC}$ a diffuse reflection is performed, otherwise a specular one is performed during a particle-surface interaction ($R_{DSMC}$:~random number with $R_{DSMC}=[0,1)$). The inflow conditions are shown in Tab.~\ref{tab:DSCM__flow} and take into account all the components in VLEO according to the NRLMSISE-00 model. 

\begin{table}[hp]
\centering
\caption{PICLas Simulations Flow Parameters based on NRLMSISE-00 Atmospheric Model.}
~\\
\label{tab:DSCM__flow}
\begin{tabular}{ p{0.1\textwidth}<{\raggedright} p{0.12\textwidth}<{\centering}  p{0.12\textwidth}<{\centering}  p{0.12\textwidth}<{\centering} p{0.18\textwidth}<{\centering}  }
\toprule
\textbf{$h$} & \textbf{$T_{in}$} & \textbf{$T_{wall}$} &  \textbf{$v_{SC}$} & \textbf{$n_{in,total}$}   \\ 
\SI{}{\kilo\meter} & \SI{}{\kelvin} & \SI{}{\kelvin} & \SI{}{\meter\per{\second}} & $\SI{}{1/\meter^3}$ \\ 
\midrule
150 & 582.0 & 300 & 7818.2 & 4.131E+16 \\
180 & 666.1 & $"$ & 7800.3 & 1.042E+16 \\
200 & 690.7 & $"$ & 7788.5 & 4.967E+15\\
220 & 703.9 & $"$ & 7776.6 & 2.560E+15\\
250 & 713.4 & $"$ & 7759.0 & 1.045E+15\\
\bottomrule
\end{tabular}
\end{table}

\section{Enhanced Funnel Design (EFD) Intake}
The Enhanced Funnel Design (EFD) intake is developed as a passive device for an ABEP system in VLEO, for the altitude range $h=120-\SI{250}{\kilo\meter}$, and optimized for the target thruster discharge channel diameter of $d_{out}=\SI{37}{\milli\meter}$ corresponding to $A_{out}=\SI{1.075E-3}{\meter^2}$. 
The cylindrical EFD with hexagonal ducts, see Fig.~\ref{fig:efdiso}, is based on BM calculations and simulated $\Theta$ for Earth application.
Instead, the squared EFD with squared ducts, see Fig.~\ref{fig:efdisosq}, is based on both BM calculations and simulated $\Theta$, as well as a full PICLas simulation, but optimised for Mars application and served to verify the BM against PICLas. Finally, the results yielded a final relative error on $\eta_c$ between BM and full PICLas simulation of $\abs{4.5}-\abs{6.5}\%$~\cite{boldini}, see details in Appendix~A,Tab.~\ref{tab:EFDVLMO}. The EFD intake designs are gold coated for diffuse reflection properties~\cite{KOVALEV2011744,RAMESH1974291}. 
\begin{table}[h]
 \caption{Cylindrical EFD with Hexagonal Ducts Geometrical Design.}
 \label{tab:EFD}
 \centering
 \begin{tabular}{cccccccc}
 \toprule
$A_{in}$ & $F_f$ & $N_{ducts}$ & $L/R_{ducts}$ & $R_{ducts}$ & $L/R_{tube}$ & $R_{tube}=R_{in}$ & $\eta_c$\\
$\SI{}{\meter^2}$ & - & - & - & $\SI{}{\centi\meter}$ & - & $\SI{}{\centi\meter}$ & -\\
\midrule
0.00538 & 0.918 & 228 & 5 & 0.25 & 10.667 & 4.14 & 0.43\\
 \bottomrule
 \end{tabular}
 \end{table} 
 \begin{figure}
 	\centering
 	\begin{subfigure}[b]{0.47\linewidth}
 		\includegraphics[width=\linewidth]{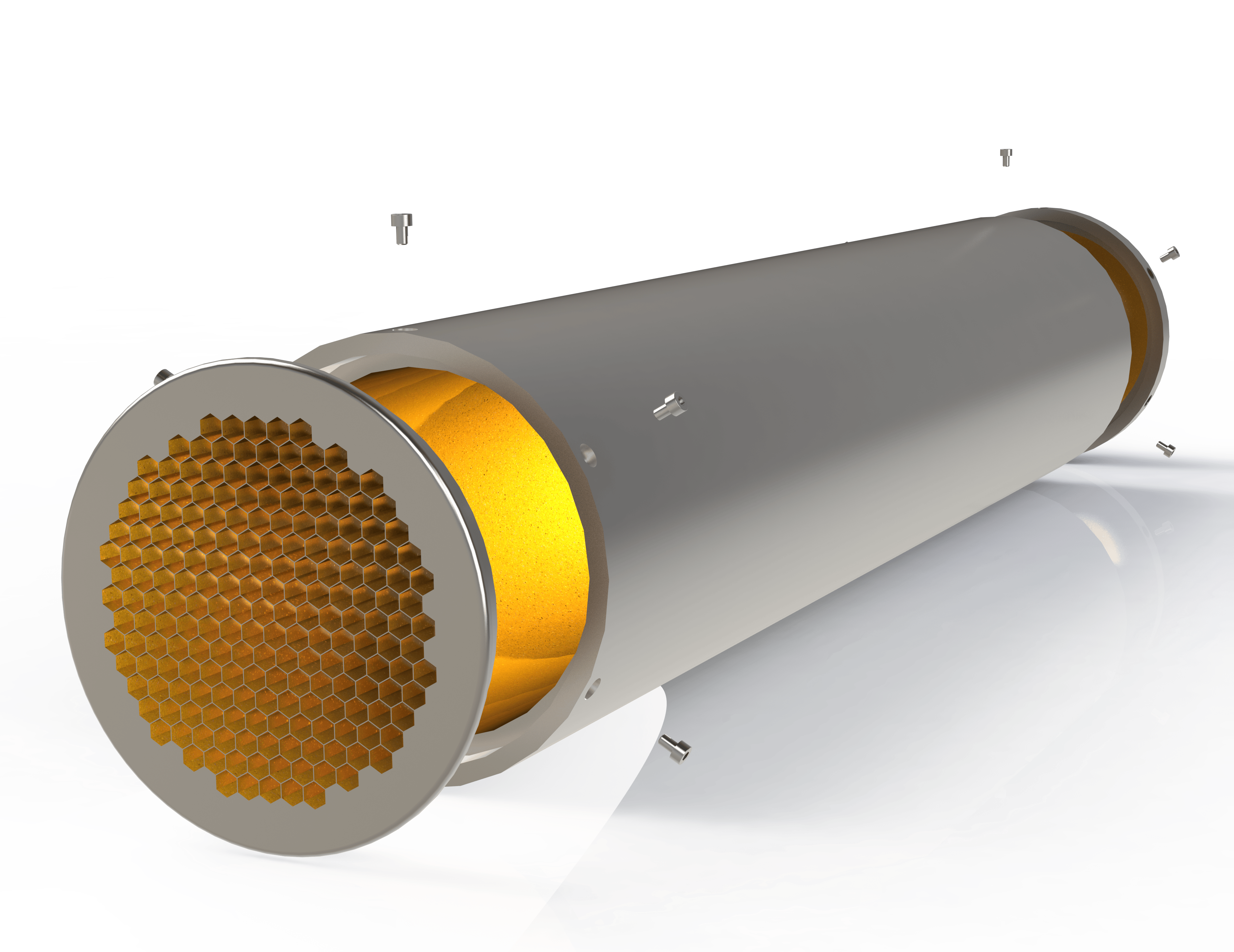}
 	\caption{Cylindrical EFD with Hexagonal Ducts.}
 		\label{fig:efdiso}
 	\end{subfigure}
 	\begin{subfigure}[b]{0.47\linewidth}
 		\includegraphics[width=\linewidth]{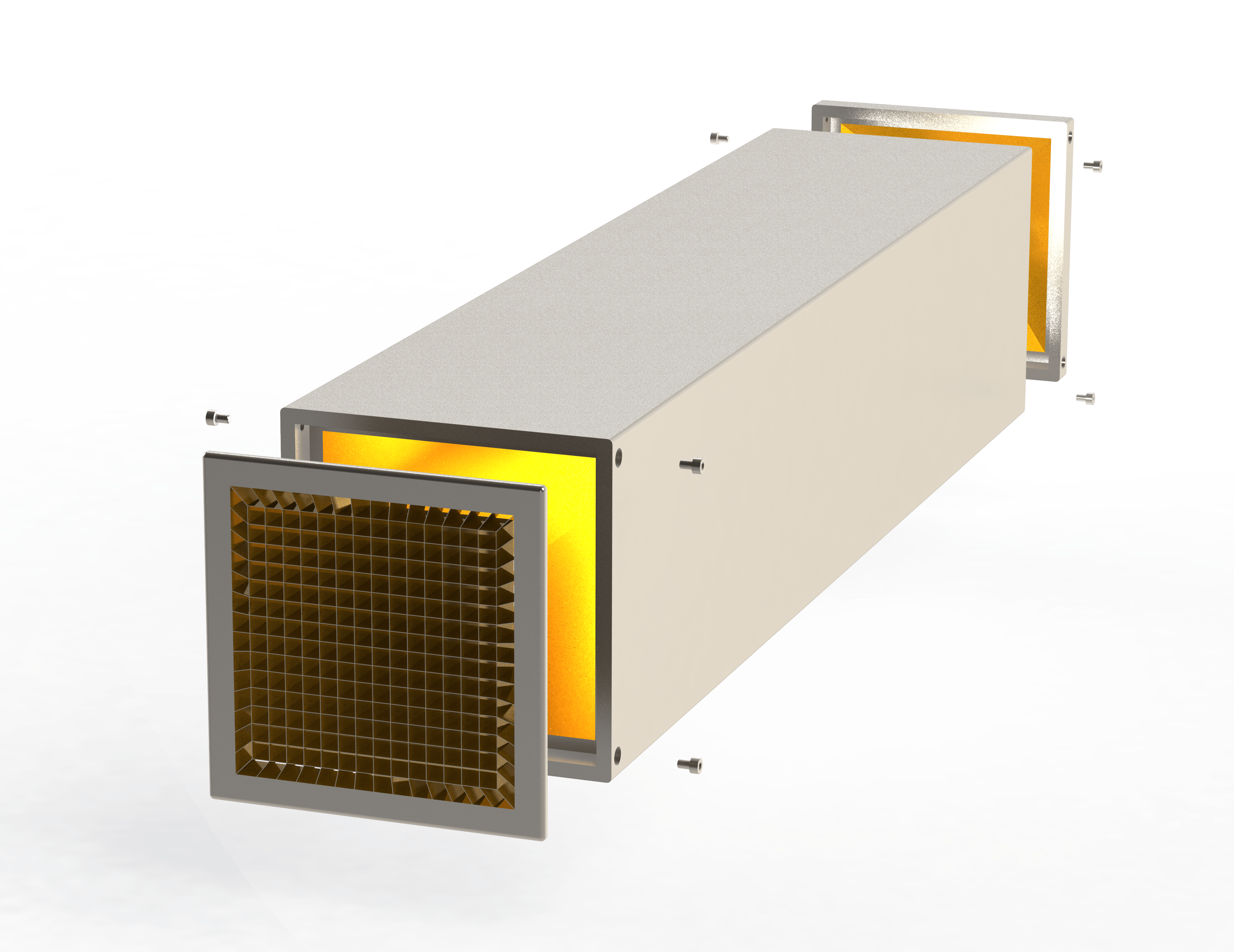}
 	\caption{Squared EFD with Squared Ducts.}
 		\label{fig:efdisosq}
 	\end{subfigure}
 	\caption{EFD Intake Designs (Exploded Views).}
 	\label{fig:EFD}
 	\end{figure}

The cylindrical EFD with hexagonal ducts achieves an intake efficiency $\eta_c=0.43$ based on the geometrical details shown in Tab.~\ref{tab:EFD}. The small ducts have hexagonal shape and a defined $L/R=5$, corresponding to $X=0.18-0.51$ for $S=7-20$, the outer section is matched to the main duct by a $\SI{45}{\degree}$ conical structure. The assembly is composed by three parts: the honeycomb structure of hexagonal ducts which is inserted into the second part, the main cylindrical duct, and the end part,  the conical section connected to the back of the main cylindrical duct, see Fig.~\ref{fig:EFD}. By changing the geometry, higher $\eta_c$ can be achieved. An important parameter is $A_{in}/A_{out}$. For example by changing to $A_{out}=\SI{6.082E-3}{\meter^2}$ (that of IPG3 at IRS~\cite{herdrich2002operational}), the same intake design can reach $\eta_c=0.623$. However, this also decreases density and pressure in the thruster’s discharge channel due to the larger cross section $A_{out}$, see Eq.~\ref{eq:effn} and Eq.~\ref{eq:pch} in Appendix~A, and, hence, volume, with the respective consequences of increased $P$, ignition condition, as well as overall performance. Finally, there is a trade-off between $\eta_c$ and $\dot{m}_{thr}$~\cite{romanosp2016}. Under the aforementioned hypothesis it is not possible to maximize both at the same time. Therefore, the design and selection must refer to the mission and thruster requirements and iterated on basis of them.
 
\section{Diffuse Intake}
The "Diffuse Intake" is presented in the following. It's design is based on the EFD. It is named "Diffuse Intake" and the outcome of PICLas simulations result in a $\eta_c<0.456$. It has $A_{in}=\SI{4E-3}{\meter^2}$ for a compact intake, and a total length of $L=\SI{0.085}{\meter}$, resulting in a $82\%$ length reduction compared to the EFD. 
\begin{figure}[H]
    	\centering
    	\includegraphics[width=1\textwidth, trim={4cm 1cm 3cm 1cm},clip]{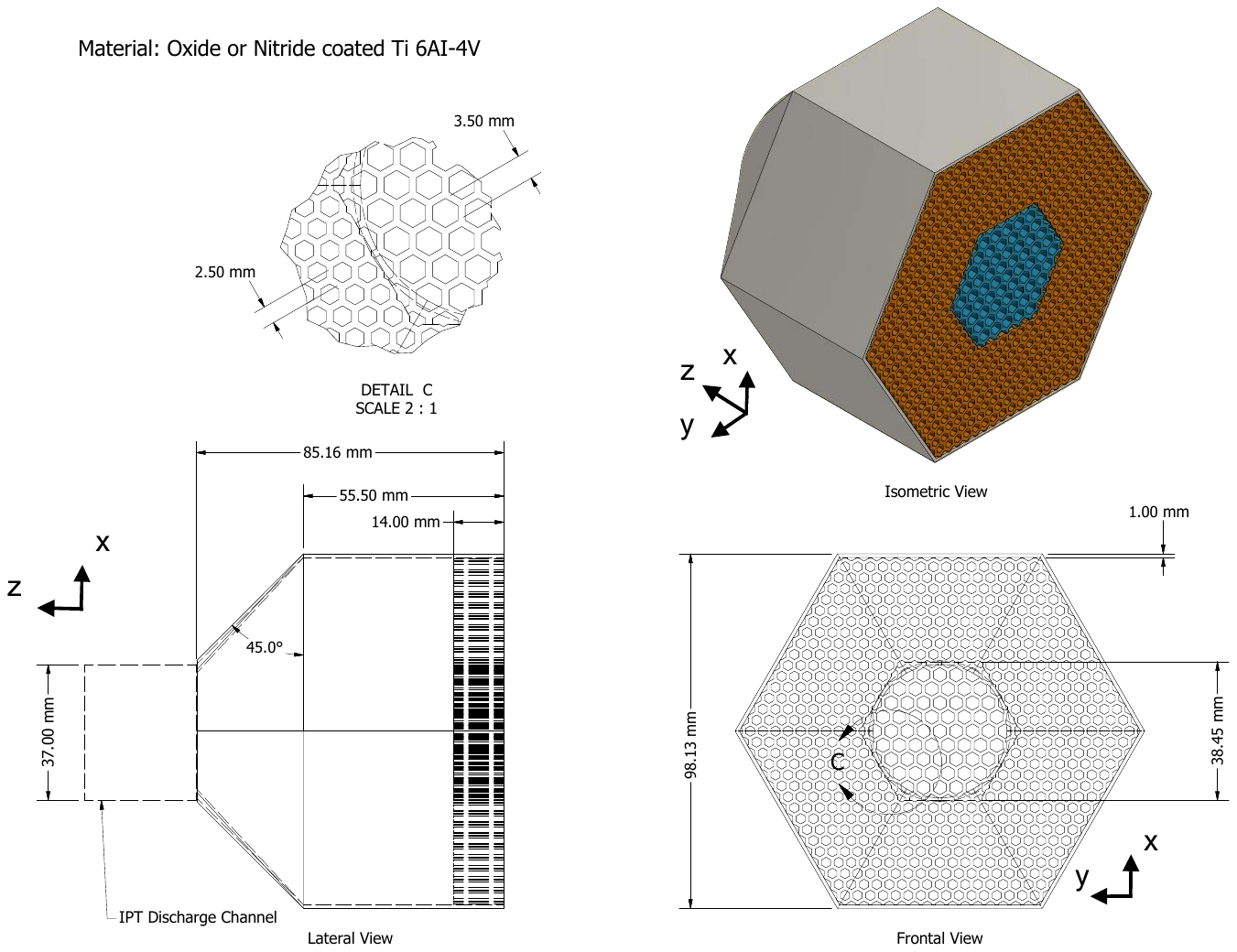}
    	\caption{Diffuse Intake Technical Schematics~\cite{espinosa}.}
    		\label{fig:intakedSchem}
\end{figure}
It has a circular cross section filled with a honeycomb structure of hexagonal ducts with multiple aspect ratios (AR), the outlet section $A_{out}$ is the thruster's discharge channel diameter $d_{out}=\SI{37}{\milli\meter}$. The technical schematics of the "Diffuse Intake" is shown in Fig.~\ref{fig:intakedSchem}, while the detailed performance is presented in Tab.~\ref{tab:diffuseintake}. In particular, $\eta_c$ slightly decreases with $h$ by $4\%$ between $h=150-\SI{250}{\kilo\meter}$ and consequently so does $\dot{m}_{thr}$ in relation to the inflow condition. Furthermore, the pressure distribution at $h=\SI{150}{\kilo\meter}$ simulated with PIClas is shown in Fig.~\ref{fig:intakedpress}. It highlights a region of homogeneous pressure $p_{ch}=0.02-\SI{0.3}{\pascal}$ within the intake, favourable for propellant injection into the discharge channel which is at a lower pressure $p$.

\begin{table}[H]
\centering
\caption{Diffuse Intake Performance vs Altitude.}
\label{tab:diffuseintake}
\begin{tabular}{cccccc}
\toprule
$h$ & $A_{in}$ & $\dot{N}_{in}$ & $\dot{N}_{out}$ & $\eta_{c}$ & $\dot{m}_{thr}$\\ 
$\SI{}{\kilo\meter}$ & $\SI{}{\meter^2}$ & $\SI{}{1/\second}$ & $\SI{}{1/\second}$ & - & $\SI{}{\milli\gram\per{\second}}$\\ 
\midrule
150 & 0.008 & 1.34E+18 & 6.11E+17 & 0.456 & 0.0240\\ 
180 & `` & 3.37E+17 & 1.49E+17 & 0.443 & 0.0053\\ 
200 & `` & 1.60E+17 & 7.14E+16 & 0.445 & 0.0023\\ 
220 & `` & 8.26E+16 & 3.51E+16 & 0.424 & 0.0011\\ 
250 & `` & 3.36E+16 & 1.40E+16 & 0.416 & 0.0004\\ 
\bottomrule
\end{tabular}
\end{table}

\begin{figure}[H]
    \centering
    \includegraphics[width=0.7\textwidth, trim={5cm 0cm 5cm 10.5cm},clip]{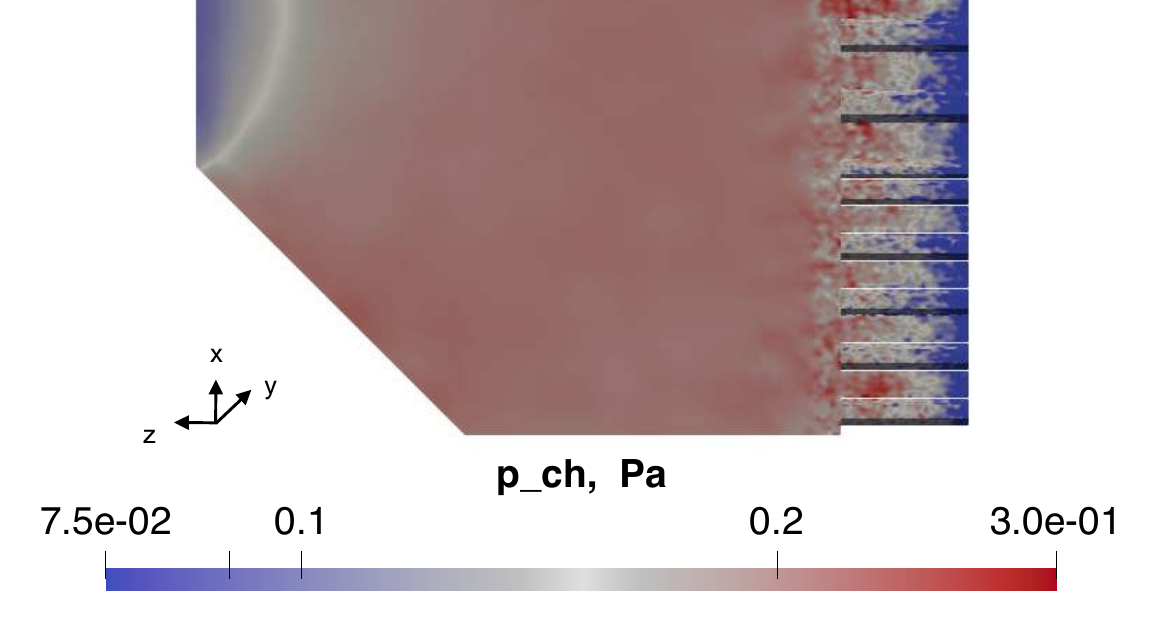}
    \caption{Diffuse Intake (Half) $p_{ch}$ Distribution, $h=\SI{150}{\kilo\meter}$~\cite{romanoacta3}.}
    \label{fig:intakedpress}
\end{figure}

The selected material, to provide diffuse reflection and resist to erosion in VLEO, is the titanium alloy \ce{Ti-6Al-4V} oxide or nitride coated~\cite{titaniumnitrideoxide,espinosa}.
Further investigation is performed to evaluate the effect of the misalignment with the incoming flow, given by the angle $\beta$ to the intake performance, see Fig.~\ref{fig:beta}.

 \begin{figure}[H]
     \centering
     \includegraphics[width=0.7\textwidth]{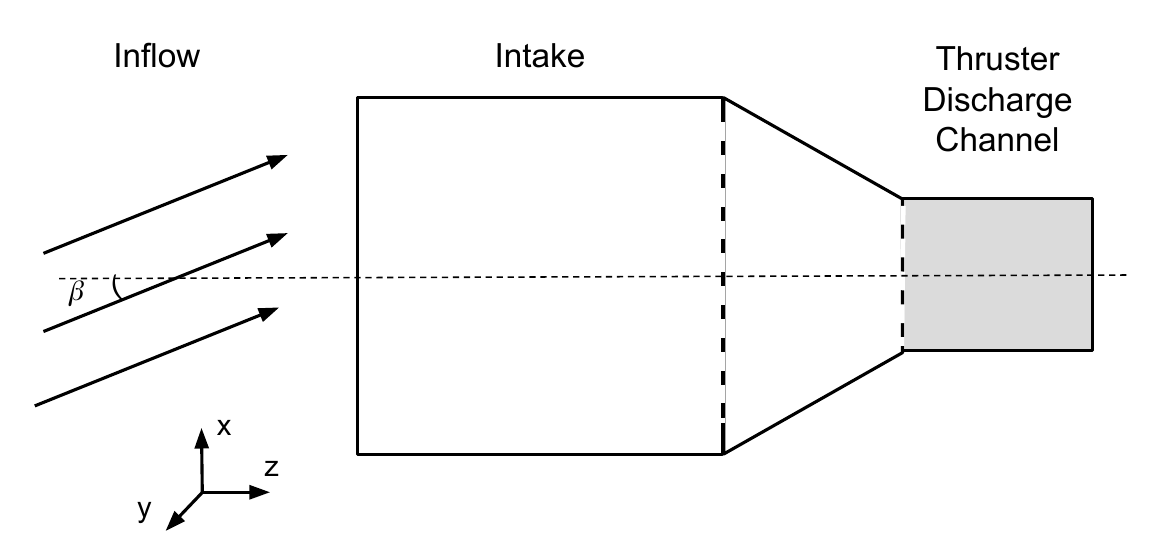}
     \caption{Misalignment Angle $\beta$ to the Flow Definition.}
     \label{fig:beta}
 \end{figure}
 
 The results from PICLas simulation is shown in Tab.~\ref{tab:diffuseintakebeta} which highlights that $\eta_c$ relative reduction $\Delta\eta_c<20\%$ for $\beta<\SI{5}{\degree}$, while for $\beta=\SI{10}{\degree}$ this is of $\Delta\eta_{c}<50\%$, where $\beta$ is the angle between the flow and the axis of symmetry of the intake, see Tab~\ref{tab:diffuseintakebeta}. This highlights a strong dependency on $\beta$ and a strict requirement for the spacecraft attitude. This also means that an assessment in terms of atmospheric winds must be included in the mission design.
 
\begin{table}[H]
 \centering
 \caption{Diffuse Intake $\beta$ Analysis,~$h=\SI{150}{\kilo\meter}$.}
 \label{tab:diffuseintakebeta}
 \begin{tabular}{ccc}
 \toprule
$\beta$ & $\eta_{c}$ & $\Delta \eta_c$\\
$\SI{}{\degree}$ & - & $\%$\\
  \toprule
0 & 0.456  & - \\ 
5 &  0.378  & $-17$\\ 
10 & 0.270 & $-41$\\ 
15 & 0.194 & $-58$\\ 
20 & 0.150 & $-67$\\ 
  \bottomrule
 \end{tabular}
 \end{table}
 
 \section{Specular Intake}
 The intake design based on specular reflecting materials is hereby presented. Fully specular reflections are represented by the energy accommodation coefficient $\alpha=0$. When a particle hits the surface, no energy is lost or transferred, and it is reflected at an angle opposite to that of incidence. An analogy to free molecular flow and $\alpha=0$, is that of optics and light rays, e.g.~telescopes. Based on this, an intake based on a parabolic design is designed and simulated, named "Specular Intake". The outlet section $A_{out}$ is the thruster's discharge channel diameter $d_{out}=\SI{37}{\milli\meter}$, while the main design drivers are $A_{in}$, and the location of the parabola's focus, used to direct the collected particles into a specific region. PICLas simulations revealed that the focus positioned inside the discharge channel volume yields the highest $\eta_c$. The technical schematics of the designed intake are shown in Fig.~\ref{fig:intakesSchem}. The distribution of $p_{ch}$ resulting from PICLas simulations is shown for $h=\SI{150}{\kilo\meter}$ in Fig.~\ref{fig:intakespress}, highlighting the area of higher pressure concentrated at the back of the intake $p_{ch}=0.02-\SI{0.34}{\pascal}$, toward the location of the focus.
 
 \begin{figure}[H]
     	\centering
     	\includegraphics[width=1\textwidth, trim={2cm 1cm 2cm 1cm},clip]{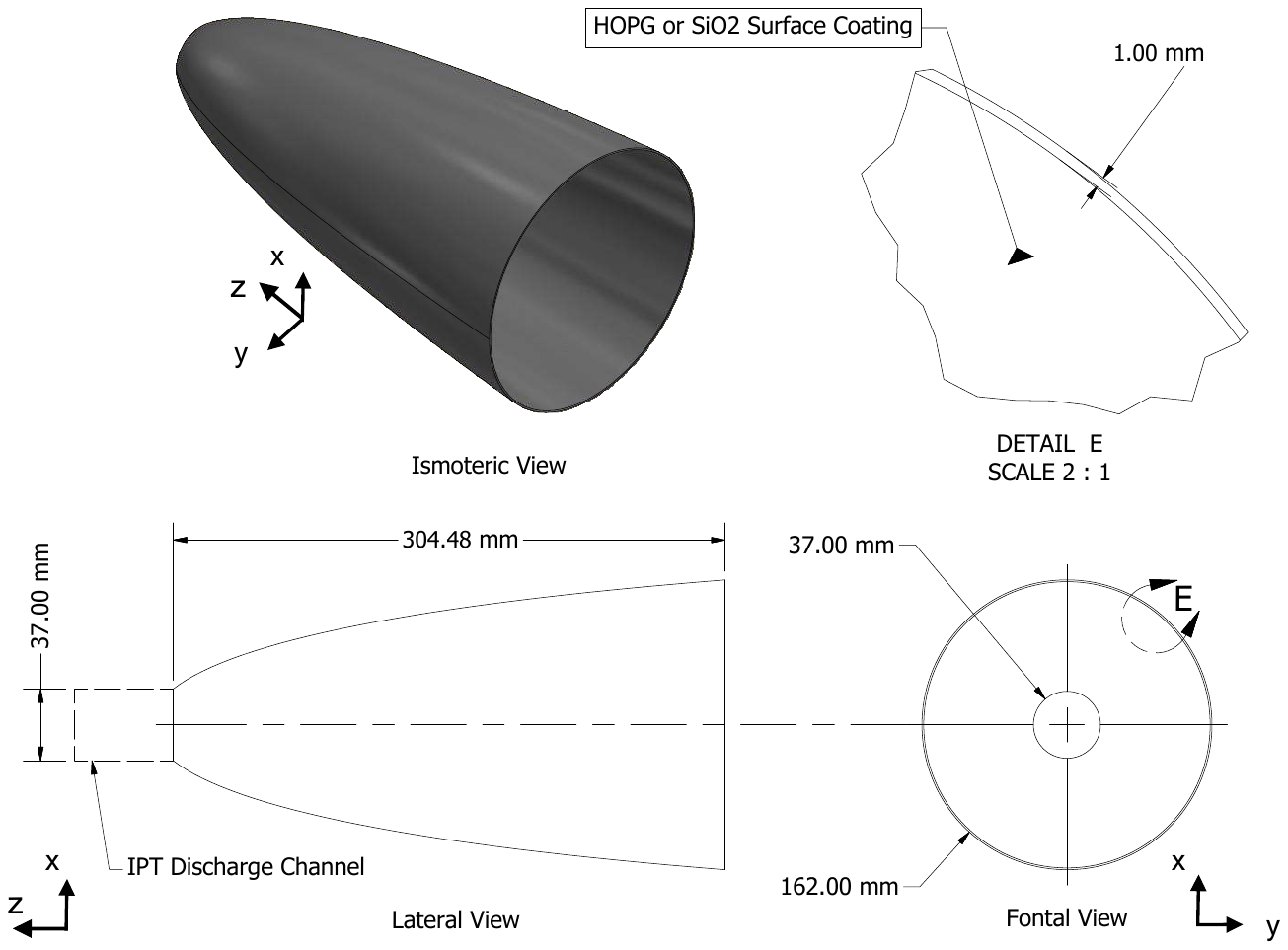}
     	\caption{Specular Intake Technical Schematics~\cite{espinosa}.}
     		\label{fig:intakesSchem}
 \end{figure}

 \begin{figure}[H]
  \centering
  \includegraphics[width=.85\textwidth, trim={0cm 1cm 0cm 3cm},clip]{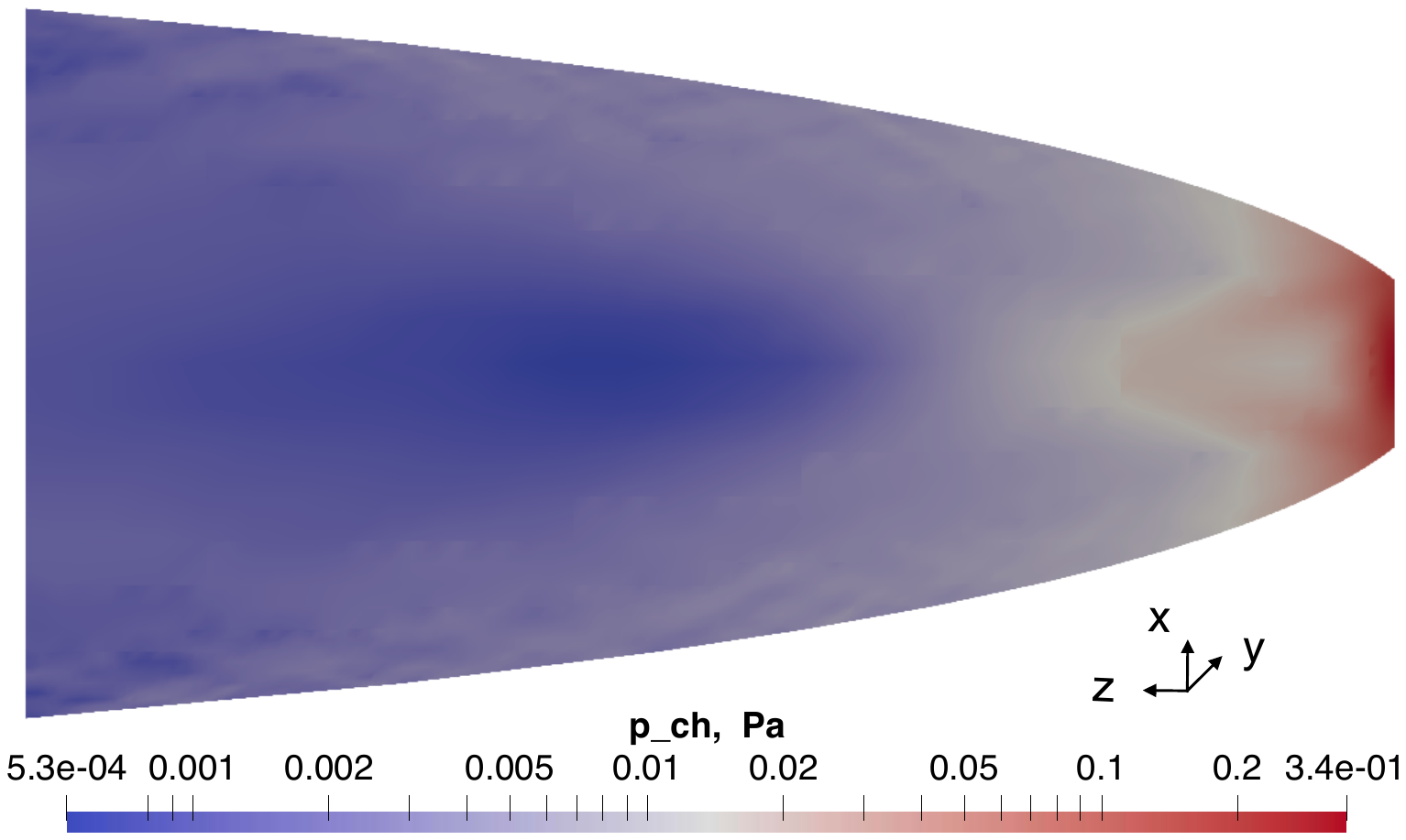}
  \caption{Specular Intake Pressure Distribution, $h=\SI{150}{\kilo\meter}$~\cite{romanoacta3}.}
  \label{fig:intakespress}
 \end{figure}

The intake performance, in particular $\eta_c$ and $\dot{m}_{thr}$, dependency on the misalignment with the flow (given by $\beta$) is shown in Fig.~\ref{fig:parabolabeta}. Hereby, the dependence is less strong than in the diffuse case. Indeed, only for $\beta>\SI{10}{\degree}$ the impact on $\eta_c$ becomes notable. In particular, by increasing $\beta=\SI{10}{\degree} \rightarrow \SI{11}{\degree}$ there is a relative reduction of $\eta_c$ by $\Delta\eta_c \sim 10\%$. For $\beta=\SI{20}{\degree}$, $\eta_c \sim 0.13$. This makes the "Specular Intake" a better candidate compared to the "Diffuse Intake" not only due a significantly higher $\eta_c$, but also due to its less sensitiveness to the flow misalignment. The proposed materials are highly oriented pyrolytic graphite (HOPG) coating or \ce{SiO2} due to their $\alpha\rightarrow0$ properties~\cite{KOVALEV2011744,HOPG}.
\begin{figure}[h]
     \centering
     \includegraphics[width=.75\textwidth]{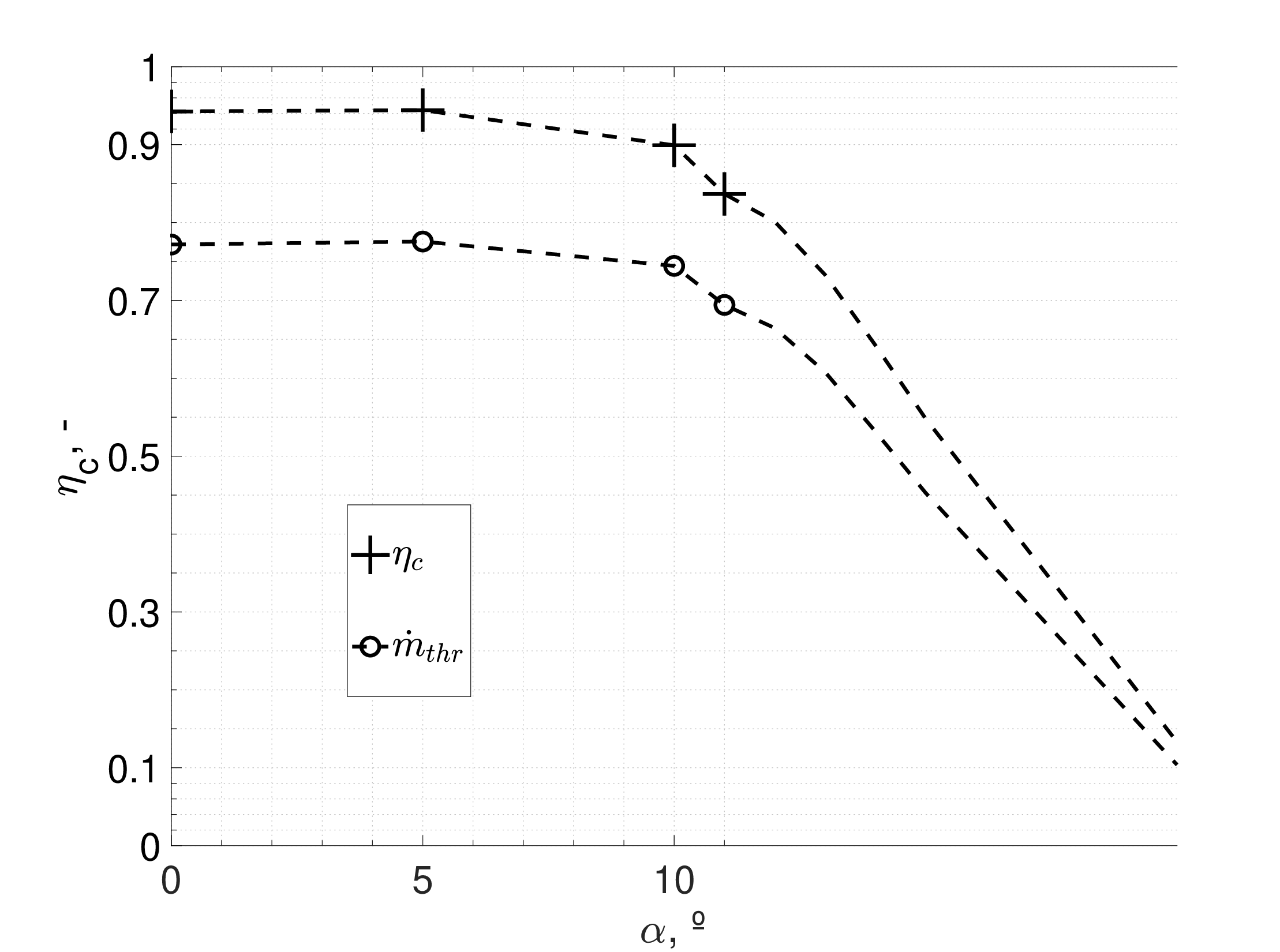}
     \caption{Specular Intake Flow Misalignment $\beta$ Analysis~\cite{romanoacta3}.}
     \label{fig:parabolabeta}
 \end{figure}
 
\section{Hybrid Intake Design}
Finally, the concept of a hybrid intake based on the synergetic use of both specular and diffuse reflecting surfaces is hereby presented. The intake is composed of a first stage based on specular reflections to effectively collect the particles, followed by a second stage, which is based on fully diffuse reflection to absorb the particle kinetic energies and have them proceeding to the thruster inlet by thermal diffusion. Given the high $\eta_c$ of the "Specular Intake", it is expected that the total $\eta_c$ of the hybrid intake is lower. Such design can be implemented once the requirements on flow conditions reaching the thruster are defined, such as pressure, density, particle velocities, and so on. The hybrid intake concept is presented in Fig.~\ref{fig:hybrid}. 
\begin{figure}[h]
  	\centering
  	\includegraphics[width=1\textwidth, trim={8cm 6.5cm 7cm 6cm},clip]{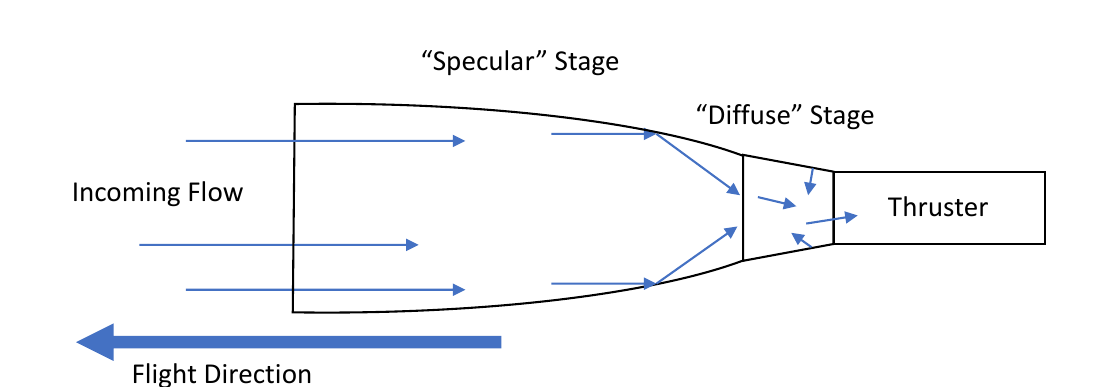}
  	\caption{Hybrid Intake Design.}
  		\label{fig:hybrid}
\end{figure}
It is expected that the hybrid intake can achieve efficiencies higher than the diffuse-based designs, but lower than a specular-based intake. Finally, a detailed analysis is required, especially Monte Carlo simulation of full designs to provide accurate intake performance estimations.

\section{Intake Designs Performance}
The intake performances are presented in Tab.~\ref{tab:intakes} in terms of $\eta_c$, and $\alpha$ with the geometrical values of intake area $A_{in}$, diameter $D_{in}$, and length $L$, as well the relative intake efficiency decrease $\Delta \eta_c$ for misalignment angles of $\beta=\SI{5}{\degree},~\SI{10}{\degree},~\SI{15}{\degree}$.
\begin{table}[h]
 \centering
 \caption{ABEP Intake Designs Comparison.}
 \label{tab:intakes}
 \begin{tabular}{ccccccccc}
 \toprule
Intake & $A_{in}$ & $D_{in}$ & $L$ & $\eta_{c}$ &  $\alpha$ & $\Delta \eta_{c, \beta=\SI{5}{\degree}}$ & $\Delta \eta_{c, \beta=\SI{10}{\degree}}$ &$\Delta \eta_{c, \beta=\SI{15}{\degree}}$\\
& $\SI{}{\meter^2}$ & $\SI{}{\meter}$ & $\SI{}{\meter}$ & - & - & $\%$ & $\%$ & $\%$\\
 \toprule
EFD & 0.005 & 0.080 & 0.480 & 0.430 & 1 & NA & NA & NA\\
Diffuse Intake & 0.008 & 0.100 & 0.085 & 0.458 & 1 & $-17$ & $-41$ & $-67$\\
Specular Intake & 0.019 & 0.162 & 0.305 & 0.943 & 0 & $\sim0$ & $-5$ & $-8$\\
\bottomrule
\end{tabular}
\end{table}

The "Specular Intake" outperforms the "EFD", and the "Diffuse Intake", independently of $A_{in}$ due to its $\eta_c$ being twice as the diffuse-based designs. The EFD and the "Diffuse Intake" are based on fully diffuse reflection. By comparing the two, the EFD intake is much longer than the "Diffuse Intake", therefore using unnecessary volume that can be saved. Finally, based on the results presented above, the chosen design is the "Specular Intake" due to its high $\eta_c$ and relative minimum sensitiveness to flow misalignment $\beta$.

As a final remark, the thruster minimum mass flow $\dot{m}_{thr,min.}$ and pressure $p_{ch,min}$ for ignition and operation are points of verification to be performed during the thruster test characterization campaign. Finally, further optimization of both intake and thruster are possible, in particular by changing the respective areas to adapt for the required values of $\dot{m}_{thr}$, $p_{ch}$, or $\eta_{c}$.

\chapter{Thruster}
This chapters deals with the thruster design, in particular for an ABEP system, and it is structured as following.
First, an introduction to plasma physics is provided that is narrowed on electromagnetic waves in the plasma and RF plasma discharges. The physics of an Helicon plasma thruster are then presented followed by an analysis on the thruster requirements and its concept. Estimation of the thruster's design requirements is then performed with the numerical tool HELIC. Furthermore, RF fundamentals are presented followed by the birdcage antenna theory. Finally, the thruster's electrical circuit and the birdcage antenna design are presented followed by experimental verification.

	\section{Plasma Fundamentals Definitions}
	\label{ch:plasma}
	Plasma is the fourth state of matter: by heating a solid, its state changes first to liquid, then to gas and, finally, to plasma. Its definition is: \textit{A plasma is a quasi-neutral gas of charged and neutrals particles which exhibits collective behaviour} \cite{thebible}.

Plasma behaves similarly to an electrical conductor. As charged particles move within the plasma, $E$- and $B$-fields are produced locally that are experienced also by charged particles in the plasma that are far away, and, therefore influence them in a collision-less manner. Plasma also reacts, locally and globally, in the presence of externally applied $E$ and $B$-fields.

The collective behaviour of the plasma means that particle motions do not depend only on the conditions of the local plasma region, but also on its state in the remote regions. 

Quasi-neutrality means that a plasma maintains a globally neutral charge: if the whole plasma is taken into account, its amount of positive and negative charges will be almost the same, so that the total charge remains neutral. If any external action is performed on the plasma to change its quasi-neutrality condition, the plasma reacts, trying to restore the quasi-neutrality condition. By inserting an electrode with positive potential into the plasma, for example, negatively charged particles immediately form a cloud around the electrode that compensates, locally, for the difference in charge. The thickness of the cloud, called sheath, is defined by the Debye length $\lambda_D$ as in Eq.~\ref{eq:debye}.
\begin{equation}
\lambda_D = \sqrt{\frac{\epsilon_0 k_B T_e}{n_e e^2}}
\label{eq:debye}
\end{equation}
Here, $T_e$ is the electron temperature, $\epsilon_0$ is the permittivity of free space, $n_e$ the electron plasma density, $k_B=\SI{1.3806503E-23}{\meter^2\kilo\gram\per{\second^2\kelvin}}$ the Boltzmann constant, and $e=\SI{1.60217646E-19}{\coulomb}$ the electron charge.
According to~\cite{thebible}, the conditions required for a gas to be defined a plasma are the following:
\begin{enumerate}
\item $\lambda_D \ll L$
\item $N_D \ggg 1$
\item $\omega\tau > 1$
\end{enumerate}
Condition~$1.$ means that an ionized  gas has to be dense enough such that this condition is realized, where $L$ is the length of the plasma boundary taken into consideration. The condition~$2.$ is for the Debye shielding to be valid, as there have to be enough particles in the cloud to realize the shielding, where $N_D$ is the number of charged particles in the Debye sphere. Condition~$3.$, finally states that if $\omega$ is the frequency of the typical plasma oscillation, and $\tau$ the mean time between collisions with neutral atoms, the condition~$3.$ must be satisfied such that the gas behaves like a plasma, rather than a neutral gas.

Plasma is defined by the \textbf{electron plasma density}, the number of electrons $n_e$ or (ionized atoms $n_i$) present in a unit of volume $\SI{}{\meter^{-3}}$. 
In the case of a singly ionized plasma, the quasi-neutrality condition can be written as in Eq.~\ref{eq:plasman}.
\begin{equation}
n_e\simeq n_i\simeq n_p
\label{eq:plasman}
\end{equation}

An important parameter is the \textbf{plasma temperature}, defined in terms of energy. A plasma in thermal equilibrium has particles with different velocities, and their most probable distribution is known as Maxwellian distribution, see Eq.~\ref{eq:max1} for the one-dimensional case.
\begin{equation}
f(u)=Y\exp\biggl(-\frac{1}{2}mu^2 / k_BT_{e,i}\biggr)
\label{eq:max1}
\end{equation}
The number of particles per $\SI{}{\meter^{-3}}$ that has a velocity between $u$ and $u+du$ is given by $f(u)$, whereas $Y$ is a constant related to the plasma density $n$, $k_B$, and the half product of $m$ and $u^2$ is the kinetic energy. 
The plasma density $n_p$ is shown in Eq.~\ref{eq:max2}, and the constant $Y$x in Eq.~\ref{eq:max3}.
\begin{equation}
n_p=\int_{-\infty}^{+\infty} f(u)du
\label{eq:max2}
\end{equation}
\begin{equation}
Y=n_p\biggl(\frac{m}{2\pi k_BT_{e,i}}\biggr)^{1/2}
\label{eq:max3}
\end{equation}
The width of the Maxwellian velocity distribution is given by the temperature $T_{e,i}$. This is defined by the average kinetic energy of the particles of the distribution $E_{av}$. By performing algebraic substitution, see~\cite{thebible}, it is possible to show that Eq.~\ref{eq:max3} can be rewritten as Eq.~\ref{eq:max4}.
\begin{equation}
E_{av}=\frac{3}{2}k_BT_{e,i}
\label{eq:max4}
\end{equation}
As $T_{e,i}$ and $E_{av}$ are closely related, plasma temperature is usually expressed by $k_BT_e,i$ with the unit of \SI{}{\electronvolt}. The conversion factor is $k_BT_{e,i}=\SI{1}{\electronvolt}=\SI{1.6e-19}{\joule}$, therefore $T_{e,i}=\SI{1}{\electronvolt}=\SI{11600}{\kelvin}$. The temperature $T_{e,i}$ can be different in the same plasma for ions, $T_i$, and for electrons $T_e$, but also for different directions.

Finally, a plasma has its own frequency due to the arising small charge separations that make the electron clouds oscillate around an equilibrium point made by the heavier steady ions, and it is called the plasma frequency $\omega_p$ defined as in Eq.~\ref{eq:plasmafreq}.
\begin{equation}
\omega_p = \sqrt{\frac{n_e e^2}{\epsilon_0 m}_e }
\label{eq:plasmafreq}
\end{equation} 

\subsection{Plasma under Uniform Electromagnetic Fields}
In this section, the behaviour of single particles in a plasma under the influence of uniform electromagnetic fields is described. This is needed to later discuss the influence of time-varying electromagnetic fields as the thruster is based on RF operation with an externally applied static magnetic field. 

\subsubsection{$\vec{E}=0,~\vec{B}=const.$}
By considering $\vec{E}=0$ and $\vec{B}=const.$, the equation of motion is given in Eq.~\ref{eq:e0}, where $\vec{v}$ is the single particle velocity vector, $m$ its mass, $t$ the time, and $q$ its charge.
\begin{equation}
m\frac{d\vec{v}}{dt}=q\vec{v} \times \vec{B}
\label{eq:e0}
\end{equation}
Solving for $\vec{B}=B\hat{z}$, where $\hat{z}$ is the unit vector in the direction of the $z$ axis in a standard $x,y,z$ reference system, and $B$ is the magnetic field intensity, the equation of a simple harmonic oscillator is obtained, with its natural frequency $\omega_c$ called the cyclotron frequency as in Eq.~\ref{eq:ecr}.
\begin{equation}
\omega_c = \frac{\abs{q}B}{m}
\label{eq:ecr}
\end{equation}
Therefore, electrons and ions rotate around the magnetic field lines with an angular frequency $\omega_c$ in an orbit of radius the Larmor radius, defined as in Eq.~\ref{eq:rl}, where the guiding centre is on the magnetic field lines.
\begin{equation}
r_L=\frac{v_{\bot}}{\omega_c}=\frac{mv_{\bot}}{\abs{q}B}
\label{eq:rl}
\end{equation}
Finally, the direction of rotation generates a $\vec{B}$-field that is always opposite to the imposed one $\vec{B}=B\hat{z}$. 

\subsubsection{$\vec{E}=const.,~\vec{B}=const.$}
If an electric field is added, the motion of the charged particles is the sum of the Larmor gyration of the particles, plus a drift of the guiding centre given by the $\vec{E}$-field in the $x-z$ plane as described in Eq.~\ref{eq:e1}.
\begin{equation}
m\frac{d\vec{v}}{dt}=q(\vec{E}+\vec{v}\times \vec{B})
\label{eq:e1}
\end{equation}
The motion looks like an inclined helix with increasing pitch (acceleration). The drift due to the $\vec{E}$-field is given in Eq.~\ref{eq:drift} for a given $\vec{E}$ and $\vec{B}$-field, and is independent of $m$, $q$, and $v_{\bot}$.
\begin{equation}
\vec{v}_E = \frac{\vec{E}\times\vec{B}}{\vec{B}^2}
\label{eq:drift}
\end{equation}

\subsection{Electromagnetic Waves in the Plasma}
Waves can develop into a plasma due to time-varying electromagnetic fields. Any periodic motion can be described using Fourier analysis theory and can be modelled as superposition of sinusoidal oscillations at given frequencies $\omega$ and wavelengths $\lambda$. 
A general oscillating quantity $A$ is expressed in Eq.~\ref{eq:oscillating}.
\begin{equation}
A = \overline{A} \exp{[i(\vec{k}\cdot\vec{r}-\omega t )]}
\label{eq:oscillating}
\end{equation}
The wave amplitude is $\overline{A}$, $\vec{r}$ is the coordinate vector, and $\vec{k}$ is the propagation constant vector.
The phase velocity $v_\phi$  is defined as $v_\phi=\frac{\omega}{k}$, and is the velocity at which a wave propagates through a medium, where $k=2\pi/\lambda$ is the spatial frequency of the wave. The $v_\phi$ can exceed the speed of light $c$ without going against the relativity theory as an infinitely long train of waves at constant amplitude does not carry any information. Instead, what cannot exceed $c$ is the group velocity $v_g$ defined in Eq.~\ref{eq:groupvel}, which is, instead, the velocity at which the envelop of the wave propagates through space.
\begin{equation}
v_g = \frac{d\omega}{dk}
\label{eq:groupvel}
\end{equation}

\subsubsection{Electron Plasma Waves}
The plasma oscillation propagated by the electron thermal motion can be called a plasma wave. The equation of motion is given by the natural plasma oscillation, defined by the plasma frequency $\omega_p$, and the thermal electron motion itself. The dispersion relation $\omega(k)$ is in Eq.~\ref{eq:elecwave}.
\begin{equation}
\omega^2=\omega^2_p + \frac{3}{2}k^2v^2_{th}
\label{eq:elecwave}
\end{equation}
The thermal velocity is $v_{th}=\sqrt{2k_BT_e/m}$, and the group velocity results in $v_g = \frac{3}{2}\frac{v^2_{th}}{v_\phi}$.

Applying a static $\vec{B}$-field to the electron waves is modelled as following. 
The $\vec{E}$-field is $\vec{E}_1=E_1\hat{x}$ along $x$ and parallel to $\vec{k}=k\hat{x}$. The static $\vec{B}$-field is $\vec{B}_0=B_0\hat{z}$ and is along $z$ and, therefore, perpendicular to $\vec{E}_1$. Finally, $k_x = k_y = E_y = E_z = 0$, leading to the dispersion relation of Eq.~\ref{eq:disp1}.
\begin{equation}
\omega^2 = \omega_p^2 + \omega_c^2 = \omega_h^2
\label{eq:disp1}
\end{equation}
The term $\omega_h$ is defined as the upper hybrid frequency, the frequency of the electrostatic electron waves across $\vec{B}$, while those along $\vec{B}$ have a frequency $\omega=\omega_p$. Therefore, the plasma oscillation is at a frequency $\omega_h>\omega_p$, and the electron gyration motion is stretched by the sum of the electrostatic force and the Lorentz force, so to form an ellipse between planes of different densities of the plasma due to the wave motion. If the waves are propagating at a given angle $\theta$ to the $\vec{B}$-field, two waves are possible: one similar to the plasma oscillation and the other like the upper hybrid one. 

\subsubsection{Sound Waves and Ion Waves}
Sound waves in plasmas are treated by applying the Navier-Stokes equations, that results in the respective dispersion relation in Eq.~\ref{eq:soundwave}, where $\gamma$ is the adiabatic index and $M$ the mass of ions.
\begin{equation}
\frac{\omega}{k}=\biggl( \frac{\gamma k_BT}{M}\biggr)^{\frac{1}{2}}=c_s
\label{eq:soundwave}
\end{equation}
The velocity of sound waves is $c_s$. This represents the sound wave propagating in a neutral gas, from one layer to another by means of collisions. If neutrals and collisions are removed, the analogue phenomena of ion waves arises, in which the vibrations are transmitted by the $\vec{E}$-field. The dispersion relation is in Eq.~\ref{eq:ionwave} where sound speed in plasma is $v_s$.
\begin{equation}
\frac{\omega}{k}=\biggl( \frac{k_BT_e + \gamma_i k_BT_i}{M} \biggr)^{\frac{1}{2}}=v_s
\label{eq:ionwave}
\end{equation}  
The ion dispersion relation is different from that of electrons. In particular, plasma oscillations due to electrons are constant-frequency waves plus a correction required for their thermal motion, while ions waves have a constant-velocity and exist only when thermal motions are present, moreover, group and phase velocity are always equal~$v_g=v_\phi$. For higher frequencies waves (shorter wavelength), ion waves become constant-frequency waves, and electron waves become constant-velocity waves. 

Ion acoustic waves develop when $\vec{k}$ is almost perpendicular to the applied $\vec{B}$-field. 
The dispersion relation for electrostatic ion cyclotron waves is given in Eq.~\ref{eq:disp2}, where $\Omega_c$ is the ion cyclotron frequency.
\begin{equation}
\omega^2 = \Omega_c^2 + k^2 v_s^2
\label{eq:disp2}
\end{equation}
Physically, ions undergo acoustic-type oscillations, but the presence of the $\vec{B}$-field, therefore the arise of the Lorentz force, makes this frequency of oscillation larger by the term $\Omega_c^2$.

If the $\vec{B}$-field is set exactly perpendicular to $\vec{k}$, the dispersion relation is given by Eq.~\ref{eq:disp3}.
\begin{equation}
\omega = (\Omega_c\omega_c)^\frac{1}{2}= \omega_l
\label{eq:disp3}
\end{equation}
The term $\omega_l$ is defined as the lower hybrid frequency, being a combination of  $\omega_c$ and $\Omega_c$, an oscillation of ions and electrons perpendicularly to the $B$-field, and only appears if $\vec{k}$ is perpendicular to $\vec{B}$.

\subsubsection{Electromagnetic Waves}
An $\vec{E}$-field parallel to an applied $\vec{B}$-field, so that $\vec{E}=E\hat{z}$, $\vec{B}=B\hat{z}$, and $\vec{k}=k\hat{x}$ is now considered. This condition can be resumed as a microwave source sending a wave through a waveguide to a magnetized plasma column. The trains of waves reach the plasma column perpendicularly, with $\vec{E}$ parallel to $\vec{B}$ and $\vec{k}$ perpendicular to both. This is the "ordinary" O-wave, described by the dispersion relation of Eq.~\ref{eq:disp4}.
\begin{equation}
\omega^2 = \omega^2_p + c^2k^2
\label{eq:disp4}
\end{equation}
For such dispersion relation, the cutoff phenomena arise. If a microwave beam is sent through the plasma at an angular frequency $\omega$, the wavelength in the plasma $\lambda = 2\pi/k$, will change according to Eq.~\ref{eq:disp4}. If $n_e$, therefore $\omega_p$, is raised within the plasma such that $\omega=\omega_p$, with $k\rightarrow0$, waves cannot propagate. Cutoff is defined as the point at which the refraction index goes to $0$, equivalent to an infinite wavelength. Therefore, there is a critical plasma density $n_c$ such that the condition $\omega=\omega_p$ is realized, defined as $n_c = m \epsilon_0 \omega^2 / e^2$. Useful methods are derived based on this property to estimate the $n_e$. 

If the $\vec{E}$-field is perpendicular to the $\vec{B}$-field, the "extraordinary" X-wave arises. Since the motion of the particles is influenced by the $\vec{B}$-field, $\vec{E}$-field is $\vec{E}=E_x\hat{x}+E_y\hat{y}$, and $\vec{k}=k\hat{x}$. The dispersion relation is given in Eq.~\ref{eq:disp5}.
\begin{equation}
\omega^2 \mp \omega\omega_c-\omega^2_p=0
\label{eq:disp5}
\end{equation}
The corresponding cutoff frequencies are $\omega_R$ and $\omega_L$, with $\omega_c$ and $\omega_p$ the electron cyclotron frequency and the plasma frequency respectively, see Eq.~\ref{eq:disp5bisR} and Eq.~\ref{eq:disp5bisL}.
\begin{equation}
\omega_R = \frac{1}{2}\biggl[ \omega_c + \sqrt{\omega^2_c+4\omega_p^2} \biggr]
\label{eq:disp5bisR}
\end{equation}
\begin{equation}
\omega_L = \frac{1}{2}\biggl[ - \omega_c + \sqrt{\omega^2_c+4\omega_p^2} \biggr]
\label{eq:disp5bisL}
\end{equation}
Due to the $\vec{E}$-field, this electromagnetic wave is longitudinal and transversal propagating to $\vec{B}$. This last dispersion relation highlights also the resonance phenomena, corresponding to an infinite index of refraction and $0$ wavelength and, precisely, at the condition where $\omega^2_h=\omega^2_p+\omega^2_c=\omega^2$. 

Therefore, a wave is fully absorbed at resonance, and fully reflected at cutoff. 

At resonance, the wave encountering the plasma converts its energy into upper hybrid oscillations $\omega^2=\omega^2_h$. The "extraordinary" X-wave is both electromagnetic and electrostatic, while at resonance it becomes electrostatic only. The cutoff frequencies are two, $\omega_R$ and $\omega_L$, right-hand and left-hand cutoff. 

Finally, electromagnetic waves that are parallel to the $\vec{B}$-field, propagating along $z$ so that $\vec{k}=k\hat{z}$ result in the dispersion relation of Eq.~\ref{eq:disp6}.
\begin{equation}
\frac{c^2k^2}{\omega^2}= \begin{cases} 1-\frac{\omega^2_p/\omega^2}{1+(\omega_c/\omega)} & \mbox{L-Wave} \\
& \\
1 - \frac{\omega^2_p/\omega^2}{1 - (\omega_c/\omega)} & \mbox{R-Wave}
\end{cases}
\label{eq:disp6}
\end{equation}
The solutions are two different waves, the R, right-hand circularly polarized waves, and L, the left-hand circularly polarized waves, meaning that the $\vec{E}$-field vector rotates in the R and L direction respectively. To better visualize movement, by looking toward the direction of $\vec{B}$ and $\vec{k}$, the R-wave rotates clockwise, and the L-wave counter-clockwise. The resonance case for $\vec{k}\rightarrow \infty$ and $\omega=\omega_c$ is the R-wave: its direction of rotation is the same as the electrons, therefore the wave loses energy by transferring it to the electrons and cannot propagate. The L-wave would have resonance with $\Omega_c$, but ion movements are neglected in the theoretical analysis due to their inertia. The cutoff frequencies are the same as for X- and O-waves. For the R-wave is $\omega_R$, while for the L-wave is $\omega_L$. For $\omega < \omega_c$ there is a band propagation in the $\omega-{v^2_\phi}/	{c^2}$ diagram called the whistler mode which will be later described in Sec.~\ref{sec:helicon}.

Finally, by analysing low frequency ion oscillations in $\vec{B}$-fields, there are two cases: the Alfv\'en and the magnetosonic waves. The Alfv\'en wave has $\vec{k}$ along $\vec{B}$ and the oscillating $\vec{B}$- and $\vec{E}$-fields $\vec{B}_1$ $\vec{E}_1$ perpendicular to each other. The main condition is that $\omega \ll \Omega_c$. These waves travel along the $\vec{B}$-field at a constant velocity, called the  Alfv\'en velocity $v_A$ as defined in Eq.~\ref{eq:alfven1}, with $\rho$ being the mass density and $\mu_0$ the permeability of vacuum.
\begin{equation}
v_A = \frac{B}{\sqrt{\mu_0 \rho}} 
\label{eq:alfven1}
\end{equation}
	In an Alfv\'en wave, the plasma is "frozen" to the line of force, due to the low frequency. This velocity can be very large, e.g. at $n=\SI{6e-21}{\meter^{-3}}$,  $v_A=\SI{\sim2.8e5}{\meter\per{\second}}$~\cite{thebible}. 

The magnetosonic wave, instead, is a low frequency wave that propagates across a steady $\vec{B}=B\hat{z}$, with $\vec{E}_1=E_1\hat{x}$, and $\vec{k}=k\hat{y}$. 
The resulting dispersion relation is given by Eq.~\ref{eq:magnetosonic}. 
\begin{equation}
\frac{\omega^2}{k^2}=c^2\frac{v^2_s+v^2_A}{c^2+v^2_A}
\label{eq:magnetosonic}
\end{equation}
In magnetosonic waves, compression and decompression are produced by $\vec{E}\times\vec{B}$ drifts across~$\vec{E}$.

\subsection{Helicon Waves}
\label{sec:helicon}
Within the family of low frequency waves belong the whistler waves: low frequency electromagnetic waves that were first discovered as result of lightning strikes on Earth. Such electromagnetic waves produce descending whistling tones that can be detected by radios, hereby the name as they were first discovered. 

Helicon waves are whistler waves (R~waves) bounded in a cylinder. For many years, they have been studied and their description is mainly understood~\cite{thebible,isayama2018review,takahashi2019helicon}, but their mechanism of power deposition is still to be fully unlocked~\cite{EPFL4}. Helicon waves are low frequency electromagnetic waves, with $\omega_l<\omega_{Helicon}<\omega_c$, that develop in plasmas confined within a physical cylinder when a static $B$-field is applied along its axis.

Helicon waves can develop in different azimuthal mode numbers $m$, mainly the modes $m=+1$, $m=0$, and $m=-1$ that describe the EM fields configuration. Within the field of electric propulsion, $m=+1$ and $m=0$ are mostly considered, and especially $m=+1$ as it delivers higher $n_e$ compared to the $m=-1$~\cite{chen2015}. Nevertheless, higher modes $m>+1$ are also achievable. 
The $m=+1$ helicon wave is a right-hand polarized EM wave that presents the $\vec{E}$-field as illustrated in Fig.~\ref{fig:heliconm01}~(a), while the $m=0$ is azimuthally symmetric, see Fig.~\ref{fig:heliconm01}~(b), and has the $\vec{E}$-field lines changing from totally solenoidal (electromagnetic) to totally radial (electrostatic) each half cycle, while a mixture of the two is present in between. \begin{figure}[h]
 	\centering
 	\includegraphics[width=14cm]{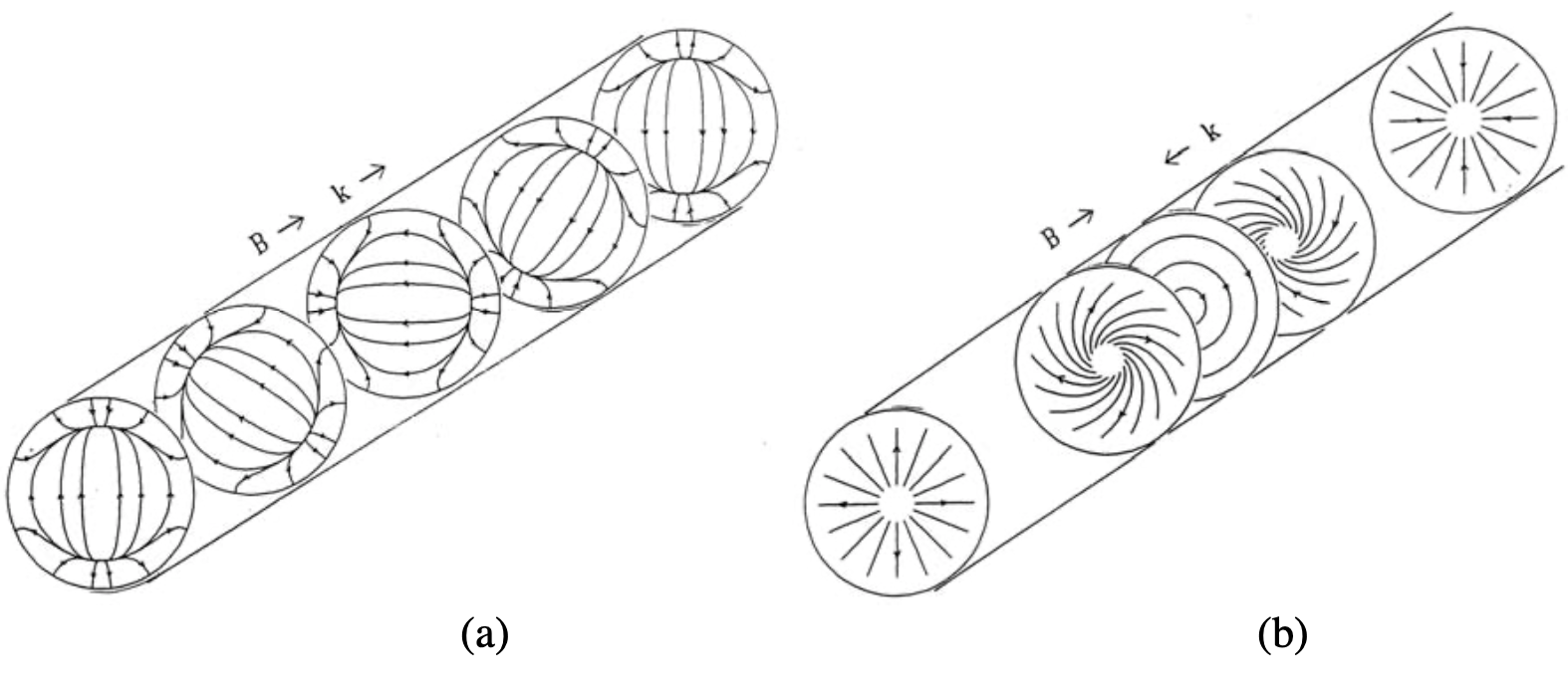}
 	\caption{Electric field patterns for the $m=+1$ mode $(a)$, and for the $m=0$ mode $(b)$ mode~\cite{chen235R}.}
 	\label{fig:heliconm01}
 \end{figure}

Helicon waves require a static $B$-field, $B_0$, along the cylinder axis to develop, and do not have a strict resonance condition which allows helicon plasma sources to be operated on a wide range of applied $B$-field strengths and $f$. 
 
When modelling helicon waves, by applying the boundary condition of the physical discharge channel to the plasma, and by taking into account a finite mass of the electrons, a second wave arises: the Trivelpiece-Gould (TG)~\cite{arnush2000role,isayama2018review}. Such a wave is an electron cyclotron wave that travels obliquely, at an angle $\theta$, to the applied $B$-field. While the TG-wave is highly damped and, therefore, confined in a thin ring close to the discharge channel wall, the helicon wave is not and develops toward the centre of the discharge channel. 

To derive the dispersion relation of helicon and TG-waves, the electron fluid equation of motion is written as in~\cite{HELIC_Theory_I}, in which the main hypotheses are of only first-order perturbation of the general type $\exp{i(m\theta+kz-\omega t)}$, a steady $B$-field aligned with the discharge channel's symmetry axis $z$, $\vec{B}_0 = B_0\hat{z}$, and all the dissipation mechanisms summarized within the collision rate $\nu$. 
\begin{equation}
-i\omega m_e \vec{v} = -e(\vec{E}+\vec{v}\times\vec{B}_0)-m_e \nu \vec{v}
\label{eq:elecfluidmot}
\end{equation}
The helicon wave dispersion relation derived from Eq.~\ref{eq:elecfluidmot}, is shown in its basic form in Eq.~\ref{eq:helicon}.
It is based on the hypotheses that $T_i << T_e$, that ion movements are neglected due the their inertia, and that the electron mass $m_e$ is neglected as well. In particular, it highlights the direct relation between $n_e$ and the applied $B$-field $B_0$. 
 \begin{equation}
 \beta=\frac{\omega}{k}\frac{n_e e \mu_0}{B_0}
 \label{eq:helicon}
 \end{equation}
The term $\beta$ is the total wave number $\beta^2=k^2_{\perp}+k^2_z$. If, instead, $m_e$ is not neglected, Eq.~\ref{eq:elecfluidmot} can be reduced and factored, see~\cite{HELIC_Theory_I} for the detailed derivation, leading to the two roots $\beta_{1,2}$ shown in Eq.~\ref{eq:heliconroot}, where $\delta=\omega/\omega_c$, and $k^2_w=\delta(\omega_p/c)^2$, with $c$ the speed of sound in plasma. 
\begin{equation}
\beta_{1,2}~\approx \frac{k}{2\delta}\biggl[1\mp\biggl(1-\frac{2\delta k^2_w}{k^2}\biggr)\biggr] 
\label{eq:heliconroot}
\end{equation}
The first root $\beta_1 \approx k_w^2/k$ leads to Eq.~\ref{eq:helicon}, the helicon wave dispersion relation, while the second root $\beta_2 \approx k/\delta$ leads to the TG-wave one of Eq.~\ref{eq:heliconTG} with $\theta$ the angle of the TG-wave respect to $\vec{B}_0$.
\begin{equation}
\beta=k_\parallel \frac{\omega_c}{\omega}=\beta\cos{\theta} \frac{\omega_c}{\omega}
\label{eq:heliconTG}
\end{equation}

To visualize how helicon and TG-wave propagates within a discharge channel of radius $r_0$, the typical helicon plasma radial density profile $n_e(r)$ with its peak in the centre, is considered. Conditions exist for both the radial position, and plasma density, at which the helicon and/or the TG-wave are triggered and can propagate, see Fig.~\ref{fig:helicon_n}~\cite{shamrai1998stable}. In the low density region where $n<n_{low}$ and $r>r_{up}$, only the TG wave can propagate. An intermediate region for $r_{up}<r<r_{low}$ and $n_{low}<n<n_{up}$ exists, in which both helicon and TG-wave can propagate. The last region, is that of $n<n_{up}$ and $r<r_{up}$ in which both waves cannot propagate and become evanescent~\cite{shamrai1998stable,isayama2018review}. Finally, the TG wave deposits the RF power from the antenna to the helicon wave which then penetrates deeper into the discharge volume~\cite{shamrai1998stable,isayama2018review}.\begin{figure}[h]
 	\centering
 	\includegraphics[width=8cm]{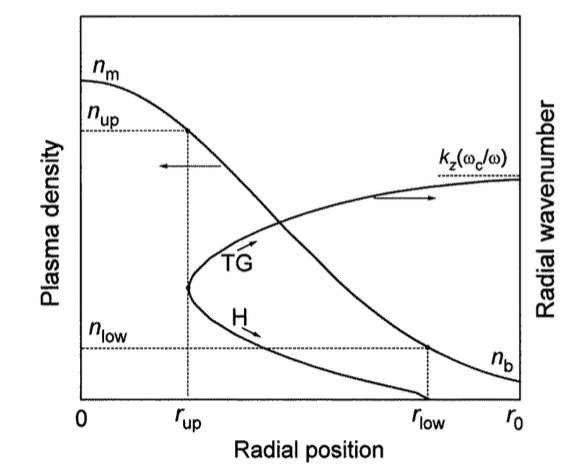}
 	\caption{Plasma density profile and radial variation of transverse wave number $k$. Helicon (H) and Trivelpiece-Gould (TG) branches of the dispersion relation~\cite{shamrai1998stable}.}
 	\label{fig:helicon_n}
 \end{figure}
The plasma in low pressure helicon discharges is in a non equilibrium condition, which means that $T_e>>T_i$ therefore $E_{k,e}>E_{k,i}$.

The capability of efficiently producing high density quasi-neutral plasma, up to $n_e=\SI{1E19}{\meter^{-3}}$~\cite{chen2015}, in a contactless manner made helicon plasma sources very interesting for plasma propulsion application by adding a magnetic nozzle to maximize thrust production. What most caught the attention of many researchers is the presence of a double layer (DL) in the plasma. This was firstly seen as a mechanism that could further increase the overall thrust~\cite{charles2006helicon}, but later found that it does accelerates ions, but does not increase the net momentum flux of the plasma itself,~\cite{fruchtman2006electric}. Finally, the helicon thruster became object of extensive research, but still, its physics are not yet fully understood~\cite{takahashi2019helicon}.

\subsection{RF Plasma Discharges}
This dissertation is focused on plasma generated by the use of radio frequency (RF) electromagnetic waves. The range of frequencies used in RF plasma sources is $f=1-\SI{500}{\mega\hertz}$~\cite{chabert_RF}. DC or low-frequency plasma sources operate at $f<\SI{1}{\mega\hertz}$, while those operating at microwave (MW) range are between $0.5 < f < \SI{10}{\giga\hertz}$, see Tab.~\ref{tab:frequencies}.\begin{table}[h]
\center
\caption{Frequency Ranges Definition~\cite{chabert_RF}}
\label{tab:frequencies}
\begin{tabular}{lc}
\toprule
Definition & Frequency Range\\
\midrule
Low-Frequency & $f<\SI{1}{\mega\hertz}$\\
Radio-Frequency & $\SI{1}{\mega\hertz}<f<\SI{500}{\mega\hertz}$\\
Microwave & $\SI{0.5}{\giga\hertz} < f < \SI{10}{\giga\hertz}$\\
\bottomrule
\end{tabular}
\end{table}

Specifically designed antennae fed by alternating current (AC) in the RF range, allow plasma generation without having a direct contact with it. Such contact-less operation has the major advantage of reducing at minimum any possible issue of erosion over time of any component in direct contact with the plasma, as it leads to performance degradation of the device. For the same reason, it enhances the flexibility in terms of working gas that can be utilised. According to~\cite{chabert_RF}, the RF range is suitable for plasma, as most of the charged particles, except the very heavy ions for the lowest frequencies, can instantaneously respond to the RF fields, while at the highest frequencies, ions are inertially constrained and only respond to the time-averaged fields. Electrons, instead, instantaneously respond within all the RF range.

In RF plasma sources, an antenna fed by AC at a given frequency in the RF range generates time-varying electromagnetic fields, that are antenna design dependent, that ionize the gas inside the discharge channel. The particles are heated up, until electrons start being ripped apart from their atoms, forming ions, and colliding with other neutrals increasing the ionization degree by ripping away more electrons from their atoms. An RF plasma discharge can develop in different modes. These modes, or regimes, define how the RF power is transferred to the plasma, which are commonly known as the $E$, $H$, and $W$ regime/mode~\cite{chabert_RF}, see Tab.~\ref{tab:regimes}.\begin{table}[h]
\centering
\caption{RF Plasma Regimes/Modes}
\label{tab:regimes}
\begin{tabular}{lcl}
\toprule
Regime/Mode & Abbreviation & Coupling\\
\midrule
Electrostatic & $E$ & Capacitive\\
Electromagnetic & $H$ & Inductive\\
Electromagnetic Wave & $W$ & Wave-based\\
\bottomrule
\end{tabular}
\end{table}

Generally, by increasing input power, the capacitive regime $E$ is achieved first, as the plasma is ignited. As the input power is increased and the plasma density also increases, there is a transition from $E$ to the inductive regime $H$~\cite{herdrich,chabert_RF}. If the plasma source is associated with an externally applied static magnetic field, the wave-based regime $W$ comes into place as a more efficient power coupling mechanism. This is involved with the formation of propagating waves within the plasma, and the $W$-regime can reach plasma densities in the order of $n_e=\SI{1E19}{\meter^{-3}}$. This regime is associated with a higher ionization degree compared to $E$ and $H$-regimes~\cite{chabert_RF,chen224}. The mode jump is commonly observed as an increment of plasma density, usually also associated with greater plasma brightness, but it can be also absent if the antenna is well designed~\cite{takahashi2019helicon}.

\section{Helicon Plasma Thruster}
In this section, the helicon plasma thruster concept, its physics of plasma discharge and plasma acceleration for thrust generation are presented, as well as the role of the Current -Free Double Layer. Finally, an outlook of possible critical thruster improvement is presented. 
\subsection{Concept}
The thruster conceptual schematics is shown within Fig.~\ref{fig:IPT}. 
\begin{figure}[H]
	\centering
	\includegraphics[width=.85\textwidth]{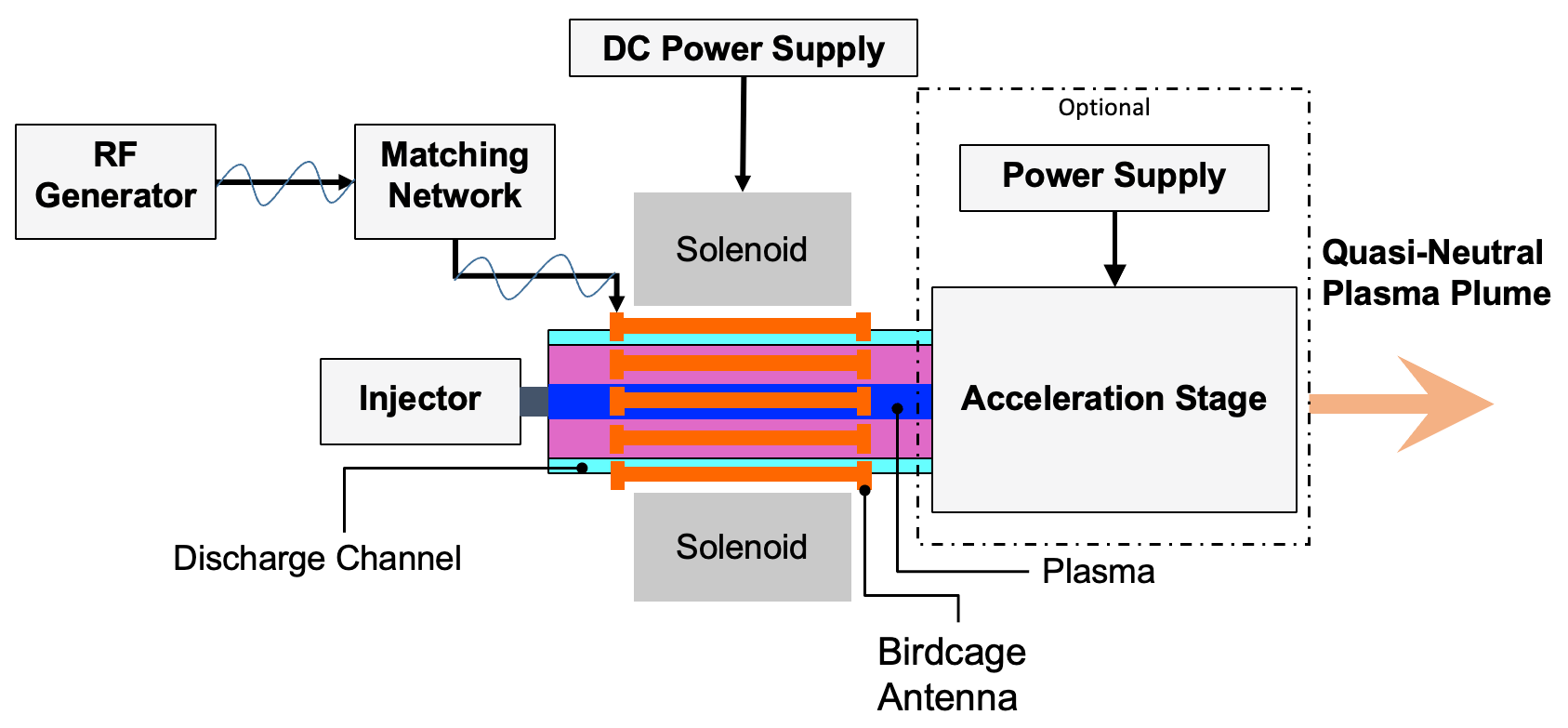}
	\caption{RF Helicon-based Plasma Thruster (IPT) Concept.}
	\label{fig:IPT}
\end{figure}
The RF generator provides the input power through the matching network to the thruster. An external solenoid fed by a DC power supply, alternatively a permanent magnet system, provides the static $B$-field required for triggering helicon wave formation and, by diverging at the outlet section of the discharge channel, aids the acceleration of the quasi-neutral plasma plume for thrust generation. This is known in literature as the (electro)magnetic nozzle. Further improvement can also be provided by combining it with an additional applied $E$-field as well, its effect on the performance are currently under investigation~\cite{chansp2021}. The injector (intake in the case of the ABEP system) delivers the propellant into the discharge channel, around which the RF-fed antenna is located. There, the propellant is ionized and accelerated by the electromagnetic fields generated by the antenna in combination with the applied $B$-field. The thruster developed within this dissertation is based on a birdcage antenna, which is treated more in detail in Chapter~\ref{ch:antenna}. An additional acceleration stage can be added if required.  The currently investigated helicon-based plasma thrusters in the world are those from the Australian National University (ANU)~\cite{charles2006helicon}, the Carlos III University of Madrid, Spain~\cite{ahedo2019helicon},  the University of Maryland, USA~\cite{vitucci2019development}, the small-enterprise T4i, Italy~\cite{manente2019regulus}, the Tohoku University, Japan~\cite{takahashi2019helicon}, the Tokyo University of Agriculture and Technology, Japan~\cite{isayama2018review}, the Washington University, USA~\cite{vereen2019recent}. Ad Astra Rocket Company also includes an helicon plasma stage for the VASIMR engine~\cite{squire2019steady}. 

%

\subsection{Thrust Generation Mechanism}
\label{sec:thrustmodel}
The mechanism of thrust generation in a magnetic nozzle is still subject to investigation as a complete understanding of its physics has not been yet achieved~\cite{takahashi2019helicon,taka2021}. The thrust model hereby presented is based on the work from~\cite{fruchtman2006electric,fruchtman2008,taka2011,taka2011b,fruchtman2012,takahashi2019helicon}. The derivation of the thrust force is based on a fluid model of an axisymmetric, magnetically expanding, and current-free plasma. 

The starting point is the equation for the thrust force $T$ as the sum of static and dynamic pressures over a cross section $A$, see Eq.~\ref{eq:thrust0}, with $\tau$ the momentum flux per unit cross section.
\begin{equation}
T=\int_{A} \tau dA = \int_{A} (p_e + m_i n_p u_z^2)dA
\label{eq:thrust0}
\end{equation}
The derivation of $\tau$ is based on the momentum equation, of ions and electrons, in a steady state condition, and it is derived based on the following hypotheses.
\begin{itemize}
\item $T_i << T_e$;
\item electron inertia negligible $m_e\sim0$;
\item quasi-neutrality $n_i\sim n_e = n_p$;
\item axisymmetric system.
\end{itemize}
The momentum equation is shown in Eq.~\ref{eq:mom}, with $j=i,e$ for ions or electrons respectively, where the left hand side is the inertial term $m_j n_j(\vec{v}_j \cdot \nabla)\vec{v}_j$ if the continuity equation is satisfied ($\nabla(n_j\vec{v}_j)=0$), and the right hand side the sum of the Lorentz force and of that given by the pressure gradient $\nabla{p_j}$. 
\begin{equation}
m_j \nabla{(n_j \vec{v}_j \vec{v}_j)}=q_j n_j (\vec{E}+\vec{v}_j\times\vec{B})-\nabla{p_j} 
\label{eq:mom}
\end{equation}

The momentum equation of Eq.~\ref{eq:mom} can be rewritten in the respective radial $r$ and axial $z$ components for velocities of electrons $v_{r,\theta,z}$, Eq.~\ref{eq:momrade},~\ref{eq:momaxe}, and of ions $u_{r,\theta,z}$, Eq.~\ref{eq:momradI},~\ref{eq:momaxI}.
\begin{equation}
-e n_p (E_r+v_\theta B_z) = \frac{\partial p_e}{\partial r}
\label{eq:momrade}
\end{equation}

\begin{equation}
-e n_p (E_z-v_\theta B_r) = \frac{\partial p_e}{\partial z}
\label{eq:momaxe}
\end{equation}

\begin{equation}
e n_p (E_r+u_\theta B_z)= 0
\label{eq:momradI}
\end{equation}

\begin{equation}
e n_p (E_z - u_\theta B_r)= \frac{1}{r} \frac{\partial{(r m_i n_p u_r u_z)}}{\partial{r}}+\frac{\partial{(m_i n_p u^2_z)}}{\partial{z}}
\label{eq:momaxI}
\end{equation}
In Eq.~\ref{eq:momradI}, the inertial terms for the ion in the radial direction is assumed to be zero for $T_i  < < T_e$ as well as for the agreement of this model with experiments~\cite{takahashi2019helicon}. By assuming $E_z$ to be negligible in the axial projections of the momentum equations Eq.~\ref{eq:momaxe} and Eq.~\ref{eq:momaxI}, and by combining Eq.~\ref{eq:momrade} and Eq.~\ref{eq:momradI} for a discharge channel radius $r_s$ and an expanding plasma radius $r_p(z)$, the momentum flux can be finally obtained as in Eq.~\ref{eq:momentumflux}
\begin{equation}
\frac{\partial \tau}{\partial z} = \frac{\partial (p_e+m_i n_p u_z^2)}{\partial z}=e n_p (v_\theta - u_\theta)B_r - \frac{1}{r}\frac{\partial {(r m_i n_p u_r u_z)}}{\partial r}
\label{eq:momentumflux}
\end{equation}

The total thrust $T$ provided by a magnetic nozzle in an helicon plasma thruster can be decomposed based on Eq.~\ref{eq:thrust0} and Eq.~\ref{eq:momentumflux} as in Eq.~\ref{eq:heliconT} in an expanding plasma $r_p(z)$~\cite{taka2011}, by the sum of the contributions of electron pressure $T_S$, Lorentz force $T_B$, and ion lost at the walls $T_W$, see Fig.~\ref{fig:magneticnozzle}.
\begin{equation}
\begin{aligned}
T& = 2\pi \int_{r_s} r p_{e,0} dr-2\pi \int_{z} \int_{r_p} r \frac{B_r}{B_z} \frac{\partial{p_e}}{\partial r} dr dz-2\pi \int_{z}\int_{r_p} \frac{\partial}{\partial r} (r m_i n_p u_r u_z) dr dz \\
& =T_S+T_B+T_W
\end{aligned}
\label{eq:heliconT}
\end{equation}
Where $T_S$ is the thrust generated by the static electron pressure $p_e$ pushing at the upstream of the plasma source, commonly the injector plate, specified in Eq.~\ref{eq:heliconTS}, where $p_{e,0}$ is the maximum electron pressure force inside the source, and $r_s$ is the discharge channel radius.\begin{figure}[h]
\centering
\includegraphics[width=.7\textwidth]{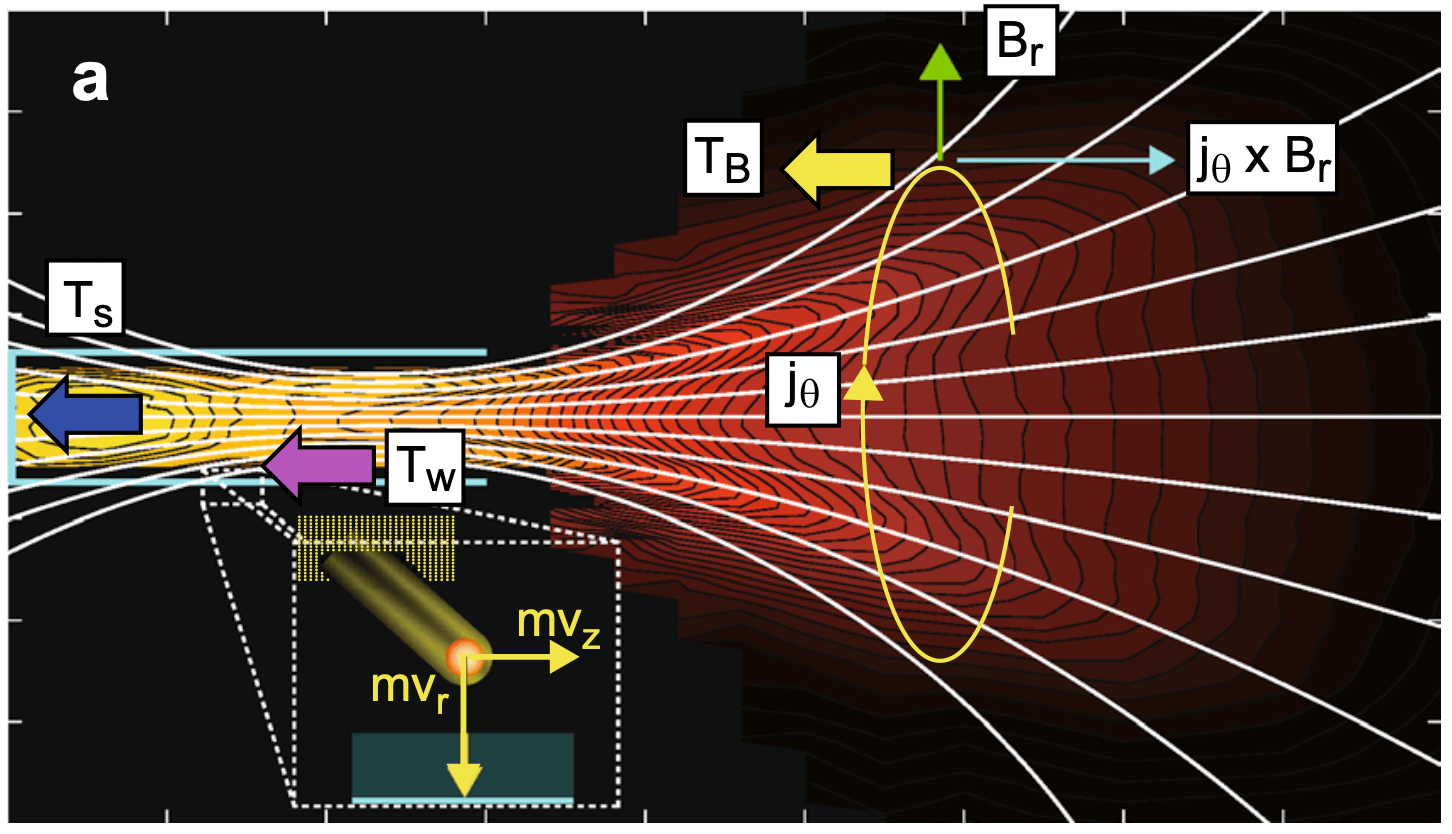}
\caption{Schematics of Thrust Production Mechanism in a Magnetic Nozzle~\cite{takahashi2019helicon}.}
\label{fig:magneticnozzle}
\end{figure}
\begin{equation}
T_S=2\pi \int_{r_s} r p_{e,0} dr
\label{eq:heliconTS}
\end{equation}
The electron pressure $p_e$ is converted upstream of the plasma source into ion dynamic momentum via the sheath acceleration. 

The second term of Eq.~\ref{eq:heliconT} is $T_B$, that is the volume integration of the Lorentz force, see Eq.~\ref{eq:heliconTB}. This is a result of the radial $B$-field $B_r$ that arises in the diverging $B$-field region, as well as the azimuthal electron-diamagnetic current $B_z^{-1}\partial{p_e}/\partial{r}$ which increases the axial momentum flux of the plasma acting on the $B$-field lines. Finally, the magnetic nozzle converts the radial $p_e$ into axial plasma momentum via the Lorentz force, and the main responsible for the energy of the magnetic nozzle are the electrons themselves.
\begin{equation}
T_B=-2\pi \int_{z} \int_{r_p} r \frac{B_r}{B_z} \frac{\partial{p_e}}{\partial r} dr dz
\label{eq:heliconTB}
\end{equation}
Finally, $T_W$ is the axial momentum delivered by the ions that are lost at the wall, which imparts some axial component to the thrust that should not be neglected, see Eq.~\ref{eq:heliconTW}.
\begin{equation}
T_W=-2\pi \int_{z}\int_{r_p} \frac{\partial}{\partial r} (r m_i n_p u_r u_z) dr dz
\label{eq:heliconTW} 
\end{equation}

\subsection{Role of the Current-Free Double Layer}
Current-free double layers (CFDL), discovered in helicon plasma thrusters~\cite{charles2007review}, are transient or stationary localized potential structures~\cite{doublelayercarles} that form in current-free expanding plasmas in diverging $B$-fields~\cite{fruchtman2006electric}, and are potential barriers with a voltage drop $\Delta V$ that accelerate the ions opposing to the flux of electrons that overcomes this potential drop do maintain charge neutrality, such that the global net current is zero. Therefore, a CFDL does not increase the net plasma momentum~\cite{fruchtman2006electric,fruchtman2010thrust}, but only affects the plasma upstream in terms of density and temperature.
Finally, a threshold for the minimum applied magnetic field that leads to the formation of the CFDL exists, and this is for a discharge channel diameter that is larger than the corresponding ion Larmor radius~\cite{takahashi2019helicon,takahashi2010double}.

\subsection{Improvement of Helicon Plasma Thrusters}
The maximum measured thruster efficiency $\eta_T$ of helicon thrusters to day (April 2021) reached $\eta_T=20\%$~\cite{takahashi2019helicon}. According to~\cite{takahashi2019helicon,taka2021}, $\eta_T$ can be improved by inhibiting cross-field diffusion and plasma losses at the wall of the discharge channel, and by improving the overall understanding of the plasma acceleration mechanisms, such as the detachment of plasma from the $B$-field lines in the magnetic nozzle region, the effect of neutrals, and the electron thermodynamics~\cite{takahashi2019helicon}. 
The thruster can be also improved in terms of circuit efficiency as well, with an improved power processing unit (PPU), but also by designing the thruster to match the required impedance as described later, in detail, within Ch.~\ref{ch:RF}. This drastically minimizes power losses and also leads to a preciser knowledge of how much of the input power is effectively used for the plasma ionization and acceleration. Furthermore, the thruster's antenna design is not only important from the point of view of a matched impedance, but also in terms of the spatial and temporal configuration of the generated electromagnetic fields~\cite{romanoacta2}. Therefore, the antenna shall be designed to be matched at the correct impedance, and also to provide an optimized electromagnetic field configuration for both efficient ionization and acceleration of the plasma. Finally, the application of the thruster on an ABEP system, means that atmospheric propellant is used, leads to the fact that the required energy of ionization can be higher, therefore requiring higher input power for ionization, when compared to \ce{Xe}-based helicon plasma thrusters~\cite{heliconABEP2008,heliconABEP}. 

	\section{Thruster Concept and Requirement Analysis}
	\label{ch:IPT}
	This section presents the design approach of the thruster. At first, top level thruster requirements are given and the respective concept is briefly described. Then, to determine the main design features such as size, input frequency, and required magnetic field, the software HELIC is utilised. It is 2-dimensional tool that highlights plasma resistance behaviour depending on, mainly, thruster geometry, plasma density, input frequency, and magnetic field.  Finally, the thruster main design parameters are selected and presented.

The thruster requirements are based on those given within the H2020 DISCOVERER project: an RF-based contactless and neutraliser-less thruster that can operate on atmospheric propellant, e.g. atomic oxygen AO and nitrogen \ce{N2} in VLEO, see Fig.~\ref{fig:rho_h_comp}, and that can efficiently cope with the respective aerodynamic drag by operating at an input power $P_{in}<\SI{5}{\kilo\watt}$.
The thruster concept is based on RF helicon wave plasma discharges, to provide higher plasma densities for a given input power than capacitive or inductive-based devices~\cite{Chen_2015}, and electromagnetic-based acceleration. The high ionization degree that can be achieved with helicon wave based discharges enables electromagnetic acceleration due to a reduced presence of neutrals, leading to potentially higher exhaust velocities. Finally, the main design drivers of the thruster are in terms of dimensions of the discharge channel, the antenna type, the operating frequency, and the required static magnetic field range over a given plasma density range. 

\subsection{Thruster Analysis with HELIC} 
\label{ch:helic}
The analysis over the main thruster design drivers is performed by using HELIC, the freed 2D numerical tool based on the helicon wave theory developed by D.~Arnush and F.~Chen~\cite{HELIC_Theory_I, HELIC_Theory_II}. 
It allows to estimate the power coupling and resulting plasma resistance $R_P$ for given geometrical (discharge channel and antenna) and plasma device input parameters (input frequency $f_0$, propellant, pressure $p$, applied $B$-field, plasma density $n$ and its respective radial profile $n_e(r)$). Five antenna types are available: 1-loop, Nagoya type III, Boswell, Half Helix, and Multi Turn (solenoid like, defined by number of turns, and length). Moreover, an applied magnetic field $B_0$ is applied along the symmetry axis of the discharge channel. Top (right) and bottom (left) plates are present that can be either conductive or insulating. The geometry of the HELIC software tool is shown in Fig.~\ref{fig:helic_Geometry}.\begin{figure}[h]
	\centering
	\includegraphics[width=9cm]{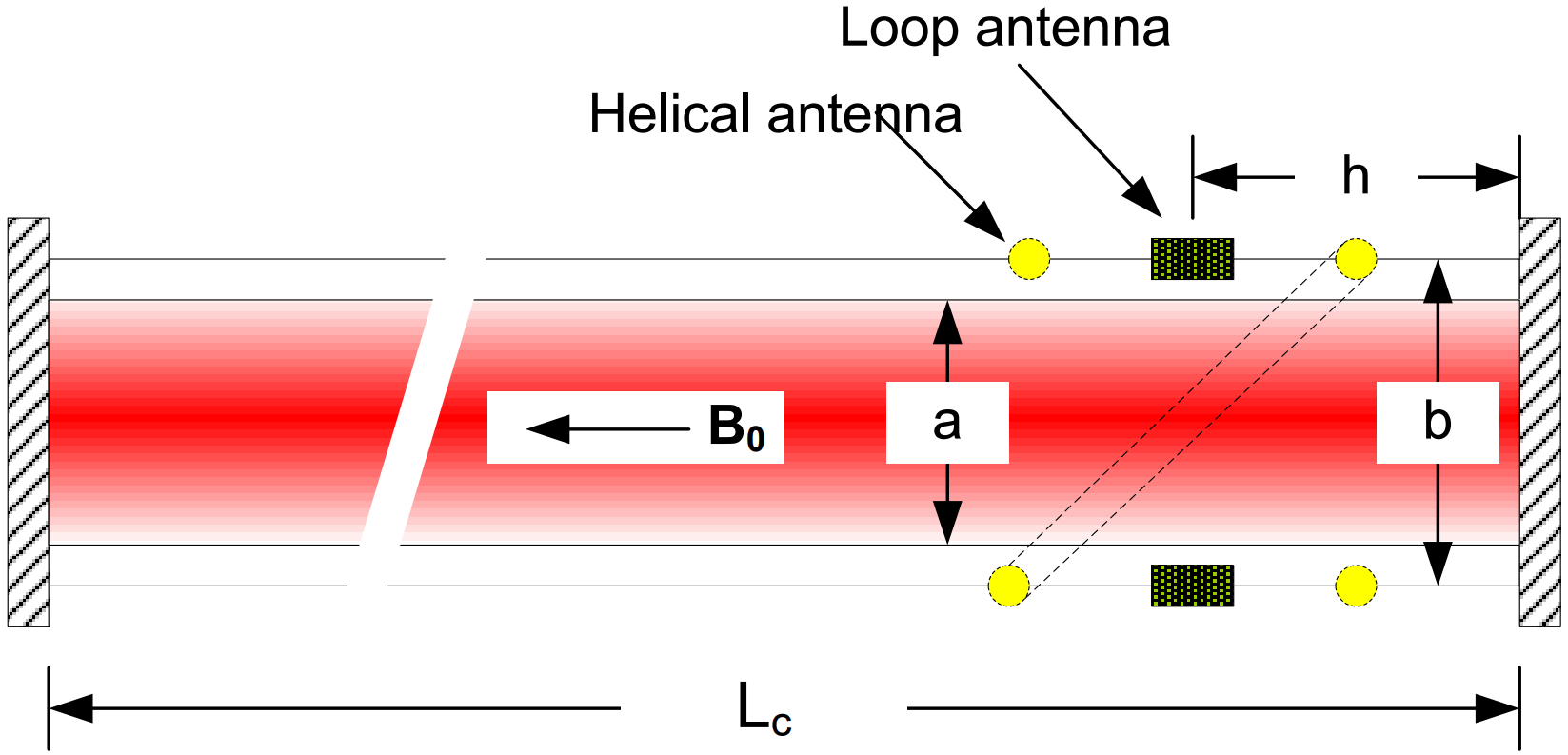}
	\caption{HELIC Geometry~\cite{chen217R}.}
	\label{fig:helic_Geometry}
\end{figure}

\subsubsection{Input}
Within this subsection the input parameters for the HELIC software are described. Three exemplary discharge channel diameters are selected based on the small-scale inductively heated plasma generator IPG6-S~\cite{romanoacta}, based on which the thruster is developed~\cite{romanorgcep,romanoiac,romanoiepc2,romanorgcep2,romanoiac2,romanosp2018,romanoacta}: $\phi=\SI{37}{\milli\meter}$, the half of it $\phi=\SI{18.5}{\milli\meter}$, and its double $\phi=\SI{74}{\milli\meter}$. The reference propellant is \ce{Ar} to simplify the design process. For the final steps, optimization is done for atmospheric propellant \ce{N2} and \ce{O2} (AO is not available in HELIC) as well. The target plasma density is that achievable by helicon-based plasma sources, up to $n=\SI{1E18}{}-\SI{1E19}{\meter^{-3}}$~\cite{chen224}. 
\begin{table}[h]
	\caption{HELIC Inputs: Plasma, Antenna, and Cavity Radii.}
	\centering
	\begin{tabular}{lcl}
		\toprule
		Quantity & Symbol & Value\\
		\midrule
	Plasma Radius & $a$ & $\SI{9.25}{\milli\meter}$\\
								& & $\SI{18.5}{\milli\meter}$\\
								& & $\SI{37.0}{\milli\meter}$\\
	Antenna Radius & $b$ & $a+t_{tube}+\frac{t_{ant}}{2}$\\
	Radius of the Conductive & & \\
	Bounding Cylinder & $c$ &$\SI{0.2}{\meter}>b$\\
	Channel Length & $L_C$ & \SI{2}{\meter}\\
		\bottomrule
	\end{tabular}
	\label{tab:input1}
\end{table}
Plasma and antenna geometry required by HELIC are given in Tab.~\ref{tab:input1}. The bounding cylinder radius $c$ is set to $c>b$, so that results are insensitive to this. The discharge channel has a thickness of $t_{tube}=\SI{1.5}{\milli\meter}$, used by HELIC to calculate the radius at which the current flows. The dielectric constant is approximated by $\epsilon_0$. Discharge channel and antenna thickness $t_{ant}$ are kept constant, so that $b=a+\SI{3}{\milli\meter}$. The cavity length $L_C=\SI{2}{\meter}$, with a small antenna this can simulate ejection of plasma, as the geometry is provided with two boundaries of conductive or insulating plates at begin and at the end of the discharge channel. The input RF frequency is based on experiments with IPG6-S of $f_0=\SI{3.3}{\mega\hertz}$~\cite{romanorgcep2}. The effect of other frequencies is also investigated by setting $f_0=\SI{13.56}{\mega\hertz}$, common industrial standard, and its higher harmonics $f_0=\SI{27.12}{\mega\hertz}$, and $f_0=\SI{40.68}{\mega\hertz}$. A wave mode of $m=+1$ is selected (and targeted in the design) as it better performs compared to $m=-1$ mode~\cite{HELIC_Theory_II}, see Tab.~\ref{tab:input2}. 

The antenna starting point is a 5-turn coil based on IPG6-S design, with a length of $L_A=\SI{0.075}{\meter}$, and targeting a $m=+1$ helicon wave mode. Moreover, the helicon wavelength of $\lambda=\SI{0.15}{\meter}$ according to~\cite{chen158R} is applied.\begin{table}[h]
	\caption{HELIC Inputs: Ion, Frequency, and Azimuthal Mode.}
	\centering
	\begin{tabular}{lcl}
		\toprule
		Quantity & Symbol & Value\\
		\midrule
		Ion Species & - & \ce{Ar}\\
		RF Frequency & $f_0$ & $3.3$,~$13.56$,~$27.12$,~$\SI{40.68}{\mega\hertz}$ \\
		Wave & $m$ & $+1$\\
		Antenna Type & type & Multi Turn:~5 turns\\
		Turns/gap & - & NA, 0\\
		Antenna Length & $L_A$ & \SI{0.075}{\meter}\\
		\bottomrule
	\end{tabular}
	\label{tab:input2}
\end{table}

The cavity type is set to bounded, so that the its size determines the range of $k$ of the waves. For finite cavities, the maximum wavelength is $2 Lc$. The maximum $k$ at which the waves can propagate depends on $n$ and $B_0$. The boundary plates of the discharge channel can be set as conductive $ei=-1$ or insulating $ei=+1$. The latter is chosen as first approach. The antenna is positioned at the centre with $ZA=0$, see Tab.~\ref{tab:input3}.\begin{table}[h]
	\caption{Wave-number Spectrum.}
	\centering
	\begin{tabular}{lcl}
		\toprule
		Quantity & Symbol & Value\\
		\midrule
		Boundary Condition & $ei$ & $+1$\\
		Antenna Location & $ZA$ & \SI{0}{\meter}\\
		\bottomrule
	\end{tabular}
	\label{tab:input3}
\end{table}

The magnetic field is set to $B_0=0.001-\SI{0.1}{\tesla}$, and the plasma density profile $n(r)$ uniform. The absence of magnetic field is evaluated by applying $B_0=\SI{0.0001}{\tesla}$ to represent a purely inductive discharge~\cite{HELIC_Theory_I,HELIC_Theory_II}. The electron temperature is set to $T_e=\SI{3}{\electronvolt}$, and the neutral pressure, $p=10~$mTorr~$=\SI{1.33}{\pascal}$. The remaining parameters are set to default.\begin{table}[h]
	\caption{Plasma Density, Magnetic Field, and Various.}
	\centering
	\begin{tabular}{lcl}
		\toprule
		Quantity & Symbol & Value\\
		\midrule
		Plasma Density & $n_{min}$ & $\SI{1e11}{{\centi\meter}^{-3}}$\\
		& $n_{max}$ & $\SI{1e13}{{\centi\meter}^{-3}}$\\
		Plasma Density Profile & - & Uniform\\
		Magnetic Field & $B_0$ & $B_0=10 - \SI{100}{\milli\tesla}$ \\
		Electron Temperature & $T_e$ & \SI{3}{\electronvolt}\\
		Pressure of Neutrals & $p$ &  10~mTorr $=\SI{1.33}{\pascal}$\\
		\bottomrule
	\end{tabular}
	\label{tab:input4}
\end{table}

\newpage
\subsubsection{Objectives}
According to~\cite{chen224}, the rate of the RF power deposition depends on the load/plasma resistance $R_P$ at the output of the matching network. To maximize the power deposited into the plasma, $R_P$ must be larger than the circuit's resistance $R_P>R_C$. The power actually reaching the plasma $P_{in}$ is given by Eq.~\ref{eq:power}.\begin{equation}
P_{in}=P_{RF}\frac{R_P}{R_P+R_C}
\label{eq:power}
\end{equation}
The parameters are later evaluated based on input frequency $f_0$, and applied magnetic field $B_0$.

\subsubsection{Results}
In the following, the numerical calculation results are presented, in terms of plasma resistance $R_P$ vs. plasma density $n$ and magnetic field $B_0$ at different RF input frequencies $f$. This is, in accordance to Chen, the most useful plot for the design of the source~\cite{chenlast}. For each simulation, the magnetic field $B_0$ has been varied from \SI{10}{\milli\tesla} to \SI{100}{\milli\tesla}.
\vspace{20pt}
\begin{figure}[H]
	\centering
	\includegraphics[width=.7\textwidth]{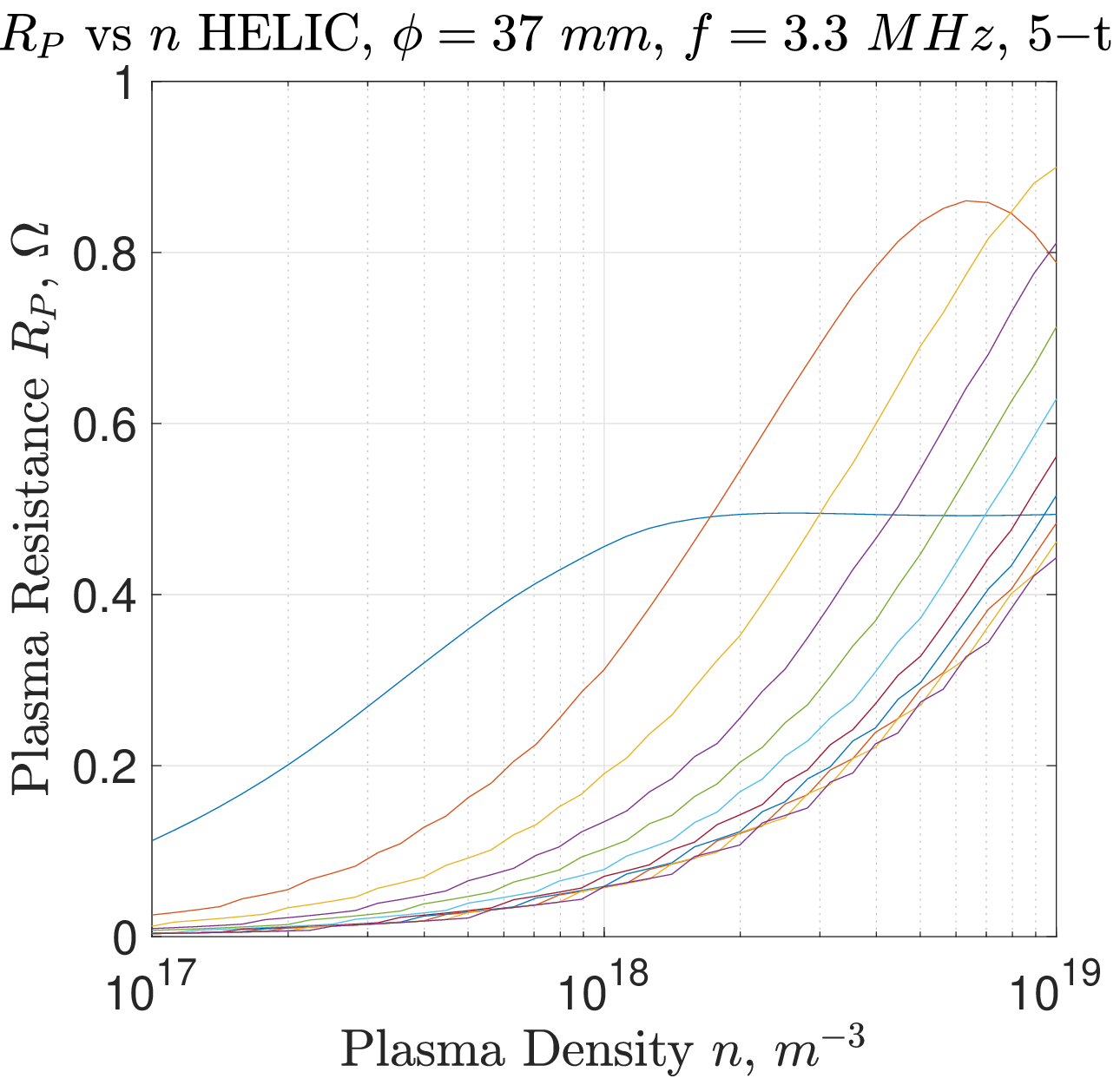}
	\caption{Plasma Resistance $R_P$ vs Plasma Density $n$ vs Applied Magnetic Field $B_0$, Input Frequency $f=\SI{3.3}{\mega\hertz}$, 5-turns Antenna.}
	\label{fig:37_33}
\end{figure}
\newpage
\begin{figure}[H]
	\centering
	\includegraphics[width=.7\textwidth]{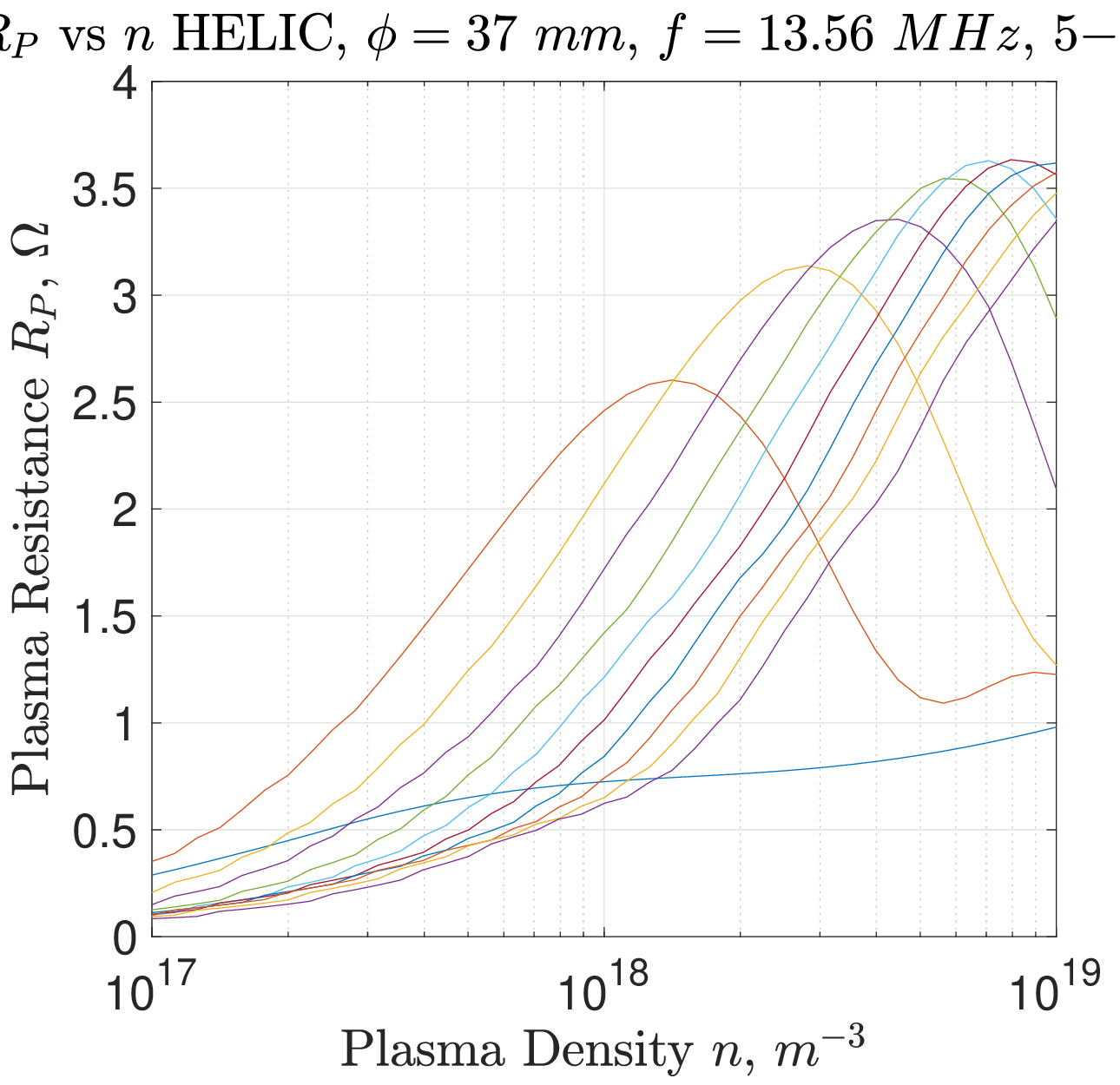}
	\caption{Plasma Resistance $R_P$ vs Plasma Density $n$ vs Applied Magnetic Field $B_0$, Input Frequency $f=\SI{13.56}{\mega\hertz}$, 5-turns Antenna.}
	\label{fig:37_1356}
\end{figure}
\begin{figure}[H]
	\centering
	\includegraphics[width=.7\textwidth]{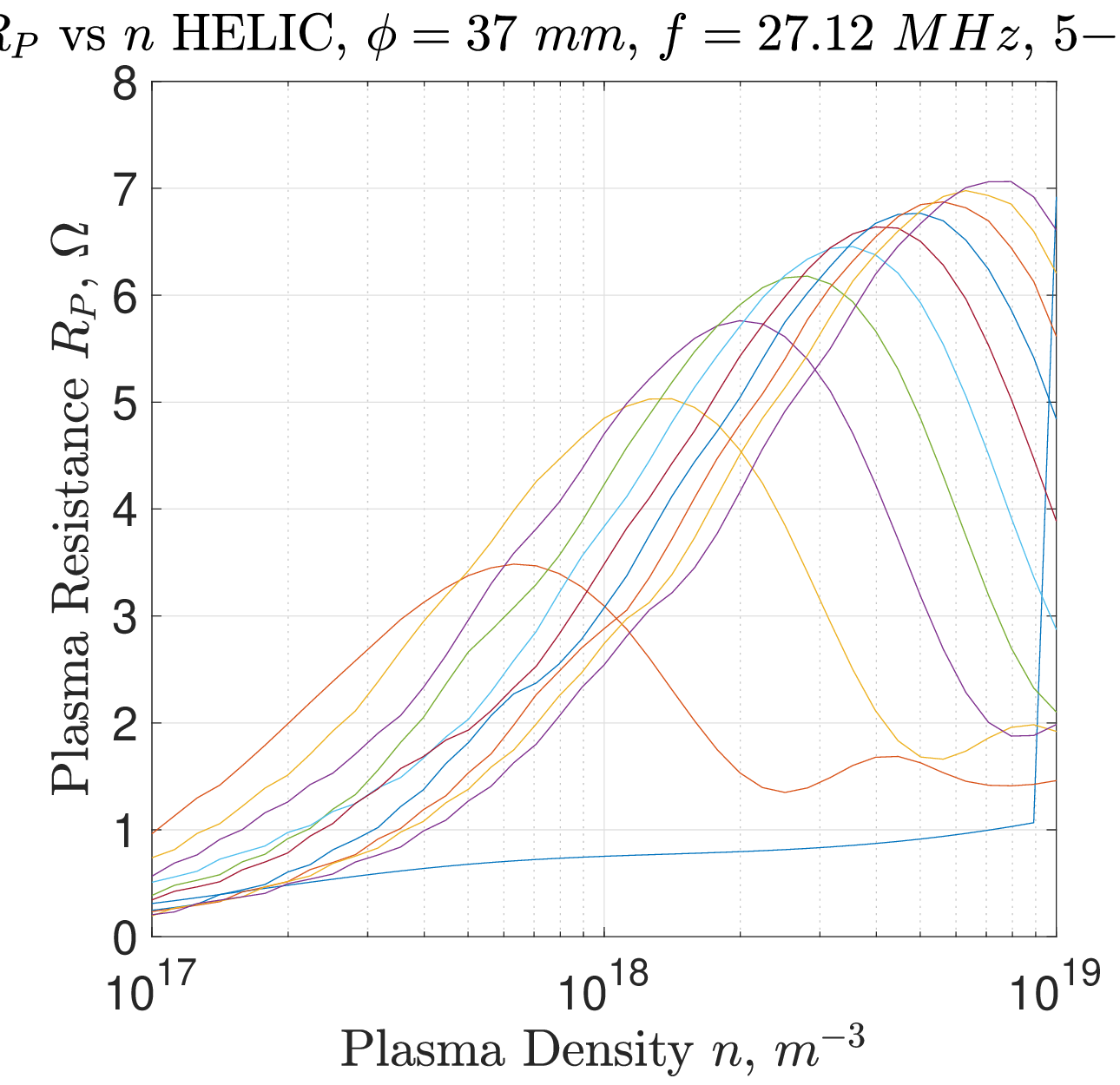}
	\caption{Plasma Resistance $R_P$ vs Plasma Density $n$ vs Applied Magnetic Field $B_0$, Input Frequency $f=\SI{27.12}{\mega\hertz}$, 5-turns Antenna.}
	\label{fig:37_2712}
\end{figure}
\begin{figure}[H]
	\centering
	\includegraphics[width=.7\textwidth]{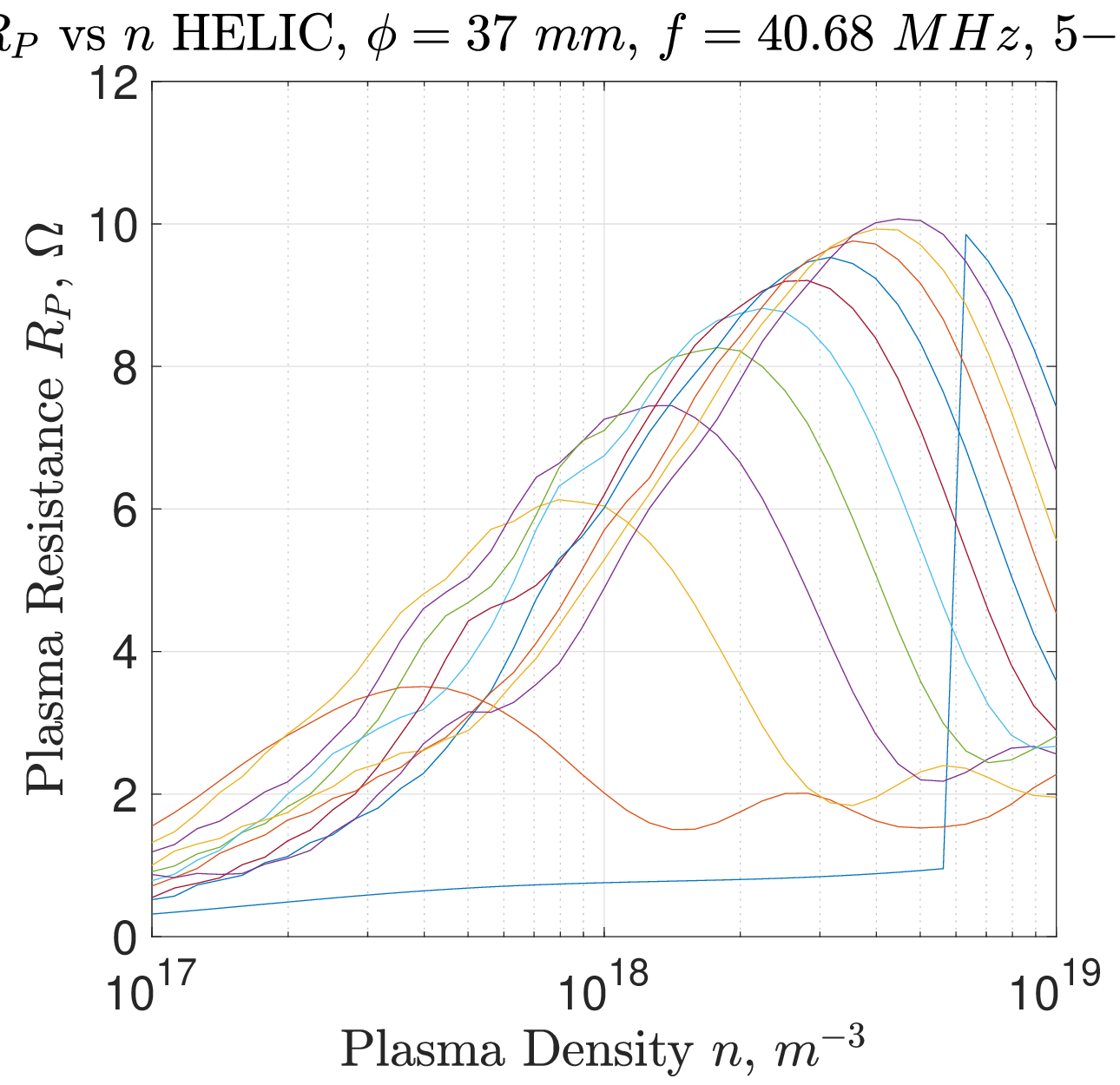}
	\caption{Plasma Resistance $R_P$ vs Plasma Density $n$ vs Applied Magnetic Field $B_0$, Input Frequency $f=\SI{40.68}{\mega\hertz}$, 5-turns Antenna.}
	\label{fig:37_4068}
\end{figure}

\subsubsection{Discussion}
At higher frequencies $f$, $R_P$ increases, and $n$ at which the highest $R_P$ is achieved, shifts to lower $n$.  As the $R_P$ peak shifts to lower $n$, $R_P$ also increases with $f$. This is an important result as $p$, therefore $n$ are expected to be low in an ABEP application, $p\sim\SI{4E-4}{}-\SI{1E-3}{\pascal}$~\cite{di2007ram}. This means that an higher $f$ is advantageous for operation and ignition at low $p$, this is also in accordance with~\cite{chen233}. Finally, with the given plasma density range $n$, to achieve $R_P>\SI{1}{\ohm}$, $f$ must be high to increase $R_P$, and the applied magnetic field $B_0$ should be tuned to match the $R_P$ peak for the expected $n$. The case at $f=\SI{3.3}{\mega\hertz}$ does not provide $R_P>\SI{1}{\ohm}$ for the given $n$ range, even with $B_0=\SI{100}{\milli\tesla}$, see Fig.~\ref{fig:37_33}. By applying $f=\SI{13.56}{\mega\hertz}$, see Fig.~\ref{fig:37_1356}, $R_P>\SI{1}{\ohm}$ can be achieved for any $B_0=10-\SI{100}{\milli\tesla}$ for $n>\SI{2e18}{\meter^{-3}}$, while for lower ones, this can be achieved given that the correct $B_0$ is applied. It can be noted the tendency of $R_P=\SI{1}{\ohm}$ without $B_0$ for $n=\SI{1e19}{\meter^{-3}}$. The case at  $f=\SI{27.12}{\mega\hertz}$ in Fig.~\ref{fig:37_2712}, shows that $R_P>\SI{1}{\ohm}$ is achieved for any $B_0=10-\SI{100}{\milli\tesla}$ for plasma densities $n>\SI{4e17}{\meter^{-3}}$, while for lower ones, this can be achieved given that the correct $B_0$ is applied. The case for $f=\SI{40.68}{\mega\hertz}$, see Fig.~\ref{fig:37_4068}, shows that $R_P>\SI{1}{\ohm}$ is achieved for any $B_0=10-\SI{100}{\milli\tesla}$ for plasma densities $n>\SI{2e17}{\meter^{-3}}$, while for lower ones, this can be achieved given that the correct  $B_0$ is applied. Finally, the amplitude of $R_P$ for the same $n,~B_0$ is larger for the $f=\SI{40.68}{\mega\hertz}$ case. At higher frequencies, the $R_P$ peaks shifts down to lower $n$ while its amplitude increase. The irregularities in the plot are due to low step size set into HELIC, set to achieve shorter simulation times, as the scope is to visualise the qualitative behaviour of $R_P$. Finally, $f=\SI{40.68}{\mega\hertz}$ is selected with an expected required $B_0=10-\SI{100}{\milli\tesla}$. The results are resumed in Tab.~\ref{tab:HELIC1}.

\begin{table}[h]
\center
\caption{HELIC Results, 5-turn Antenna, $R_P>\SI{1}{\ohm}$.}
\label{tab:HELIC1}
\begin{tabular}{ccc}
\toprule
$f$ & $B_0$ & $n$ \\
\SI{}{\mega\hertz} & \SI{}{\milli\tesla} & $\SI{}{1\per{\meter^3}}$\\
\midrule
$3.3$ & $>100$ & $>1\times10^{19}$\\
$13.56$ & $10-100$  & $>2\times10^{18}$ \\
$27.12$ & $10-100$  & $>4\times10^{17}$\\
$40.68$ & $10-100$ & $>2\times10^{17}$ \\
\bottomrule
\end{tabular}
\end{table}

\subsubsection{Refined Inputs}
Further analysis is performed by applying a finite length discharge chamber and setting the position of the antenna $ZA$ from the centre of the discharge channel. The discharge channel is set to that of IPG6-S, $L_C=\SI{180}{\milli\meter}$ and $ZA=L_C/2-L_A/2=\SI{0.0525}{\meter}$, to have the antenna end corresponding to that of the discharge channel. The range of $B$-field is set according to the results of the precedent section. The HELIC outputs are shown in Fig.~\ref{fig:37mm_refined} and Fig.~\ref{fig:37mm_refined_conducting}.

\begin{figure}[hp]
	\centering
	\includegraphics[width=.7\textwidth]{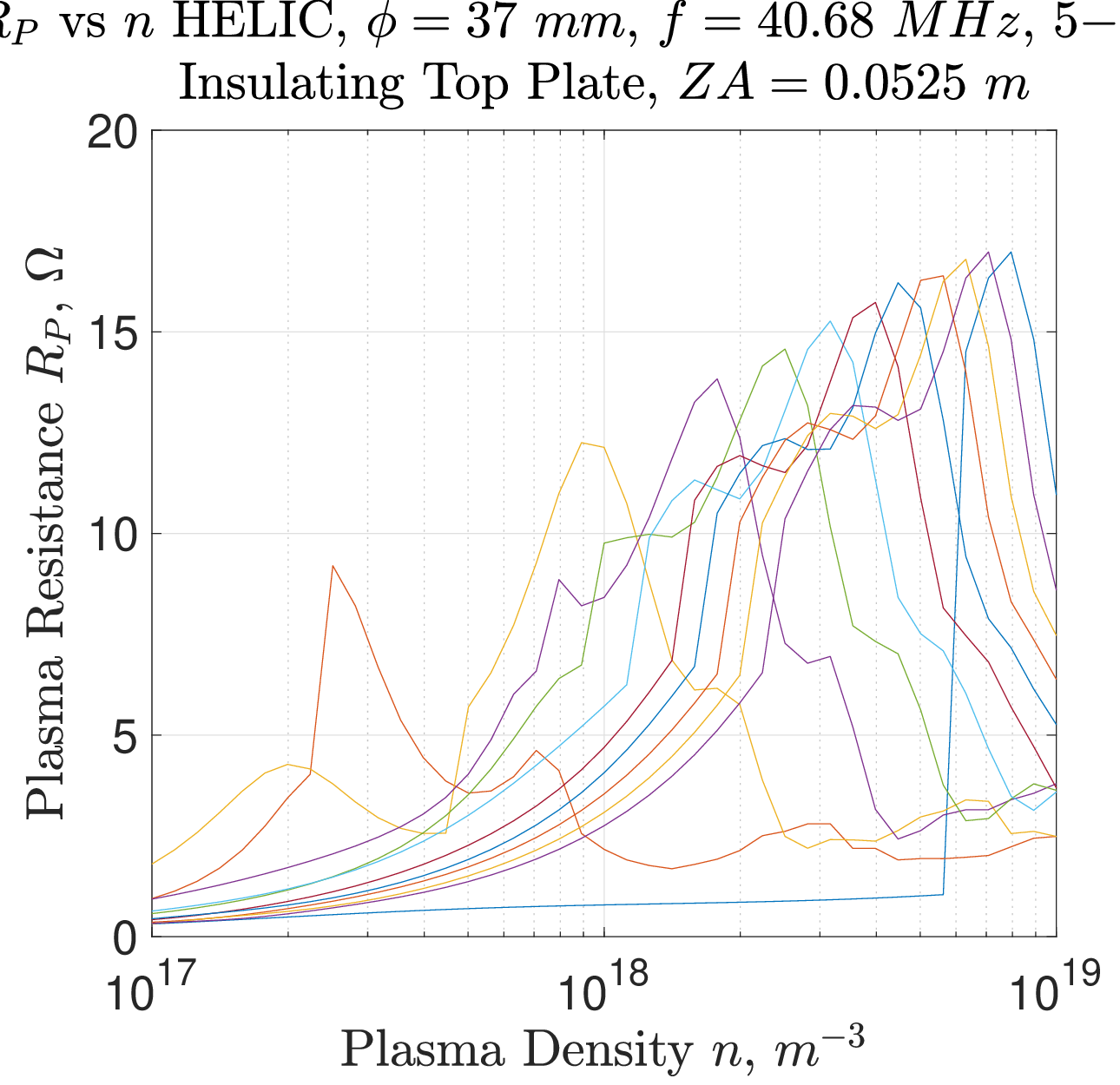}
	\caption{$R_P$ vs $n$ vs $B_0$, Insulating Top Plate, at $f=\SI{40.68}{\mega\hertz}$.}
	\label{fig:37mm_refined}
\end{figure}
\begin{figure}[hp]
	\centering
	\includegraphics[width=.7\textwidth]{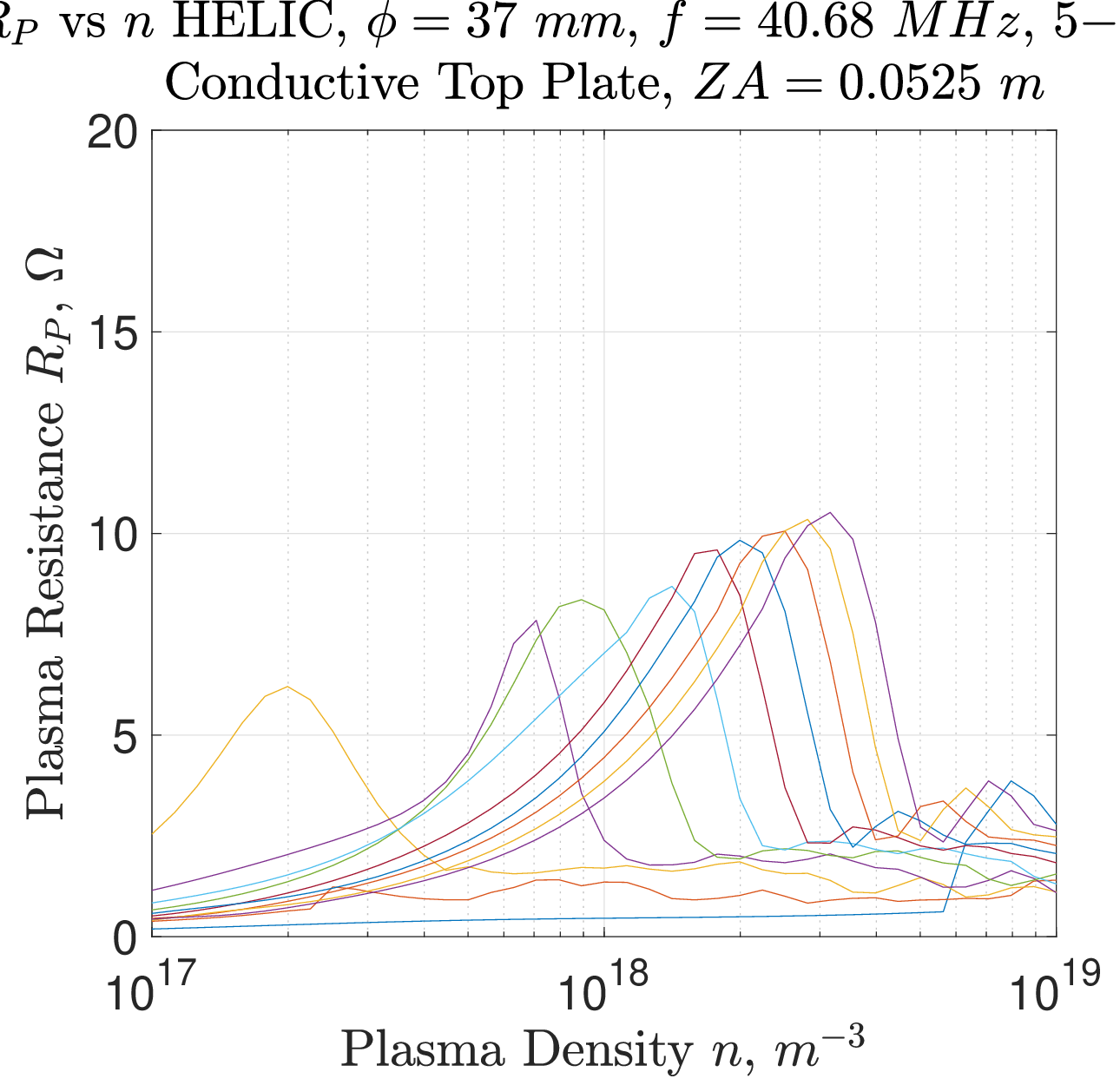}
	\caption{$R_P$ vs $n$ vs $B_0$, Conductive Top Plate, at $f=\SI{40.68}{\mega\hertz}$.}
	\label{fig:37mm_refined_conducting}
\end{figure}

The conductive top plate reduces $R_P$, creates a sharper shape of the peaks, and shifts them to lower $n$. Notable is the increase in magnitude of the low density peak of $R_P$ for $B_0=\SI{20}{\milli\tesla}$. For comparison, the same analysis is performed for $f=\SI{13.56}{\mega\hertz}$. The results are presented in Fig.~\ref{fig:37_1356_good},~\ref{fig:37_1356_good_c}. The same behaviour as of the $f=\SI{40.68}{\mega\hertz}$ case can be observed, but at generally much lower $R_P$ magnitudes.

\begin{figure}[hp]
	\centering
	\includegraphics[width=.7\textwidth]{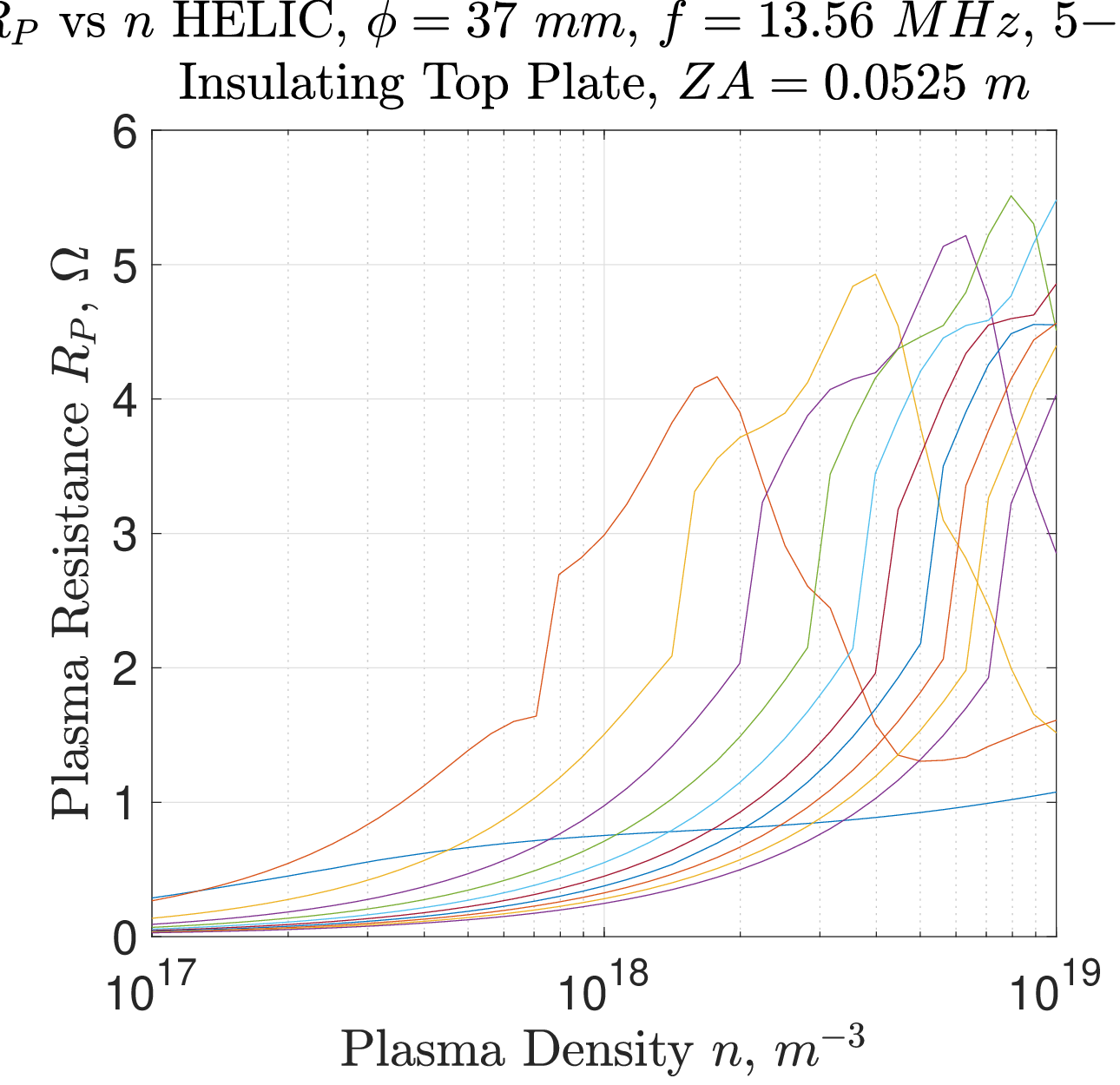}
	\caption{$R_P$ vs $n$ vs $B_0$, Insulating Top Plate, at $f=\SI{13.56}{\mega\hertz}$.}
	\label{fig:37_1356_good}
\end{figure}

\subsubsection{Discussion}
The numerical results for the different parameters provides a clearer view of their effects on $R_P$. Such analysis converges to $f=\SI{40.68}{\mega\hertz}$ for the discharge channel diameter of $\phi=\SI{37}{\milli\meter}$ as resulting in high $R_P$ for given $n$ compared to lower $f$. The required $B$-field is $B_0\sim 20-\SI{30}{\milli\tesla}$ for $n=\SI{1e18}{\meter^{-3}}$ to follow the $R_P$ peak. However, lower $B$-fields can be applied while still resulting in $R_P>\SI{1}{\ohm}$. If a lower $n$ is chosen, the required $B_0$ decreases, and a conductive top plate can be used to improve performance, especially by the low density peak~\cite{chen233}. The case at $f=\SI{13.56}{\mega\hertz}$ is analysed as well, and shows the possibility of delivering $R_P>\SI{1}{\ohm}$ for $n>\SI{4e18}{\meter^{-3}}$ and $B_0 \geq \SI{0.02}{\tesla}$. Finally, the resulting $R_P$ at  $f=\SI{40.68}{\mega\hertz}$ is more than double compared to that at $f=\SI{13.56}{\mega\hertz}$.

\subsubsection{Plasma Density Profile at 40.68 MHz}
For more precise evaluation, the typical helicon-discharge plasma density profile $n(r)$, represented by a Gaussian-like profile in which the peak is reached at the centre of the discharge channel~\cite{chen224}, is applied to the final $f=\SI{40.68}{\mega\hertz}$ case derived previously. The HELIC parameters are $s=2$, $t=4$, $fa=0.1$. The discharge channel has $L_C=\SI{180}{\milli\meter}$, $\phi=\SI{37}{\milli\meter}$, wall thickness of \SI{1.5}{\milli\meter}, the antenna is a 5-turns with $L_A=\SI{80}{\milli\meter}$ providing $m=+1$ mode. The antenna ends with the end section of the discharge channel, therefore $ZA=\SI{0.050}{\meter}$. Following, $R_P$ is displayed in Fig.~\ref{fig:last40ins} and Fig.~\ref{fig:last40cond}.
\begin{figure}[hp]
	\centering
	\includegraphics[width=.7\textwidth]{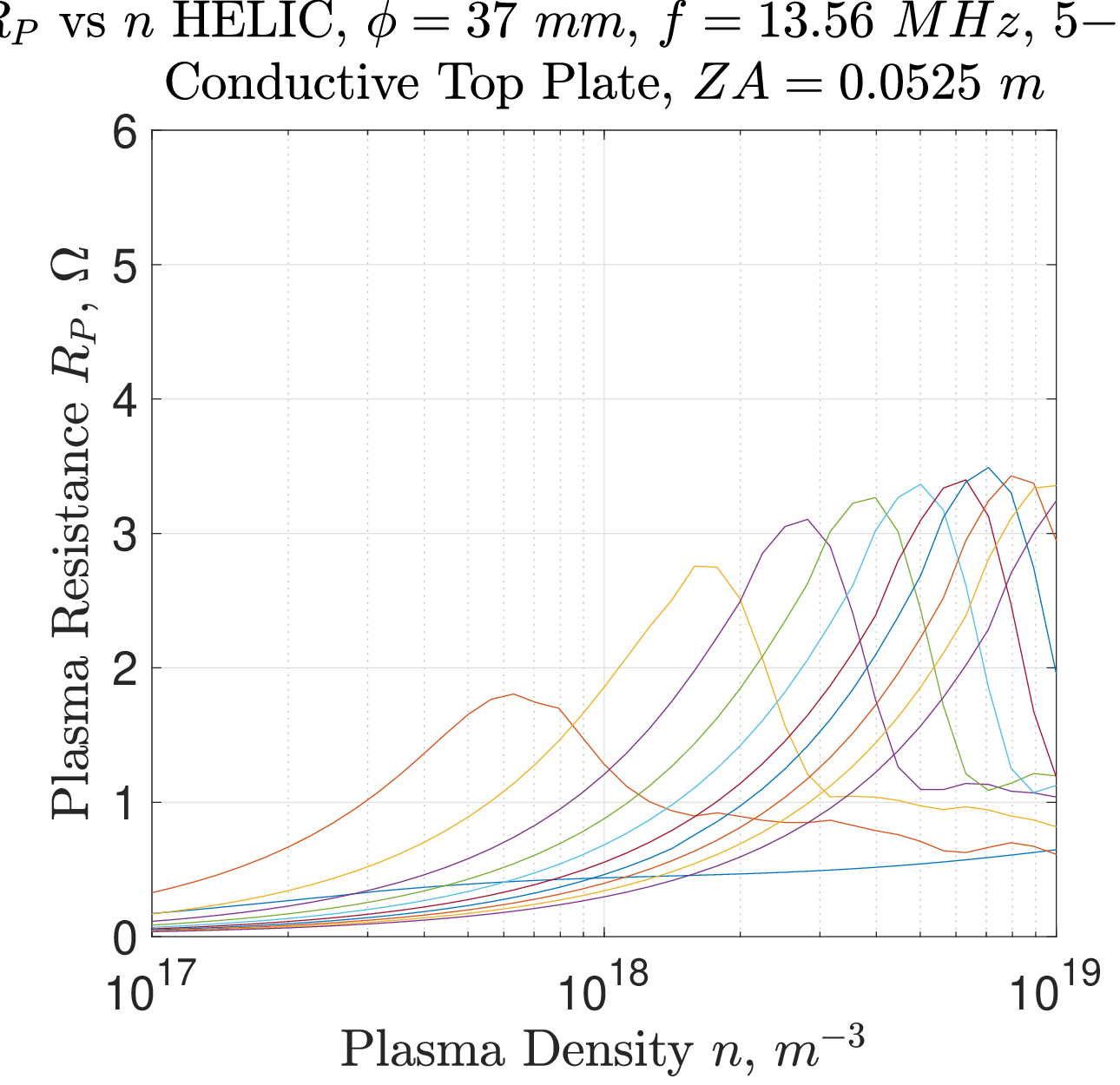}
	\caption{$R_P$ vs $n$ vs $B_0$, Conductive Top Plate, at $f=\SI{13.56}{\mega\hertz}$.}
	\label{fig:37_1356_good_c}
\end{figure}

\begin{figure}[hp]
	\centering
	\includegraphics[width=.7\textwidth]{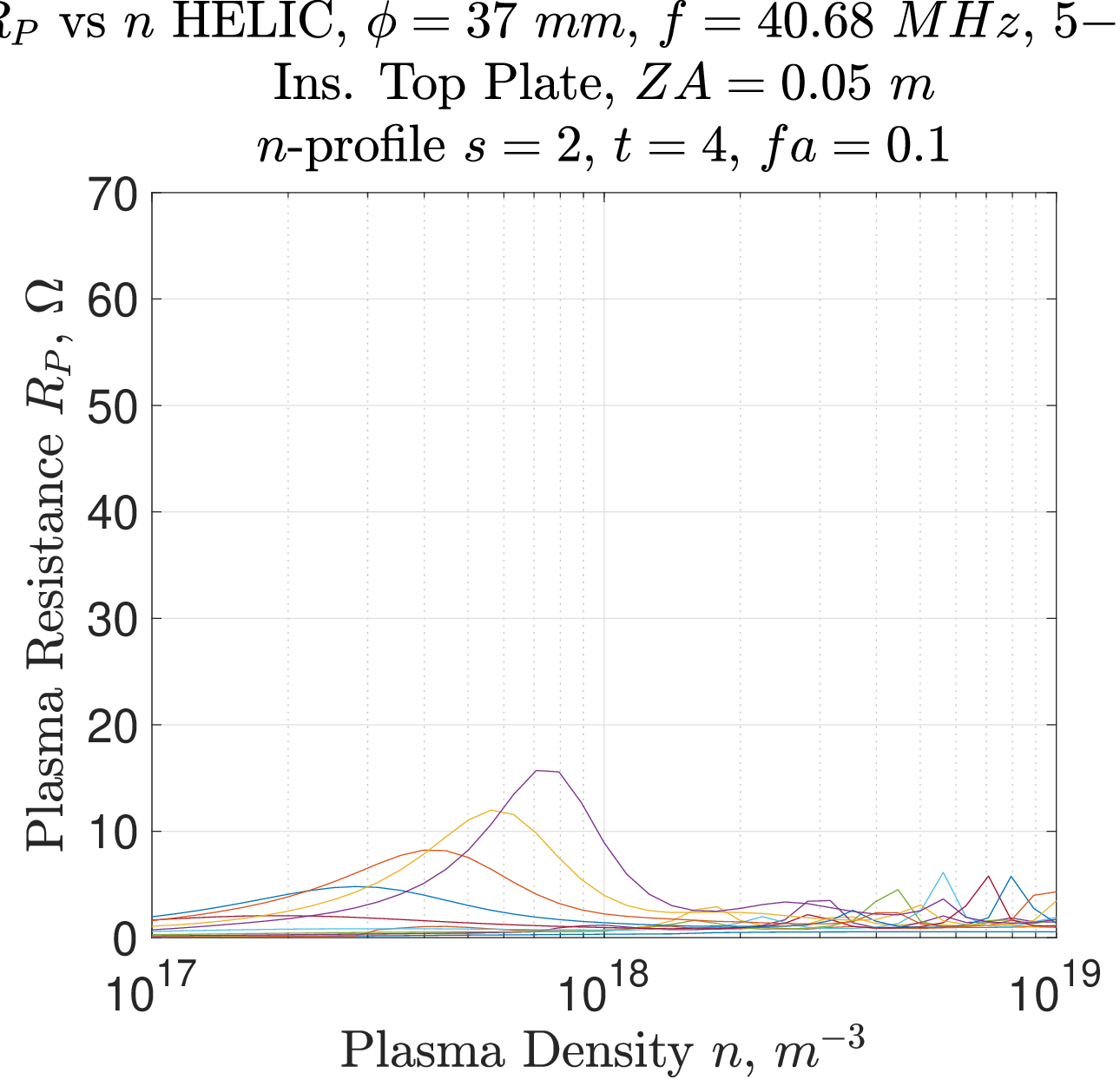}
	\caption{$R_P$ vs $n$ vs $B_0$, Insulating Top Plate, at $f=\SI{40.68}{\mega\hertz}$, \ce{Ar}, $n$-profile.}
	\label{fig:last40ins}
\end{figure}
\begin{figure}[hp]
	\centering
	\includegraphics[width=.7\textwidth]{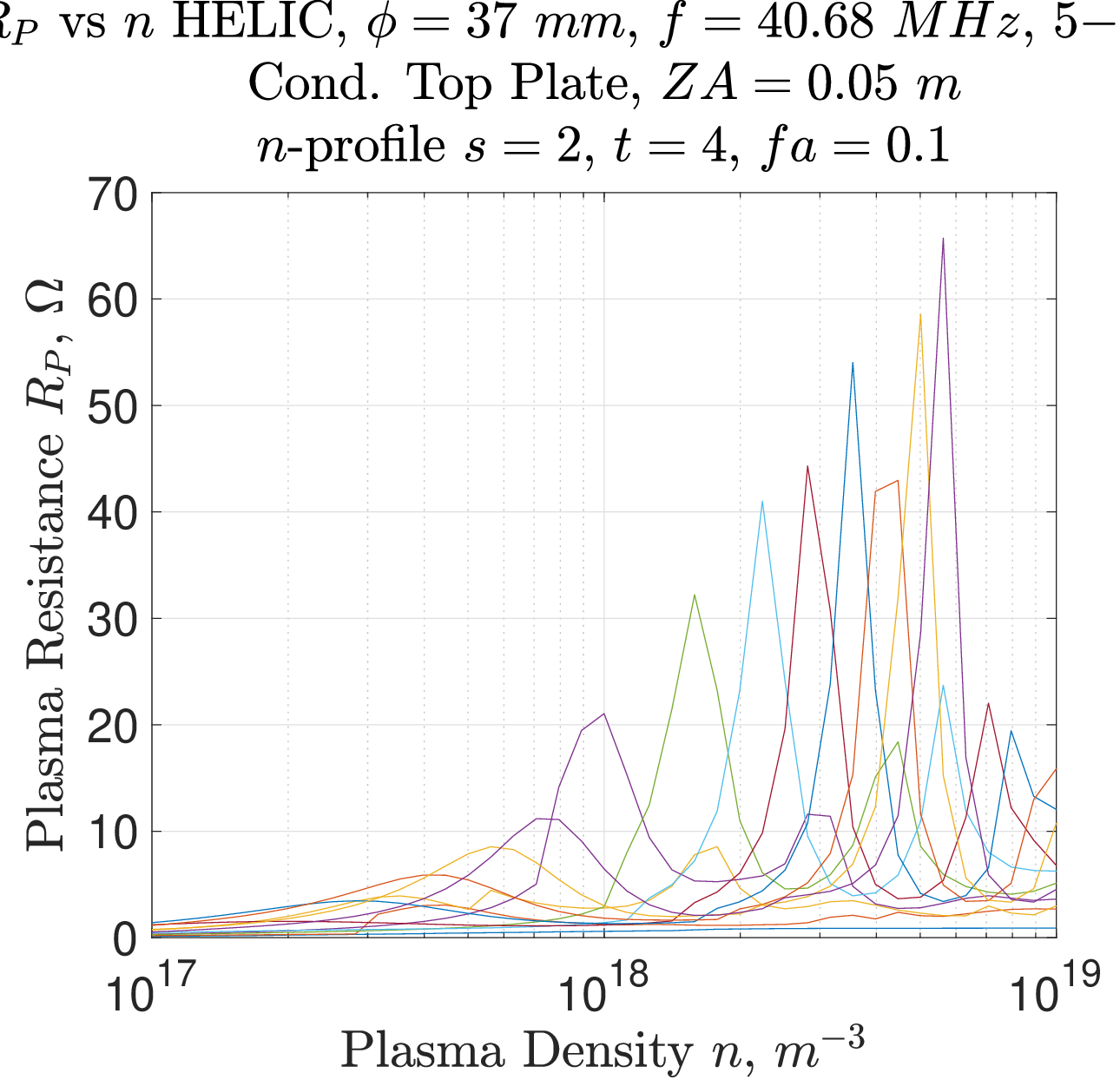}
	\caption{$R_P$ vs $n$ vs $B_0$, Conductive Top Plate, at $f=\SI{40.68}{\mega\hertz}$, \ce{Ar}, $n$-profile.}
	\label{fig:last40cond}
\end{figure}
Fig.~\ref{fig:last40ins} and Fig.~\ref{fig:last40cond} show that $R_P>\SI{1}{\ohm}$ can be achieved. With an insulating plate, the $R_P$ peaks are between $n\sim2\times10^{17}-\SI{8e17}{\meter^{-3}}$, whereas a conductive top plate would generate a much larger $R_P$ peaks at higher $n$, therefore widening the plasma density region of high $R_P$ for $B_0=30-\SI{100}{\milli\tesla}$. The analysis is performed also for \ce{N2} and it is shown in Fig.~\ref{fig:last40insN2} and Fig.~\ref{fig:last40condN2}. The general behaviour is similar to the \ce{Ar} case, but resulting in lower $R_P$.
\begin{figure}[hp]
	\centering
	\includegraphics[width=.7\textwidth]{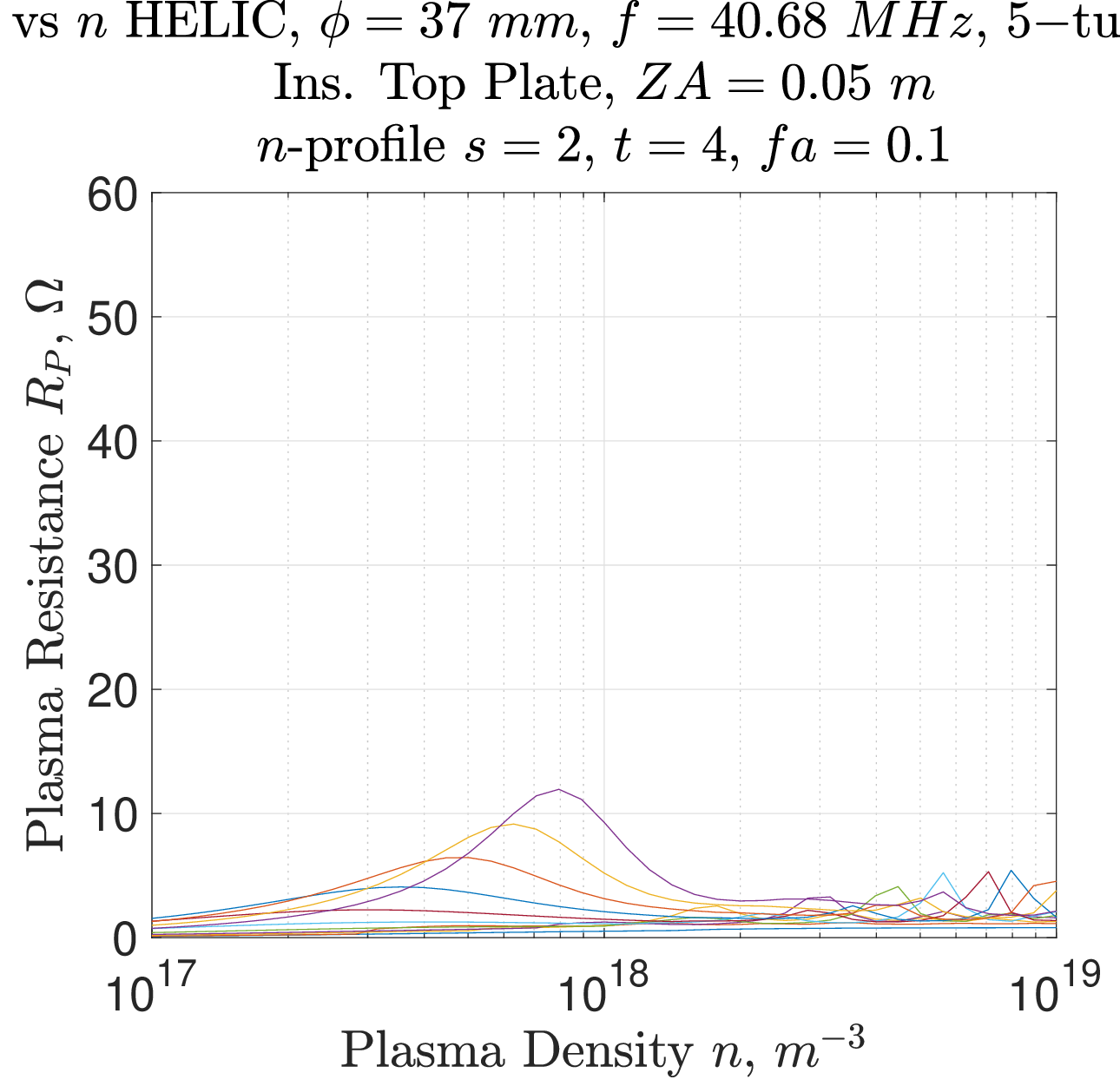}
	\caption{$R_P$ vs $n$ vs $B_0$, Insulating Top Plate, at $f=\SI{40.68}{\mega\hertz}$, \ce{N2}.}
	\label{fig:last40insN2}
\end{figure}
\begin{figure}[H]
	\centering
	\includegraphics[width=.7\textwidth]{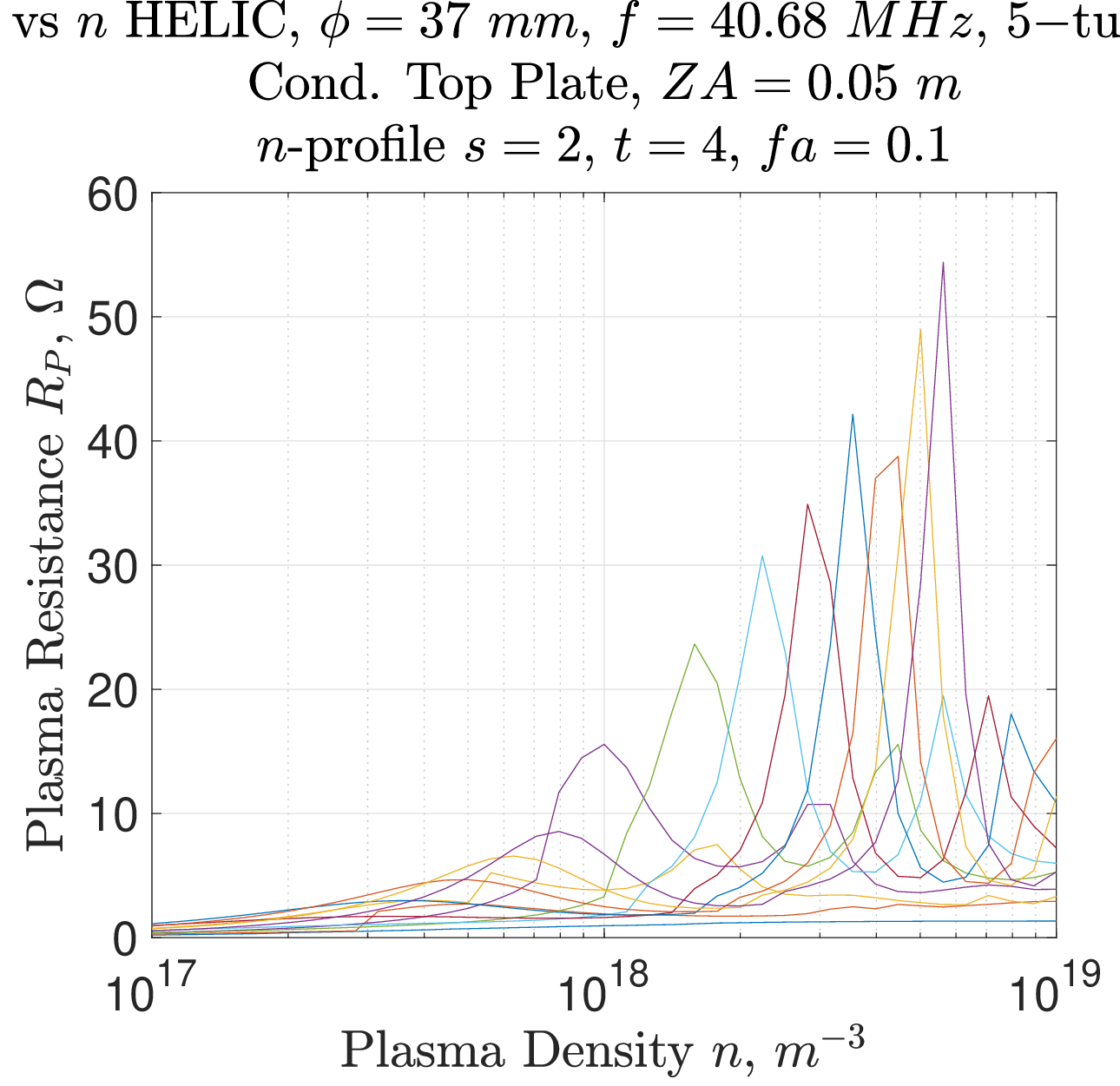}
	\caption{$R_P$ vs $n$ vs $B_0$, Conductive Top Plate, at $f=\SI{40.68}{\mega\hertz}$, \ce{N2}.}
	\label{fig:last40condN2}
\end{figure}

\subsection{Conclusions and Final Remarks}
Finally, the behaviour of $R_P$ has been qualitatively analysed, and the requirements quantitatively defined. An operating input frequency of $f=\SI{40.68}{\mega\hertz}$, a discharge channel of $\phi=\SI{37}{\milli\meter}$ diameter, $\SI{1.5}{\milli\meter}$ thickness, and $L=\SI{180}{\milli\meter}$ length are selected. The applied magnetic field shall provide an intensity between $B_0=10-\SI{100}{\milli\tesla}$ depending on the achieved plasma density $n$, that shall lay in the typical helicon plasma sources range of $n=1\times10^{17}-\SI{1e19}{\meter^{-3}}$~\cite{Chen_2015}. The magnetic field strength must be finally also tuned in terms of the desired confinement and the requirement for the magnetic nozzle in a thruster application. By using \ce{N2} as injected gas, the overall $R_P$ is slightly lower, in general $5-\SI{10}{\ohm}$ less compared to the \ce{Ar} case. In HELIC, this depends on the gas-dependent collision frequency that lead to different propagation of the waves, while in the real application one must take into account the different higher energy required for ionization. Finally, the behaviour stays qualitatively the same. 
Finally, HELIC does not account for the imaginary part of the impedance $Z$, the reactance $X$ of the plasma and of the antenna. Indeed, the plasma, and any electrical component in an AC circuit will have its resistance $R$, and its reactance $X$. The latter is given by the sum of the capacitive $X_C$ and the inductive term $X_L$. Especially $X_L=2\pi f L$, where $L$ is the inductance, is directly related to the input frequency. The higher the frequency, the higher its contribution, independently of the plasma itself, especially if a coil antenna with $N=5$ turns is implemented, each turn adds on $X_L$ and increases the mismatch, therefore the power loss, if this is not compensated by $X_C$. This leads to researching for a more optimized antenna that can provide a partially matched load. To discuss this further, the basics of AC circuitry must be introduced, see Section~\ref{ch:RF}.

\newpage
\subsection{The Case of IPG6-S}
A brief description of an experimental test campaign with the inductively-heated plasma generator IPG6-S~\cite{dropmann} operating with an applied $B$-field compared to the HELIC results is presented, as of being a representative case that aided to the development of the new thruster.
IPG6-S is a water cooled ICP-based plasma source with a discharge channel of $\phi = \SI{37}{\milli\meter}$ operating at $f=\SI{3.3}{\mega\hertz}$ based on a $5.5$-turns coil antenna, and it has been used as test-bed for the development of the ABEP-based thruster~\cite{romanorgcep,romanoiac,romanoiepc2,romanorgcep2,romanoiac2,romanosp2018,romanoacta}, see Fig.~\ref{fig:ipg6s}.
\begin{figure}[H]
\centering
\includegraphics[width=.8\textwidth, trim={1cm 1cm 0cm 0cm},clip]{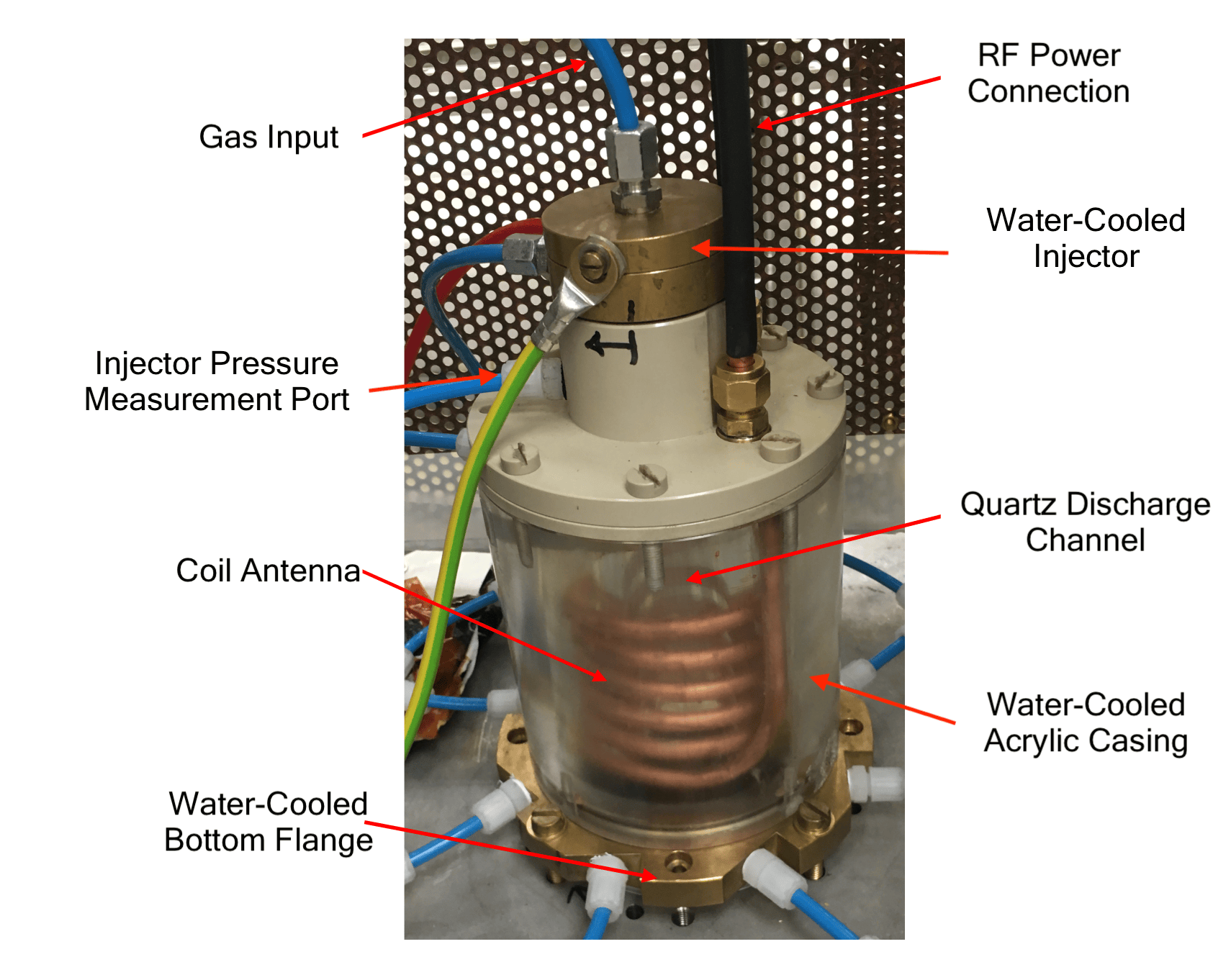}
\caption{Inductively-heated Plasma Generator IPG6-S.}
\label{fig:ipg6s}
\end{figure}

An applied magnetic field $B_0=5-\SI{66.5}{\milli\tesla}$ has been applied to IPG6-S, along the symmetry axis of the discharge channel, for various gas flows of~\ce{Ar},~\ce{N2}, and~\ce{O2}. As the specific power supply provides as output the absorbed power by the load $P_{RF}$, instead of the forward power $P_{in}$, the relative change in power absorption can be evaluated. Indeed, the following is observed for all gases:
\begin{itemize}
\item Relatively small $B$-fields, $B_0\sim 5-\SI{10}{\milli\tesla}$ trigger ignition at low $P_f$ and $\dot{m}$;
\item $B_0$ can tune the reflected power $P_r$;
\item $B_0$ leads to large variations in the absorbed power $P_{abs}$;
\item $B_0$ leads to a visually confined plasma.
\end{itemize}

In more detail, plasma ignition can be triggered at low powers and mass flows if a small $B_0$ is applied. The $P_r$ is indirectly observed in the oscillation of the $P_{abs}$ output analogue signal, and can be tuned, either reduced or amplified by tuning the amplitude of $B_0$. In particular the case of $\dot{m}=\SI{4.36}{\milli\gram\per{\second}}$~\ce{O2} shows that an increase of $B_0=5 \rightarrow \SI{27}{\milli\tesla}$ triplicates $P_{abs}$ as well as the pressure in the discharge channel measured by $p_{inj}$, see Fig.~\ref{fig:IPG6-SO2}, where $P_{abs}=P_{RF}$, $P_{gen,cool}$ is the water cooling power of IPG6-S, $I_A$ is the input current, and $p_{inj}$ the pressure measured just after the injector head, Fig.~\ref{fig:ipg6s}. Instead, $B_0$ could not be recorded along the other parameters, and it is therefore shown by single values at given times shown by vertical arrows. 
\begin{figure}[H]
\centering
\includegraphics[width=.9\textwidth]{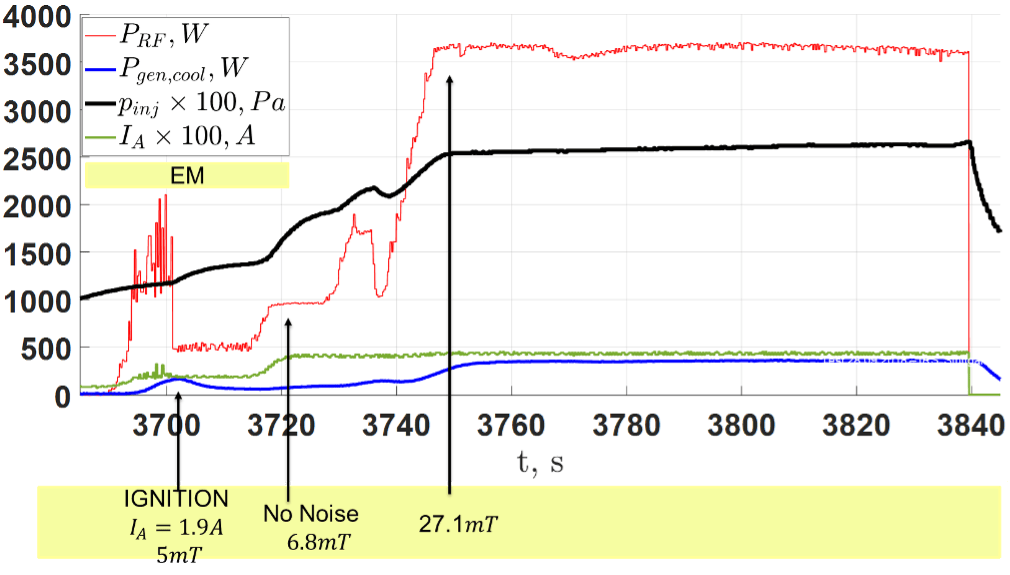}
\caption{IPG6-S Test Results: \ce{O2} at $\dot{m}=\SI{4.36}{\milli\gram\per{\second}}$~\cite{romanorgcep2}.}
\label{fig:IPG6-SO2}
\end{figure}
 Instead, when operating at $\dot{m}=\SI{1.38}{\milli\gram\per{\second}}$ it is shown that by raising $B_0$, $P_{abs}$ is also increased but showing large oscillations possibly indicating large $P_r$, a sudden plasma disruption happens once $P_f$ is slightly increased. This might indicate a shift in the $Z$ of load and plasma to a condition in which $R_P$ becomes too low compared to $R_C$.
Based on the previous analysis with HELIC, in particular Fig.~\ref{fig:37_33}, and on the experiments performed with IPG6-S, more in detail reported in~\cite{romanorgcep2}, the direct influence of $B_0$ on $P_{abs}$ is shown. It indirectly highlights the change in $R_P$ as in the HELIC analysis, and the possibility to have developed a W-regime that increased coupling. Also, the improvement due the applied magnetic field can be also due to the enhanced plasma confinement that is reported to improve the coupling in E and~H regimes~\cite{boswell1984very}. Finally, the positive effect of $B_0$ on power absorption, tuning, and ignition at low pressure, the latter an important parameter for an ABEP-based thruster, are experimentally shown.

\newpage
	\section{Radio-Frequency}
	\label{ch:RF}
	This section provides a brief introduction of alternate current (AC) theory with the most required notions necessary to understand the RF thruster design.

\subsection{Basic Definitions and Fundamentals}
In AC, the flow of electric charge periodically reverses its direction, so that both current and voltage variate over time in the form of a sinusoidal-like curve. The number of oscillations per unit of time of current and voltage is the frequency, in particular RF is the range between $f=\SI{3}{\hertz} - \SI{300}{\giga\hertz}$~\cite{chen2016handbook}.
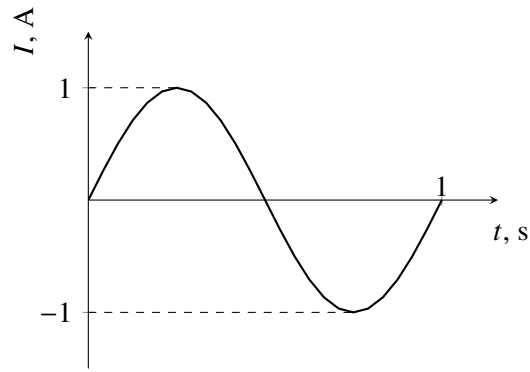
\begin{figure}[H]
	\centering
	\begin{tikzpicture}[scale=1]
	\centering
	\begin{axis}[
	axis lines=middle,
	x label style={at={(axis description cs:1.1,.4)},anchor=east},
	y label style={at={(axis description cs:-0.1,1)},rotate=90,anchor=south},
	xmin=0,xmax=2*pi+1,ymin=-1.5,ymax=1.5,
	ylabel={$I$, \SI{}{\ampere}},
	xlabel={$t$, \SI{}{\second}},
	xtick={0, 2*pi},
    xticklabel style={font=\normalfont, above},
	xticklabels={0, 1},
	]
	\addplot+[thick,grid,no marks,domain=0:2*pi,black]{sin(deg(x))};
	\addplot+[dashed,grid,no marks,domain=0:pi/2,black]{1};
	\addplot+[dashed,grid,no marks,domain=0:2*pi-pi/2,black]{-1};
	\end{axis}
	\end{tikzpicture}
\caption{Current Waveform in AC, at \SI{1}{\hertz}.}
\label{fig:accurrent}
\end{figure}

The main differences in AC compared to direct current (DC) are:\begin{itemize}
	\item Current and voltage periodically vary over time;
	\item Current flows only within an external thin layer of the conductor, the thickness of this layer is defined as the skin depth;
	\item At high frequencies, the alternating fields cause emission of electromagnetic waves from the conductor to the surrounding space;
	\item Resistance has an extended definition, with both real and imaginary part: the impedance;
	\item Any circuit in AC with resistors, capacitors and inductors has its resonance frequencies.
\end{itemize}

\subsubsection{Skin Depth}
The skin depth $\delta$, see Eq.~\ref{eq:skindepth}, is defined as the thickness of the layer of the conductor where the current flows, precisely until where the current density falls to $1/e$ of the initial value.
\begin{equation}
	\delta=\sqrt{\frac{\rho}{\pi f\mu}}
	\label{eq:skindepth}
\end{equation}
Here, $\rho$ is the resistivity, and $\mu$ the magnetic permeability of the conductor. Considering copper at the design frequency of the thruster, $f=\SI{40.68}{\mega\hertz}$, $\delta=\SI{0.0115}{\milli\meter}$. In the practical case, it is assumed that $98\%$ of the current flows within a layer that is a few times $\delta$~\cite{chen2016handbook}. 

\subsubsection{Proximity Effect}
The flow of AC current in any exposed conductor causes it to emit electromagnetic waves that interact with any conductive material located nearby. In particular, the resulting changing magnetic field flux $\Phi_B=\iint_{A} \vec{B}(t)\cdot d\vec{A}$ being the surface integral of the $B$-field through the cross section $A$, induces electromotive forces $\varepsilon$ on such nearby materials, according to Faraday's law, see Eq.~\ref{eq:faraday}.
\begin{equation}
\varepsilon=-\frac{d\Phi_B}{dt}
\label{eq:faraday}
\end{equation}
The currents deriving from $\varepsilon$ are called eddy currents and flow in closed paths to produce a magnetic field that is opposite to the one inducing them. Those currents can distort the applied electromagnetic fields, and potentially cause power losses and Joule heating. Such effects are to be taken into account for the thruster design, as many components with more or less conductive parts are at close distance to the antenna.
 
\subsubsection{Impedance, Inductance, Capacitance}
In AC, the resistance $R$ definition is extended to the impedance $Z$, a phasor composed by the resistance $R$ plus the imaginary component named the reactance $X$. The phasor $Z$ rotates on the real-imaginary plane forming an angle $\theta$ with the real axis, see Eq.~\ref{eq:impedance} and Fig.~\ref{fig:phasor}.
\begin{equation}
	\vec{Z}=\abs{Z}e^{j\theta}=\vec{R}+\vec{X}i
	\label{eq:impedance}
\end{equation}
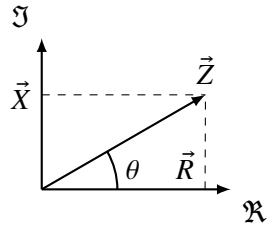
\begin{figure}[H]
	\centering
\begin{tikzpicture}[>=latex]
\draw[style=help lines] (0,0) (3,2);
\coordinate (vec2) at (30:2.5);
\coordinate (vec3) at (0:2.5);
\coordinate (vec4) at (90:2);
\coordinate (vec5) at (270:2);
\coordinate (vec6) at (180:2);
\draw[->,thick,black] (0,0) -- (vec2) node[above] {$\vec{Z}$};
\draw[->,thick,black] (0,0) -- (vec3) node [below right] {$\Re$};
\draw[->,thick,black] (0,0) -- (vec4) node [above left] {$\Im$};
\draw[dashed](0,1.25) node [left] {$\vec{X}$} -- (vec2);
\draw[dashed]  (2.16,0) node [above left] {$\vec{R}$} -- (vec2);
\draw [black, thick ] (1.0,0) arc (0:30:1cm)    node [midway, right] {$\theta$};    
\end{tikzpicture}
\caption{Impedance $Z$ Phasor.}
\label{fig:phasor}
\end{figure}

The reactance $X$ is the sum of the inductive $X_L$ and the capacitive term $X_C$, see Eq.~\ref{eq:inductance} and Eq.~\ref{eq:capacitance}.
\begin{equation}
	X_L =2\pi f L
	\label{eq:inductance}
\end{equation}
\begin{equation}
X_C =-\frac{1}{2\pi f C}
\label{eq:capacitance}
\end{equation}
The equivalent capacitance and inductance of the circuit are $C$ and $L$, respectively. In phasorial coordinates, $\theta=\arctan{(X/R)}$.

\subsection{Radio-Frequency Generation}
The basic circuit required to describe the operating principle of RF generators is the \textit{RLC} one. It consists of a DC source, a resistor $R$, an inductor $L$, and a capacitor $C$, see Fig.~\ref{fig:RLC} and Fig.~\ref{fig:RLC2}.
\begin{figure}[H]
	\centering 
	\begin{tikzpicture}
	\draw[step=0.5,very thin,black!20] (-1,-0.5);
	\path (0,0) coordinate (ref_gnd);
	\draw
	(ref_gnd) to[american voltage source=\(V\)] ++(0,2)
	to[R=\(R\)] ++(3,0) 
	to[L=\(L\)] ++(2,0) 
	to[C=\(C\)] ++(0,-2) 
	-- (ref_gnd);
	\end{tikzpicture}
\caption{Series RLC.}
	\label{fig:RLC}
		\vspace{10pt}
		\begin{tikzpicture}
	\draw[step=0.5,very thin,black!20] (-1,-1);
	\path (0,0) coordinate (ref_gnd);
	\draw
	(ref_gnd) to[american voltage source=\(V\)] ++(0,2)
	-- ++(2,0)   to[R=\(R\)] ++(0,-2) -- (ref_gnd) 
	++(2,2)   -- ++(1.5,0) to[L=\(L\)] ++(0,-2) -- ++(-1.5,0)
	++(1.5,2) -- ++(1.5,0) to[C=\(C\)] ++(0,-2) -- ++(-1.5,0);
	\fill[color=black] (ref_gnd)++(2,0)   circle[radius=0.08];
	\fill[color=black] (ref_gnd)++(3.5,0) circle[radius=0.08];
	\fill[color=black] (ref_gnd)++(2,2)   circle[radius=0.08];
	\fill[color=black] (ref_gnd)++(3.5,2) circle[radius=0.08];
	\end{tikzpicture}\vspace{-10pt}
	\caption{Parallel RLC.}
	\label{fig:RLC2}
\end{figure}
Once the DC voltage source is switched on, the energy flows through $R$,  $L$, and $C$. Due to the circuit configuration, the energy flow periodically reverses its direction, therefore creating an AC signal. If the system is run at its resonant frequency $\omega_r$ see Eq.~\ref{eq:resonancedef}, and by considering there are no losses ($R=\SI{0}{\ohm}$ in the superconductive regime), the DC source could be disconnected and the system would run for an infinite time.
\begin{equation}
	\omega_r= 2\pi f_R = \frac{1}{\sqrt{LC}}
	\label{eq:resonancedef}
\end{equation}


\subsubsection{Power in AC}
The power in AC is of two kinds: active and reactive. The active power $P$ is that transferred into one direction, while the reactive power $Q=P_r$ is that reflected back to the power source. If the load is matched to exactly the same $\vec{Z}$ of the source, then $P_r=\SI{0}{\watt}$ and the respective phase shift of the voltage $V$ is $\varphi=\SI{0}{\degree}$. If the phase difference between the two is $\varphi=\SI{90}{\degree}$, then the load is purely reactive and all the power is reflected back. The apparent power $\abs{S}$ is the one sent from the AC generator. The current and voltage root mean squares are respectively $I_{rms}$ and $V_{rms}$. By definition:
\begin{itemize}
\item Active power $P$:
\begin{equation}
P=\abs{S}\cos{\varphi}=V_{rms}I_{rms}\cos{\varphi}=\frac{\vec{\abs{V}}^2}{\vec{\abs{Z}}^2} R
\end{equation}

\item Reactive power $Q=P_r$:
\begin{equation}
Q=P_r=\abs{S}\sin{\varphi}=\frac{\vec{\abs{V}}^2}{\vec{\abs{Z}}^2} X
\end{equation}

\item Complex power $\vec{S}$:
\begin{equation}
\vec{S}=\vec{V}\vec{I}^*=\vec{\abs{I}}^2 \vec{Z}=\frac{\vec{\abs{V}}^2}{\vec{Z}^*}
\end{equation}

\item Apparent power $\abs{S}=P_f$:
\begin{equation}
\abs{S}=\sqrt{P^2+Q^2}=V_{rms}I_{rms}
\end{equation}
%
\end{itemize}
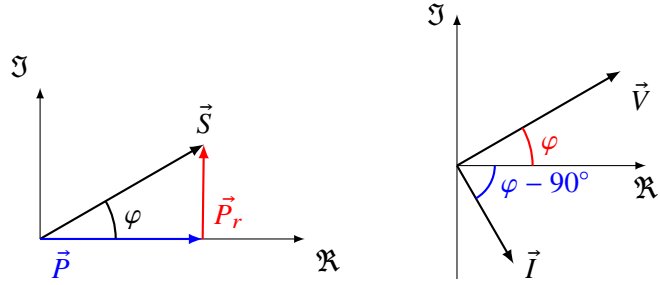
\begin{figure}[H]
	\centering
	\begin{tikzpicture}[>=latex]
	\draw[style=help lines] (0,0) (3,2);
	\coordinate (vec2) at (30:2.5);
	\coordinate (vec3) at (0:3.5);
	\coordinate (vec4) at (90:2);
	\coordinate (vec5) at (270:2);
	\coordinate (vec6) at (180:2);
	\draw[->,thick,black] (0,0) -- (vec2) node[above] {$\vec{S}$};
	\draw[->,black] (0,0) -- (vec3) node [below right] {$\Re$};
	\draw[->,black] (0,0) -- (vec4) node [above left] {$\Im$};
	\draw[->, thick, blue](0,0) node [below right] {$\vec{P}$} -- (2.15,0);
	\draw[->, thick, red] (2.15,0) node [above right] {$\vec{P_r}$} --  (vec2);
	\draw [black, thick ] (1.0,0) arc (0:30:1cm)    node [midway, right] {$\varphi$};    
	\end{tikzpicture}
	\qquad
	\begin{tikzpicture}[>=latex]
		\draw[style=help lines] (0,0) (3,2);
		\coordinate (vec1) at (300:1.5); 
		\coordinate (vec2) at (30:2.5);
		\coordinate (vec3) at (0:2.5);
		\coordinate (vec4) at (90:2);
		\coordinate (vec5) at (270:2);
		\coordinate (vec6) at (180:2);
		\draw[->,thick,black] (0,0) -- (vec1) node[right] {$\vec{I}$};
		\draw[->,thick,black] (0,0) -- (vec2) node[below right] {$\vec{V}$};
		\draw[->,black] (0,0) -- (vec3) node [below] {$\Re$};
		\draw[->,black] (0,-1.5) -- (vec4) node [left] {$\Im$};
		\draw [red, thick ] (1.0,0) arc (0:30:1cm)    node [midway, right] {$\varphi$};    
		\draw [blue, thick] (0.5,0) arc (0:-60:0.5cm) node [midway, right] {$\varphi-\ang{90}$};  
		\end{tikzpicture}
		\caption{Power phasor (left), $\vec{I}$ and $\vec{V}$ phasor (right) represented on the complex plane.}
		\label{fig:fasoreIV}
\end{figure} 
Losses and reflections of power in the circuit make $I$ and $V$ not having the peak at the same instant of time, resulting in a phase difference $\varphi$, see Fig.~\ref{fig:fasoreIV}. Inductive circuits have $\varphi >\SI{0}{\degree}$, vice versa capacitive circuits have $\varphi <\SI{0}{\degree}$. 

\subsection{Impedance Matching}
\label{sec:matching}
Impedance matching~\cite{davis2001radio} is of crucial importance for an efficient power transmission in AC (RF), from the source, the RF generator, to the load, in this case the thruster. The maximum power transfer from source to load is achieved only if the impedance of the source $Z_S$ is "matched" to the conjugate ($^*$) of the load, corresponding to $Z_{S}=Z^*_{L}$ that means $R_{S}+jX_{S}=R_{L}-jX_{L}$. This can be easily seen in the reflection coefficient $\Gamma$ definition in Eq.~\ref{eq:reflection}.
\begin{equation}
\Gamma=\frac{Z_L-Z_S}{Z_L+Z_S}
\label{eq:reflection}
\end{equation}
RF generators (source), usually deliver power at a given impedance $Z_S=50+j0\SI{}{\ohm}$ that is purely resistive (common industrial standard). To transfer all the power from source to load, the condition is that $Z_L=50+j0\SI{}{\ohm}$, also corresponding to $\varphi=\SI{0}{\degree}$. Concerning the thruster circuit, the load is the thruster (IPT) itself, and presents a variable value that can be written as $Z_{L}=Z_{IPT,circuit}+Z_{plasma}$.
The plasma is a variable impedance $Z_{plasma}$ because of variations that are due to applied $B$-field, propellant flow rate and composition variation, and input power. Therefore, the RF generator must cope with a variable $Z_L$. To protect it from damage, a second device, the matching network, is installed in the circuit between the source and the load. This device is made of variable capacitors and inductors that are either manually or automatically operated with the aim of forming a circuit that incorporates $Z_L$ with a resulting impedance matched to $Z_S$. This ensures that the RF generator always sees the same $Z$, and therefore the $P_r$ reaching it is minimized. Finally, $Z_L$ is only seen by the matching network, and the reflected power is partly dissipated as Joule heating $Q_{diss}$ within the matching network itself. The conceptual scheme of the thruster circuit is shown in Fig.~\ref{fig:RF_circ}.
\begin{figure}[H]
	\centering
	\includegraphics[width=0.9\textwidth]{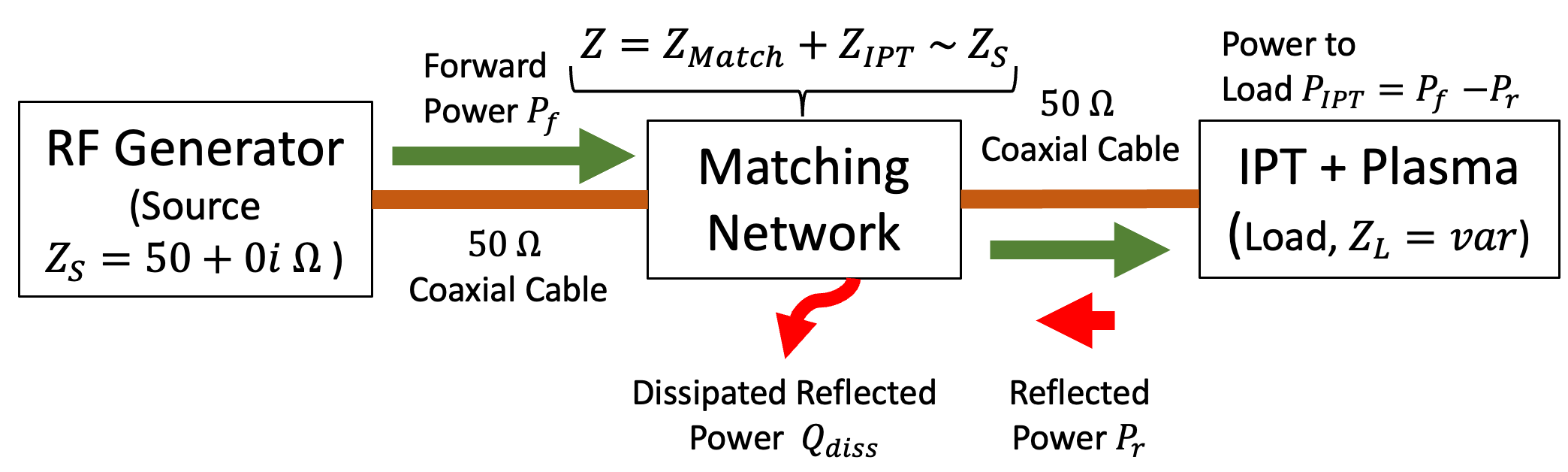}
	\caption{Simplified RF Circuitry with Thruster (IPT) as Load.}
	\label{fig:RF_circ}
\end{figure}

Finally, the matching network does not improve the load itself, but it maximizes the power transfer from one branch of the circuit to another. That is, forming a RLC circuit with an impedance $Z=50+j0\SI{}{\ohm}$ composed by the load plus the matching network, in which the power flows. The load represented by $Z_L$, therefore antenna and plasma, is not changed by the matching network.

Fundamental is that, at resonance, $Z$ is totally resistive, $Z=R$, hence, $X=\SI{0}{\ohm}$. This is the case that is later analysed within this dissertation concerning the plasma thruster design, as it leads to potential minimized power losses, finally leading to a partially optimized circuitry and a maximized, from the electrical circuit point of view, high efficiency plasma thruster. 

\subsection{RF Antennae}
An antenna is defined as the interface between propagating electromagnetic waves through a medium and electric currents travelling within a conductor. An antenna can work as a transmitter or as a receiver. In the field of EP, antennae are used to transmit the EM waves to the gas contained within the discharge channel to ionize it and, eventually, accelerate it to produce thrust. An antenna can have different shapes depending on the application, and each geometry delivers a certain configuration of EM fields. To ensure an efficient power transmission, antennae needs to be matched to the correct $Z$, and tuned to operate at the $f$ of the supplied signal. An important reference value is the $S_{11}$ parameter. The $S_{11}$ is one of the element of the S-matrix composed by the scattering parameters, see Eq.~\ref{eq:smatr}.
\begin{equation}
S=
\begin{pmatrix}
S_{11} & S_{12} \\
S_{21} & S_{22} \\
\end{pmatrix}
\label{eq:smatr}
\end{equation} 
It defines how a signal sent from port number 1 reaches port number 2. The reverse voltage gain is $S_{12}$, while $S_{21}$ is the forward voltage gain. The output port voltage reflection coefficient is $S_{22}$, while $S_{11}$ is the input port voltage reflection coefficient~\cite{balanis}. The $S_{11}$ is equal to the reflection coefficient $\Gamma$ and provides information on how efficiently a signal is transmitted at a certain $f$ to the antenna. In particular, the resonance frequencies of an antenna can be found by plotting the $S_{11}$ parameter over $f$. Its definition based on reflected and absorbed power is $\abs{S_{11}}=10 \log{ (P_r/P_{in})}$, measured in $\SI{}{\decibel}$, where $P_{r}$ is the reflected power and $P_{in}$ the incident/input power in $\SI{}{\watt}$, some common $S_{11}$ levels are shown in Tab.~\ref{tab:s11pr}.
\begin{table}[H]
	\centering
	\caption{$S_{11}$ vs Power Ratio.}
	\label{tab:s11pr}
	\begin{tabular}{rccc}
		\toprule
		$S_{11}$ & $P_{in}$ & $P_{abs}$ & $P_{r}$\\
		\SI{}{\decibel} & $\%$ & $\%$ & $\%$\\
		\midrule
		$-20$ & $100$ & $99$ & 1\\
		$-10$ & " & $90$ & 10\\
		$-6$ &  " &$75$ & 25\\
		$-3$ &  " &$50$ & 50\\
		$-1$ &  " &$20$ & 80\\
		$0$ &  " &$0$ & 100\\
		\bottomrule
	\end{tabular}
\end{table}

For $S_{11}=\SI{0}{\decibel}$, $0\%$ of the power is absorbed and $100\%$ is reflected back. 
At $S_{11}=\SI{-20}{\decibel}$, $99\%$ of the power fed at port 1 is absorbed. 

Finally, the closer $Z_L$ is to $Z_S$, the less power is reflected back, therefore increasing the power absorption and, consequently, the power ratio. 
The best case, coinciding with $S_{11}<\SI{-20}{\decibel}$ is when the condition $Z_S=Z^*_{L}$ is satisfied, see Subsection~\ref{sec:matching}. 

\subsection{Transmission Lines}
As the unknown variable is the plasma impedance, the RF circuit can be a priori designed to minimize losses. This is composed of RF generator, matching network, thruster, cabling, and connectors. As stated in Sec.~\ref{sec:matching}, any mismatch in $Z$ results in power reflected that need to be somehow dissipated, finally reducing the overall efficiency of the plasma thruster system. The circuit can be divided in two branches, see Fig.~\ref{fig:RF_circ}. The first branch is made by the RF generator that sees the second branch, composed by the matching network and the load, as a matched load. 
To reduce mismatches and losses the coaxial cable characteristic must be carefully selected, in particular:
\begin{itemize}
	\item Coaxial cable impedance must be the same as the source's;
	\item Coaxial cable length shall be shorter than the wavelength $\lambda+\frac{c}{f}$, with $c$ the speed of light.
\end{itemize}
More in detail, if the coaxial length is comparable to $\lambda$, the cable has to be considered as a "transmission line" and must be carefully integrated into the circuit. Indeed, its length leads to impedance transformations. The second branch is composed of the matching network and the load (plasma and thruster), but also of the connections between matching network and the thruster's antenna. The second branch shall have a total impedance $Z$ equal to $Z_S$ in ideal conditions. The plasma possess its own impedance, $Z_{plasma}$, that changes during operation and it is generally very difficult to predict. The impedance of the thruster, $Z_{IPT}$, instead, is defined by its geometry and electrical design. The connections between the matching network and the thruster can be realized in two ways, see Fig.~\ref{fig:balunbal}, by using balanced or unbalanced lines.
\begin{figure}[H]
	\centering
	\includegraphics[width=.5\textwidth]{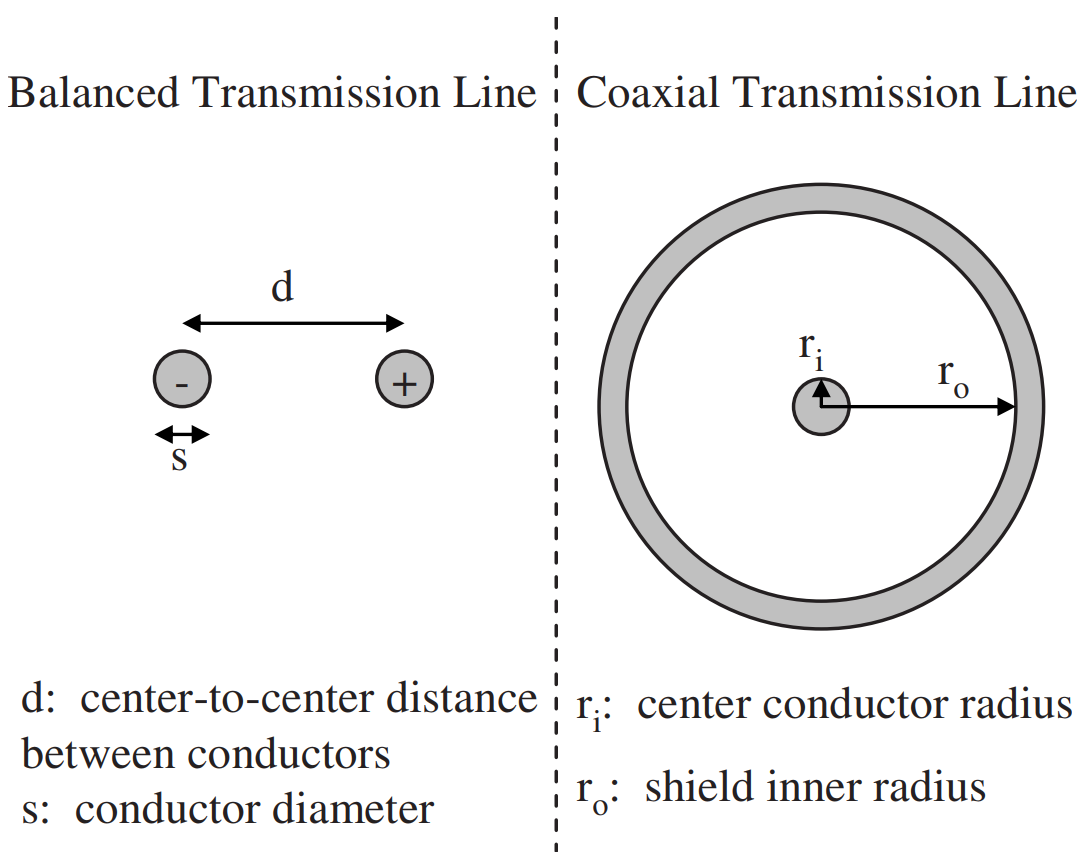}
	\caption{Balanced and Unbalanced (Coaxial) Transmission Lines~\cite{walkerRF}.}
	\label{fig:balunbal}
\end{figure}

In a balanced line, the electrical connections from the matching network to the thruster's antenna can be composed by a pair of conductors placed parallelly at an equal distance. Voltage and current are equal and opposite to the ones flowing in the other conductor. Their disadvantages is first, that the spacing between them must be kept exactly constant, and second, as they are not shielded, they emit EM waves, therefore interacting with the environment close to them, finally leading to losses. These lines can be shielded and the required constant distance can be achieved by using solid conductor rods, for example. An unbalanced line, instead, uses a single conductor that is provided with a shield that is the ground: a coaxial cable. Such lines do not couple power of nearby surfaces and the impedance is constant along their length. It is important to select the correct coaxial cable, in particular its impedance, maximum power, length $L$, and velocity factor VF, defining the speed of propagation of the signal through the coaxial cable with respect to the speed of light and leading on an effective wavelength that is $\lambda_{eff}=\lambda$VF. 

Finally, an unbalanced line is chosen for the thruster system, for being more robust and easier to implement for a laboratory model. In particular, the total impedance $Z$ of a coaxial cable of $Z_{cable}$ being a transmission line and considered lossless, is shown in Eq.~\ref{eq:trans}~\cite{walkerRF}. It highlights how its length $L$ related to the wavelength $\lambda$ can transform the $Z$ of the system.
\begin{equation}
	Z=Z_{cable}\frac{Z_{load}+i Z_{cable}\tan{(2\pi L/\lambda)}}{Z_{cable}+i Z_{load}\tan{(2\pi L/\lambda)}}
	\label{eq:trans}
\end{equation} 
The detail of the select connections are shown in Sec.~\ref{sec:iptrf}

\subsection{Thruster Optimized RF Circuit}
\label{sec:iptrf}
In the following, the details on the selected components of the thruster RF circuit, the respective optimization choice, and their implementation are presented. The IPT thruster facility is composed by:
\begin{itemize}
	\item RF Generator: Advanced Energy CESAR RF 4040;
	\item Auto-Matching Network: Advanced Energy Navigator 4040-L70;
	\item Coaxial cables $RG393/U$ \SI{50}{\ohm} VF$=0.695$;
	\item Thruster IPT;
	\item Rogowski Coil: CWT Mini 50HF/3.
\end{itemize}

The RF generator provides an input power of up to $P_{in,max}=\SI{4}{\kilo\watt}$ at $f=\SI{40.68}{\mega\hertz}$ $\pm 0.005\%$ with a $Z_S=50+j0\SI{}{\ohm}$ and it is water and air cooled. 

The auto-matching network is an L-type, designed to operate at $f=\SI{40.68}{\mega\hertz}$, match to a $Z_S=50+j0\SI{}{\ohm}$, and can cope up to $I_{rms}=\SI{70}{\ampere}$ and a $V_{peak}=\SI{5}{\kilo\volt}$, the tuning time is $t_{tuning}<\SI{1}{\second}$ according to the user manual.

The coaxial cables are all of the $RG393/U$~\SI{50}{\ohm} with VF$=0.695$ to cope with the expected power, voltages, and currents levels, and are provided with double shielding for both safety and increased robustness of the system to external EM disturbances. 

The coaxial cable connecting the RF Generator to the auto-matching network has $7/16"$~\SI{50}{\ohm} connectors and a calibrated length of $L=\lambda_{eff}/4$ to allow a further filtering of any reflected power back to the RF generator. 

The coaxial cable connecting the auto-matching network to the thruster, instead, has N-type~\SI{50}{\ohm} connectors and a calibrated length of $L=\lambda_{eff}/2$ to ensures that there is no $Z$ transformation due to the transmission line between the matching network and the thruster. 

According to Eq.~\ref{eq:trans}, by choosing $L=\lambda_{eff}/2$ the imaginary component in Eq.~\ref{eq:trans} disappears, and the auto-matching network must transform only $Z_{load}$, while the cable is totally invisible in terms of $Z$.

The thruster is hereby designed to provide a targeted $Z_{IPT,circuit}=50+j0\SI{}{\ohm}$ in absence of plasma at $f=\SI{40.68}{\mega\hertz}$. The presence of the plasma will change $Z_{IPT}$ and is to be transformed by the auto-matching network, see Fig.~\ref{fig:RF_circ2}.
\begin{figure}[H]
	\centering
	\includegraphics[width=1\textwidth]{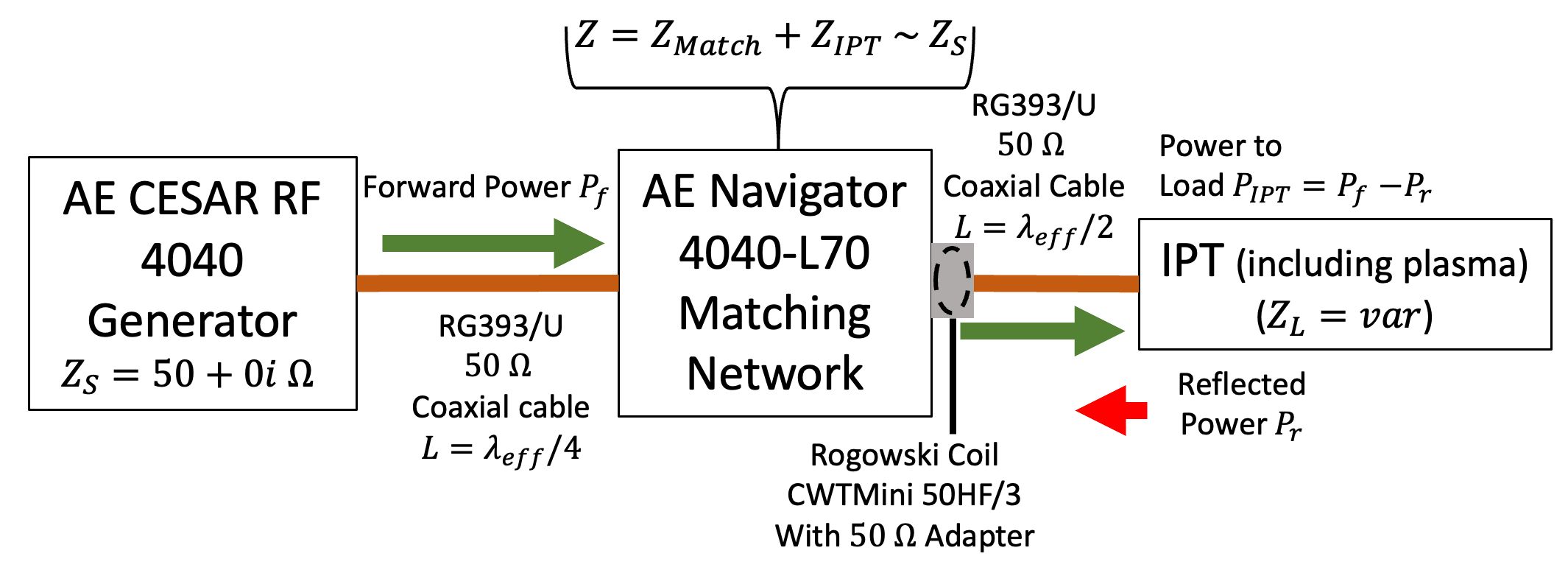}
	\caption{RF Optimized Circuitry with Thruster (IPT).}
	\label{fig:RF_circ2}
\end{figure} 

Moreover, a special adapter is designed, built, and installed at the auto-matching network RF-output plug. The adapter is matched at $50+j0\SI{}{\ohm}$ and allows the insertion of the CWT Mini 50HF/3 Rogowski coil to measure the current flowing from the matching network to the thruster. Simultaneously, it converts the non-standard RF-output plug to a N-type socket for easier assembly. The rendered view is shown in Fig.~\ref{fig:rogoada}, a picture of the Rogowski coil positioned inside the outer shell of the adapter is shown in Fig.~\ref{fig:adaptorinsertion}, while the complete assembly integrated is shown in Fig.~\ref{fig:rogoadapic}.
\begin{figure}[H]
		\centering
		\begin{subfigure}[b]{0.39\linewidth}
\includegraphics[width=\textwidth]{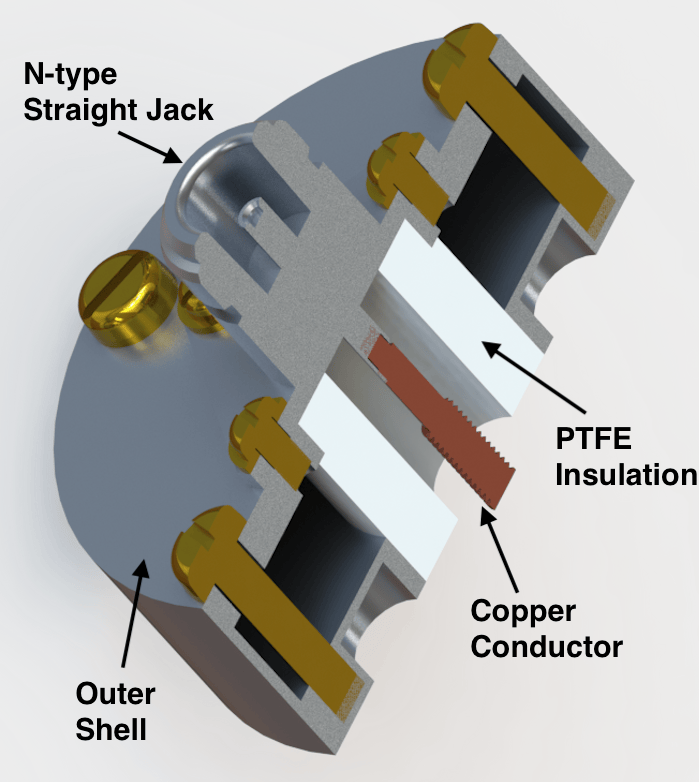}
\caption{Sectioned Render View.}
\label{fig:rogoada}
		\end{subfigure}
		\begin{subfigure}[b]{0.465\linewidth}
\includegraphics[width=\textwidth]{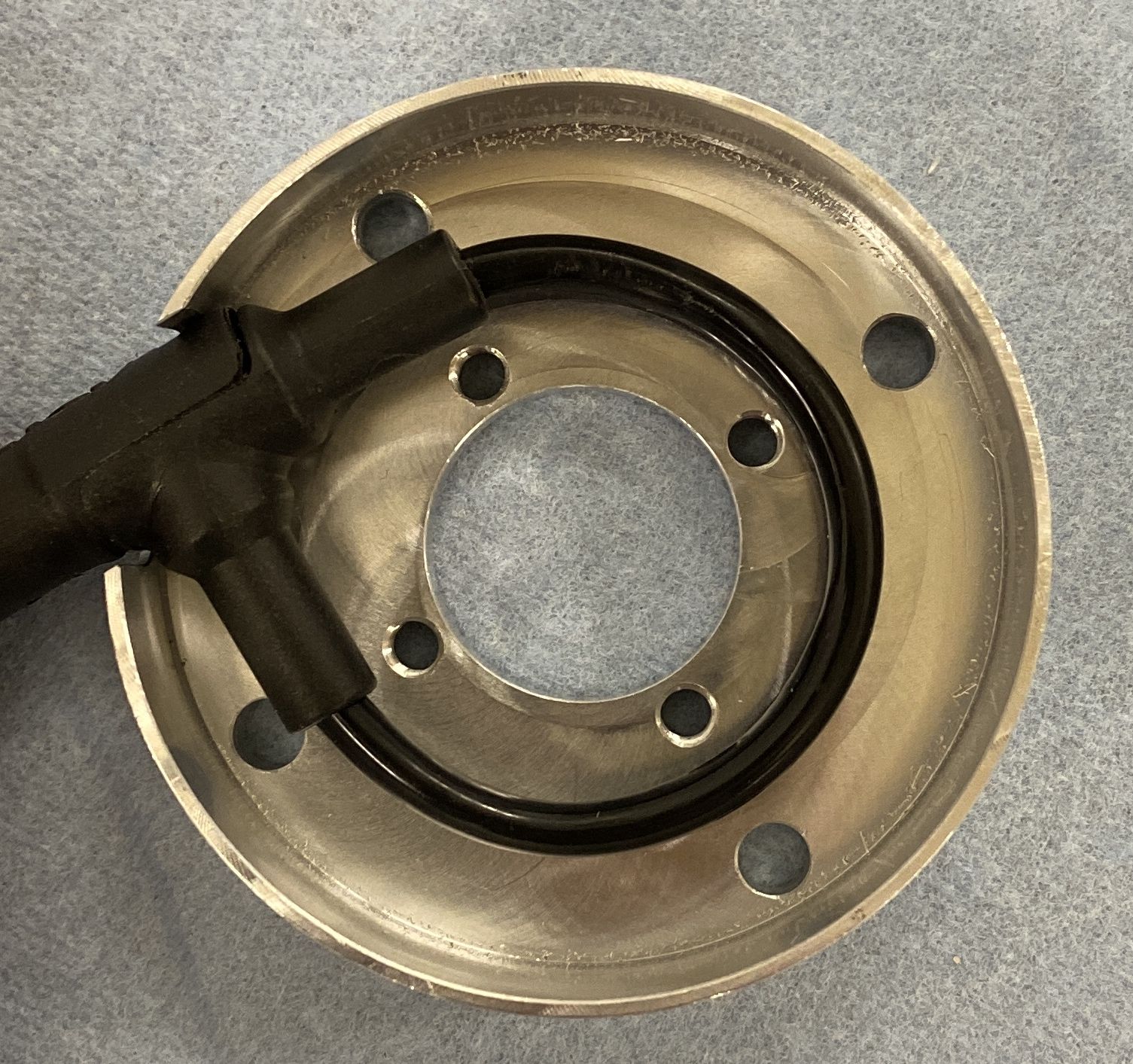}
\caption{Rogowski Coil into the Outer Shell.}
\label{fig:adaptorinsertion}
		\end{subfigure}
			\caption{Rogowski Adapter.}
\end{figure}

\begin{figure}[H]
\centering
\includegraphics[width=.75\textwidth]{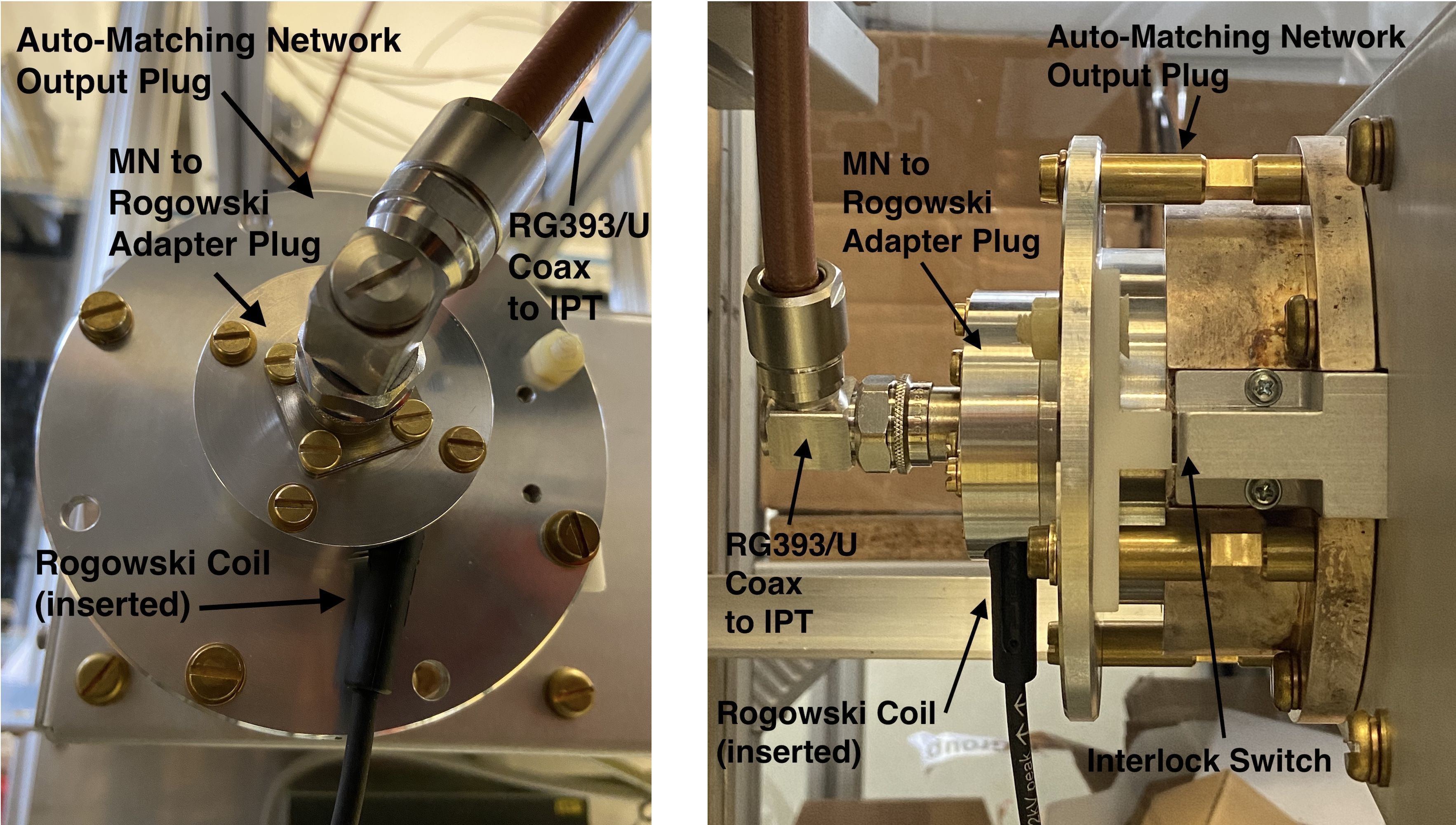}
\caption{View of the Auto-Matching Network Output Plug, Rogowski Adapter Integrated.}
\label{fig:rogoadapic}
\end{figure}

Given the aforementioned optimization process, the last contribution is that of the thruster itself. Having such an optimized circuit results in a minimum amount of losses and uncertainties in the knowledge in terms of power and current that are effectively coupled into the load. 

	\section{Birdcage Antenna Design}
	\label{ch:antenna}
	This section describes the thruster's antenna, the birdcage antenna, starting from the theoretical principle of operation, to the design procedure, simulation and verification, and ending with the engineering model design.

\subsection{Antenna for Electric Propulsion}
In an RF plasma thruster, its antenna produces the alternating EM fields within the discharge channel that excite the gas transforming it to the plasma state and, eventually, accelerate it.
Concerning the thruster development, the first approach has been of following the heritage at IRS on inductively coupled plasmas~\cite{georg1,georg2} and the literature available on both inductive and helicon plasma sources and thrusters. Nagoya-type and half-helical antennae have been also considered, but the "t-turn" design, finally, resulted to that providing higher $R_P$, see Sec.~\ref{ch:IPT}. The first approach has been the development of a "t-turn" antenna design. The geometry of any antenna determines its $Z$ and provides an estimation of the absorbed and reflected power by the antenna only. The resistance $R$ only depends on the material properties and its total surface, while the reactance $X$ is directly dependent on the frequency and self and mutual inductances determined by the antenna geometry and its surroundings. In general, RF plasma thrusters and sources, utilize the industrial standard frequency of $f=\SI{13.56}{\mega\hertz}$~\cite{charles2006helicon,ahedo2019helicon,vitucci2019development,takahashi2019helicon,isayama2018review}. By operating the same "t-turn" antenna at $f=\SI{40.68}{\mega\hertz}$ an intrinsic higher reactance is expected, in particular due to the inductance term $X_L=2\pi f L$. The question raised on whether it is possible or not to develop an antenna that provides minimum reactance $X$, therefore removing a substantial part of the "a priori" reflected power. The answer has been found in the medical sector, from the devices for magnetic resonance imaging (MRI). Such machines utilize a very strong static $B$-field combined with a RF resonant antenna. The latter provides an impulse of very homogeneous magnetic field perpendicular to the applied one to visualize the variation of spin of the hydrogen atoms and get the MRI images~\cite{HAYES1985622}. Such an antenna is called "birdcage" due to its shape and operates at one of its resonance frequencies providing $X\sim\SI{0}{\ohm}$, therefore reducing the power reflection and the matching requirements, making it a partially matched load. Due to this feature and for the configuration of the resulting EM fields that will be described later, the "t-turn" antenna has been discarded and, instead, the birdcage antenna has been implemented and developed. 

\subsection{Birdcage Antenna Theory}
\label{sec:birdcagetheory}
Birdcage antennae operate on the principle that a sinusoidal current distribution over a cylindrical surface induces a homogeneous transversal $B$-field within the cylindrical volume itself~\cite{birdcage1985,chen2016handbook}. Depending on the chosen resonance mode and the antenna's feeding, the resulting $B$-field can be linearly or circularly polarized. Such antennae are made by two end-rings, connected by $N$ equally spaced legs. The legs and/or the end-rings have capacitors in between to adjust the birdcage antenna resonance frequency to the one required by the application. Depending on the capacitors installation, those can be designed as low-pass, high-pass, or band-pass frequency response, see Fig.~\ref{fig:birdcage}. \begin{figure}[H]
	\centering
	\includegraphics[width=\textwidth]{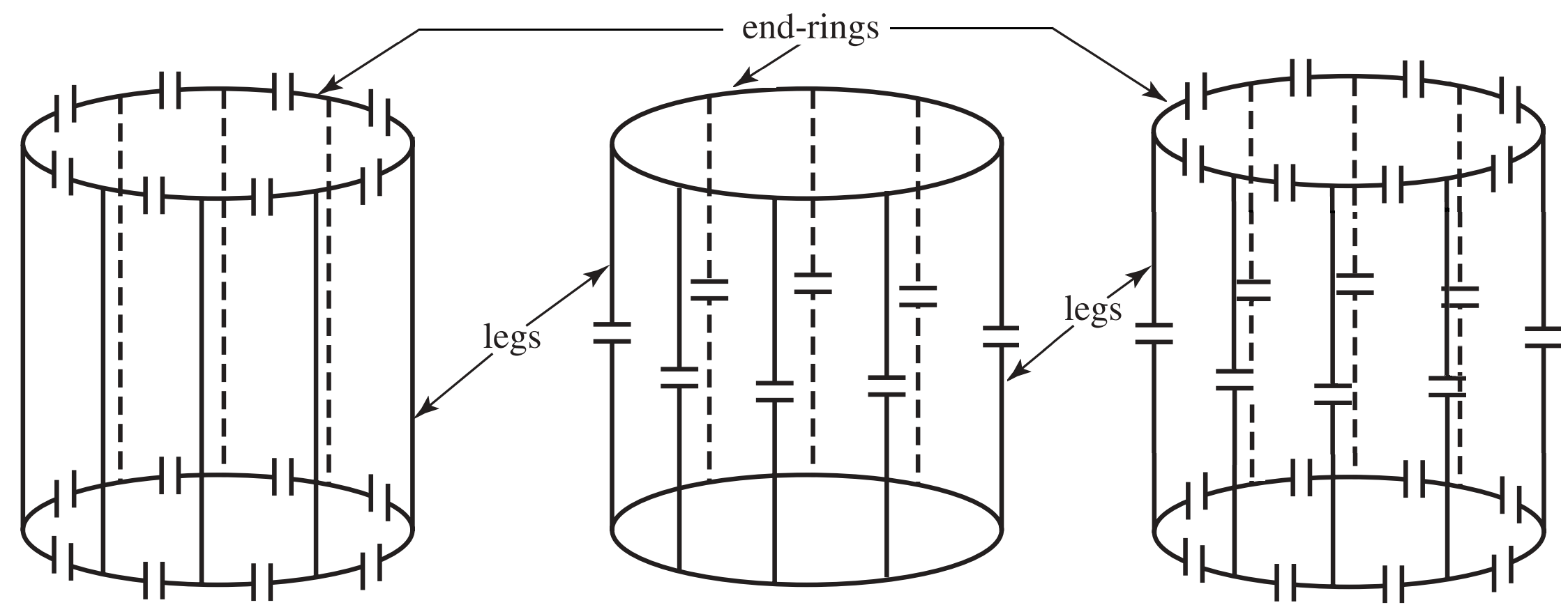}
	\caption{Birdcage antenna: High pass (left), Low pass (middle), Band pass (right), adapted from~\cite{balanis}.}
	\label{fig:birdcage}
\end{figure}
From the electrical point of view, the birdcage is modelled by self and mutual inductances of legs and end-rings, and by the applied capacitances.
The current distribution along the antenna has to follow the law described in Eq.~\ref{eq:current_birdcage}, where $I_{jk}$ is the normalized current at the $j$-th loop, made by two adjacent legs and sections of end-rings, for the $k$ mode of a birdcage antenna with $N$ legs~\cite{balanis}, see Fig.~\ref{fig:birdcageloops}.

\begin{figure}[H]
	\centering
	\includegraphics[width=\textwidth]{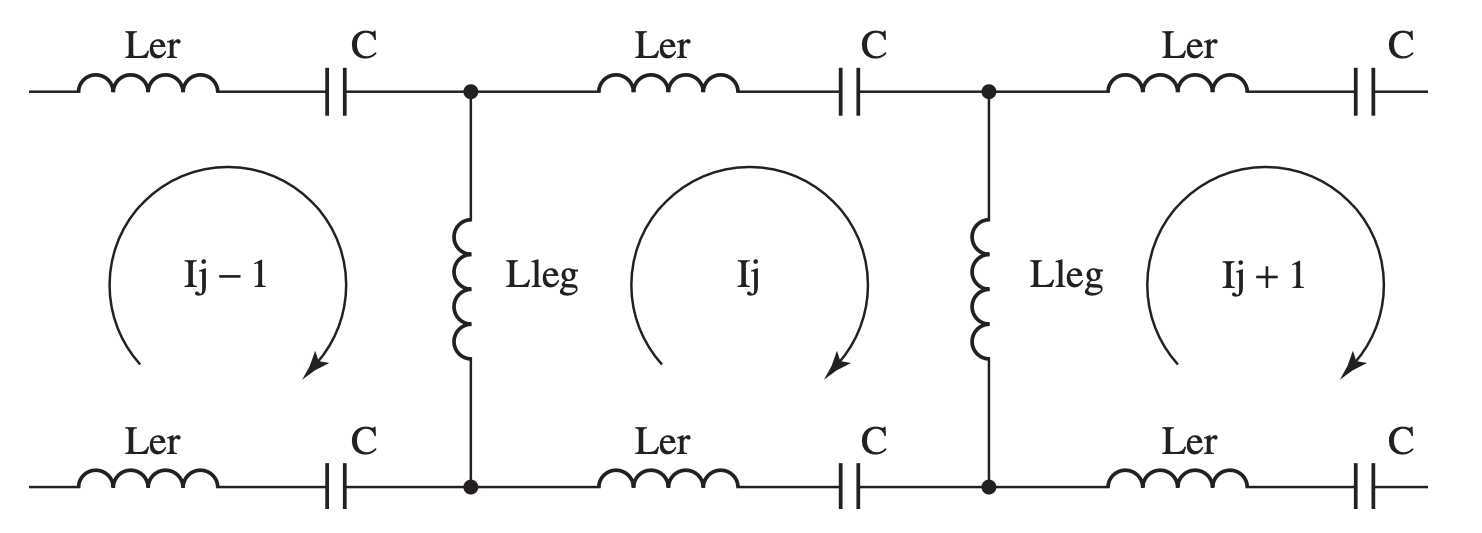}
	\caption{Equivalent Circuit of a Birdcage Antenna in a High Pass Configuration~\cite{balanis}.}
	\label{fig:birdcageloops}
\end{figure}

\begin{equation}
\label{eq:current_birdcage}
I_{jk} = 
\begin{cases}
\cos{\biggl(\frac{2\pi j k}{N}\biggr)};~k=0,1,2,...,N/2\\ \sin{\biggl(\frac{2\pi j k}{N}\biggr)};~k=1,2,...,(N/2-1)
\end{cases}
\end{equation}

At resonance, the current amplitude along a single leg follows a sinusoidal behaviour as described in Eq.~\ref{eq:current_birdcage}, as it goes from $-I_{max}$ to $0$ up to $+I_{max}$, see Fig.~\ref{fig:currentbirdcage} normalized to $\abs{I_{max}}=\SI{1}{\ampere}$. The sinusoidal current distribution is achieved along the circumference of the cylinder, as the currents along each leg are separated by the same phase shift, see Fig.~\ref{fig:currentbirdcage}. This means, the more the number of legs $N$, the more the current distribution profile along the circumference of the cylinder matches that of a sinusoidal curve.

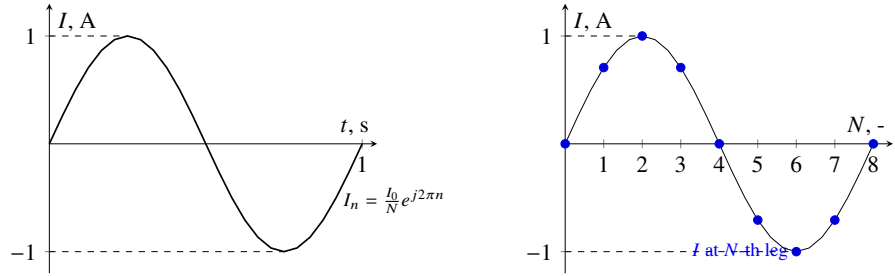
\begin{figure}[H]
\centering
	\begin{tikzpicture}[scale=0.8]
	\centering
	\begin{axis}[
	axis lines=middle,clip=false,
	xmin=0,xmax=2*pi+0.3,ymin=-1.2,ymax=1.3,
	ylabel={$I$, \SI{}{\ampere}},
	xlabel={$t$, \SI{}{\second}},
	xtick={0, 2*pi},
	xticklabels={0, 1}
	]
	\addplot+[thick,grid,no marks,domain=0:2*pi,black]{sin(deg(x))}
	node[right,pos=0.9,font=\footnotesize]{$I_n=\frac{I_0}{N}e^{j2\pi n}$};
	\addplot+[dashed,grid,no marks,domain=0:pi/2,black]{1};
	\addplot+[dashed,grid,no marks,domain=0:2*pi-pi/2,black]{-1};
	\end{axis}
	\end{tikzpicture}
\qquad
	\begin{tikzpicture}[scale=0.8]
	\centering
	\begin{axis}[
	axis lines=middle,clip=false,
	xmin=0,xmax=2*pi+0.4,ymin=-1.2,ymax=1.3,
	ylabel={$I$, \SI{}{\ampere}},
	xlabel={$N$, -},
	xtick={0, pi/4, 2*pi/4, 3*pi/4, 4*pi/4, 5*pi/4, 6*pi/4, 7*pi/4, 2*pi},
	xticklabels={0, 1, 2, 3, 4, 5, 6, 7, 8}
	]
	\addplot+[grid,only marks,domain=0:2*pi,samples=9]{sin(deg(x))}
	node[left,pos=0.8,font=\footnotesize]{$I$~at~$N$-th~leg};
	\addplot+[grid,no marks,domain=0:2*pi,black]{sin(deg(x))};
	\addplot+[dashed,grid,no marks,domain=0:pi/2,black]{1};
	\addplot+[dashed,grid,no marks,domain=0:2*pi-pi/2,black]{-1};
	\end{axis}
	\end{tikzpicture}
\caption{$I(t)$ on a Leg (left), $I(N)$ Distribution on a 8-Leg Birdcage (right).}
\label{fig:currentbirdcage}
\end{figure}
Each birdcage antenna with $N$ legs has a spectrum of $k=N/2$ frequencies at which it resonates, called resonance modes. 
For a low pass birdcage design, the resonance mode frequencies are defined as in Eq.~\ref{eq:resonantLP}.
\begin{equation}
	\omega_{k_{LP}} = \biggl[C\biggl(L_{ER}+0.5 L_{Leg} \sin^2{\frac{\pi k}{N}}\biggr)\biggr]^{-\frac{1}{2}},~(k=0,1,2,...,N/2)
	\label{eq:resonantLP}
\end{equation}
For a high pass design, instead, the resonant mode frequencies are given by Eq.~\ref{eq:resonantHP}.
\begin{equation}
\omega_{k_{HP}} = \biggl[C\biggl(L_{ER}+2 L_{Leg} \sin^2{\frac{\pi k}{N}}\biggr)\biggr]^{-1/2},~(k=0,1,2,...,N/2)
\label{eq:resonantHP}
\end{equation}
Here, $C$ is the capacitance of the capacitors and $L_{ER}$ and $L_{Leg}$ are the inductances of the end ring and the leg, respectively. In Eq.~\ref{eq:resonantLP} and Eq.~\ref{eq:resonantHP} the mutual inductances are neglected, but their contribution must be included in the final design of the birdcage. 
\begin{figure}[H]
	\centering
	\includegraphics[width=10cm]{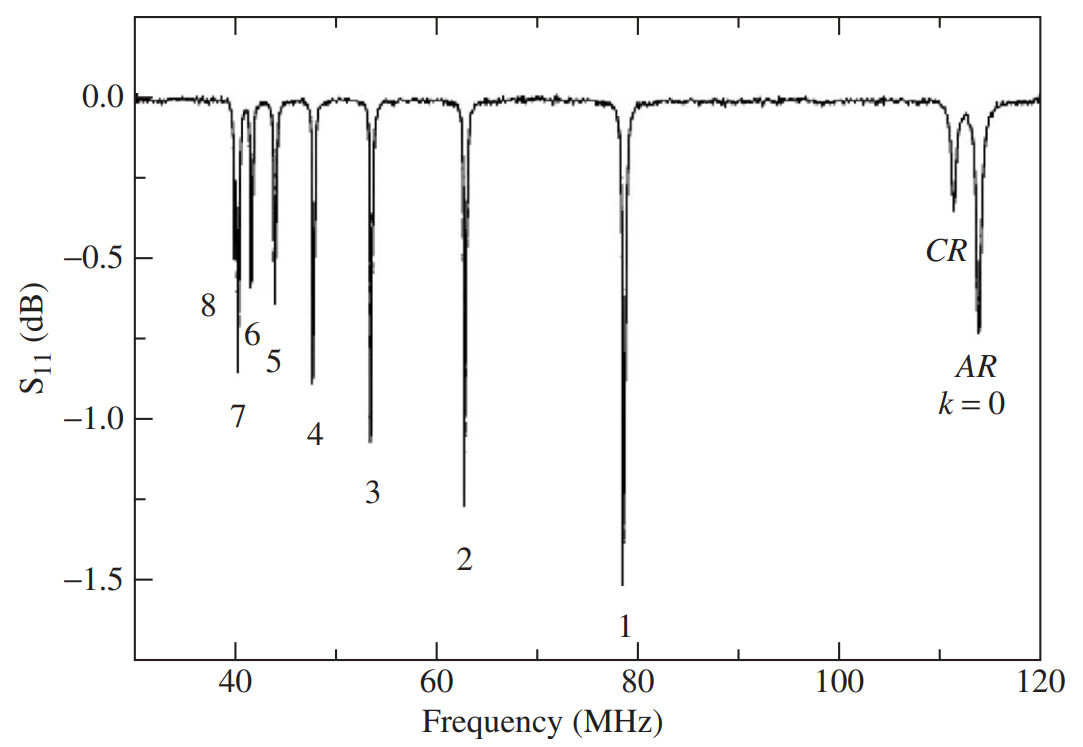}
	\caption{Example of Resonance Spectrum of a High Pass Birdcage with 16 Legs~\cite{balanis}.}
	\label{fig:resonanceex}
\end{figure}
What is of interest is the homogeneous resonance mode, as it is the only one providing a homogeneous $B$-field within the cross section of the cylindrical volume. The other modes, instead, provide a small $B$-field at the centre, but high near the legs~\cite{ozen2017novel}. The homogeneous resonant mode is reached at the highest resonant mode,  $\omega_{HP,k=1}$, for the high pass mode $k=1$, and at the lowest resonant mode, $\omega_{LP,k=0}$, for the low pass mode $k=0$. Two more resonant modes are present for the high pass configuration at high frequency, corresponding to $k=0$ and are the anti-rotating (AR) and the co-rotating (CR). They present a $B$-field aligned with the axis of symmetry of the cylinder. An example of the resonance spectrum of a high pass 16-legs birdcage antenna is shown in Fig.~\ref{fig:resonanceex}.

Finally, the frequency of the required resonance mode has to match the input one~\cite{MRI1}. Once the homogeneous resonance mode is set, antenna feeding plays an important role: while this does not influence the resonance frequency itself, it is still important in particular for the homogeneous mode. The feeding can be realized, mainly, either at one port, or at two ports, e.g. the base of a leg:
\begin{itemize}
	\item Feeding at 1 port creates a linearly polarized $B$-field;
	\item Feeding at 2 ports with quadrature, $\SI{90}{\degree}$ phase shift and geometrical separation, creates a circularly polarized $B$-field always at maximum amplitude, signal-to-noise-ratio (SNR) increased by a factor of $\sqrt{2}$, and half power consumption compared to 1 port feeding. 
\end{itemize}
For the thruster designed within this dissertation, the high-pass mode is selected, see Eq.~\ref{eq:resonantHP}. It is indeed easier to apply the capacitors at the end-rings rather then splitting legs in two halves. Moreover, this also avoids them receiving direct plasma radiation. The birdcage antenna is fed at one point (for easier first design) at the resonance mode $k=1$ at $f=\SI{40.68}{\mega\hertz}$ leading to a linearly polarized homogeneous transversal $B$-field. Consequently, also the $E$-field is linearly polarized, homogeneous, and perpendicular to the B-field within the cross section of the enclosed cylindrical volume. 

\subsection{Birdcage Antennae for Plasma Applications}
At the time of writing this dissertation, April 2021, the only known device in which cylindrical birdcage antennas are used for plasma generation, is the Resonant Antenna Ion Device (RAID)~\cite{EPFL1,EPFL2,EPFL3}, and it is used to research on negative hydrogen ions sources for the ITER fusion reactor project~\cite{ITER}. The investigation on cylindrical birdcage antennae for plasma generation was approached in 2005, for the development of a birdcage-based helicon plasma thruster~\cite{EPFL_helicon_2005}. The RAID device is capable to operate up to $P_{in}=\SI{10}{\kilo\watt}$ at a frequency of $f=\SI{13.56}{\mega\hertz}$. The discharge channel is made of aluminium oxide and it is internally water cooled. The birdcage antenna has 9 legs, each water cooled from the inside. The production of helicon waves has been confirmed by means of B-dot probe measurements~\cite{EPFL4}. The test conditions use \ce{H2} as operating gas, $P_{in}=\SI{3}{\kilo\watt}$, and $B=\SI{20}{\milli\tesla}$. The electron density along the axis is in the order of $n=\SI{1E18}{\meter^{-3}}$ at a chamber pressure of $p=\SI{0.3}{\pascal}$.

\subsection{EM Fields Configuration}
A preliminary theoretical approach on the electromagnetic (EM) fields created by a birdcage antenna with one feeding port, at the correct resonance and its influence to charged particles is hereby presented.
This analysis considers the EM fields generated by the birdcage antenna, plus the applied $B$-field applied along the axis of symmetry $z$, see Fig.~\ref{fig:birdcagefields}. The charged particles are enclosed within the birdcage antenna, ions and electrons, at a given initial condition. Such charged particles are assumed to not create any reacting EM fields. 

\begin{itemize}
	\item EM fields generated by the birdcage antenna $\vec{E_1}=(\pm E_1,0,0)$ and $\vec{B_1}=(0,\pm B_1,0)$ perpendicular to each other in the $x-y$ plane;
	\item Applied $B$-field (provided by the solenoid) $\vec{B_0}=(0,0,B_0)$.
\end{itemize} 
\begin{figure}[h]
	\centering
	\includegraphics[width=.9\textwidth]{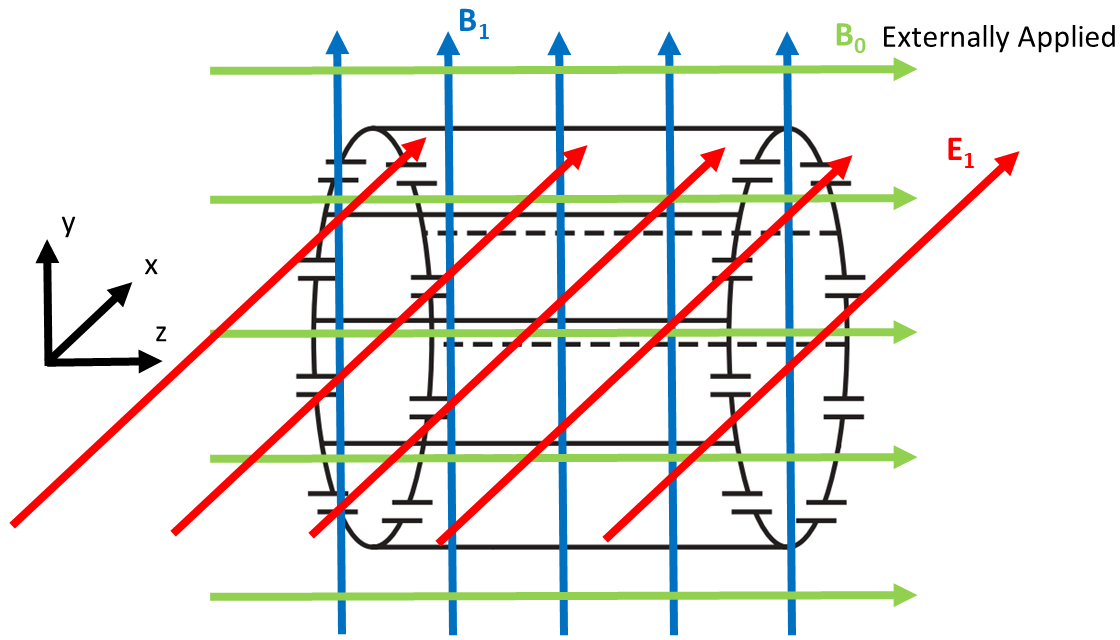}
	\caption{Birdcage EM Fields Configuration.}
	\label{fig:birdcagefields}
\end{figure}
%
%
%
%
%
%
The magnetic field $\vec{B}$ is the sum of $\vec{B_0}$ and $\vec{B_1}$. The fields generated by the birdcage are $\vec{E_1}=(E_1,0,0)$ and $\vec{B_1}=(0,B_1,0)$, therefore $\vec{B}=\vec{B_0}+\vec{B_1}=(0,B_1,B_0)$.
If one calculates the velocity drift $\vec{v}_E$ due to the $\vec{E}\times\vec{B}$ product as in~\cite{thebible}, the results is shown in Eq.~\ref{eq:vdrift} and Eq.~\ref{eq:vdrift2}: 
\begin{equation}
\vec{v}_E = \frac{\vec{E}\times\vec{B}}{\vec{B}^2}
\label{eq:vdrift}
\end{equation}
\begin{equation}
\vec{v}_E = \frac{1}{\vec{B}^2} \begin{Vmatrix}
\hat{x} & \hat{y} & \hat{z} \\
E_1 & 0 & 0 \\
0 & B_1 & B_0 \\
\end{Vmatrix} = \frac{1}{{B_0}^2+{B_1}^2}\begin{Bmatrix} 0 \\ -E_1 B_0 \\ E_1 B_1 \end{Bmatrix}
\label{eq:vdrift2}
\end{equation}
The $\vec{v}_E$ can be written in polar coordinates $(r, \theta, z)$ as in Eq.~\ref{eq:vdriftb}:
\begin{equation}
\begin{array}{l}
\vec{v}_E = \frac{1}{\vec{B}^2} \begin{Vmatrix}
\hat{r} & \hat{\theta} & \hat{z} \\
E_1 \cos{\theta} & -E_1 \sin{\theta} & 0 \\
B_1\sin{\theta} & B_1\cos{\theta} & B_0 \\
\end{Vmatrix}\vspace{15pt} = \\ 
\hspace{17pt}=\frac{1}{{B_1}^2\sin^2{\theta}+{B_1}^2\cos^2{\theta}+{B_0}^2}\begin{Bmatrix} -E_1 B_0 \sin{\theta} \\ -E_1 B_1\cos{\theta} \\ E_1 B_1 \end{Bmatrix}
\end{array}
\label{eq:vdriftb}
\end{equation}
The results for both coordinate systems lead to $\vec{v_E}$ along the $z$-axis that is, over the complete RF cycle, always along the same direction for both ions and electrons, charge independent, as also $\vec{B_1}$ and $\vec{E_1}$ maintain the same angle between them (linearly polarized), leading to an always positive product even if a rotation is involved (quadrature feeding at 2 ports). Finally, the motion is a rotation around the $z$-axis in the $x-y$ plane, and a drift along $z$.

This results is a solid starting point for plasma thruster development. Theoretically, some thrust can be already provided by both ion and electrons, due to $\vec{v_E}$ created by the resulting birdcage EM fields.
Furthermore, as mentioned earlier in Sec.~\ref{sec:thrustmodel}, the divergence of the applied $B$-field in the exhaust region leads to a magnetic nozzle acceleration of the quasi-neutral plasma. The $\vec{v_E}$ can be further increased by an extra acceleration stage that acts on both ions and electrons at the same time. This would enhance thrust while maintaining the plasma exhaust quasi-neutral and, therefore, removes the need of having a neutraliser, reducing complexity along with being an important advantage compared to conventional EP systems. It must be pointed out that the presence of the plasma will lead to EM fields and currents arising within the plasma itself and, consequently, to a different EM field configuration.

\subsection{Birdcage Antenna Resonance}
The resonance frequency, as described in Sec.~\ref{ch:RF}, can be generally expressed in the form of Eq.~\ref{eq:resonance}. The angular resonance frequency is $\omega_R=2\pi f_R$, and $L_{circ}$ and $C_{circ}$ are the total inductance and capacitance of the circuit respectively.\begin{equation}
	\omega_R = 2\pi f_R = \frac{1}{\sqrt{L_{circ}C_{circ}}}
	\label{eq:resonance}
\end{equation}
 
\begin{equation}
	f_R = f_{RF}
	\label{eq:resonancec}
\end{equation}
The hereby designed birdcage has to resonate at $f=\SI{40.68}{\mega\hertz}$ at the $k=1$ mode to provide the required homogenous and linearly polarized EM field configuration. The thruster characteristic impedance $Z_{IPT}$ at resonance has, therefore, $X=\SI{0}{\ohm}$. This ensures to reduce the reflected power and the respective matching requirements.
Furthermore, plasma represents a variable load $Z_{plasma}$ with both real $R_{plasma}$ and imaginary $X_{plasma}$ contribution, and adding it to the circuit will results in an up-shift of the resonance frequency $f_R$ and in a component of reflected power. A method to compensate such frequency shift is required so that the condition of Eq.~\ref{eq:resonancec} is respected. Based on the $RLC$ circuit theory, it is possible to act on the two sides of Eq.~\ref{eq:resonancec}, by changing $f_{RF}$, or $f_{R}$. Moreover, the uncertainties that arise as the sum of the manufacturing and assembly tolerances always introduce deviations from the simulation results and has to be compensated. Therefore, a fine tuning mechanism is required for the thruster at its whole. Tuning the input frequency requires a variable frequency RF generator, or a signal generator, that are currently not available at IRS. Moreover, this would also potentially change the behaviour of the plasma discharge as shown in Sec.~\ref{ch:helic}. Therefore, it is opted to work on the other side of Eq.~\ref{eq:resonancec}, by tuning resonance frequency $f_{R}$, see Eq.~\ref{eq:resonance}. To achieve the desired resonant condition, one has to act on the load directly: antenna and plasma. To do so, the following actions to adjust the resonant frequency $f_R$ can theoretically be taken:
\begin{itemize}
	\item Change $C_{circ}$ with variable capacitors on the antenna;
	\item Change $L_{circ}$ with movable conducting plate(s) that interact with the EM field lines;
	\item Change $L_{circ}$ with an external $B$-field that changes the overall field configuration;
	\item Change $L_{circ}$ by varying the antenna geometry;
	\item Change the plasma condition (gas flow/pressure) to change $C_{circ}$ and $L_{circ}$.
\end{itemize}
The use of movable conductive plates and variable capacitors is already performed in MRI machines to cope with the frequency shift caused by the introduction of the patient into the birdcage antenna~\cite{balanis,chen2016handbook,ozen2017novel,qian2012volume,MRI1}. In the hereby presented thruster case, the approach is of applying a variable external $B$-field plus a movable injector head, which is built out of an electrically conductive material, i.e. brass. 

\subsection{Thruster Birdcage Antenna Design and Verification}
The design of the birdcage antenna is based on the following requirements.\begin{itemize}
	\item Geometry: quartz tube of $D_{int}=\SI{37}{\milli\meter}$,~$D_{ext}=\SI{40}{\milli\meter}$, available length $L<\SI{100}{\milli\meter}$;
	\item Maximum design RF power input $P_{in}<\SI{1.5}{\kilo\watt}$;
	\item RF-generator frequency $f=\SI{40.68}{\mega\hertz}$;
\end{itemize}

To aid the design, the software XFdtd\textsuperscript{\textregistered} 7.8.1 from Remcom~Inc. is used. It is a 3D electromagnetic simulation software that uses the finite-difference time-domain method (FDTD)~\cite{xfdtd}.
This is required especially for the not trivial calculation of the mutual inductances between legs and end rings of the birdcage. The first step has been of getting a reliable frequency spectrum response of the birdcage antenna by tuning the simulation parameters. From the obtained results, the desired resonance peak of $k=1$ is identified by visualizing the reactance $X$ over frequency $f$, counting the peaks, as well as by extrapolating the 3D EM fields representation over time and verify it is the desired linearly polarized $\vec{E}\times\vec{B}$ configuration. The peak is then shifted by applying different capacitance $C$ values until the resonance peak matches the input frequency $f$.
\begin{figure}[h]
	\centering
	\includegraphics[width=\textwidth]{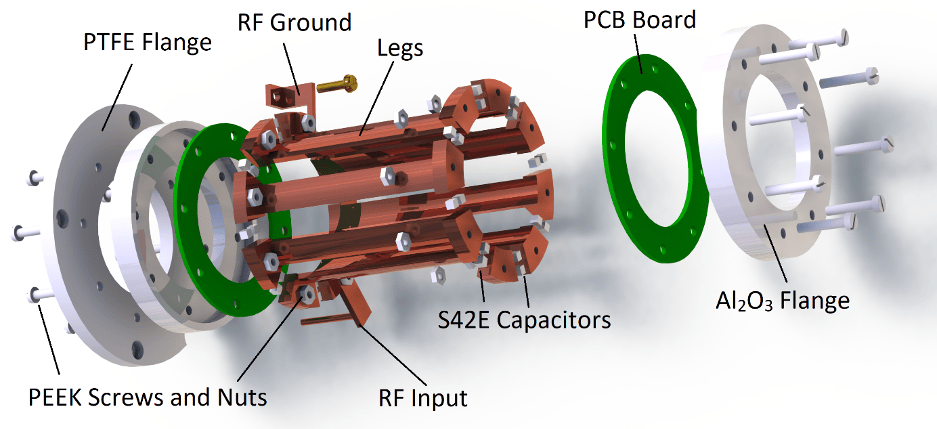}
	\caption{Exploded View of the IPT Birdcage Antenna.}
	\label{fig:IPT_bcage}
\end{figure}

The starting geometry, given the requirements, is a high pass 8 legs birdcage with an inner circumference of $D=\SI{44}{\milli\meter}$. The width of legs and end rings is chosen to cope with a maximum current per leg of $I_{max,rms}>\SI{15}{\ampere}$, deriving from $P_{in,max}$, at the given frequency.
Moreover, it is designed to provide a larger usable surface facing the plasma. In such configuration $N/2=4=k$ resonant modes are expected plus one AR and one CR at $k=0$ according to the theory of Sec.~\ref{sec:birdcagetheory}. The required for the homogeneous mode in a high pass configuration is $k=1$, as the only one producing a linearly polarized homogeneous EM fields within the $x-y$ plane.
The visualization of the $S_{11}$ parameter vs $f$ and $C$ is applied to find the desired resonance for $k=1$, see Fig.~\ref{fig:resonanceex}. A rough sweep of capacitance has been performed to visualize the peaks and extract a first value of $C_{ER}$ to start with, and fine tune the other parameters.An important parameter for the simulation is the cell size of the mesh. Such value has to be at least $1/10$ or better $1/20$ of the shortest wavelength $\lambda$ (highest frequency) of the signal sent to the antenna. Therefore, during the frequency spectrum analysis, the target cell size must be set small enough for the highest frequency of interest, when the maximum for the analysis is set to \SI{500}{\mega\hertz}, the target cell size has to be smaller than \SI{25.5}{\milli\meter}. also considering a velocity factor $V_F \sim 1$ for pure copper. For the analysis at the final fixed frequency the cell size has to respect such conditions for the wavelength at $\SI{40.68}{\mega\hertz}$, see Eq.~\ref{eq:lambda40}.
\begin{equation}
\lambda=\frac{c}{f} V_F=\frac{3\times10^8}{40.68\times10^6} V_F=\SI{7.374}{\meter}
	\label{eq:lambda40}
\end{equation} 
Therefore, at each simulation the target cell size is checked to be smaller than $1/20$ of the shortest wavelength in the simulation. Important is also the size of the control grid that surrounds the antenna. This must be at least $\frac{1}{10}\lambda$. Moreover, the software implements a special meshing technique for cylindrical shapes named XACT that is required to be active for the birdcage design due to the respective geometry.
Integrating part of the IPT is the Faraday shield: a conductive element surrounding the birdcage antenna that isolates it from the outside. External disturbances are removed ensuring for a better operation of the birdcage antenna itself by providing a defined enclosed volume that is independent of the outer environment. This also result in minimizing the leak of RF waves to the outside reducing interferences with external electronics. The birdcage antenna of the IPT is shown in the exploded view of Fig~\ref{fig:IPT_bcage}.
A PTFE flange is fixed on the closure of the IPT and it is connected through an \ce{Al2O3} flange and the printed circuit board (PCB), on which the capacitors are soldered, to the birdcage legs. The structure is held together by PEEK screws. The PEEK material is chosen as it is an insulator, it can withstand high operational temperatures $T=239-\SI{260}{\kelvin}$, and has a reasonable mechanical strength, up to $\SI{1.03e8}{\pascal}$ of tensile strength. On the other side, the same configuration is applied, but with a brass flange on top of the \ce{Al2O3} to increase thermal capacity. The RF ground is made of copper and is connected to the top IPT closure by a brass screw. The RF input connected to the antenna side is made of copper as well, and is soldered to an the N-type flanged connector that is fixed on the outer side of the IPT brass closure. The soldered copper central connector goes through the PTFE and the brass IPT closure, and its geometry is such as to provide a \SI{50}{\ohm} interface to minimize losses.

\begin{figure}[h]
	\centering
	\begin{subfigure}{.49\textwidth}
	\includegraphics[height=1.2\textwidth, trim={2cm 0cm 2cm 0cm},clip]{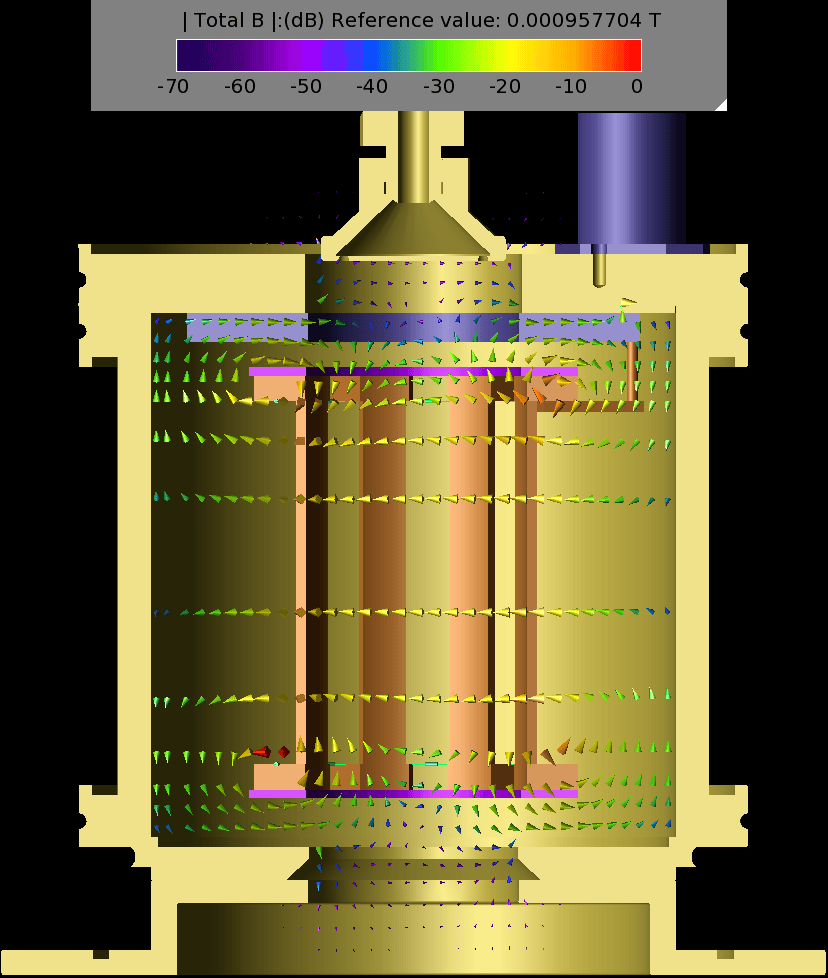}
	\caption{$\vec{B_1}$ $1st$ Half Cycle.}
	\label{fig:XFDTD_B1}
	\end{subfigure}
	\begin{subfigure}{.49\textwidth}
	\centering
	\includegraphics[height=1.2\textwidth, trim={2cm 0cm 2cm 0cm},clip]{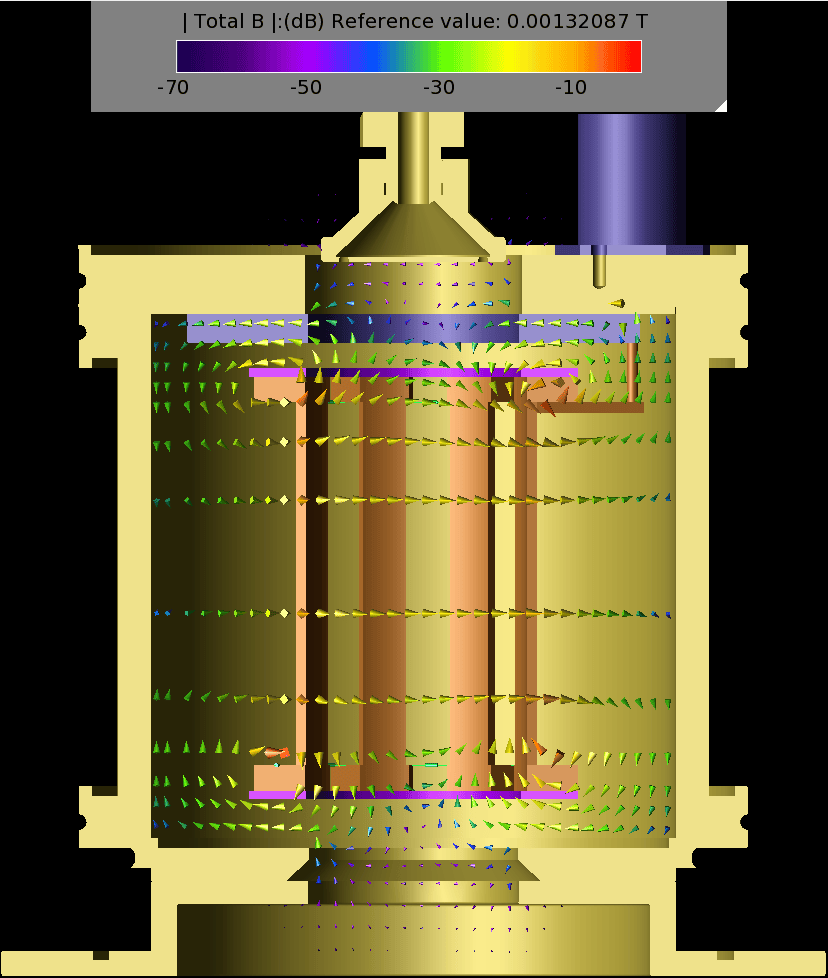}
	\caption{$\vec{B_1}$ 2nd Half Cycle.}
	\label{fig:XFDTD_B2}
	\end{subfigure}
	\label{eq:XFDTD_B_Transversal}
	\caption{IPT Birdcage $\vec{B_1}$ in the Transversal View, XFdtd\textsuperscript{\textregistered} Simulation.}
\end{figure}
\subsubsection{Resonance Verification through Electromagnetic Field Visualization}
The electromagnetic fields simulated with XFdtd\textsuperscript{\textregistered} are extracted to verify the correct resonance to be achieved. The results are shown in Fig.~\ref{fig:XFDTD_B1} and Fig.~\ref{fig:XFDTD_B2} for the transversal view of the IPT showing the vector field indicating the $B$-field direction highlighting a homogeneous $B$-field distribution within the birdcage antenna volume at the first and second half of the RF cycle. Therefore, the $B$-field $\vec{B_1}$ flips direction at each RF half cycle. The ceramic holding plates are not shown for better image's clarity.
Concerning the cross-sectional configuration of electromagnetic fields, the results are shown in Fig.~\ref{fig:XFDTD_EB_CrossSec} at the middle plane of the antenna, and in Fig.~\ref{fig:XFDTD_EB_CrossSec_Ex} at the end section of the discharge channel, the exhaust. The results highlight a uniform distribution of both $E$- and $B$-fields within the discharge channel section, and their linear polarization at both middle and exhaust plane. The slightly less uniform distribution at the exhaust region is due to the presence of electrically conductive surfaces that change the fields outside the antenna region. Those results do not take into account the applied static magnetic field form the solenoid as well as the plasma it self as those cannot be simulated with the available software.
\begin{figure}[h]
	\centering
	\begin{subfigure}{.45\textwidth}
	\includegraphics[width=6cm]{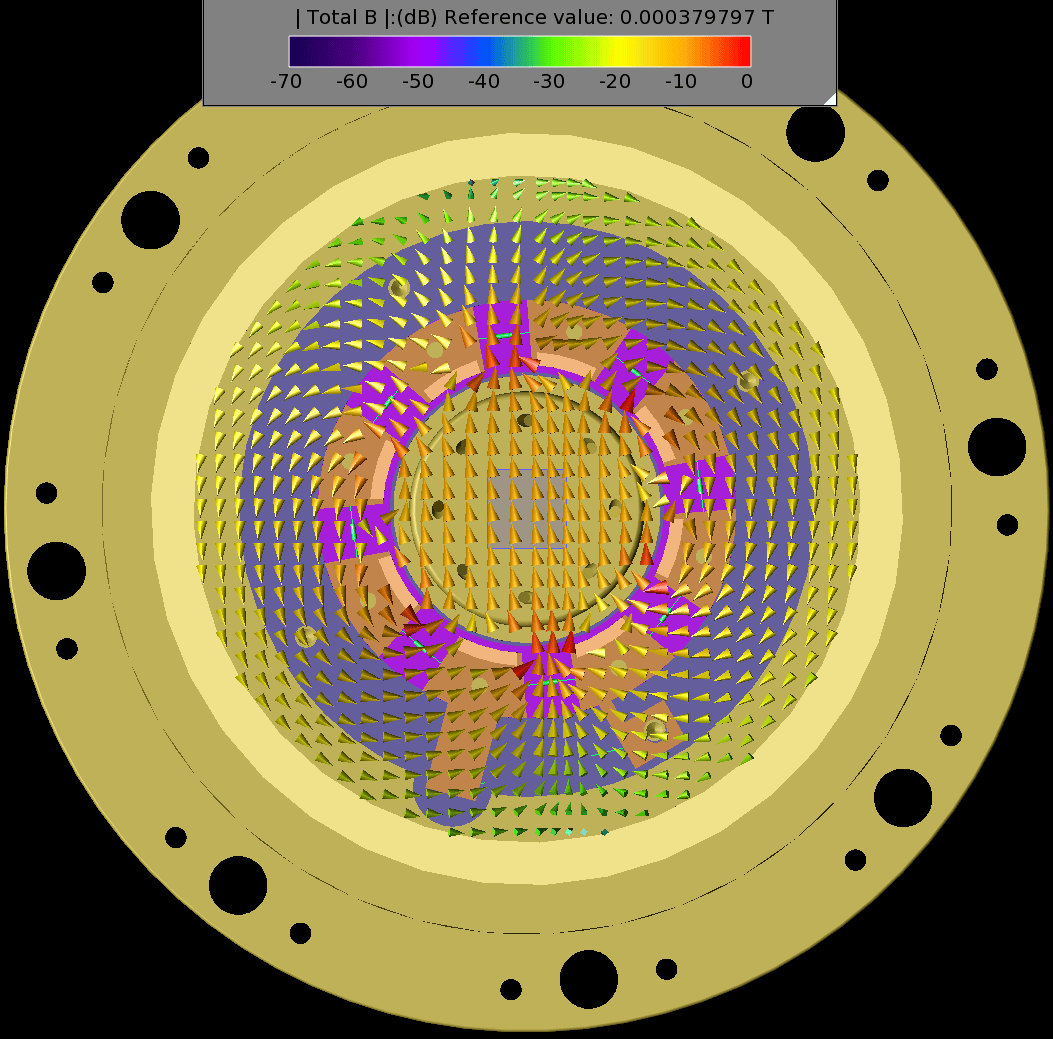}
	\caption{$\vec{B_1}$, $1st$ Half Cycle.}
	\label{fig:XFDTD_B1_c}
	\end{subfigure}
	\begin{subfigure}{.45\textwidth}
	\centering
	\includegraphics[width=6cm]{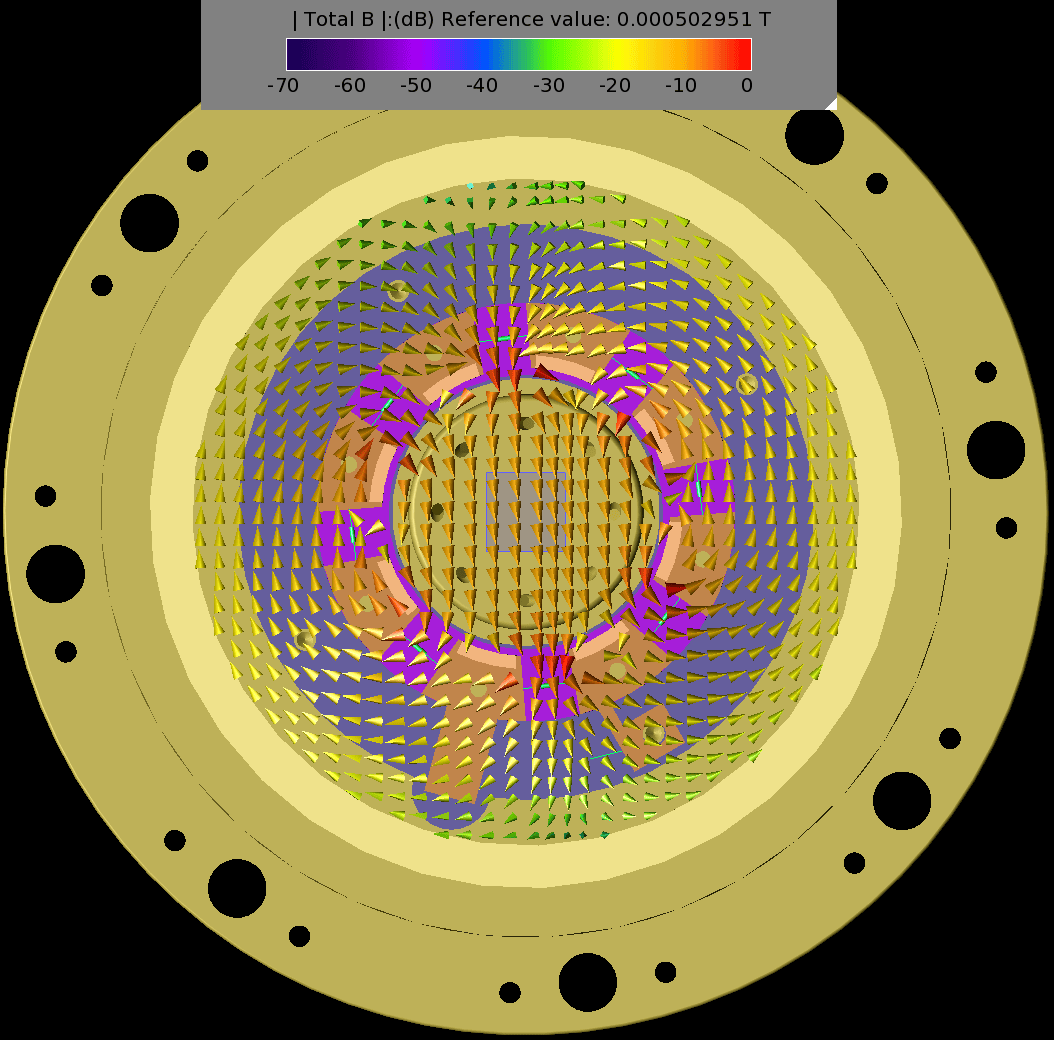}
	\caption{$\vec{B_1}$, 2nd Half Cycle.}
	\label{fig:XFDTD_B2_c}
	\end{subfigure}
	\begin{subfigure}{.45\textwidth}
	\includegraphics[width=6cm]{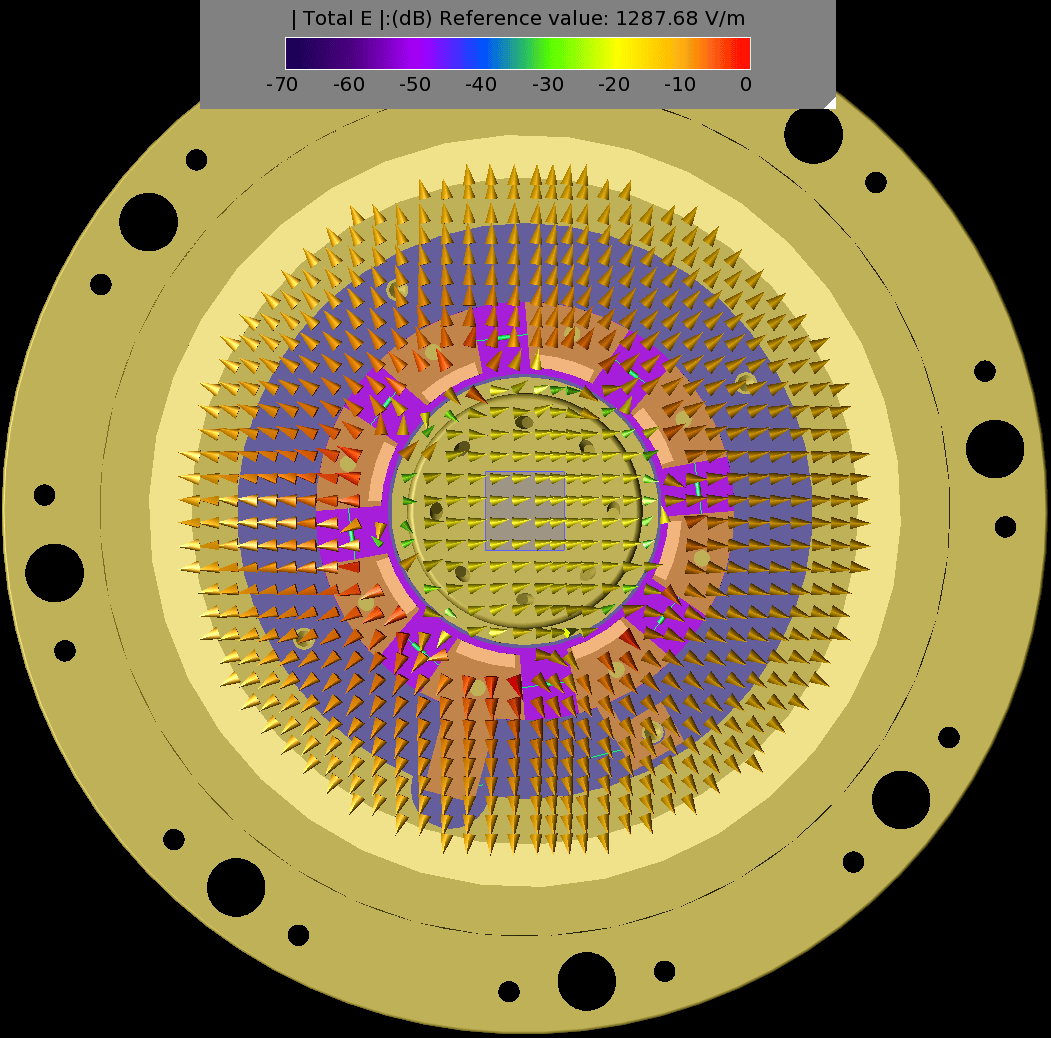}
	\caption{$\vec{E_1}$, $1st$ Half Cycle.}
	\label{fig:XFDTD_E1_c}
	\end{subfigure}
	\begin{subfigure}{.45\textwidth}
	\centering
	\includegraphics[width=6cm]{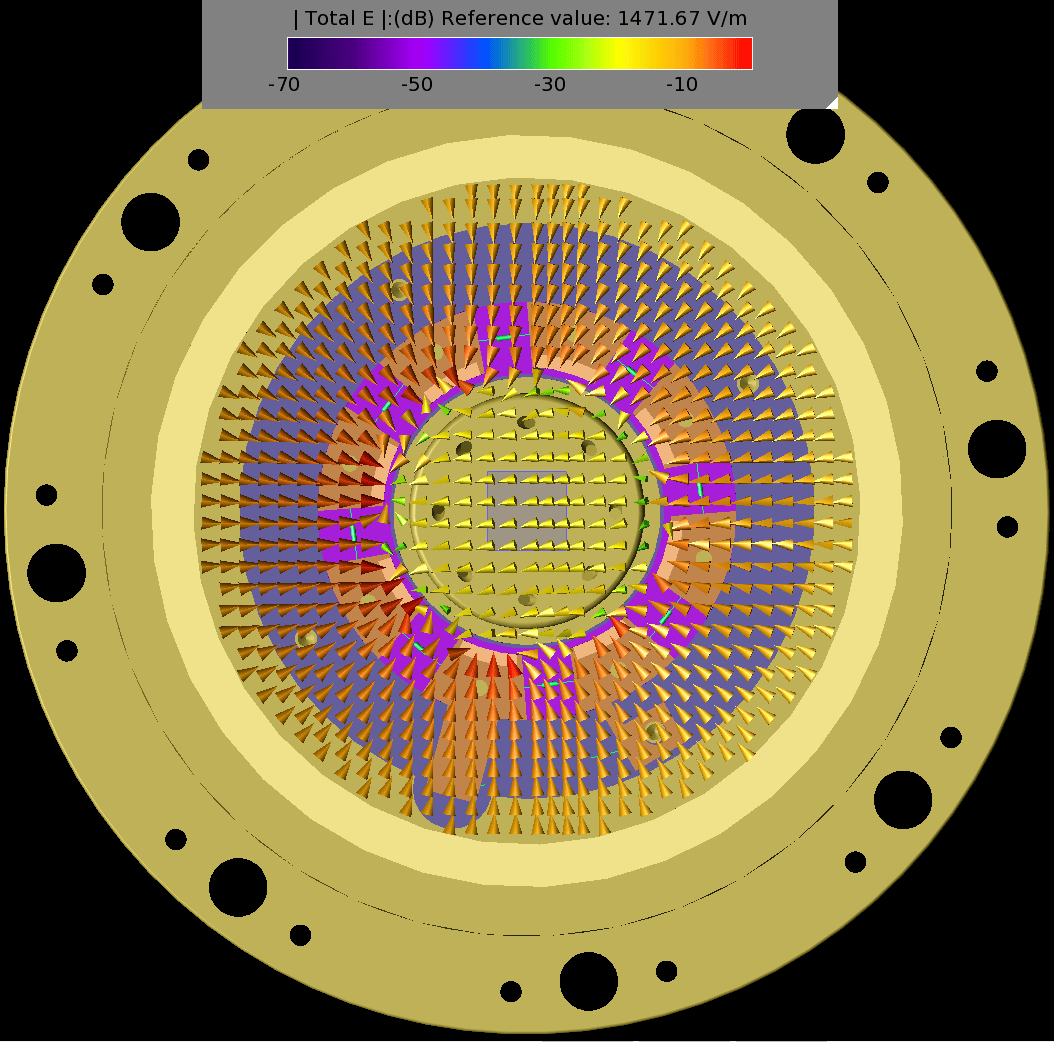}
	\caption{$\vec{E_1}$, 2nd Half Cycle.}
	\label{fig:XFDTD_E2_c}
	\end{subfigure}
	\caption{IPT Birdcage Electric $\vec{E_1}$- and Magnetic $\vec{B_1}$- Vector Fields in the Cross Sectional View, Middle Position, seen from the Exhaust Side, XFdtd\textsuperscript{\textregistered} Simulation.}
	\label{fig:XFDTD_EB_CrossSec}
\end{figure}

\begin{figure}[h]
	\centering
	\begin{subfigure}{.45\textwidth}
	\includegraphics[width=6cm]{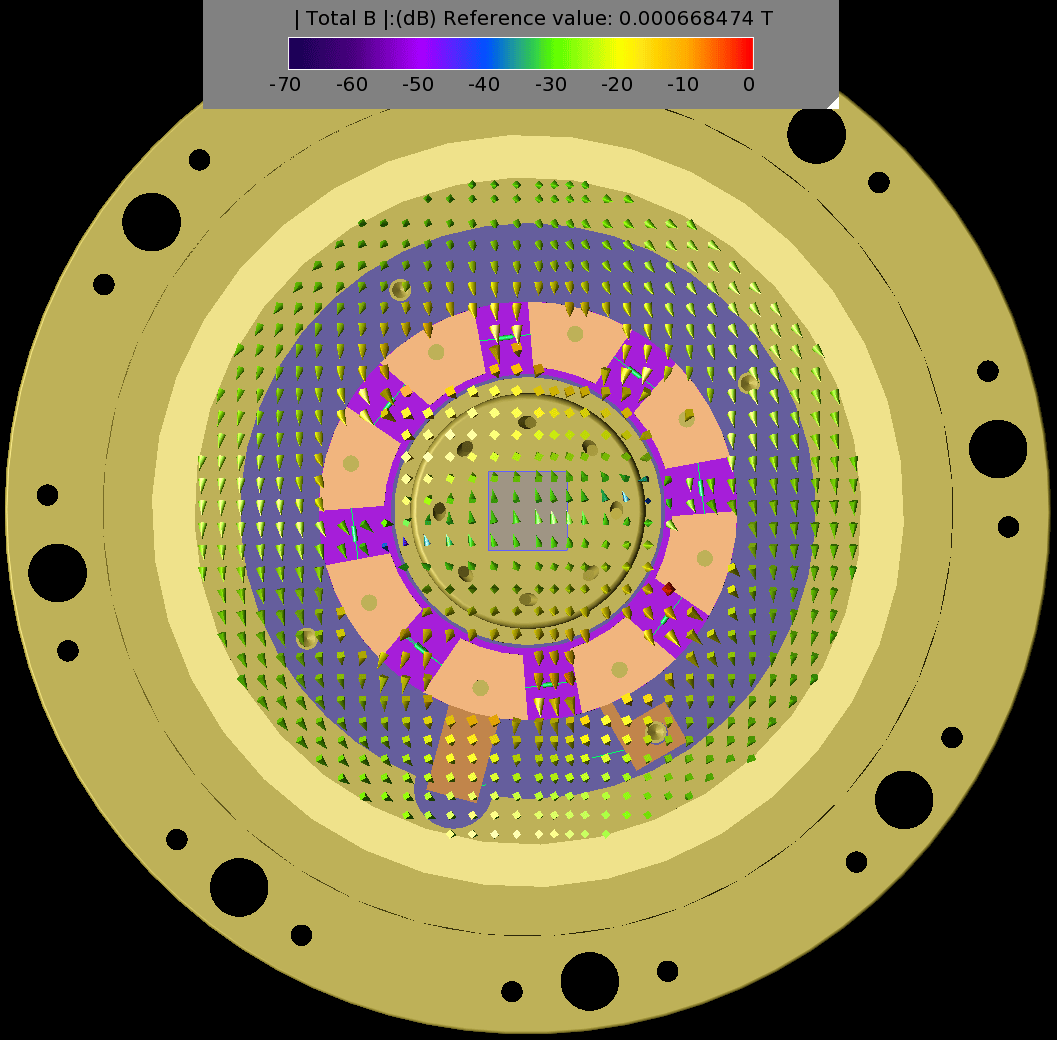}
	\caption{$\vec{B_1}$, $1st$ Half Cycle.}
	\label{fig:XFDTD_B1_ex}
	\end{subfigure}
	\begin{subfigure}{.45\textwidth}
	\centering
	\includegraphics[width=6cm]{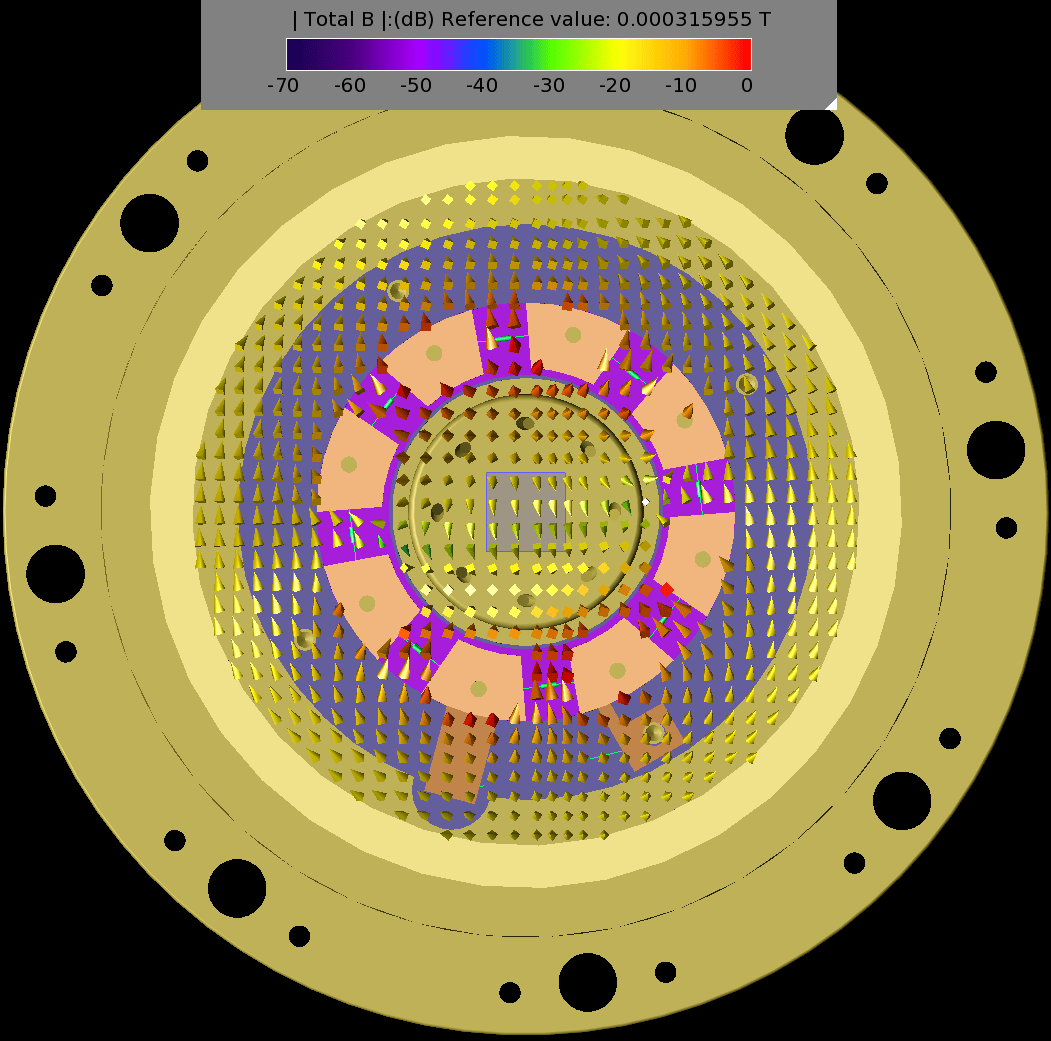}
	\caption{$\vec{B_1}$, 2nd Half Cycle.}
	\label{fig:XFDTD_B2_ex}
	\end{subfigure}
	\begin{subfigure}{.45\textwidth}
	\includegraphics[width=6cm]{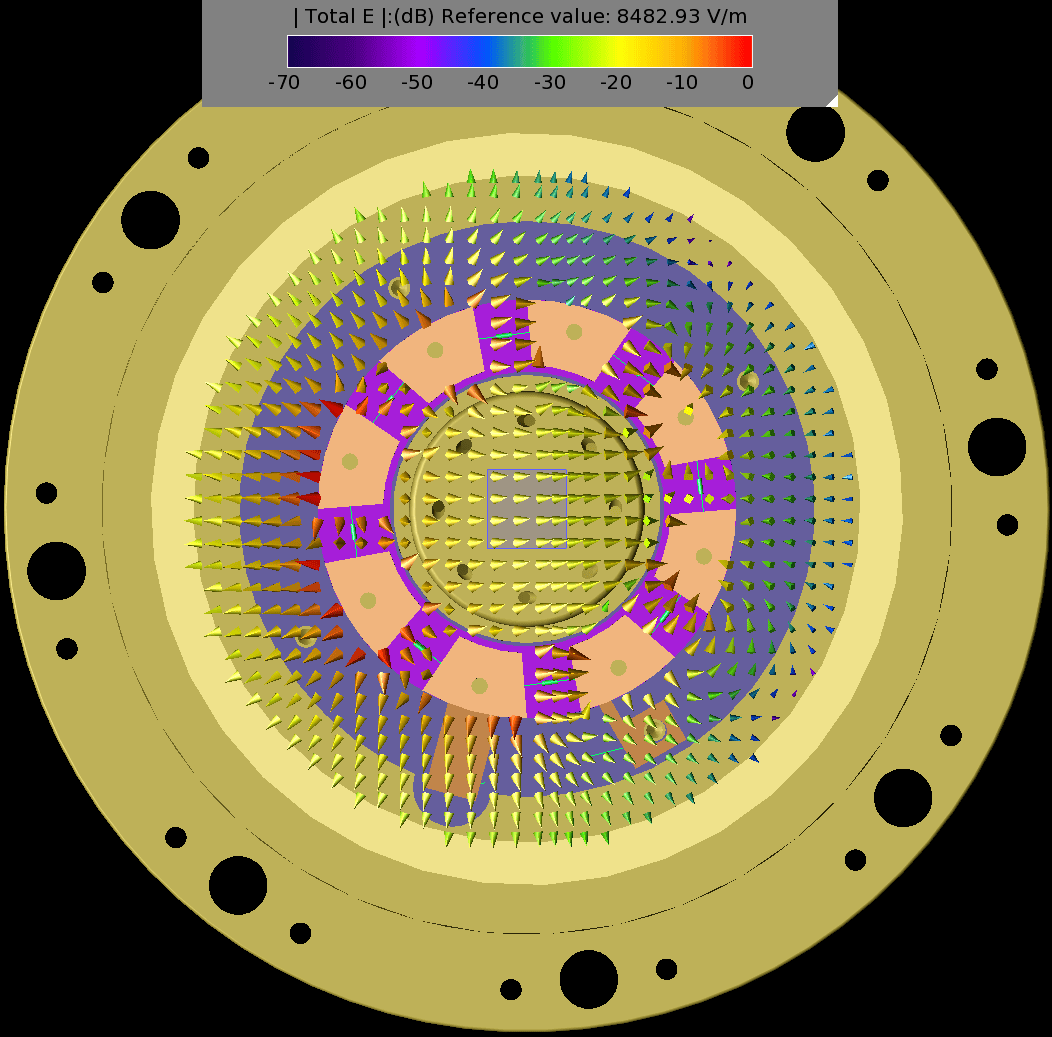}
	\caption{$\vec{E_1}$, $1st$ Half Cycle.}
	\label{fig:XFDTD_E1_ex}
	\end{subfigure}
	\begin{subfigure}{.45\textwidth}
	\centering
	\includegraphics[width=6cm]{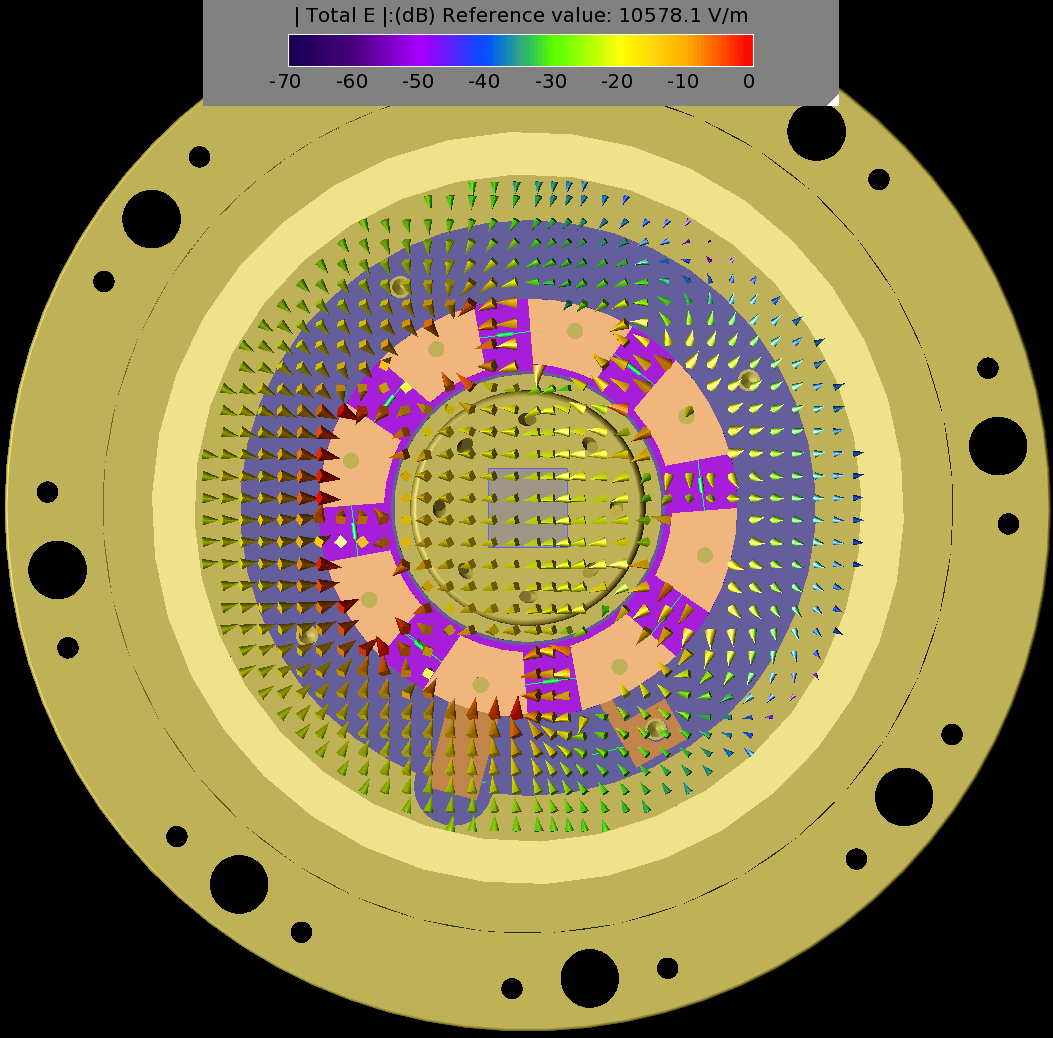}
	\caption{$\vec{E_1}$, 2nd Half Cycle.}
	\label{fig:XFDTD_E2_ex}
	\end{subfigure}
\caption{IPT Birdcage Electric $\vec{E_1}$- and Magnetic $\vec{B_1}$- Vector Fields in the Cross Sectional View, Exhaust Position, seen from the Exhaust Side, XFdtd\textsuperscript{\textregistered} Simulation.}
\label{fig:XFDTD_EB_CrossSec_Ex}
\end{figure}

\subsubsection{Achieving the Desired Resonance Mode}
The results from XFdtd\textsuperscript{\textregistered} for the IPT design, shows a required capacitance of $C=\SI{785.51}{\pico\farad}$, see the $S_{11}$ parameter vs frequency Fig.~\ref{fig:S11_f_785_51_pF} and the corresponding impedance $Z$ in Fig~\ref{fig:Z_f_785_51_pF}.
\begin{figure}[ht!]
	\centering
	\includegraphics[width=.6\textwidth]{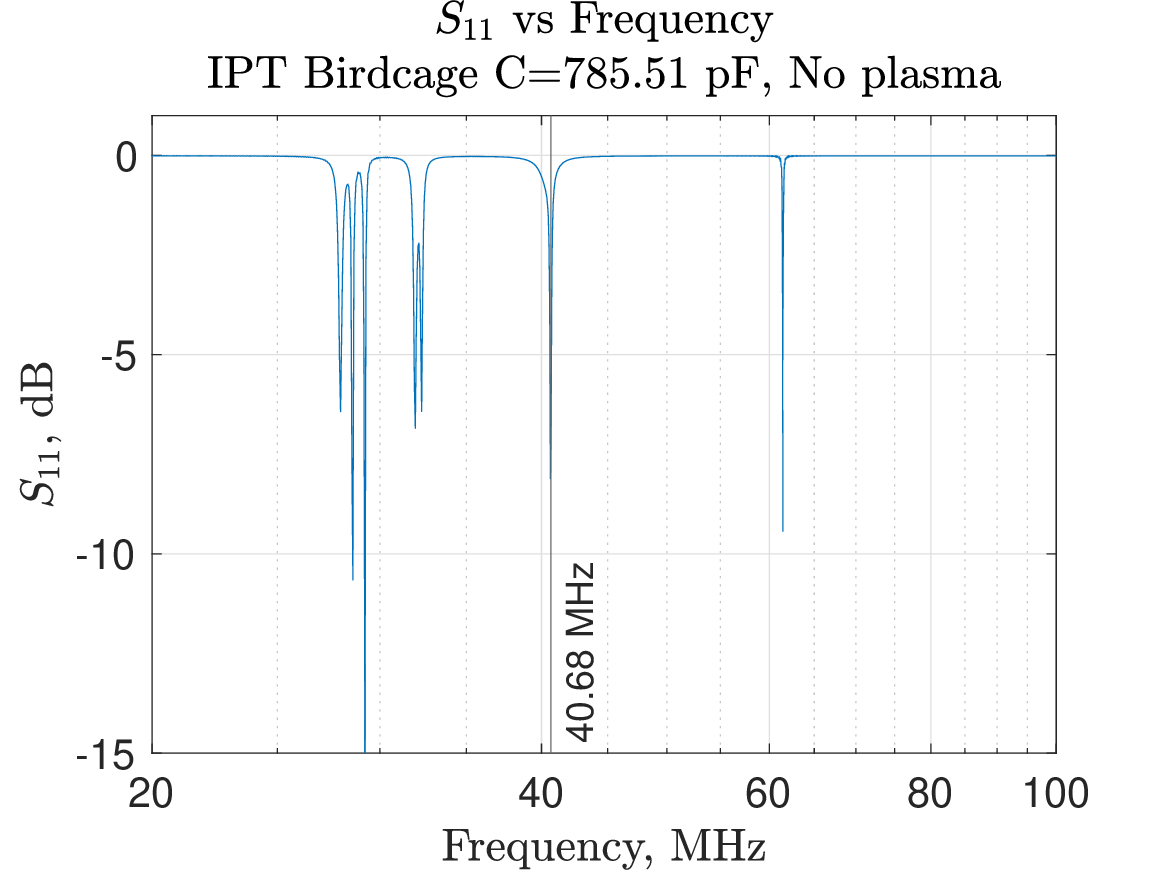}
	\caption{$S_{11}$ vs $f$ IPT Design, XFdtd\textsuperscript{\textregistered} Simulation.}
	\label{fig:S11_f_785_51_pF}
\end{figure}
\begin{figure}[hb!]
	\centering
	\includegraphics[width=.6\textwidth]{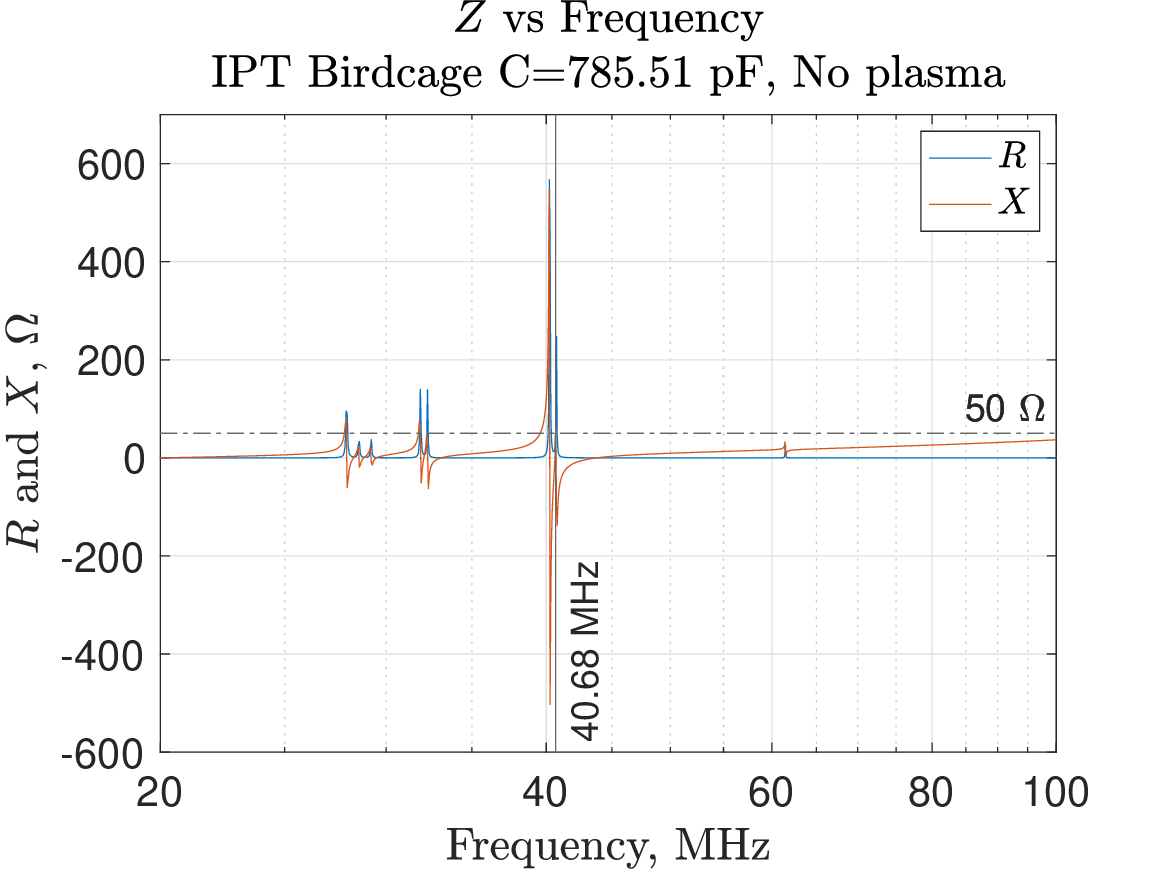}
	\caption{$Z$ vs $f$ IPT Design, XFdtd\textsuperscript{\textregistered} Simulation.}
	\label{fig:Z_f_785_51_pF}
\end{figure}
The available commercial capacitors are of either $C=\SI{360}{\pico\farad}$ or $C=\SI{390}{\pico\farad}$. The chosen ones can sustain up to $V=\SI{500}{\volt}$ and cope with a maximum of $I_{rms}=\SI{15}{\ampere}$ current. The capacitors are mounted in parallel, two per gap, at the end rings. As shown in Fig.~\ref{fig:Z_f_785_51_pF} the value at resonance of $Z$ differs from $Z_S=50+j0\SI{}{\ohm}$, hence the corresponding $S_{11}$ parameter is also small in module. The ideal case is of $S_{11}<-\SI{20}{\decibel}$ to obtain $>99\%$ of power transferred, corresponding to $Z_{IPT} \rightarrow Z_S$. The presence of plasma corresponds to an up=shift of the resonance frequency and in a drop of $Z$ that is difficult to predict~\cite{EPFL1}. Finally, the selected approach, due to the long delivery time of the $C=\SI{390}{\pico\farad}$ capacitors, has been to assembly the birdcage antenna with the $C=\SI{360}{\pico\farad}$ available capacitors with $\pm 5\%$ tolerance and verify the tuning range. The capacitors are singularly measured, and most of are of $C<\SI{360}{\pico\farad}$. By selecting those with larger $C$, an average of $C=\SI{717.475}{\pico\farad}$ per gap is achieved. The simulation results from XFdtd\textsuperscript{\textregistered} are shown in Fig.~\ref{fig:S11_f_717_pF} and Fig.~\ref{fig:Z_f_717_pF}.
\begin{figure}[ht!]
	\centering
	\includegraphics[width=.6\textwidth]{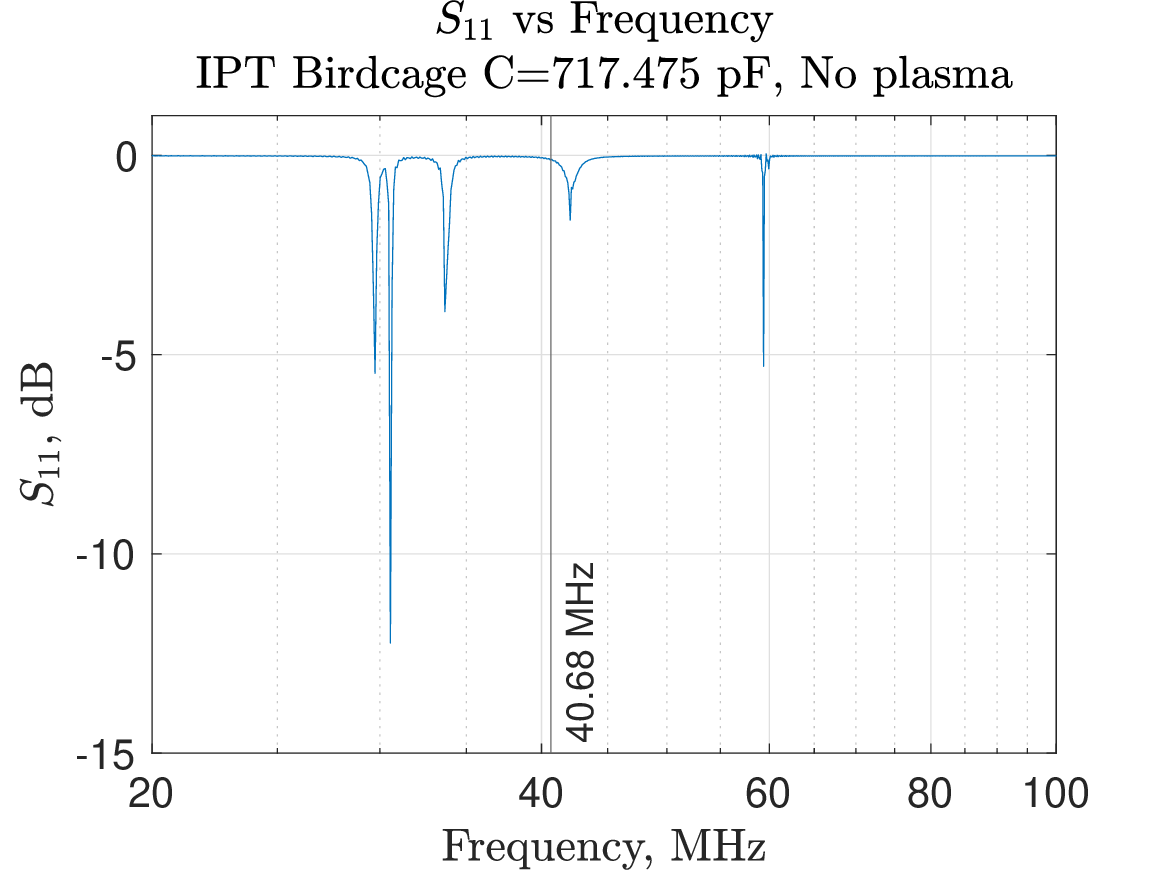}
	\caption{$S_{11}$ vs $f$ IPT Design, XFdtd\textsuperscript{\textregistered} Simulation.}
	\label{fig:S11_f_717_pF}
\end{figure}
\begin{figure}[hb!]
	\centering
	\includegraphics[width=.6\textwidth]{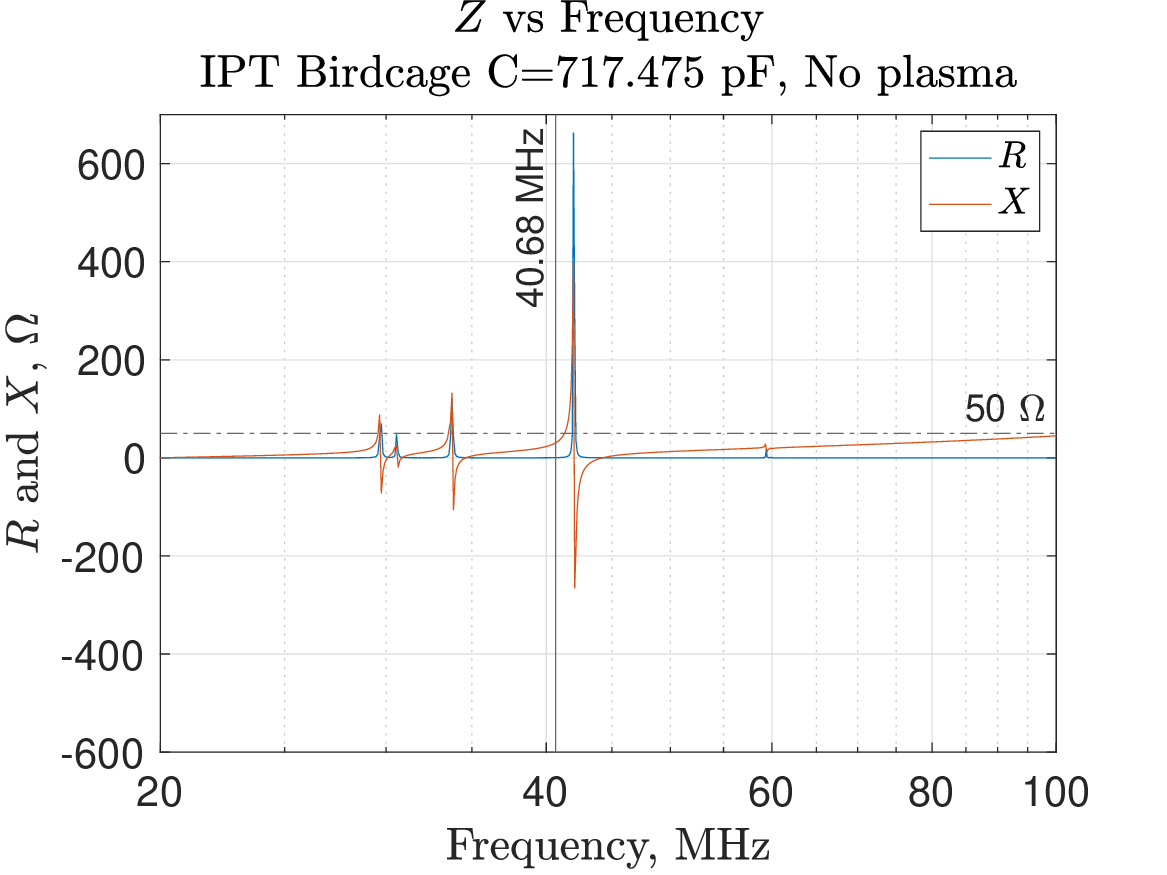}
	\caption{$Z$ vs $f$ IPT Design, XFdtd\textsuperscript{\textregistered} Simulation.}
	\label{fig:Z_f_717_pF}
\end{figure}

The IPT is assembled and tested with the FSH4 spectrum analyser from Rohde \& Schwarz that operates in the range of $f=\SI{100}{\kilo\hertz} - \SI{3.6}{\giga\hertz}$. The results show a good agreement with the XFdtd\textsuperscript{\textregistered} simulations, see Fig.~\ref{fig:S11_f_717_pF_comparison}: the relative difference in terms of frequency for the desired peak is of $3.84\%$, $f_{sim}=\SI{42.55}{\mega\hertz}$ vs $f_{meas}=\SI{44.25}{\mega\hertz}$ measured. The $S_{11}$ based on simulations is of $S_{11,sim}=\SI{-8.4}{\decibel}$ while for the measured one is of $S_{11,meas}=\SI{-6.0}{\decibel}$ resulting in a difference of $\Delta S_{11}=\SI{2.4}{\decibel}$ corresponding to $\sim 28.5\%$. The difference is acceptable due to the multiple uncertainties that are listed later, furthermore, the final antenna tuning is to be performed manually plus a mechanism of fine tuning is included into the IPT design. In particular, such deviation is due to the following aspects:\begin{itemize}
\item PCB hosting the capacitors is not included into XFdtd\textsuperscript{\textregistered} simulations;
\item Capacitors are considered ideal in the XFdtd\textsuperscript{\textregistered} simulations;
\item XFdtd\textsuperscript{\textregistered} simulations take into account an average value for $C$ equal for all capacitors;
\item Integration of the birdcage is not ideal $\Rightarrow$ tolerances in the birdcage legs manufacturing, their alignment, as well as their contact surfaces with the PCB differ from the ideal case.
\end{itemize}
\begin{figure}[h]
	\centering
	\includegraphics[width=\textwidth]{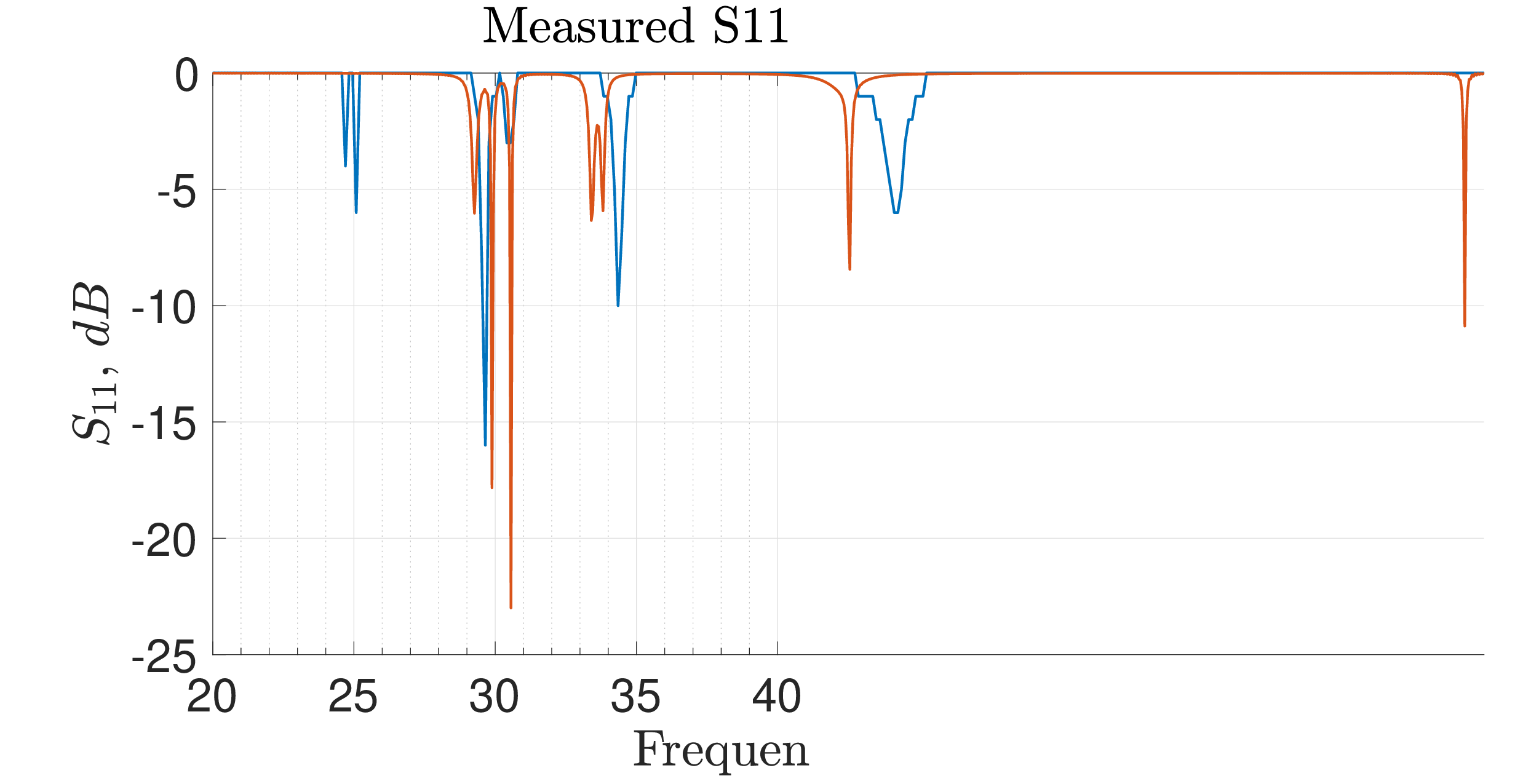}
	\caption{$S_{11}$ vs $f$ Simulation and Measurement (FSH4) IPT Design.}
	\label{fig:S11_f_717_pF_comparison}
\end{figure}
Preliminary experimenting with the IPT have shown how the resonance frequencies can be shifted upward by up to \SI{1}{\mega\hertz} by moving the injector toward the exhaust of the IPT, as well as by introducing a conductor in the discharge channel, as well as by decreasing the shield size, and or changing the leg length, this confirms the capability of later tuning once the antenna is defined. Moreover, as a plasma has densities much lower than those of a conductor, a very small shift upward in the resonance frequency is expected, \SI{0.01}{\mega\hertz} in the case of~\cite{EPFL1}. 
Finally, after testing with different capacitors and capacitances, it has been chosen to get as close as possible to the desired resonance frequency by using commercially available capacitors, and then final tune the IPT such that the injector is as far as possible from the volume enclosed by the antenna. This has been achieved with a final average capacitance of $C=\SI{813.69}{\pico\farad}$ per gap. 
Once the desired $f$ is reached, $Z$ must be matched, corresponding to a large amplitude of $S_{11}$. Firstly, a proper integration of the antenna itself must be ensured, secondly, large variations of $Z$ in a cylindrical birdcage antenna are achieved by modifying the relative position of the RF input and ground connector. To optimize $Z$, the factor $D$ defined as in Eq.~\ref{eq:D} according to~\cite{EPFL5}, is utilized, where $N_g$ is the leg number at which the ground is connected, $N_f$ is the one at which the RF input is connected, $N$ is the total number of legs, and $k$ is the desired resonance mode. By changing the respective locations, $D$ can be maximized, in particular $\abs{D}^2$ can be maximized up to $\abs{D}^2\rightarrow4$, corresponding to a matched $Z$, therefore an optimized coupling. For the current design, $N_f=1$ and $N_g=8$, resulting in $\abs{D}^2=3.85<4$.
\begin{equation}
D=\cos{\bigg[\bigg(N_g - \frac{1}{2}\bigg)\frac{k\pi}{N}\bigg]}-\cos{\bigg[\bigg(N_f - \frac{1}{2}\bigg)\frac{k\pi}{N}\bigg]}
\label{eq:D}
\end{equation}
Finally, the correct resonance $f$ is achieved at the desired $k=1$ mode, at matched $Z$ that corresponding to a maximized power coupling. The resulting IPT performance is measured by a NanoVNA v2 network analyser is shown before fine tuning in Fig.~\ref{fig:beforetuning} resulting in $f=\SI{40.55}{\mega\hertz}$ with $S_{11}=\SI{-21.7}{\decibel}$ at $Z_{IPT}=42.42-j0.99\SI{}{\ohm}$, and after fine tuning, by shifting the injector position, within Fig.~\ref{fig:aftertuning} with a final resonance frequency of $f=\SI{40.68}{\mega\hertz}$ with $S_{11}=\SI{-24.8}{\decibel}$ at $Z_{IPT}=44.56+j0.17\SI{}{\ohm}$. In both figures, the red line represents the reactance $X$, crossing the $X=\SI{0}{\ohm}$ at resonance, and the blue line representing the $S_{11}$ parameter over the calibrated range $f=40-\SI{41}{\mega\hertz}$, on the top left the resulting $Z$ is also  shown. Finally, also the measured $S_{11}$ peak width ensures to stay within the operating range once plasma is ignited	, according to an expected maximum $+\SI{1}{\kilo\hertz}$ frequency shift~\cite{EPFL1}. The final assembled birdcage antenna is shown in Fig.~\ref{fig:birdcageantenna}, and a photograph of the assembled thruster is shown in Fig.~\ref{fig:assembledIPT}. 
\begin{figure}[h]
	\centering
	\includegraphics[width=\textwidth]{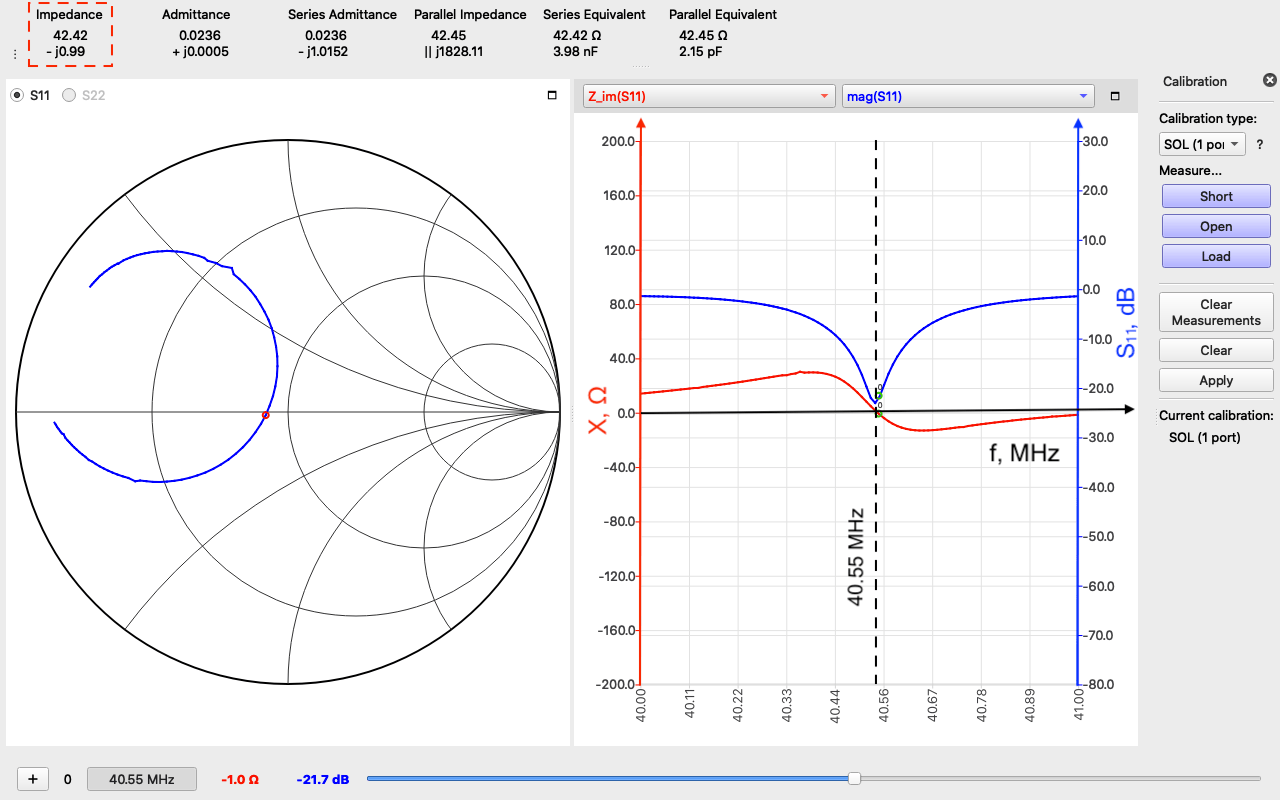}
	\caption{$S_{11}$ vs $f$ Measurement of IPT before Fine Tuning, NanoVNA v2.}
	\label{fig:beforetuning}
\end{figure}

\begin{figure}[h]
	\centering
	\includegraphics[width=\textwidth]{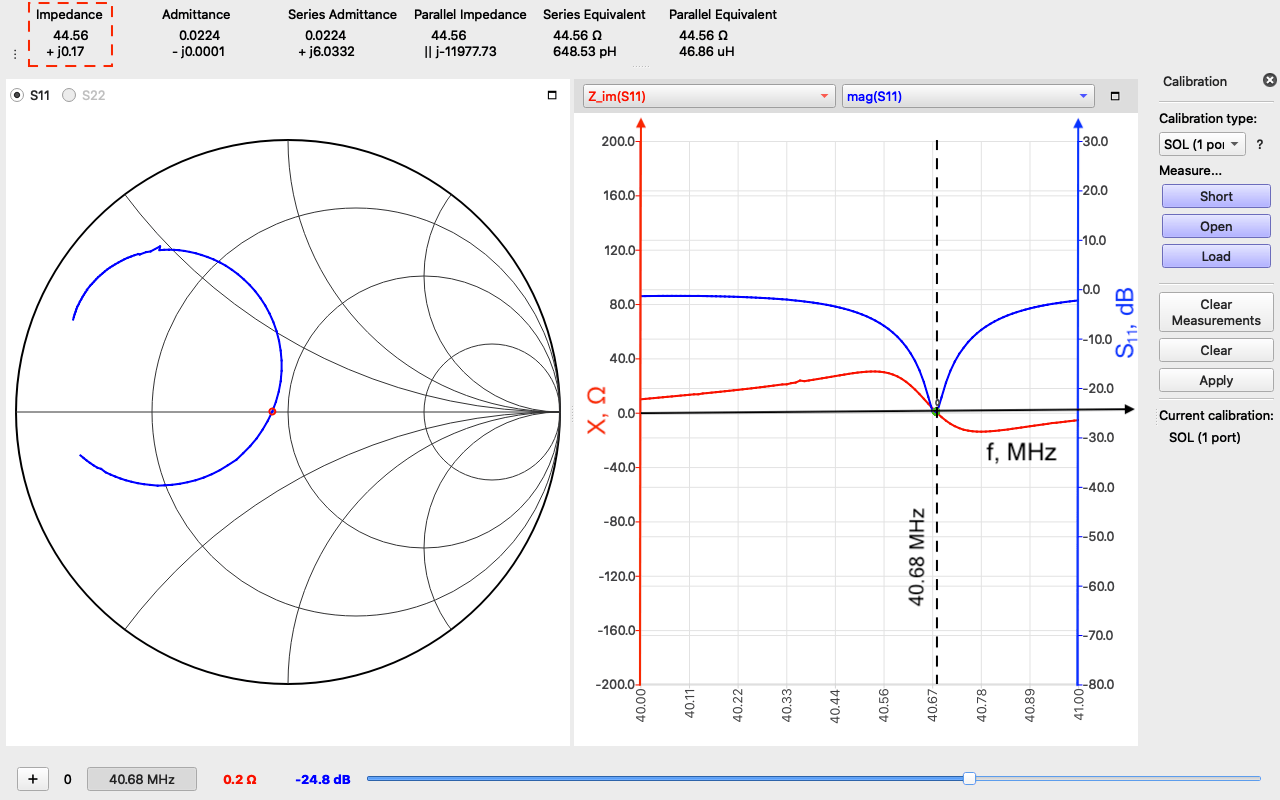}
	\caption{$S_{11}$ vs $f$ Measurement of IPT after Fine Tuning, NanoVNA v2.}
	\label{fig:aftertuning}
\end{figure}

\begin{figure}[h]
	\centering
	\includegraphics[width=\textwidth, trim={15cm 1cm 5cm 15cm},clip]{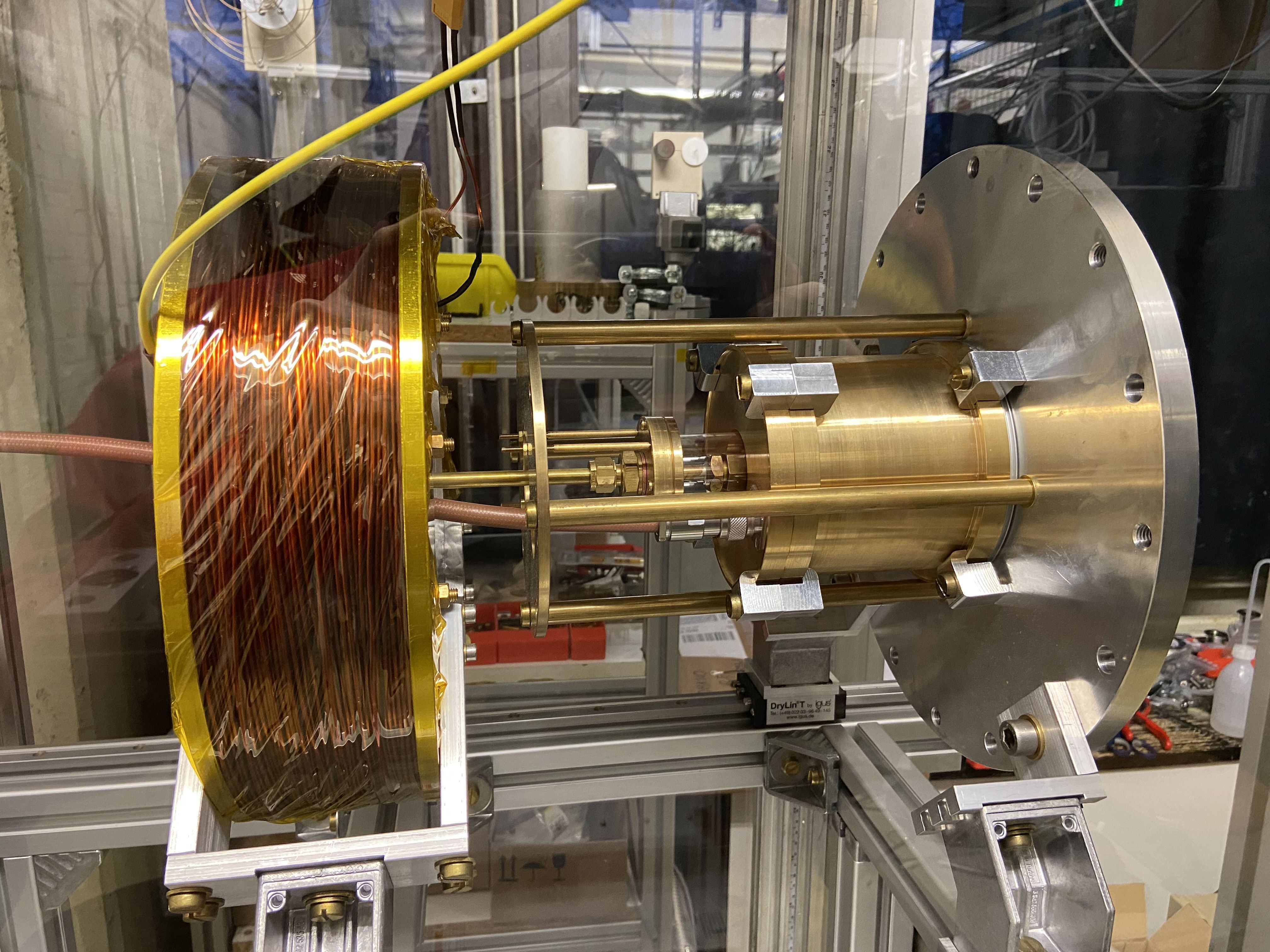}
	\caption{Assembled RF Helicon-based Plasma Thruster (IPT), Solenoid on the Left.}
	\label{fig:assembledIPT}
\end{figure}

\begin{figure}[h]
	\centering
	\includegraphics[width=.6\textwidth]{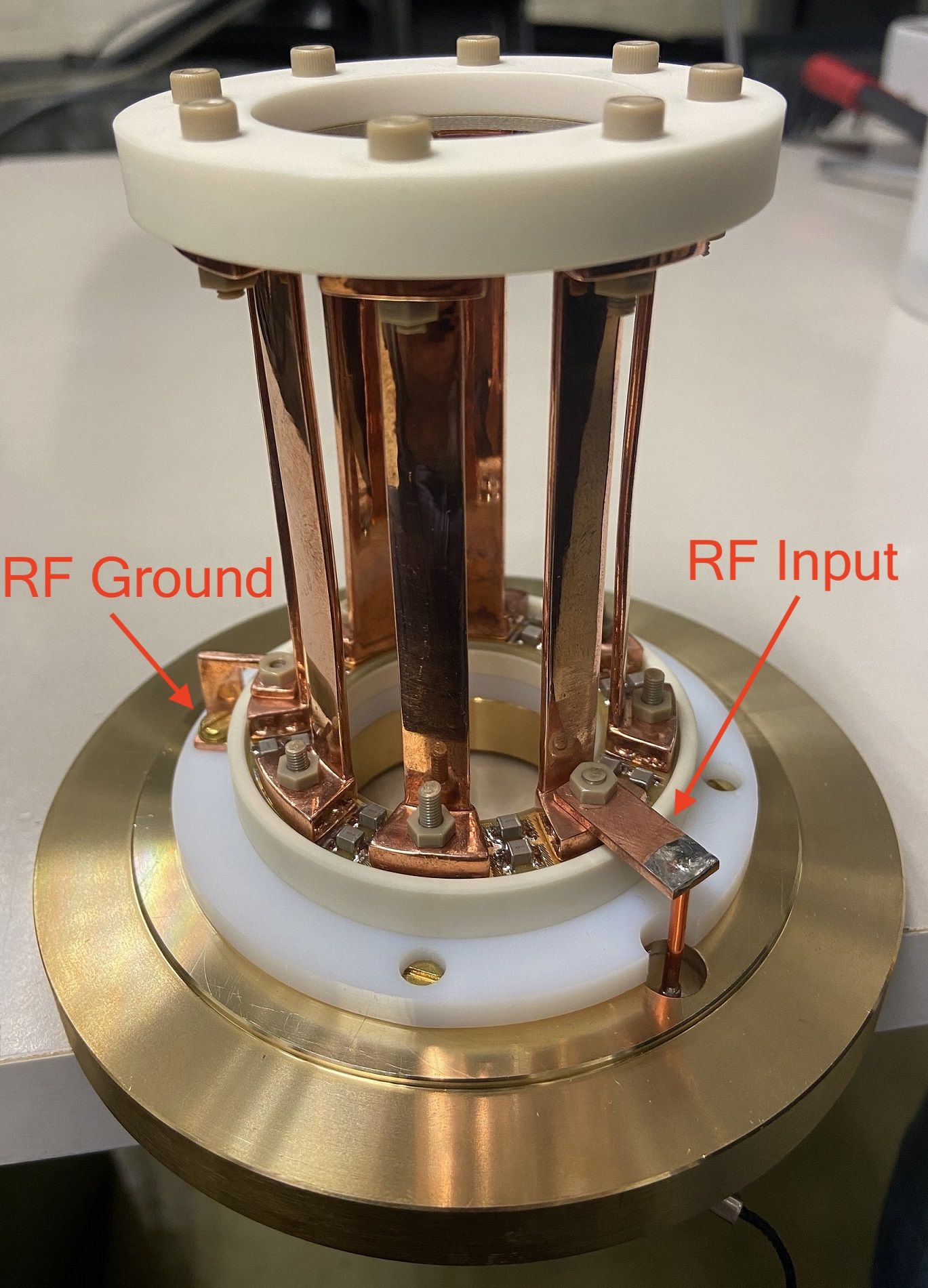}
	\caption{Assembled Birdcage Antenna.}
	\label{fig:birdcageantenna}
\end{figure}

\chapter{Experimental Set-Up}
\label{ch:fac}
\label{sec:facility}
This chapter describes the facility and the equipment that have been used The thruster and its respective power supply system are implemented into a movable rack that is attached to a vacuum chamber. In particular, this work has been performed at the vacuum chamber no.~12 at IRS. This is chosen as being the IRS EP facility as it is provided with the required pressure environment and state-of-the-art plasma diagnostics developed by Chan~\cite{chan2017characterization,chana2018breakthrough,chandevelopment,CHAN2019482,chan2019back}.

\section{Vacuum Chamber No.~12 Facility}
Vacuum Chamber No.~12 facility is equipped to test low to medium power ($P < \SI{1}{\kilo\watt}$) plasma source/electric propulsion devices, see Fig.~\ref{fig:tank12}, therefore requiring the respective pumping system and, for the plasma devices power and gas supply as well monitoring and measurement systems. The facility displayed in Fig.~\ref{fig:tank12} is provided with the following subsystems:
\begin{itemize}
\item Vacuum chamber and pumping system;
\item Data acquisition (DAQ);
\item Gas supply;
\item DC power supply.
\end{itemize}
The thruster and its respective power supply subsystem are built in a separated rack, namely RF power station (IPT) in Fig.~\ref{fig:tank12}, that is attached to the Vacuum Chamber No.~12. In the following subsections, each subsystem is described providing the respective technical specifications.

\subsection{Vacuum Subsystem}
The vacuum chamber has a diameter of $D=\SI{1}{\meter}$ and a length of $L=\SI{2.75}{\meter}$, it has multiple ports that provide optical access to the chamber as well as gas and electrical feedthroughs, both for commanding, measuring or power feeding. The vacuum is provided by a three-stage pumping system. The first stage is a Leybold Heraeus DK100 rotary piston vacuum pump with a suction rate of $\dot{V}=\SI{115}{\cubic\meter\per{\hour}}$. The second stage is a Leybold RUVAC WS 501 roots vacuum pump with a suction rate of $\dot{V}=\SI{505}{\cubic\meter\per{\hour}}$. Finally, the last stage is a Pfeiffer HiPace 700 turbomolecular pump, with a suction rate of $\dot{V}=\SI{2400}{\cubic\meter\per{\hour}}$. The final base pressure in the chamber (without gas flow) is of $p_{ch}<\SI{0.01}{\pascal}$. The suction rates and final base pressure are based on \ce{N2} gas. All the three vacuum pump operate simultaneously during the operation of the facility, the turbomolecular pump is directly mounted at the rear of the vacuum chamber through a manual gate valve.
\begin{figure}[h]
\centering
\includegraphics[width=1\textwidth, trim={1cm 0.5cm 2cm 1cm},clip]{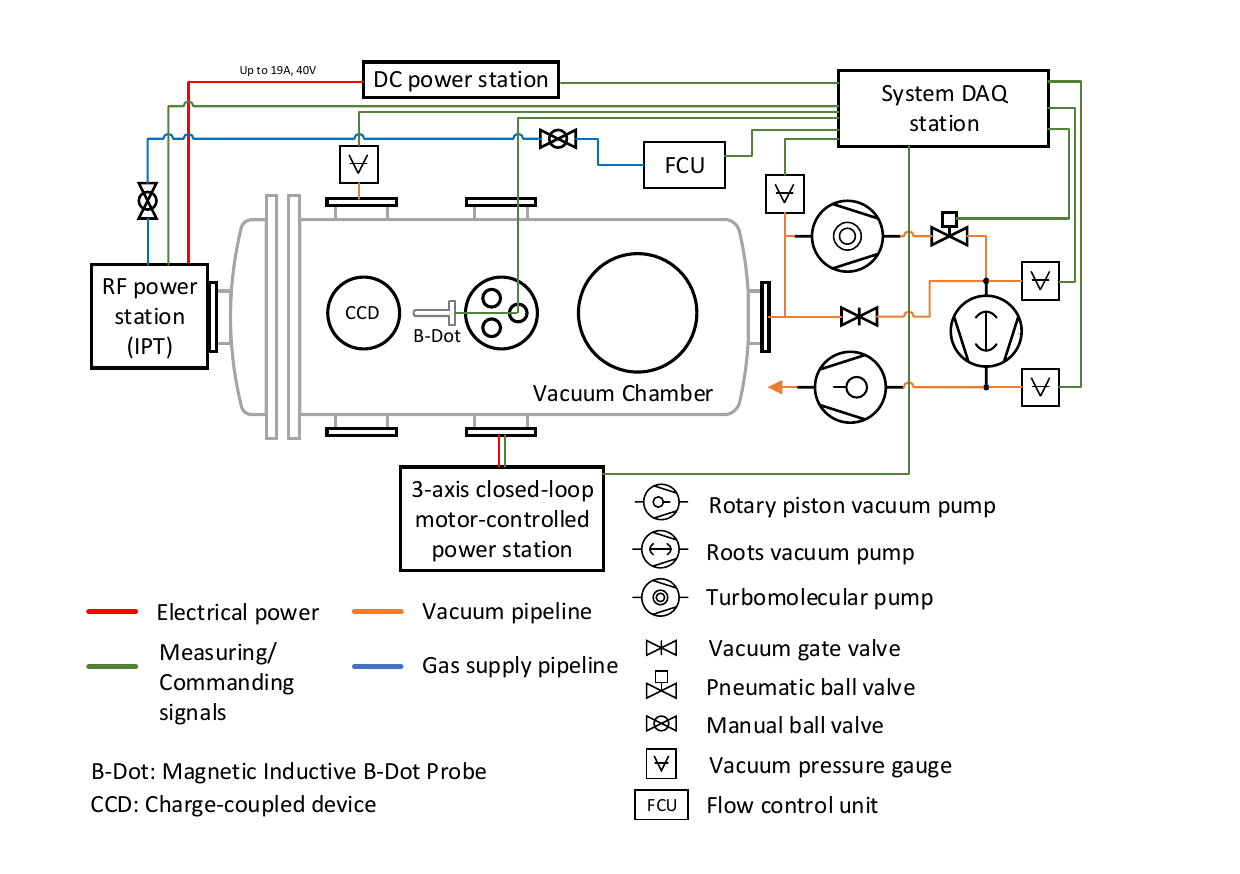}
\caption{Vacuum Chamber No.~12 Facility Schematics.}
\label{fig:tank12}
\end{figure}

\subsection{Gas Supply Subsystem}
\label{sec:gas}
Gas is supplied by a system of two mass flow controllers (MFC) Bronkhorst FG-201CV-RBD-33-V-DA, each capable of managing a wide variety of gases. Each of them can provide a mass flow of \ce{Ar} in the range between $\dot{m}=0.02-\SI{1}{\milli\gram\per{\second}}$. Bi-gas mixture can be provided at a total maximum of $70~sccm$. The system is equipped with check valves, and the gas flows through Swagelok\textsuperscript{\textregistered} pipelines to the injector head of the thruster. Gas is provided by gas bottles, containing respectively \ce{Ar}, \ce{O2}, and \ce{N2}, with a purity grade that is $\geq 5.0$. 

\subsection{Plasma Diagnostics}
The facility is also equipped with a 3-axis close-loop remotely controlled platform. The system has a position feedback-control mechanism which enables the auto-correction of the quasi-real-time position. It provides two translational and one rotational controlled axes. The stroke of the translational axes are of \SI{1200}{\milli\meter} and \SI{400}{\milli\meter}  with respective resolutions of $\pm\SI{0.025}{\milli\meter}$ and $\pm\SI{0.01}{\milli\meter}$. Both have an accuracy $< \SI{0.1}{\milli\meter}$. The rotational axis has a total sweeping angle of \SI{170}{\degree} with a resolution of \SI{0.5}{\degree}. On the platform, plasma diagnostics can be mounted. Currently Langmuir and Farady probes are available but they need the respective RF compensation circuit to be used for measuring RF thruster performance~\cite{sudit1994rf,chen2009}.
\begin{figure}[h]
\centering
	\begin{subfigure}{.5\textwidth}
\includegraphics[width=1.05\textwidth, trim={5.5cm 1cm 3.5cm 1cm},clip]{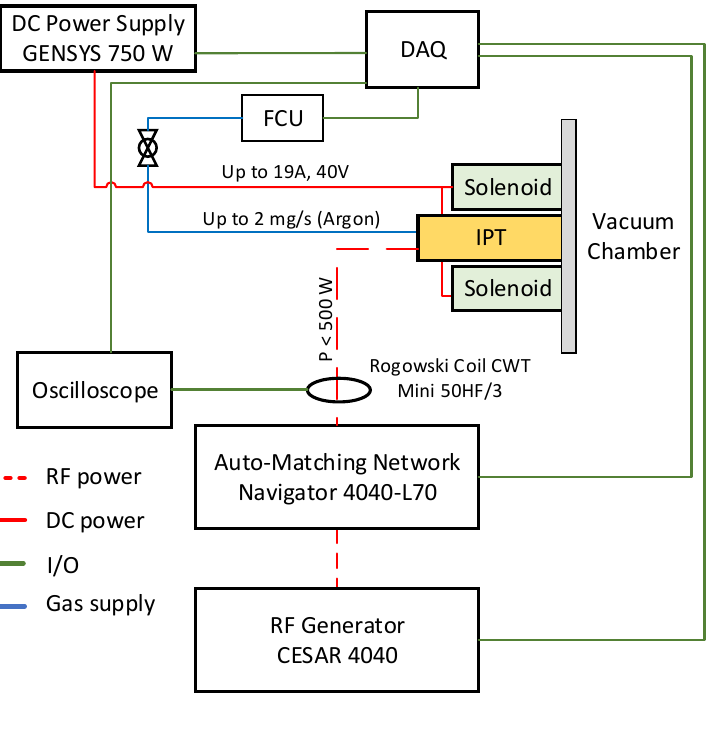}
	\caption{Thruster Power Station Schematics.}
	\label{fig:fac_schem}
	\end{subfigure}
	\begin{subfigure}{.45\textwidth}
	\centering
\includegraphics[width=.9\textwidth]{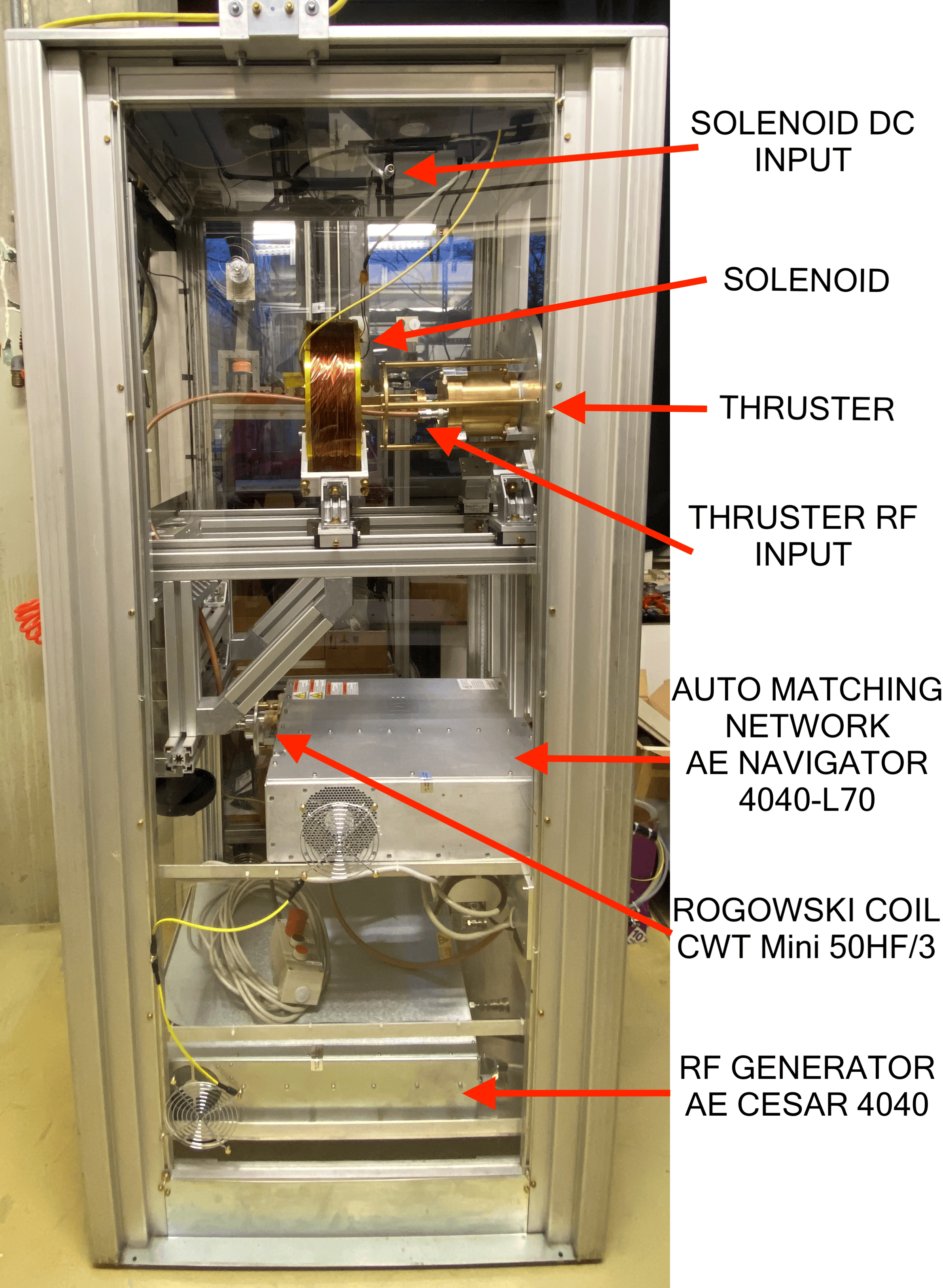}
	\caption{Thruster Power Station Photograph.}
	\label{fig:fac_schem_photo}
	\end{subfigure}
	\label{eq:fac_schem_gen}
	\caption{IPT Thruster Power Station.}
\end{figure}

\subsection{DAQ Subsystem}
The DAQ subsystem is composed of a main computer operating LabVIEW connected to a Datascan system in which the facility parameters are recorded. The power supply system and the DAQ schematics is shown within Fig.~\ref{fig:fac_schem}.

\subsection{Thruster Power Supply Subsystem}
The power supply system of the thruster is composed by the RF generator, the auto-matching network, the Rogowski coil to measure the RF current, and the DC power supply for the solenoid. The RF generator is the Advanced Energy CESAR~4040, that delivers up to $P_f<\SI{4}{\kilo\watt}$ at a frequency of $f=\SI{40.68}{\mega\hertz}$. This is connected to the Advanced Energy Navigator 4040-L70 auto-matching network that matches the thruster's impedance $Z_{IPT}$ to the standard matching impedance of this system of $Z_S=50+j0\SI{}{\ohm}$. At the output plug of the auto-matching network, the Rogowski coil CWT Mini 50HF/3 is installed to measure the RF current reaching the thruster, therefore the birdcage antenna during operation, it can operate up to $f<\SI{50}{\mega\hertz}$ with a \SI{12.5}{\nano\second} rise time. The DC power supply is the GENSYS \SI{750}{\watt} that delivers up to $I_S=\SI{19}{\ampere}$ current at the voltage of $V_S=\SI{40}{\volt}$ for the solenoid to produce the required static magnetic field $B_0$. The tuning of the thruster before each test is measured with the NanoVNA v2 Network Analyser shown in Fig.~\ref{fig:nvna}. It operates in a frequency range between \SI{50}{\kilo\hertz} and \SI{3}{\giga\hertz} with a system dynamic range (calibrated) \SI{70}{\decibel} up to \SI{1.5}{\giga\hertz} and a sweep rate of $80-100$ points$/s$, calibration is applied before each measurement is conducted. 

\begin{figure}[h]
\centering
\includegraphics[width=0.45\textwidth]{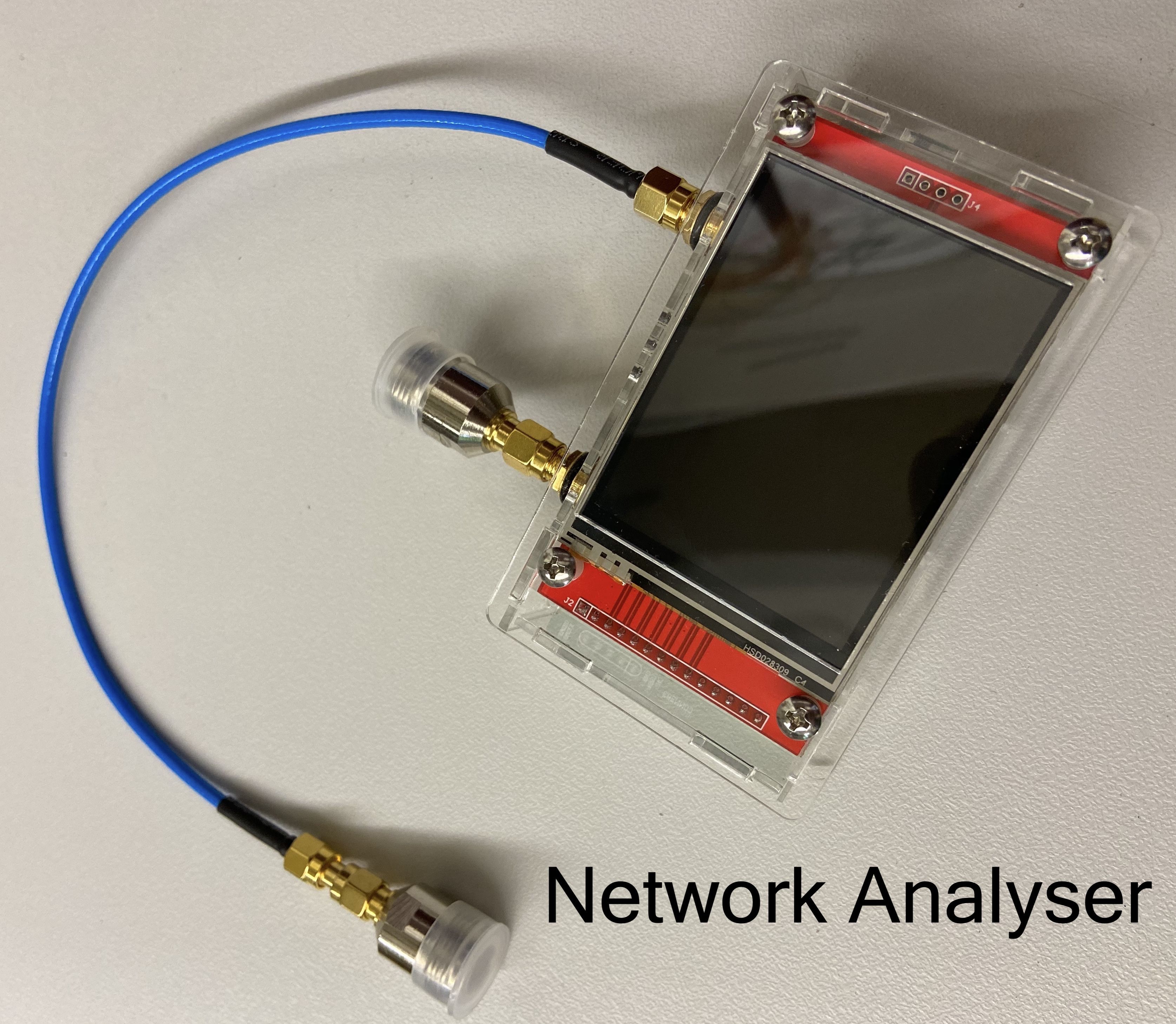}
\caption{Network Analyzer NanoVNA v2.}
\label{fig:nvna}
\end{figure}

\section{Thruster Impedance Model}
\label{ch:impedance}
The IPT RF circuit is represented in Fig.~\ref{fig:IPT_RF_circuit}. The first branch on the left is the \textbf{RF Generator} one, that includes the generator itself as the source with a fixed impedance of $Z_{S}=50+j0\SI{}{\ohm}$ and the coaxial cable that connects it to the automatic matching network. The second branch is in the middle, the \textbf{matching network}. This is composed by the internal inductors $L_1=\SI{100}{\nano\henry}$ and $L_2=\SI{350}{\nano\henry}$, the internal variable capacitors $C_1=8-\SI{250}{\pico\farad}$ and $C_2=4-\SI{100}{\pico\farad}$, and the transmission line (\SI{50}{\ohm}) coaxial cable that connects it to the IPT, that is the third branch.  The matching network's task is to optimize the impedance by changing capacitor's positions so that the circuit composed by the matching network itself and the IPT, contained within the thick line rectangle, is matched to $Z_{S}=50+j0\SI{}{\ohm}$.
\begin{figure}[h]
\centering
\begin{circuitikz}
      \draw (0,-2)
      to[sV=$Z_{S}$] (0,0)
       to[TL=$\lambda/4$](0,2);
      \draw (0,-2)
      to node[ground] {} (0,-2.5);
      \draw (0,2)
      to[L=$L_1$] (3,2)
      to[vC=$C_1$] (3,-1.75)
      to node[ground] {} (3,-2.75);
      \draw
      (3,2) to[short] (3.75,2) 
      to[L=$L_2$] (5.5,2)
      to[vC=$C_2$] (7,2)
      to[TL=$\lambda/2$] (9,2)
      to[short](10,2)
      to[short](10,0);
      \draw (10,0)
      to node[mslstub, rotate=90]{$Z_{IPT}$} (10,-2.5); 
      \draw (10,-2.25) to node[ground] {} (10,-2.25);
      \draw[black, dashed] (-2.5,-3) rectangle (0.5,3) node[above,xshift=-1.5cm,yshift=-0.5cm] {\textbf{RF Generator}};
      \draw[black, dash dot] (0.7,-3) rectangle (9,3) node[midway,xshift=0.5cm,yshift=-2.7cm] {\textbf{Matching Network}};
      \draw[black, dashed] (9.2,-3) rectangle (10.5,3) node[above,xshift=-0.6cm,yshift=-0.5cm] {\textbf{IPT}};
      \draw[black, thick] (0.6,-3.2) rectangle (10.7,3.2) node[above,xshift=-5cm] {\textbf{Matched to $Z_{S}$}};
\end{circuitikz}
\caption{Thruster System Impedance Model}
\label{fig:IPT_RF_circuit}
\end{figure}
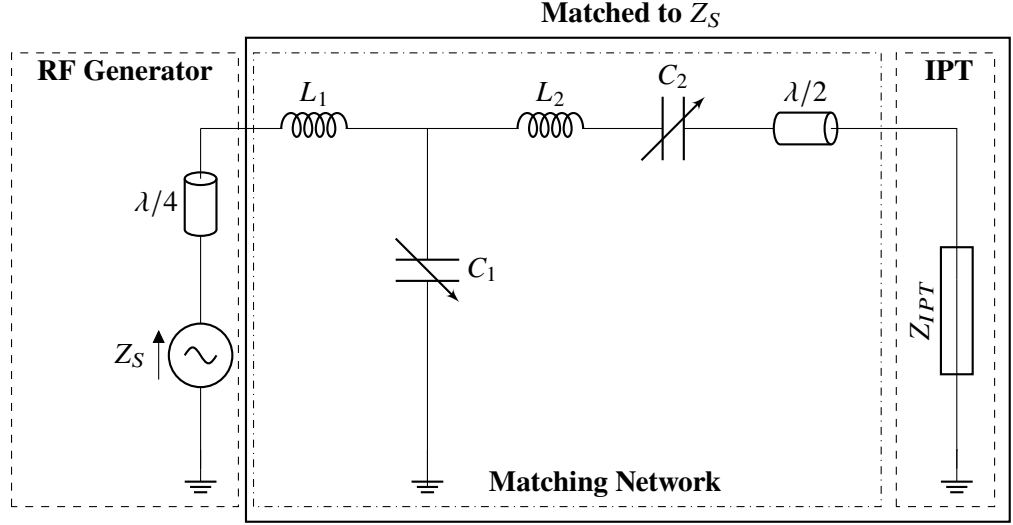
The estimation of the IPT impedance during operation with the current setup can be derived from the respective capacitor's positions given that the rest of the circuit is known. The electric circuit is shown in Fig.~\ref{fig:IPT_RF_circuit}, and the main hypothesis that the load is completely matched by the matching network, so that $Z_{S}=50+j0\SI{}{\ohm}=Z_{MN}+Z_{IPT}$, where $Z_{MN}$ is the impedance of the matching network. The transmission line of length $\lambda/4$ is ignored in the circuit, as well as the transmission line of length $\lambda/2$ as they do not lead to any $Z$ transformation.
The basic equation of the circuit is shown in Eq.~\ref{eq:RF_circuit_eq}.
\begin{equation}
Z_{S}=j\omega L_1 + \frac{1}{\frac{1}{(j\omega C_1)^{-1}}+\frac{1}{(j\omega C_2)^{-1}+j\omega L_2+Z_{IPT}}}
\label{eq:RF_circuit_eq}
\end{equation}
Given that $Z_{S}=50+j0\SI{}{\ohm}$ and $\omega = 2\pi f$, with $f=\SI{40.68}{\mega\hertz}$, are fixed, $Z_{IPT}=R_{IPT}+jX_{IPT}$ can be calculated if the capacitor's position is known and if the matching network is operating within its envelope. After some iteration, Eq.~\ref{eq:RF_circuit_eq} becomes Eq.~\ref{eq:RF_circuit_eq2}.
\begin{equation}
Z_{IPT}=R_{IPT}+jX_{IPT}=\frac{\bigg[\frac{L_1}{C_1}+\frac{L_1}{C_2}+\frac{L_2}{C_1}-\frac{1}{\omega C_1 C_2}\bigg]+j\biggl[Z_{S}\biggl( \omega L_2 - \frac{1}{\omega C_1} - \frac{1}{\omega C_2} \biggr)\biggr]} {Z_{S}+j\biggl( \frac{1}{\omega C_1} - \omega L_1 \biggr)}
\label{eq:RF_circuit_eq2}
\end{equation} 
Eq.~\ref{eq:RF_circuit_eq2} is in the form $(A+jB)/(C+jD)$, therefore the rule of division between complex numbers is applied to obtain finally Eq.~\ref{eq:RF_circuit_eq3} in which the changing capacitance $C_1$ and $C_2$ is inserted to evaluate the corresponding $Z_{IPT}$.

\begin{equation}
\begin{split}
Z_{IPT} & =\frac{\biggl[ \biggl( \frac{L_1}{C_1}+\frac{L_1}{C_2}+\frac{L_2}{C_1}-\frac{1}{\omega C_1 C_2} + \biggl( \omega L_2 - \frac{1}{\omega C_1} - \frac{1}{\omega C_2}\biggr) \biggl( \frac{1}{\omega C_1} - \omega L_1 \biggr)\biggr]Z_S} {Z^2_S+\biggl( \frac{1}{\omega C_1} - \omega L_1 \biggr)^2} + \\
& \quad + j \left[\frac{Z^2_S\biggl( \omega L_2 - \frac{1}{\omega C_1} - \frac{1}{\omega C_2} \biggr) - \biggl( \frac{L_1}{C_1}+\frac{L_1}{C_2}+\frac{L_2}{C_1}-\frac{1}{\omega C_1 C_2} \biggr) \biggl( \frac{1}{\omega C_1} - \omega L_1 \biggr) }{Z^2_S+\biggl( \frac{1}{\omega C_1} - \omega L_1 \biggr)^2}\right]
\end{split}
\label{eq:RF_circuit_eq3}
\end{equation}
The final equation that can be used to evalute $Z_{IPT}$ in its two component are shown in Eq.~\ref{eq:RF_circuit_eq_RE} for the real part of $Z_{IPT}$, and in Eq.~\ref{eq:RF_circuit_eq_IM} for the imaginary part of $Z_{IPT}$.
\begin{equation}
R_{IPT}=\frac{\biggl( 2\frac{L_1+L_2}{C_1} + 2\frac{L_1}{C_2} - \frac{1}{\omega^2 C^2_1} - \omega^2 L_1 L_2 - \frac{1}{\omega^2 C_1 C_2} \biggr)Z_S}{Z^2_S+\frac{1}{\omega^2 C^2_1}+\omega^2 L^2_1-2\frac{L_1}{C_1}}
\label{eq:RF_circuit_eq_RE}
\end{equation}

\begin{equation}
X_{IPT}=\frac{Z^2_S\biggl( \omega L_2 - \frac{1}{\omega C_1} - \frac{1}{\omega C_2} \biggr) - \biggl( \frac{L_1}{C_1}+\frac{L_1}{C_2}+\frac{L_2}{C_1}-\frac{1}{\omega C_1 C_2} \biggr) \biggl( \frac{1}{\omega C_1} - \omega L_1 \biggr) }{Z^2_S+\biggl( \frac{1}{\omega C_1} - \omega L_1 \biggr)^2}
\label{eq:RF_circuit_eq_IM}
\end{equation}

The result of those equations are valid only under the hypothesis that the auto-matching network is operating within its operational regime. 
Moreover, even though the whole circuit design is to fit the \SI{50}{\ohm} condition, the influence on $Z$ of the plug containing the Rogowski coil cannot not be measured currently and therefore is not included into the modelling. 

Finally, if the capacitor positions are known during the thruster operation, it is possible to extract $Z_{IPT}$ in its two real and imaginary components. However, many unknowns can effect the final extracted value. To conclude, the insertion of a direct $Z$ measurement during operation between the matching network and the thruster can ensure a more precise knowledge of $Z_{IPT}$ and therefore of the power coupling. A possibility would be the insertion of a network analyer at the output plug of the matching network.

\chapter{B-dot Probe Design}
\label{ch:b-dot}
This chapter describes the design of a three-axis B-dot probe, also known as inductive magnetic probe, B probe, or magnetic probe~\cite{Bdot1, Bdot2, EPFL4}. It is a diagnostic tool to sense time-varying magnetic fields. A B-dot probe is hereby designed and built to asses the presence of helicon waves within the plasma plume of the IPT. This chapter describes the working principle of the B-dot probe, the design, and the integration.

\section{Principle of Operation}
The B-dot probe operation is based on the Faraday's law of induction: a time-varying magnetic field induces an electromotive force on a conductive material nearby. In formula, this is represented by Eq.~\ref{eq:faraday}, where $\epsilon$ is the electromotive force and $\phi_{B}$ is the magnetic flux. \begin{equation}
\epsilon = -\frac{d \phi_{B}}{d t}
\label{eq:faraday}
\end{equation}

The head of the probe is the sensing element, it is composed by multiple loops of conductive wire which axis is aligned with the direction of the time-varying magnetic field to be measured~\cite{Bdot1, Bdot2, EPFL4}. Such field induces a voltage on the probe that is read and used to estimate the amplitude of the $B-$field. To correctly estimate it, the area enclosed by the loop wire $A$ must be well known, as well as the number of loops $N$. The voltage $V$ produced by the time-varying magnetic field $dB_{tot}/dt$ is given in Eq.~\ref{eq:Bdot}, where $\omega=2\pi f$ is the angular frequency at $f=\SI{40.68}{\mega\hertz}$.
\begin{equation}
V=-NA\abs{\frac{dB_{tot}}{dt}}=-NA \omega \abs{B}
\label{eq:Bdot}
\end{equation}
Finally, $NA$ is the calibration factor of the probe and is a priori measured. The probe itself should be as small as possible while at the same time producing the highest signal-to-noise ratio (SNR) with low parasitic signals~\cite{EPFL4}. The $N$ is selected to be a maximum of $N=5$ as each loop produces a counteracting magnetic field $B_{loop}$ once a current $I$ is flowing into it, that will reduce the time responsiveness of the probe itself, see Eq.~\ref{eq:Bdot1}.
\begin{equation}
V=-NA\abs{\frac{dB_{ext}}{dt}+\frac{dB_{loop}}{dt}}=V_B-L_c\frac{dI}{dt}
\label{eq:Bdot1}
\end{equation}	

A DC pick-up voltage arises in the measurement due to the $V$ difference between the probe and the plasma~\cite{EPFL4}. To remove such signal, RF power combiners, one per axis, that operate in a frequency range that includes that of the IPT input are used. In such way, the DC signal is removed from the probe output signal and only the AC one is extracted. As AC signals are measured, an accurate selection and inclusion in the measurement circuit of the respective cabling, connectors, and feedthroughs must be accounted~\cite{Bdot1}.

\section{Design}
The designed B-dot probe has three axis of measurement. The head is made of resin and provides the structure on which the three copper wire loops, each of $N=5$, are winded. The head is inserted on top of a tube made of PEEK mounted on a PEEK holder, that also functions as connecting element to the three-axis moving platform available within the Vacuum Chamber No.~12 at IRS. Inside the PEEK tube, six RG-178 \SI{50}{\ohm} coaxial cables are installed, the outer shield is grounded to the back plate of the probe, and each inner conductor is connected to one end of each of the three head loops. In such way, each wire is already shielded and does not require an extra outer shield, nor any $Z$ match within the B-dot probe shaft. The rear of the B-dot probe is made of brass, this also provides the mounting threads for 6 SMA flange adaptors. On top of the B-dot probe head and along the PEEK pipe until the holder, a borosilicate glass tube with closed tip is installed to reduce thermal loads by minimizing direct contact of the B-dot probe assembly with the plasma. At the back of the B-dot probe, six RG-178\SI{50}{\ohm} coaxial cable are connected at the SMA interfaces on one end, and to the three RF power combiners (one per axis) on the other end. Three RG-178 \SI{50}{\ohm} coaxial cables with a corresponding $\lambda/2$ length are connected to the vacuum coaxial \SI{50}{\ohm} feedthrough, in such way the impedance is not transformed and the signal loss is minimized, see Eq.~\ref{eq:trans}. For the same reason, three more RG-178 \SI{50}{\ohm} coaxial cables with a corresponding $\lambda/2$ length are connected between the vacuum coaxial \SI{50}{\ohm} feedthrough to the oscilloscope.
\begin{figure}[H]
	\centering
	\includegraphics[width=14cm]{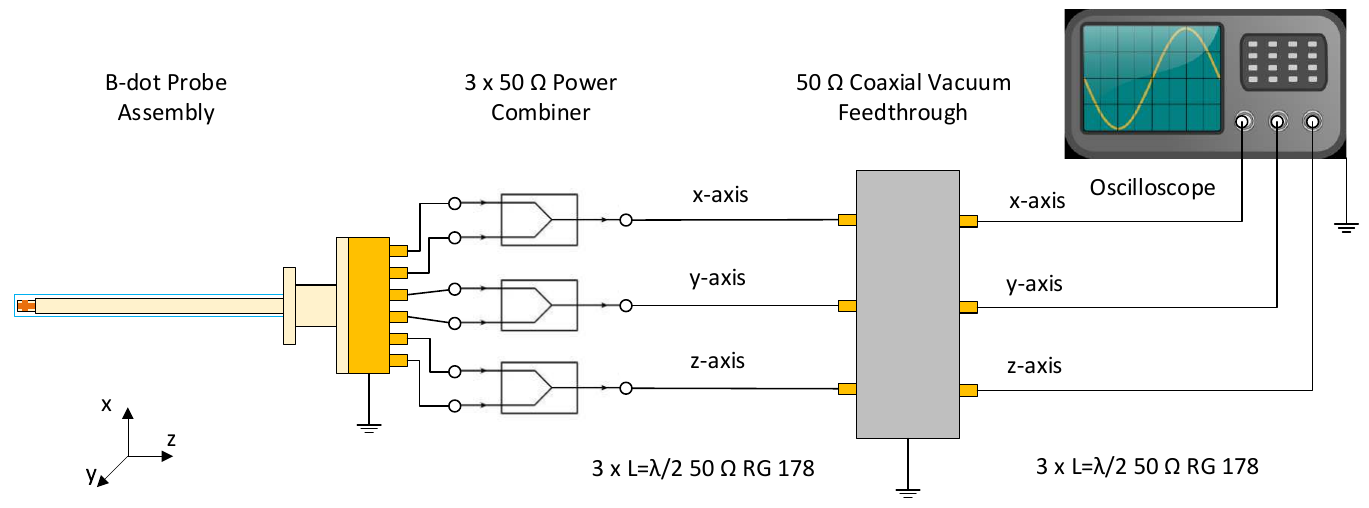}
	\caption{B-dot probe electrical circuit schematics.}
	\label{fig:bdotschematics}
\end{figure}
 The lengths of the coaxial cables are calculated based on the thruster's operating frequency and corrected by the VF of the cables. Once assembled, each cable is tested to verify the correct length and impedances. The electrical schematics of the B-dot probe is shown in Fig.~\ref{fig:bdotschematics}, while a render of the designed B-dot probe is shown within Fig.~\ref{fig:bdotrender}, and its assembled head in Fig.~\ref{fig:bdothead}.

\begin{figure}[h]
		\begin{subfigure}{.65\textwidth}
		\centering
		\includegraphics[width=1\textwidth]{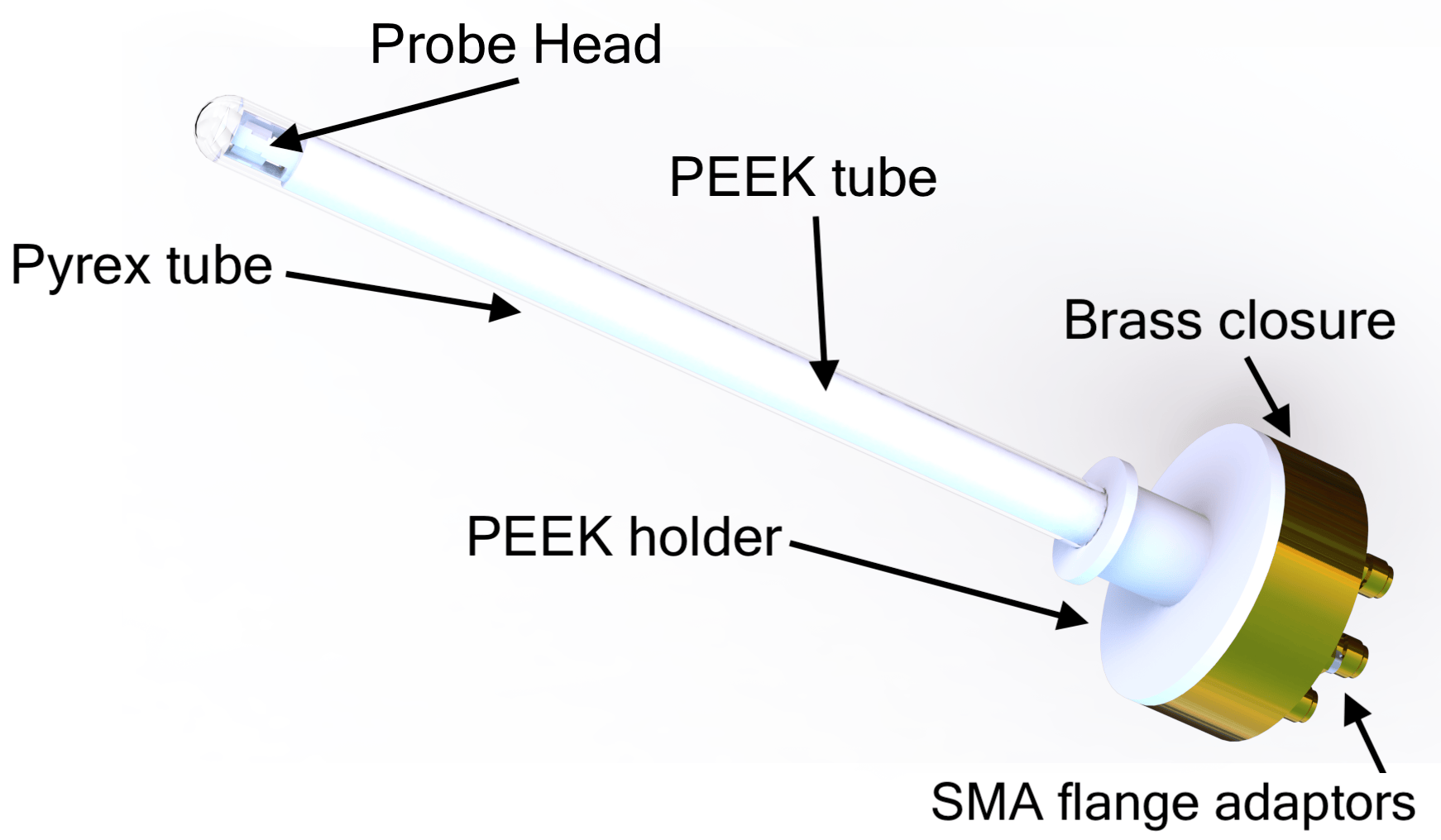}
		\caption{B-dot Probe Render.}
		\label{fig:bdotrender}
		\end{subfigure}
		\begin{subfigure}{.3\textwidth}
		\centering
		\includegraphics[height=1\textwidth, trim={3.8cm 1.5cm 3.4cm 0cm},clip]{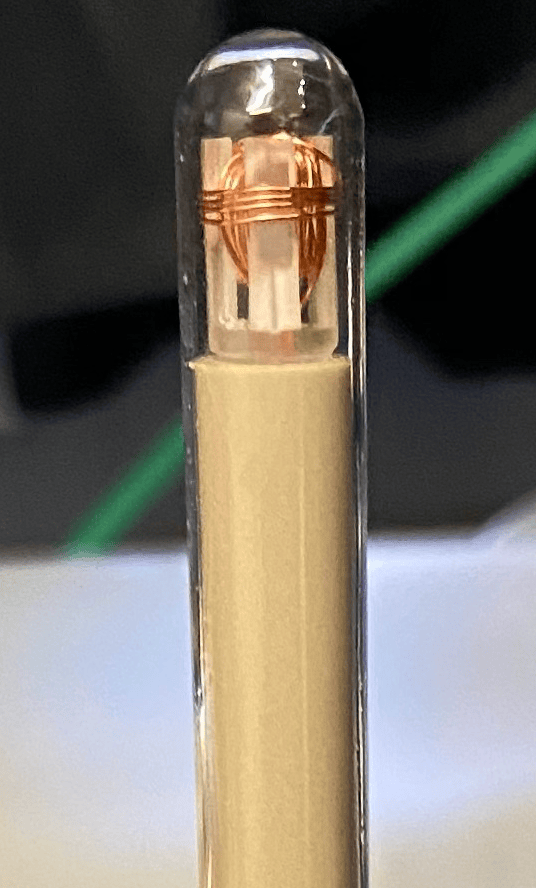}
		\caption{Assembled B-dot probe.}
		\label{fig:bdothead}
		\end{subfigure}
	\caption{Rendering of the designed B-dot probe and Assembled Head.}
	\label{fig:bdotrenderandhead}
\end{figure}


\section{Calibration}
The calibration of the B-dot probe, requires at first the calibration factor $NA$, which is the product of the number of turns $N$ and the area of each wire loop, of the  cross section magnetic wire $A_w=\SI{0.2}{\square\milli\meter}$. For the wire loop around the $x-$axis $A_x =\SI{16}{\milli\meter^2}$, around the $y-$axis $A_y =\SI{25}{\milli\meter^2}$, and around the $z$-axis $A_z =\SI{36}{\milli\meter^2}$, the precision of the manufacturing of the probe head is of $\SI{47}{\micro\meter}$. The number of turns has been fixed to $N_{x,y}=5$, while the $z$ axis has $N_z=4$ providing the respectively calibration factors of $NA_x=8.0\times10^{-5}$ turns$-\SI{}{\meter^2}$, $NA_y=12.5\times10^{-5}$ turns$-\SI{}{\meter^2}$, and $NA_z=14.4\times10^{-5}$ turns$-\SI{}{\meter^2}$.
To calibrate the probe two steps are necessary, for the low frequency calibration, Helmoltz coils are to be used~\cite{Bdot1,Bdot2}, while for the high frequencies, according to~\cite{EPFL4}, the use of the thruster itself is suggest, as the produced $B$-field is homogeneous within the thruster discharge channel and it is linearly polarized. The thruster is to be fed by means of a function generator at the resonance frequency of the antenna, $f=\SI{40.68}{\mega\hertz}$, for a given set of input voltages $V$. The known direction of the produced $B$-field serves to calibrate probe at different angles to the $B$-field produced by the thruster. By determining $I$ and $V$ flowing into the thruster's antenna, the amplitude of the produced $B$-field can be calculated, for example with XFdtd\textsuperscript{\textregistered}, and the calibration of the amplitude for high frequencies can be performed as well. The calibration set-up is shown in Fig.~\ref{fig:Bdotcal}, however due to issues in the probe final integration, the calibration could not be concluded yet. 
\begin{figure}[h]
	\centering
		\includegraphics[width=1\textwidth]{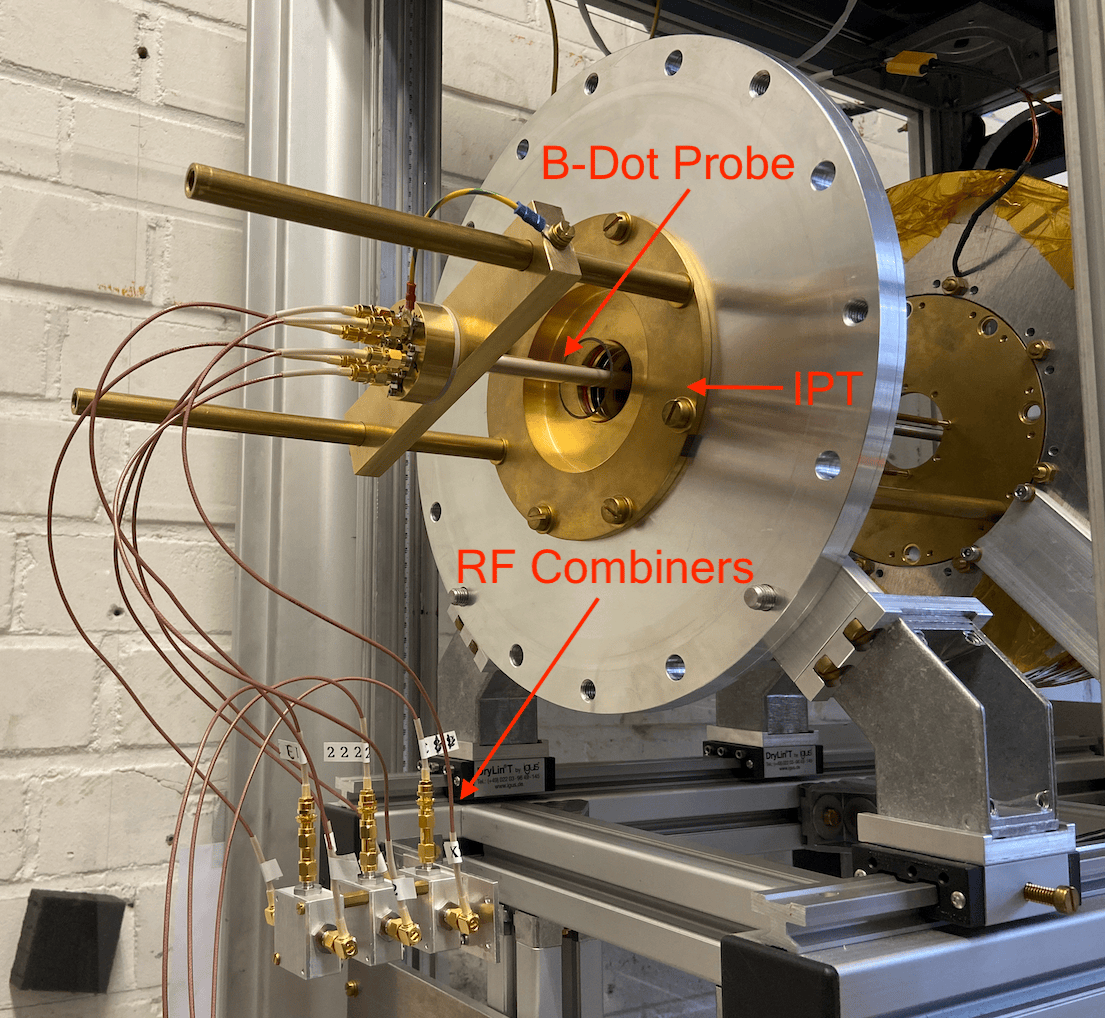}
		\caption{B-dot Probe Calibration Set-Up.}
		\label{fig:Bdotcal}
\end{figure}

\chapter{Thruster Discharge Test Campaign}
\label{ch:test}
This chapter describes the discharge characterization test campaign of the IPT. The aim is to evaluate the working envelope of the IPT on different propellants, respective particle fluxes $\dot{N}$, applied magnetic fields $B_0$, and input powers $P_f$. The test campaign is performed with \ce{Ar}, \ce{N2}, and \ce{O2} as propellant. The applied $\dot{N}$ is in the range of $\dot{N}=20.305-\SI{2.538}{1\per{\second}}$, see Tab.~\ref{table:ArN2O2}. This is based on the range of operation of the gas supply subsystem, see Sec.~\ref{sec:gas}, as well as to fit within the rage of to the ABEP estimated collectible $\dot{m}_{thr}$ for specular, diffuse, and EFD-based intakes designed in Ch.~\ref{ch:intake}. The pressure in the vacuum chamber during operation of the IPT is maintained between $p_{ch}=0.12-\SI{0.3}{\pascal}$. 
\begin{table}[H]
\centering
\caption{IPT Test Mass Flows and Particle Fluxes for \ce{Ar}, \ce{O2}, and \ce{N2}.}
\label{table:ArN2O2}
\begin{tabular}{cccc}
$\dot{m}$, \ce{Ar} & $\dot{m}$, \ce{O2} & $\dot{m}$, \ce{N2} & $\dot{N}$ \\
\toprule
\SI{}{\milli\gram\per{\second}} &\SI{}{\milli\gram\per{\second}}  &\SI{}{\milli\gram\per{\second}} & $\SI{}{1\per{\second}}$\\
\midrule
0.800 & 0.650 & 0.568 & 20.305\\
0.600 & 0.487 & 0.426 & 15.228\\
0.400 & 0.323 & 0.284 & 10.152\\
0.200 & 0.162 & 0.142 & 5.076\\
0.100 & 0.081 & 0.071 & 2.538\\
\bottomrule
\end{tabular}
\end{table}

\section{Test General Set-Up and Procedure}
The test set-up is composed by the IPT mounted at Vacuum Chamber No.~12 with no plasma diagnostic applied. The IPT is fine-tuned before the test campaign by moving the injector, while the performance, in terms of $S_{11}$ and $Z$, is visualized with the NanoVNA v2 calibrated between $40-\SI{41}{\mega\hertz}$. The measurement is performed before and after each set of tests to evaluate eventual variations due to thermal expansion. The tuning required condition is set to $S_{11}<-\SI{20}{\decibel}$ at \SI{40.68}{\mega\hertz}, so that $> 99\%$ of the $P_f$ is coupled to the antenna and $<1\%$ reflected. The solenoid is placed at a fixed position, at the minimum distance of \SI{35}{\milli\meter} from the main flange as shown in Fig.~\ref{fig:EM_location}.
\begin{figure}[H]
\centering
\includegraphics[width=.6\textwidth]{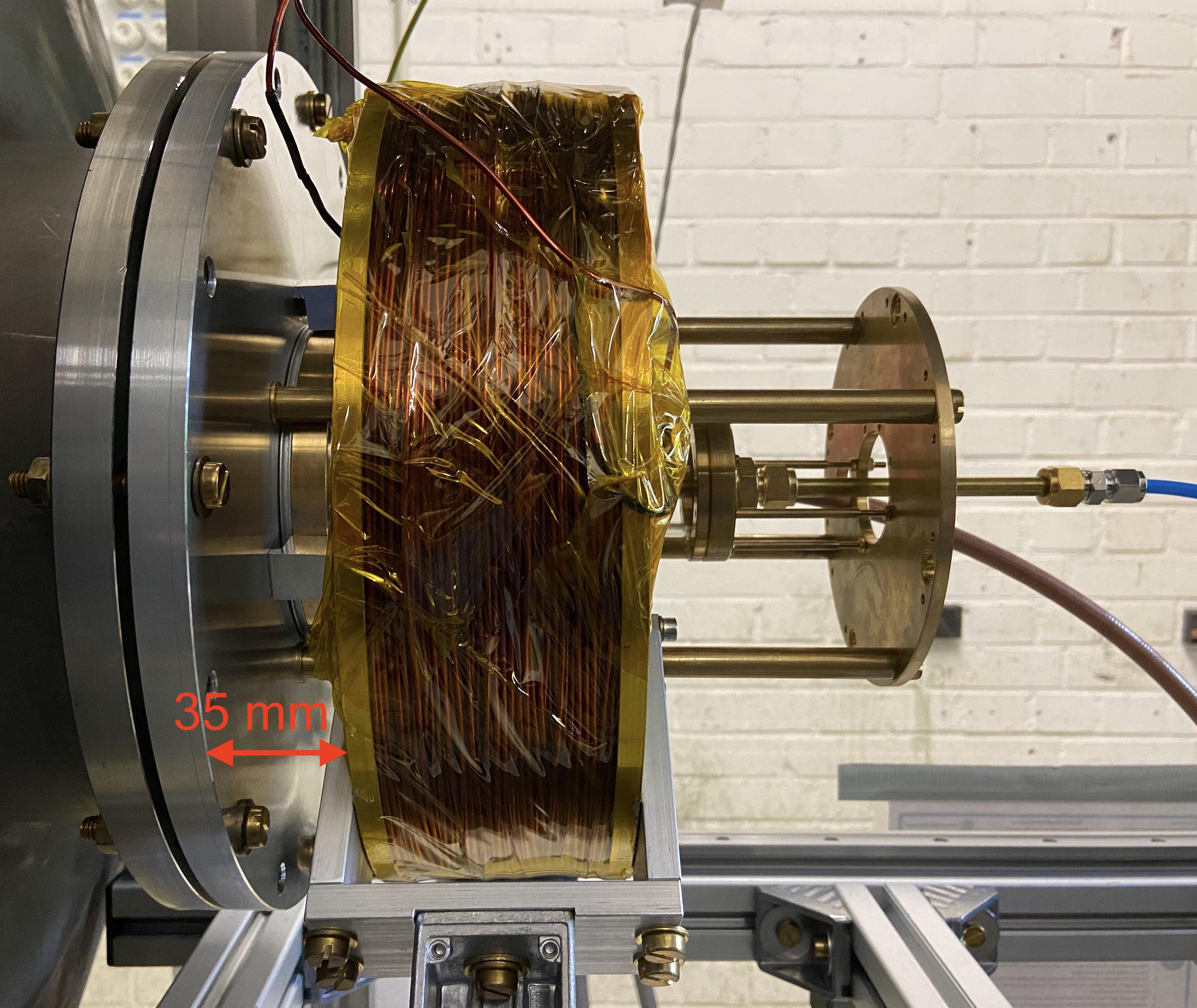}
\caption{Solenoid Location during Test.}
\label{fig:EM_location}
\end{figure}

The control parameters are the input mass flow $\dot{m}$, the forward and the reflected power $P_f$ and $P_r$, both read at the generator, and the solenoid current $I_S$. Photographs are taken through an optical window. The settings of focal length, aperture, and exposure time are set to maximize the comparability of photographs at different conditions. Finally, $P_r$ indicates the amount of power that the matching network is not able to compensate and that is finally reflected back to the RF generator. Therefore, the coupled power $P_c$ can be defined as $P_c=P_f-P_r$, and is the power flowing between the matching network and the IPT, therefore it can be re-written as $P_c=P_{IPT}+P_{loss}$ resulting in $P_{IPT}<P_f$.This finally highlights that the power that is really required for thruster operation is less than the forward power $P_{IPT}<P_f$. 
The solenoid output $B$-field $B_0$ is measured at its centre is based on the calibration curve of Fig.~\ref{fig:solenoidB0}~\cite{romanorgcep2}.
\begin{figure}[hb!]
\centering
\includegraphics[width=.8\textwidth]{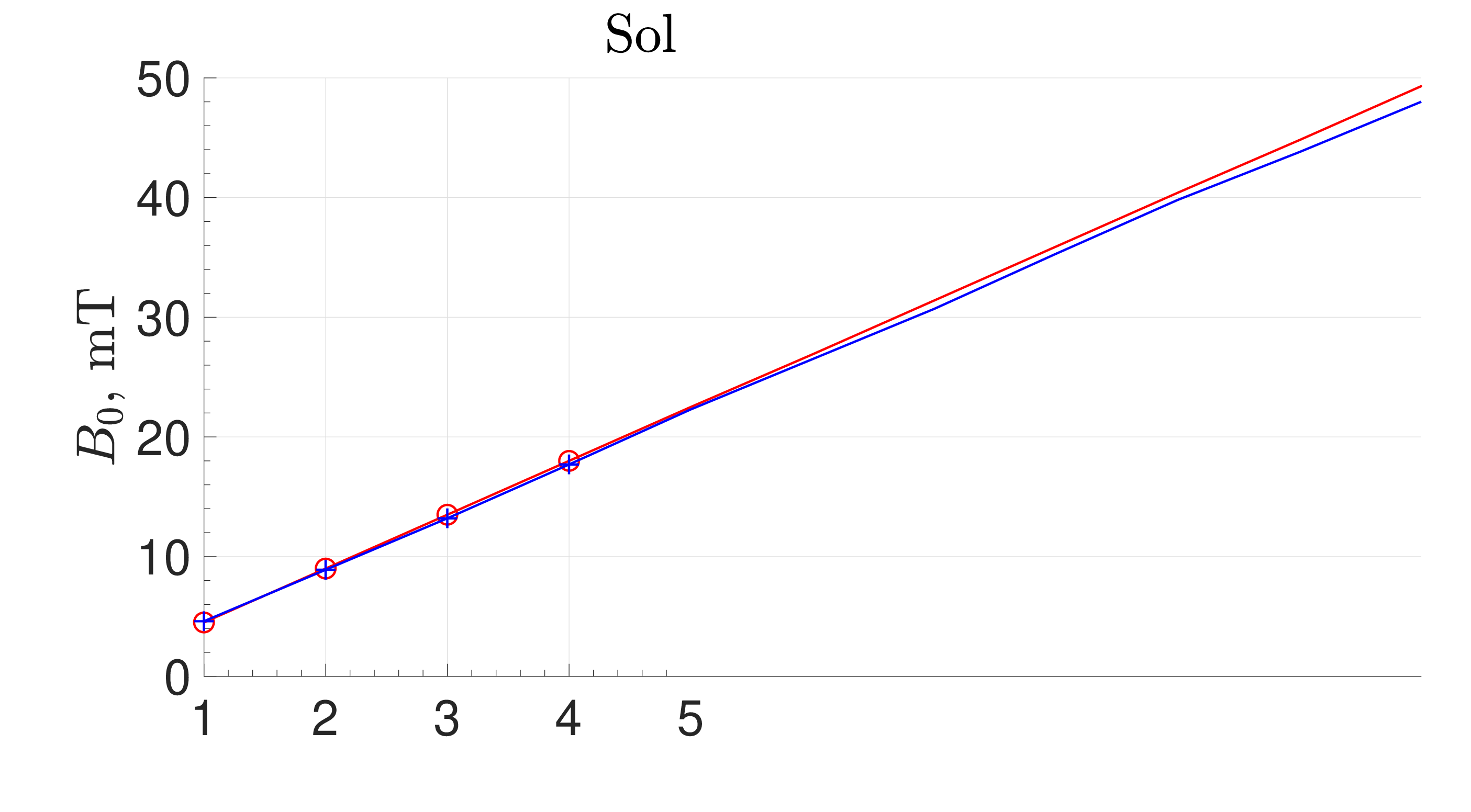}
\caption{Solenoid Output $B$-Field vs $I_S$, Theoretical and Measured.}
\label{fig:solenoidB0}
\end{figure}
 
\newpage
\section{IPT operating on \ce{Ar}} 
The IPT performance before the test in terms of impedance is of $Z=45.07+j0.43\SI{}{\ohm}$ and $S_{11}=-\SI{25.7}{\decibel}$ at $f=\SI{40.68}{\mega\hertz}$ measured with the NanoVNA v2, see Fig.~\ref{fig:resonance_Ar}. 
\begin{figure}[h]
\centering
\includegraphics[width=.9\textwidth]{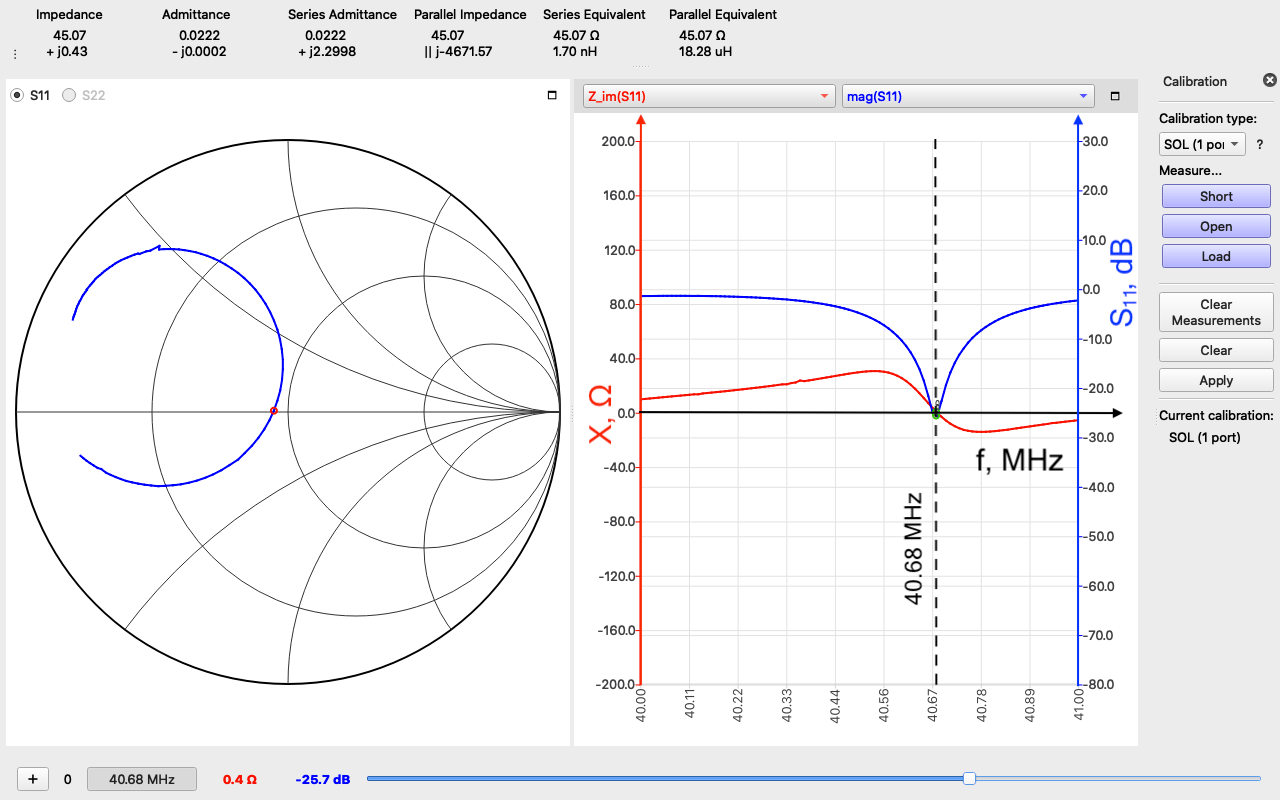}
\caption{IPT $Z_{IPT}$ and $S_{11}$ before \ce{Ar} test.}
\label{fig:resonance_Ar}
\end{figure}
\vspace{-11pt}

The test initiates with a total of $\dot{m}=\SI{0.8}{\milli\gram\per{\second}}$ of \ce{Ar}. Forward power $P_f$ is injected, while no magnetic field is applied, $B_0=\SI{0}{\milli\tesla}$. Once ignition is achieved, $B_0$ is applied.

Ignition is achieved for $P_f\sim50-\SI{60}{\watt}$ in the range $\dot{m}=0.2- \SI{0.8}{\milli\gram\per{\second}}$. For $\dot{m}=\SI{0.1}{\milli\gram\per{\second}}$, $P_f\sim\SI{85}{\watt}$ is required, while the discharge could be still maintained at $P_f=\SI{60}{\watt}$ after ignition. 
The forward power could be reduced down to $P_f=\SI{10}{\watt}$ at $\dot{m}=\SI{0.4}{\milli\gram\per{\second}}$ while having the plasma still ignited. 

After ignition, the forward power is set to $P_f=\SI{60}{\watt}$, and the solenoid is switched on at $I_S=\SI{10.07}{\ampere}$ resulting in $P_r=\SI{0}{\watt}$ for all $\dot{m}$. By decreasing it to $I_S=6.5-\SI{6.7}{\ampere}$, a visually more collimated jet is observed, while $P_r$ increases, see Fig.~\ref{fig:Ar_1} and Fig.~\ref{fig:Ar_2}. 

During the \ce{Ar} test campaign, to seek for sudden brightness jumps as observed in many helicon devices~\cite{chen220}, $P_f$ has been injected up to $P_f=\SI{300}{\watt}$ at $\dot{m}=\SI{0.6}{\milli\gram\per{\second}}$: a brighter plasma is observed with a final $P_r=30-\SI{50}{\watt}$, see Fig.~\ref{fig:Ar300}. No sudden brightness jumps have been observed while increasing $P_f$.
In general, it is observed that the reflected power $P_r$ is reduced by tuning the applied magnetic field $B_0$. Different plume configurations are achieved by varying $B_0$. A set of tests to visualize the effect of decreasing the $B_0$ has been performed, see Fig.~\ref{fig:Ar_Bfield}, a fixed input power of $P_f=\SI{60}{\watt}$ is injected for a $\dot{m}=\SI{0.6}{\milli\gram\per{\second}}$, and $I_S$ decreased in seven steps, respectively $I_S=10.07,~6.7,~6.5,~6.3,~6.0,~5.7,~\SI{3.5}{\ampere}$. In general, it must be noted that the guaranteed minimum reliable input power for RF generator operation is of $P_f=\SI{50}{\watt}$. Therefore, to reliably evaluate the IPT performance below this threshold, another power supply shall be used.

\begin{figure}[H]
\centering
\includegraphics[width=.65\textwidth, trim={12cm 0cm 2cm 0cm},clip]{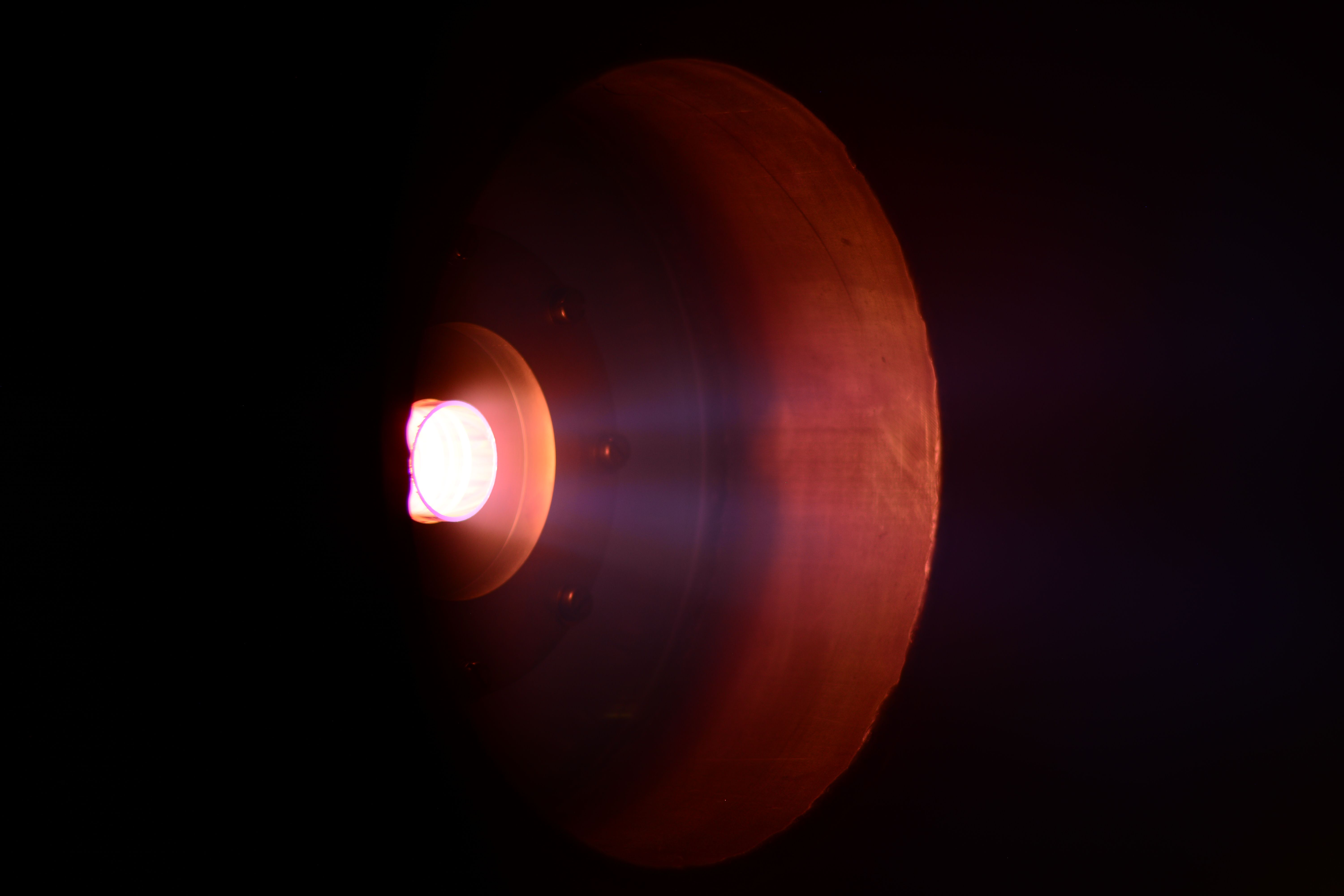}
\caption{IPT \ce{Ar} at $\dot{m}=\SI{0.8}{\milli\gram\per{\second}}, P_f=\SI{60}{\watt}$,  $P_r=\SI{4}{\watt}$,  $I_{S}=\SI{10.07}{\ampere}$, Focal length \SI{50}{\milli\meter}, Aperture $f/4$, Exposure time $1/5\SI{}{\second}$.}
\label{fig:Ar_1}
\end{figure}
\begin{figure}[H]
\centering
\includegraphics[width=.65\textwidth, trim={12cm 0cm 2cm 0cm},clip]{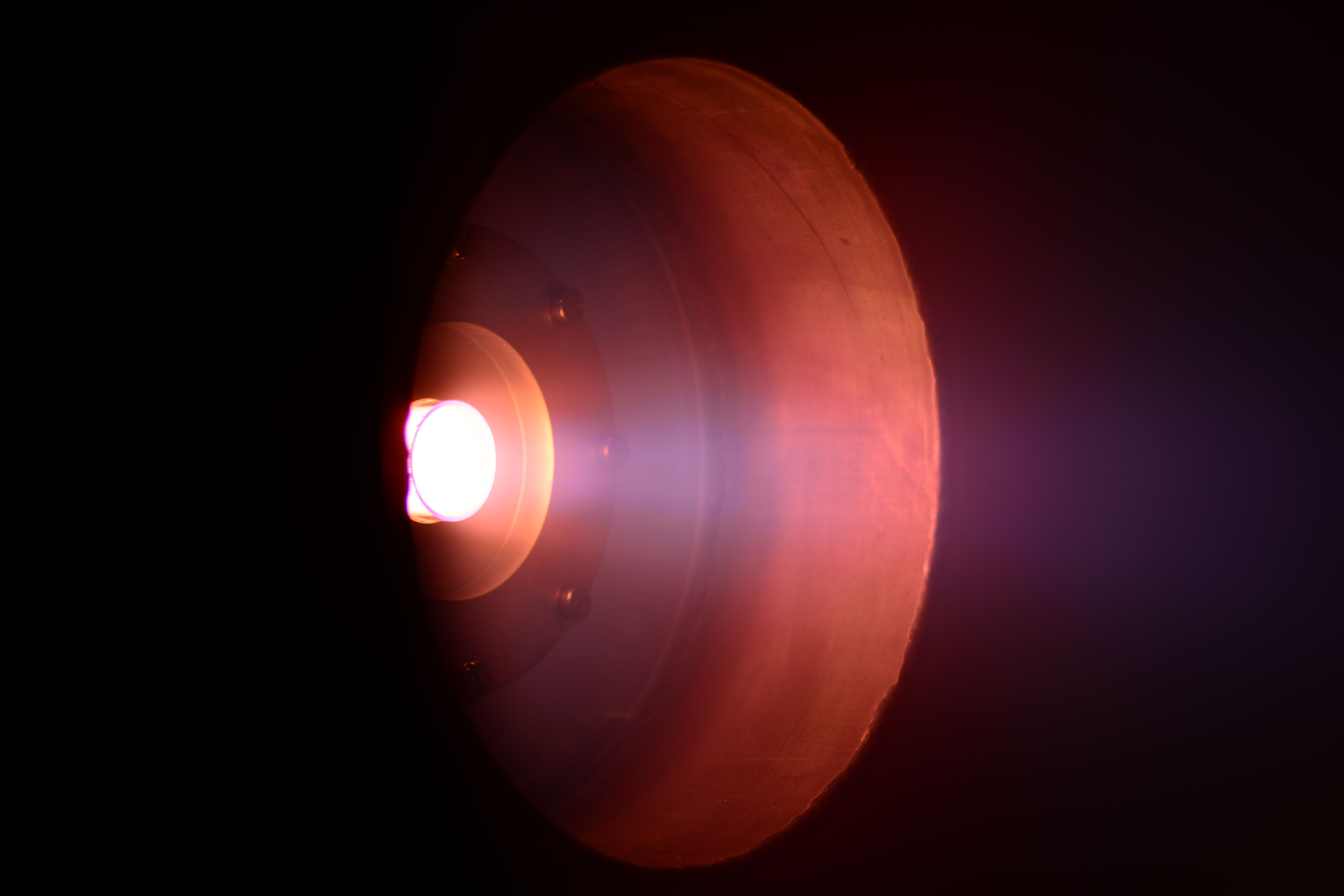}
\caption{IPT \ce{Ar} at $\dot{m}=\SI{0.8}{\milli\gram\per{\second}}, P_f=\SI{60}{\watt}$,  $P_r=15-\SI{20}{\watt}$,  $I_{S}=\SI{6.5}{\ampere}$, Focal length \SI{50}{\milli\meter}, Aperture $f/4$, Exposure time $1/5\SI{}{\second}$.}
\label{fig:Ar_2}
\end{figure}

\begin{figure}[H]
 \centering
 \includegraphics[width=.7\textwidth, trim={12cm 0cm 2cm 0cm},clip]{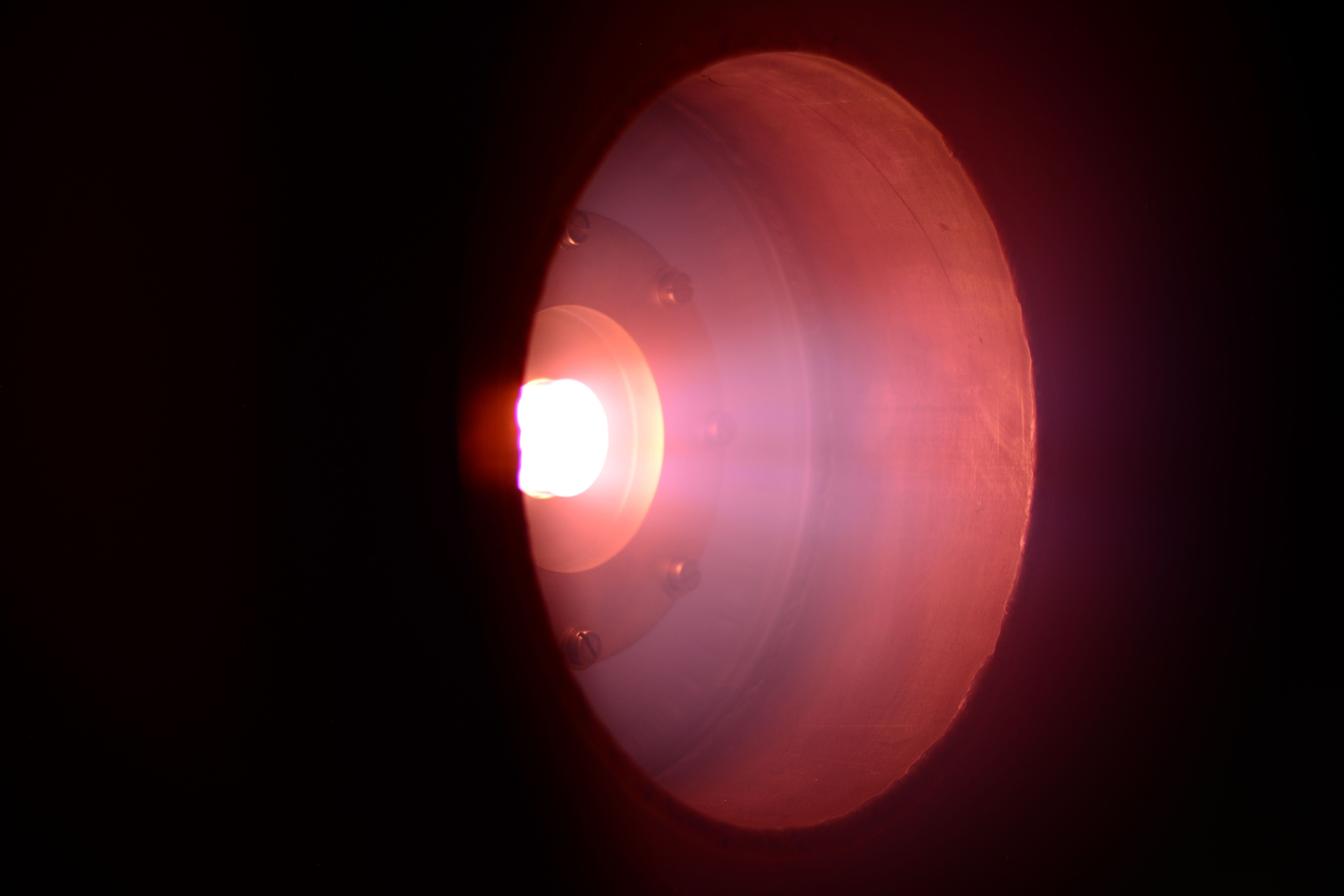}
 \caption{IPT \ce{Ar} at $\dot{m}=\SI{0.6}{\milli\gram\per{\second}}, P_f=\SI{300}{\watt}$,  $P_r=30-\SI{50}{\watt}$,  $I_{S}=\SI{10.04}{\ampere}$, Focal length \SI{50}{\milli\meter}, Aperture $f/4$, Exposure time $1/5\SI{}{\second}$.}
 \label{fig:Ar300}
 \end{figure}

 \begin{figure}[H]
     \centering
     \begin{subfigure}[b]{0.5\textwidth}
         \centering
         \includegraphics[width=.7\textwidth, trim={12cm 0cm 2cm 0cm},clip]{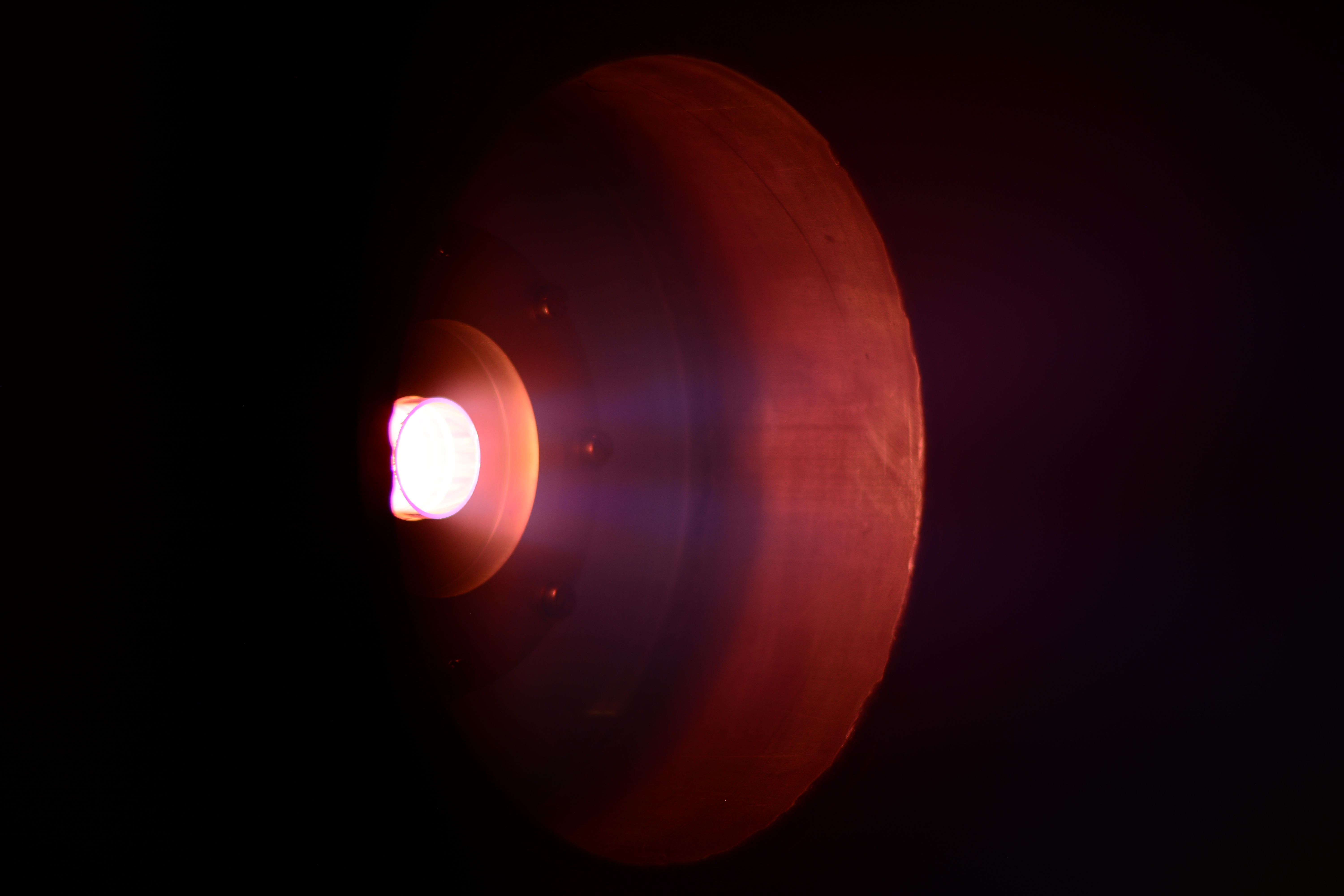}
         \caption{$I_S=\SI{10.07}{\ampere}$, $P_r=8-\SI{18}{\watt}$}
     \end{subfigure}%
     ~ 
     \begin{subfigure}[b]{0.5\textwidth}
         \centering
         \includegraphics[width=.7\textwidth, trim={12cm 0cm 2cm 0cm},clip]{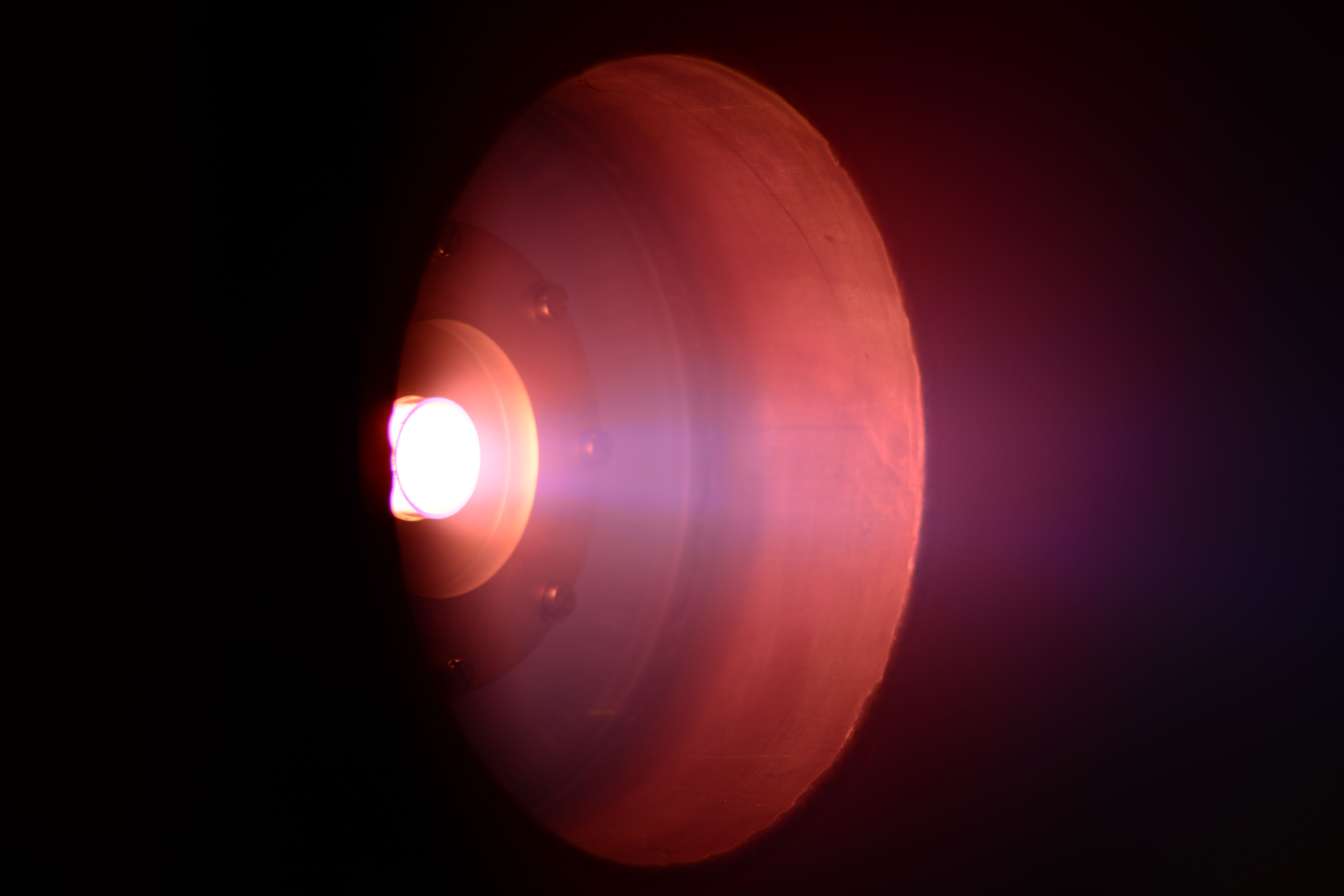}
         \caption{$I_S=\SI{6.7}{\ampere}$, $P_r=9-\SI{18}{\watt}$}
     \end{subfigure}
     ~
       \begin{subfigure}[b]{0.5\textwidth}
             \centering
             \includegraphics[width=.7\textwidth, trim={12cm 0cm 2cm 0cm},clip]{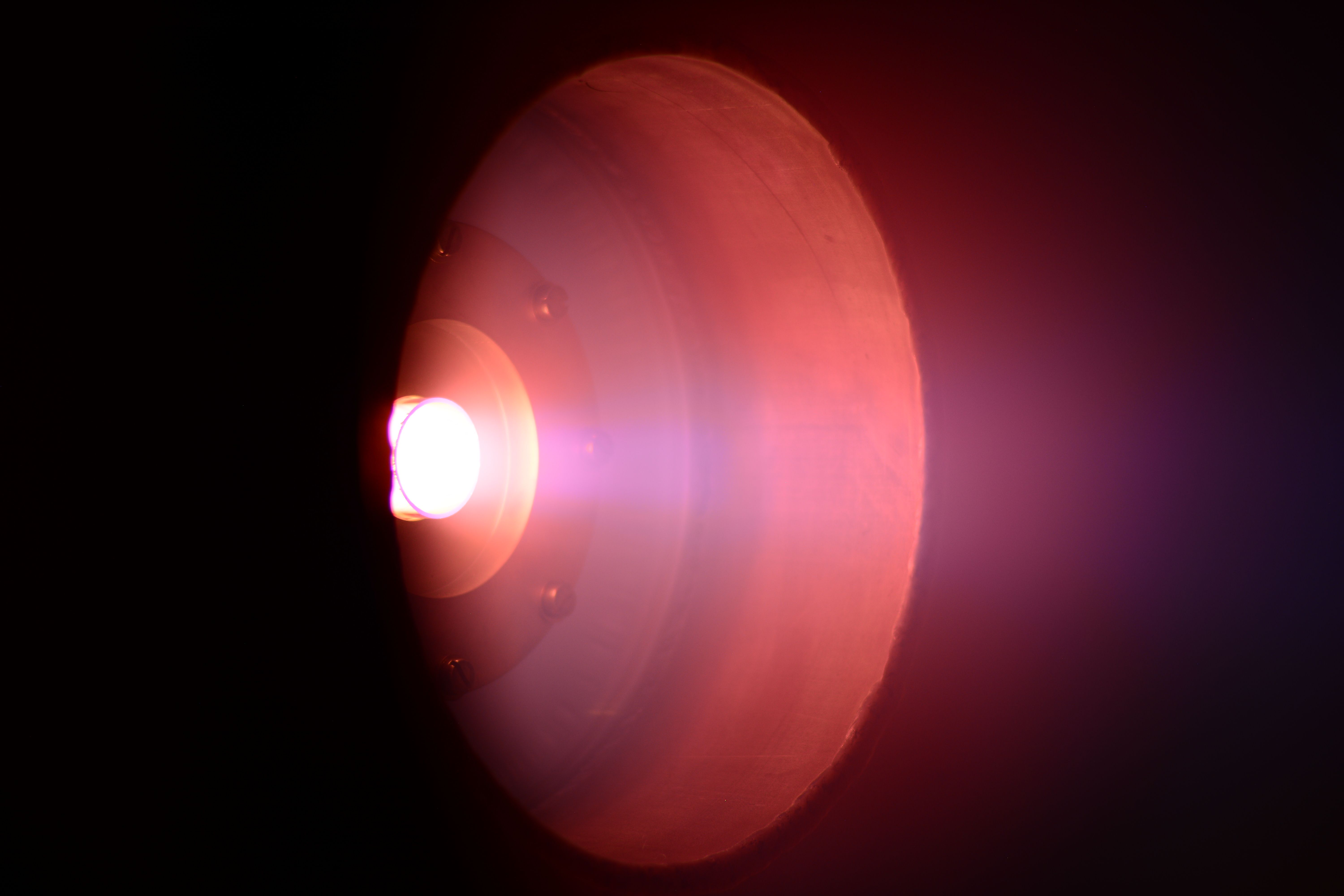}
             \caption{$I_S=\SI{6.5}{\ampere}$, $P_r=10-\SI{20}{\watt}$}
       \end{subfigure}%
       ~ 
       \begin{subfigure}[b]{0.5\textwidth}
                 \centering
                 \includegraphics[width=.7\textwidth, trim={12cm 0cm 2cm 0cm},clip]{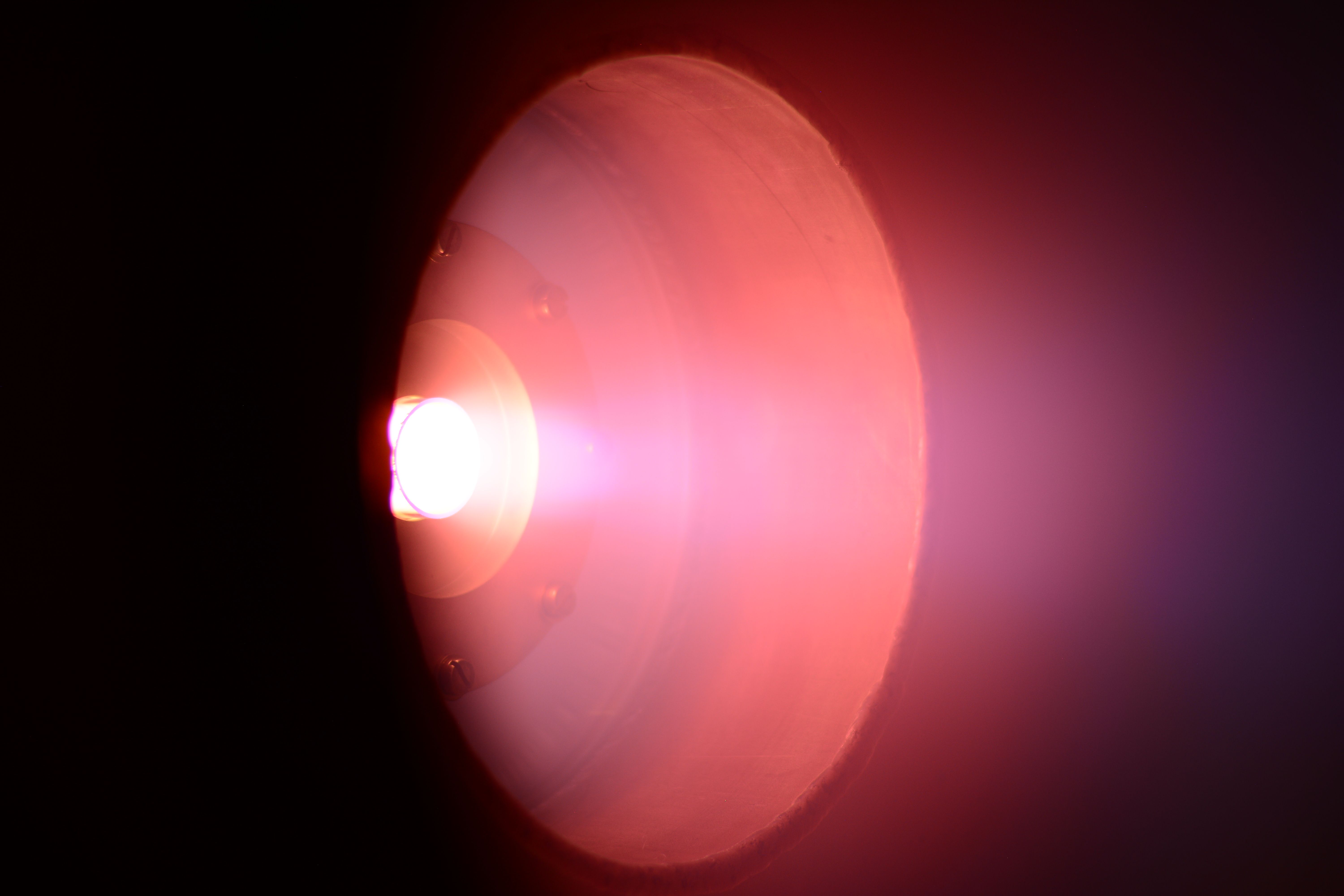}
                 \caption{$I_S=\SI{6.3}{\ampere}$, $P_r=15-\SI{20}{\watt}$}
       \end{subfigure}%
       ~\\
       \begin{subfigure}[b]{0.5\textwidth}
                     \centering
                     \includegraphics[width=.7\textwidth, trim={12cm 0cm 2cm 0cm},clip]{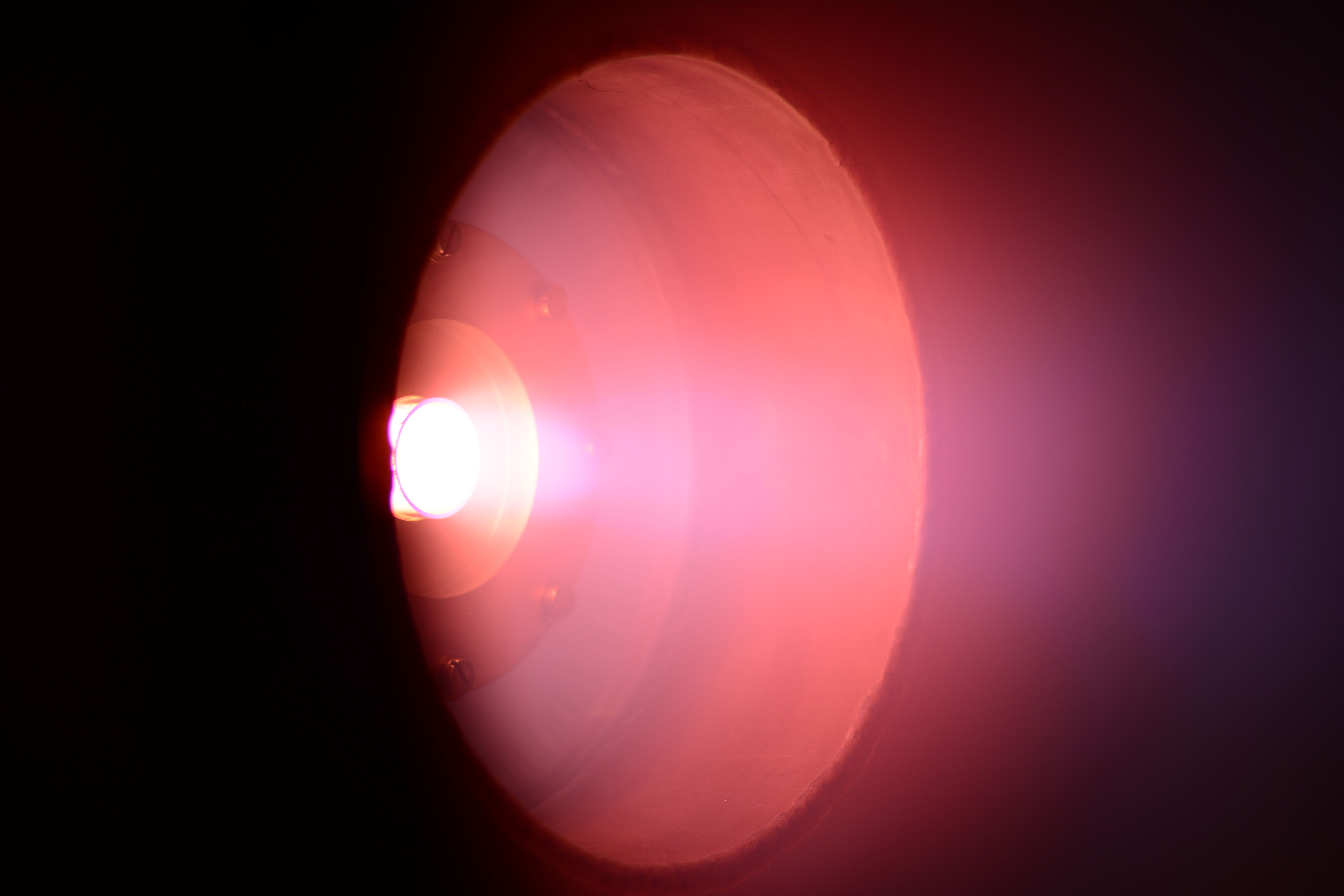}
                     \caption{$I_S=\SI{6.0}{\ampere}$, $P_r=16-\SI{25}{\watt}$}
       \end{subfigure}%
                 ~ 
       \begin{subfigure}[b]{0.5\textwidth}
                         \centering
                         \includegraphics[width=.7\textwidth, trim={12cm 0cm 2cm 0cm},clip]{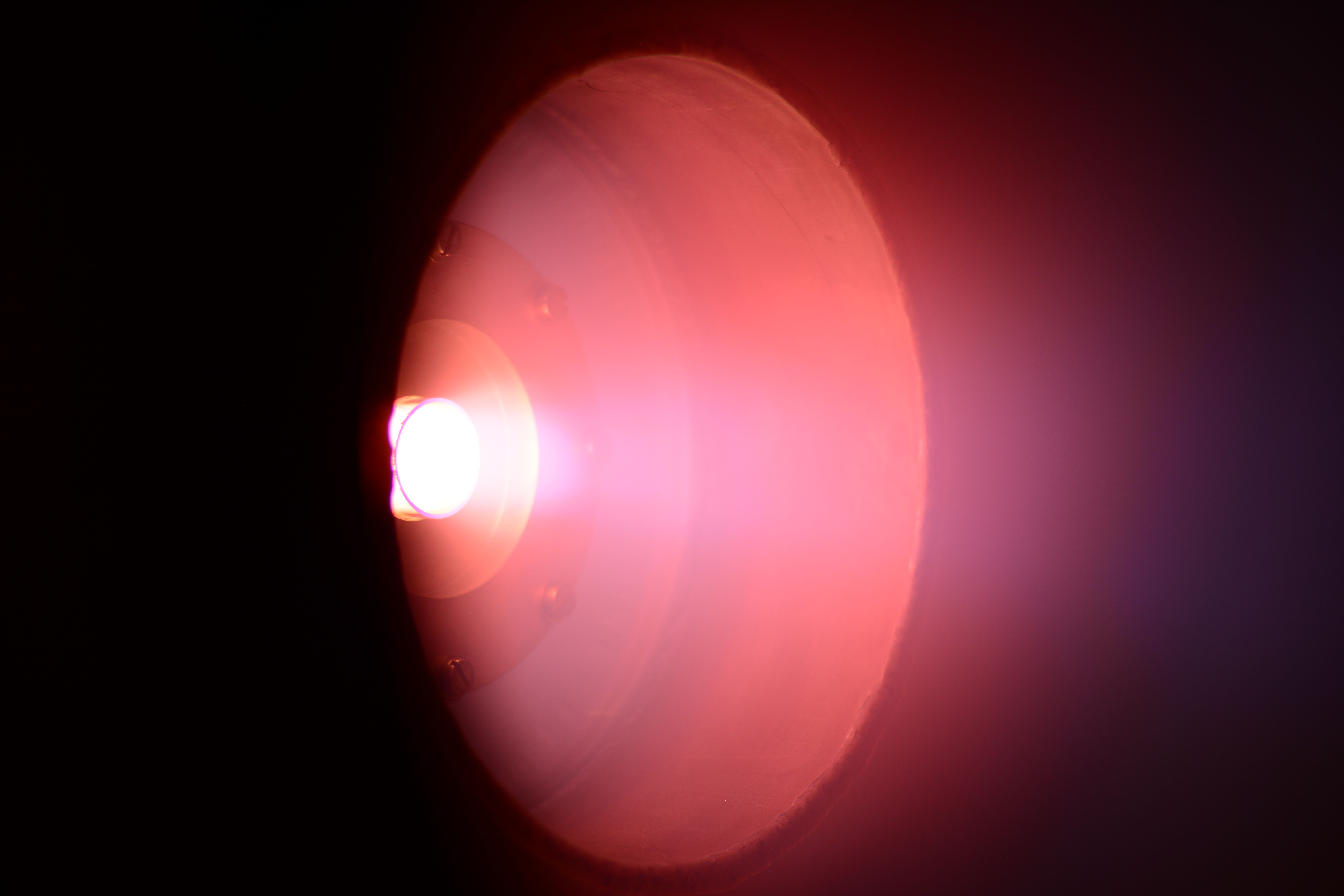}
                         \caption{$I_S=\SI{5.7}{\ampere}$, $P_r=27-\SI{29}{\watt}$}
       \end{subfigure}%
                     ~ \\
       \begin{subfigure}[b]{0.5\textwidth}
                        \centering
                        \includegraphics[width=.7\textwidth, trim={12cm 0cm 2cm 0cm},clip]{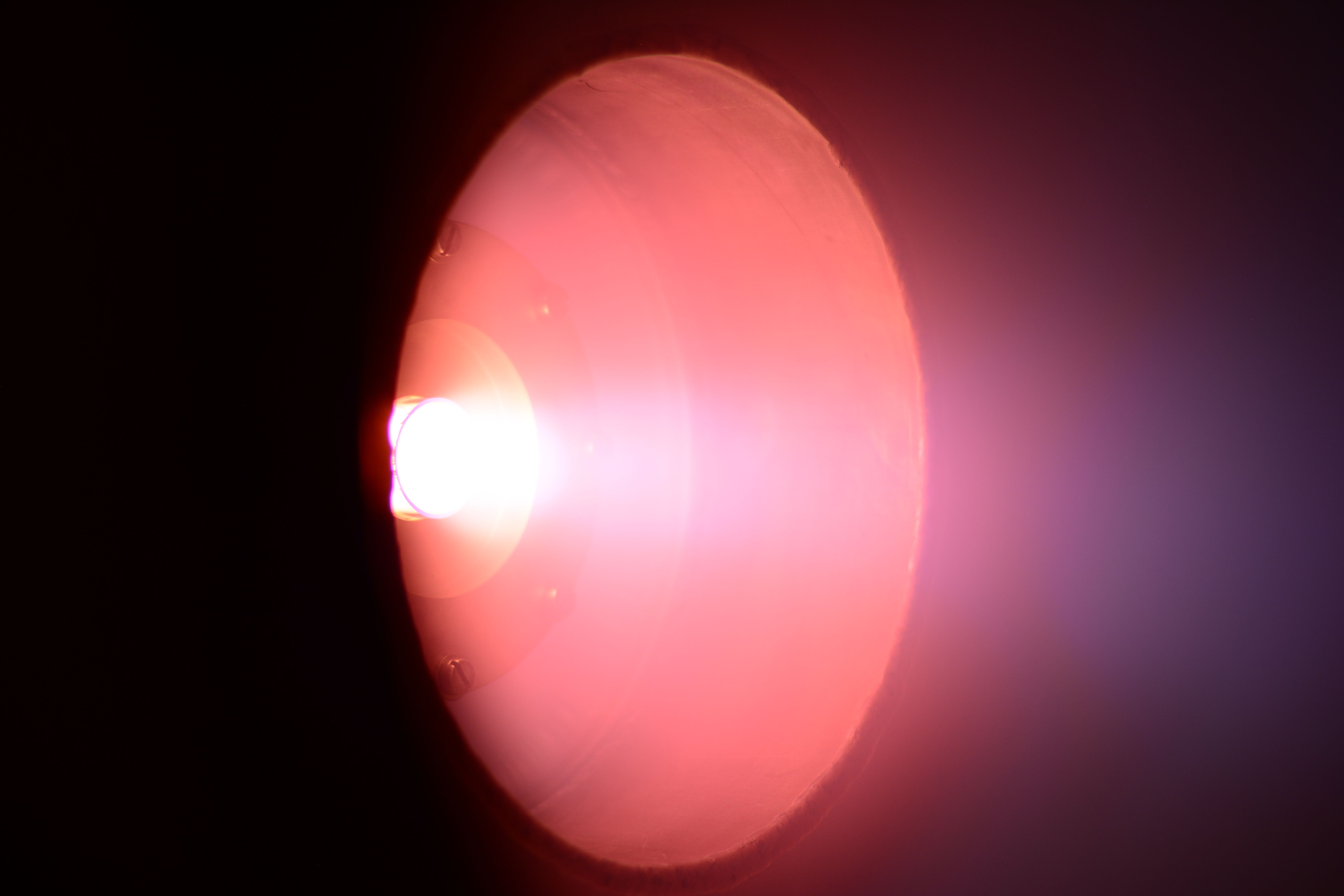}
                        \caption{$I_S=\SI{3.5}{\ampere}$, $P_r=20-\SI{30}{\watt}$}
       \end{subfigure}%
                         ~
 \caption{IPT \ce{Ar}, $P_f=\SI{60}{\watt}$,~$\dot{m}=\SI{0.6}{\milli\gram\per{\second}}$, Focal length \SI{50}{\milli\meter}, Aperture $f/4$, Exposure time $1/5\SI{}{\second}$, Solenoid at \SI{35}{\milli\meter}.}
 \label{fig:Ar_Bfield}
 \end{figure}
  
 By comparing the photographs in Fig.~\ref{fig:Ar_Bfield} with the same exposure time, it can be observed how the brightness at the outlet section of the IPT and the respective plasma jet increases with decreasing $B_0$. In particular, at larger $B_0$ a less collimated plasma jet with a conical ring structure of bluish plasma is observed. Such structure seems to converge to a single one at the centre for lower $B_0$. This could indicate a too strong $B_0$ in the first case and also, possibly, further ionization outside of the discharge channel, while for lower $B_0$ a focusing of the plasma in the axis region. After testing, the IPT performance is measured again with the NanoVNA2 and, for the \ce{Ar} test, $S_{11}$ and $Z$ at $f=\SI{40.68}{\mega\hertz}$ are reported identical to the condition before ignition.

\section{IPT operating on \ce{N2}}
The IPT discharge analysis test operating on \ce{N2} is performed by injecting the same particle flux as that of the corresponding \ce{Ar} test, see Tab.~\ref{table:ArN2O2}. With no externally applied magnetic field $B_0$, $P_f$ is injected, ignition happens at $P_f\sim\SI{60}{\watt}$, power is then fixed to $P_f=\SI{60}{\watt}$, and the solenoid is switched on. 
Ignition is achieved at any propellant flow condition between $\dot{m}=0.568 - \SI{0.142}{\milli\gram\per{\second}}$ at $P_f\sim50-\SI{70}{\watt}$, while for the $\dot{m}=\SI{0.071}{\milli\gram\per{\second}}$ case $P_f=\SI{100}{\watt}$ is required, while the operation could can be still maintained after reducing it to $P_f=\SI{60}{\watt}$. Furthermore, $P_f$ can be reduced down to $P_f=\SI{20}{\watt}$ at $\SI{0.4}{\milli\gram\per{\second}}$ while having the plasma still ignited. 
The input powers have been tested from an initial condition of $P_f=\SI{60}{\watt}$ with an applied solenoid current of $I_S=\SI{10.07}{\ampere}$ based on $P_r=\SI{0}{\watt}$ as of experimenting with \ce{Ar}. Similarly, by lowering the solenoid current to  $I_S=6.5-\SI{6.7}{\ampere}$, a visually more collimated jet and an overall brighter plasma is observed while $P_r$ increase. The most representative cases are shown in Fig.~\ref{fig:N2_1} and Fig.~\ref{fig:N2_2}, showing the difference between an applied $B_0$ with a solenoid current of $I_S=\SI{10.07}{\ampere}$ and of $I_S=\SI{6.7}{\ampere}$. 
The $P_r$ can be reduced by tuning the applied magnetic field $B_0$ similarly to the \ce{Ar} test. Different plume configurations can be achieved by varying the applied magnetic field $B_0$ intensity. The test to visualize the effect of the decreasing applied magnetic field $B_0$ is performed with the same set of $I_S$ as of the \ce{Ar} test, see Fig.~\ref{fig:N2_Bfield} and the behaviour is visually similar to the \ce{Ar} case.
By comparing the photographs having the same exposure time, it can be observed how the brightness at the outlet section of the IPT and the respective plasma jet increases with decreasing $B_0$. The behaviour is similar to that of the \ce{Ar} case. After testing, the IPT performance slightly changed with a decrease of the resonance frequency and an increase in $S_{11}$, while the condition for $S_{11}\leq\SI{-20}{\decibel}$ at $f=\SI{40.68}{\mega\hertz}$ is still respected, see Fig.~\ref{fig:resonance_N2}.
\begin{figure}[H]
\centering
\includegraphics[width=.65\textwidth, trim={12cm 0cm 2cm 0cm},clip]{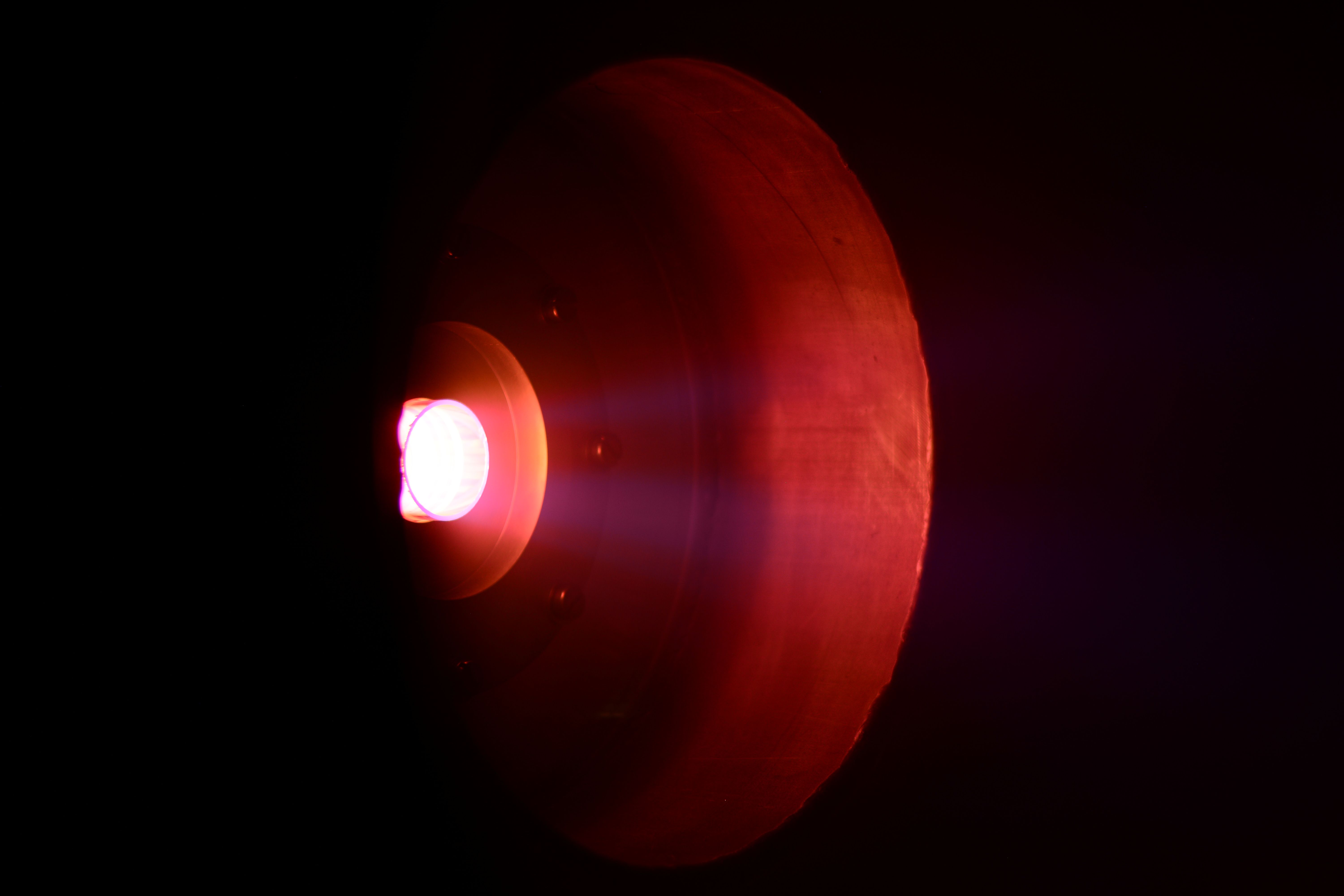}
\caption{IPT \ce{N2} at $\dot{m}=\SI{0.568}{\milli\gram\per{\second}}, P_f=\SI{60}{\watt}$,  $P_r=\SI{0}{\watt}$,  $I_{S}=\SI{10.07}{\ampere}$, Focal length \SI{50}{\milli\meter}, Aperture $f/4$, Exposure time $1/5\SI{}{\second}$.}
\label{fig:N2_1}
\end{figure}

\begin{figure}[H]
\centering
\includegraphics[width=.65\textwidth, trim={12cm 0cm 2cm 0cm},clip]{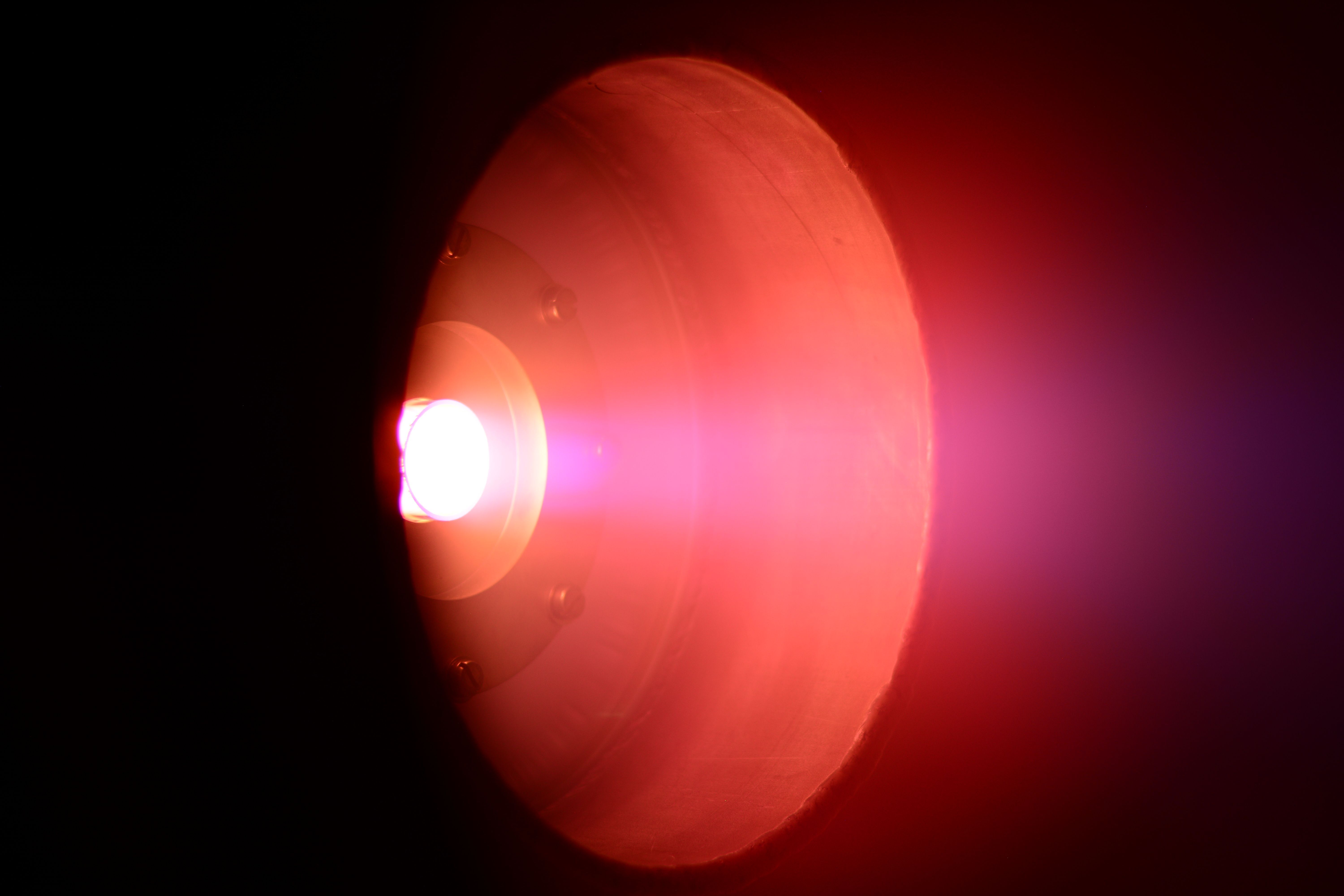}
\caption{IPT \ce{N2} at $\dot{m}=\SI{0.568}{\milli\gram\per{\second}}, P_f=\SI{60}{\watt}$,  $P_r=1-\SI{13}{\watt}$,  $I_{S}=\SI{6.7}{\ampere}$, Focal length \SI{50}{\milli\meter}, Aperture $f/4$, Exposure time $1/5\SI{}{\second}$.}
\label{fig:N2_2}
\end{figure}

\begin{figure}[H]
    \centering
    \begin{subfigure}[b]{0.5\textwidth}
        \centering
        \includegraphics[width=.7\textwidth, trim={12cm 0cm 2cm 0cm},clip]{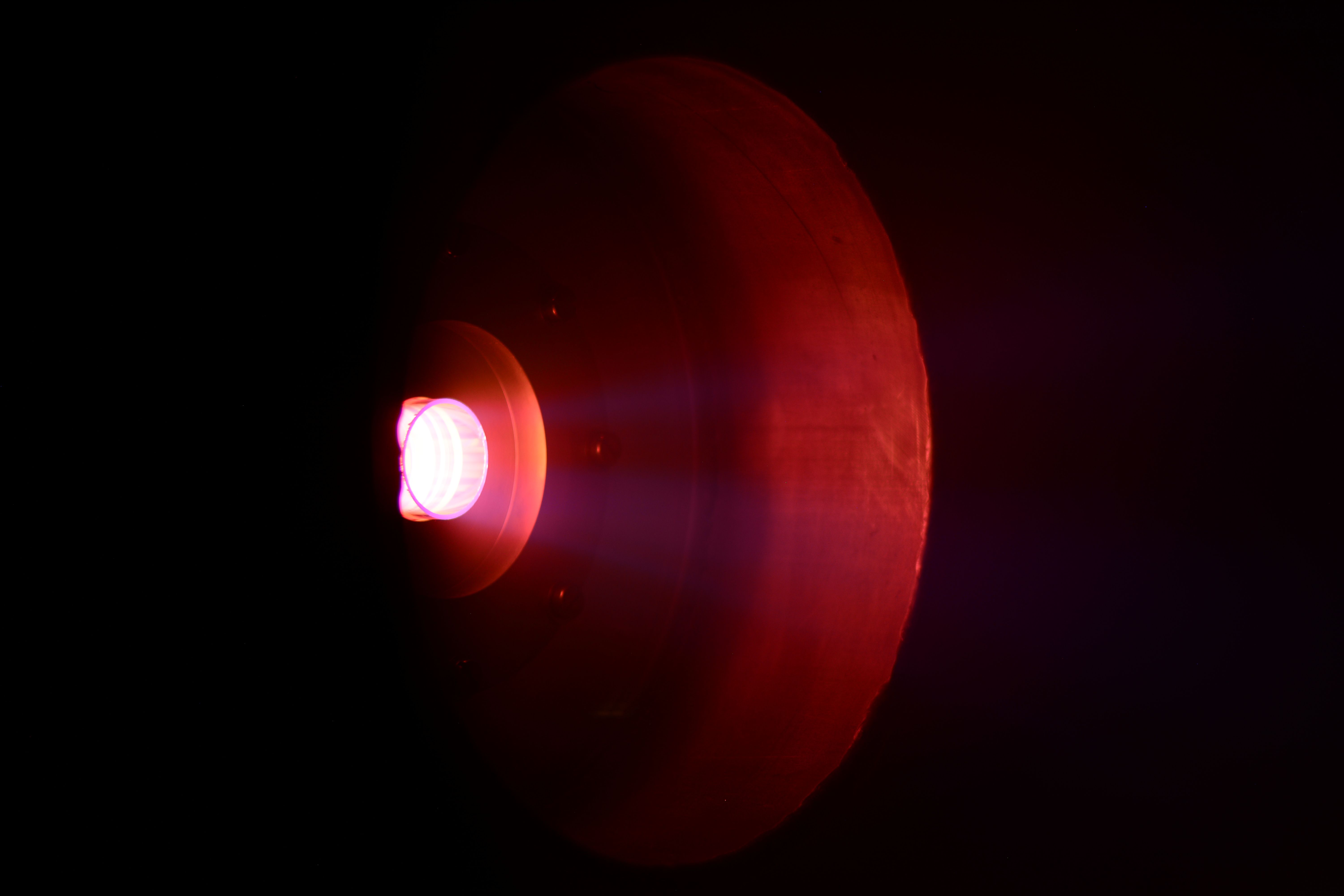}
        \caption{$I_S=\SI{10.07}{\ampere}$, $P_r=\SI{0}{\watt}$}
    \end{subfigure}%
    ~ 
    \begin{subfigure}[b]{0.5\textwidth}
        \centering
        \includegraphics[width=.7\textwidth, trim={12cm 0cm 2cm 0cm},clip]{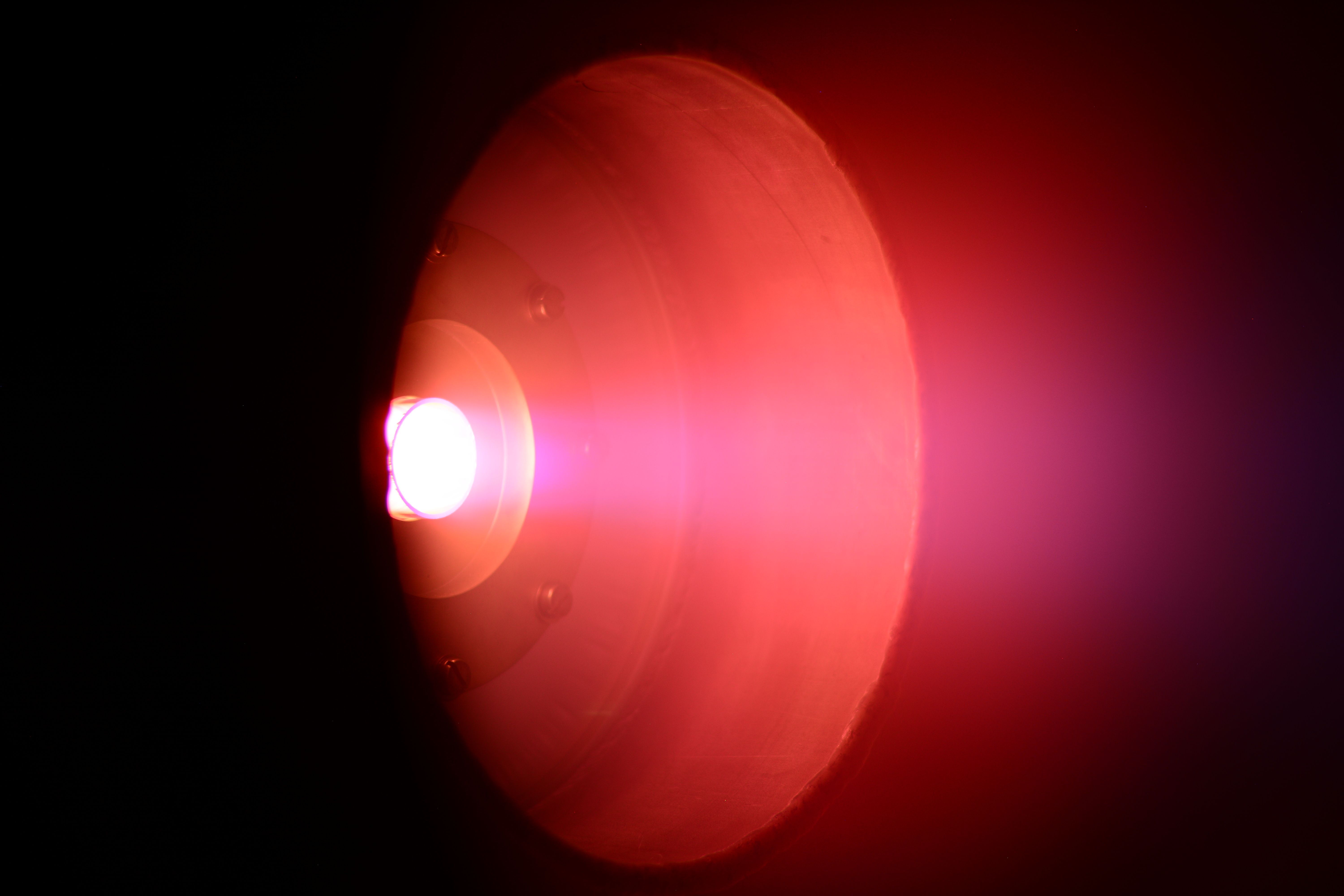}
        \caption{$I_S=\SI{6.7}{\ampere}$, $P_r=1-\SI{8}{\watt}$}
    \end{subfigure}
    ~
      \begin{subfigure}[b]{0.5\textwidth}
            \centering
            \includegraphics[width=.7\textwidth, trim={12cm 0cm 2cm 0cm},clip]{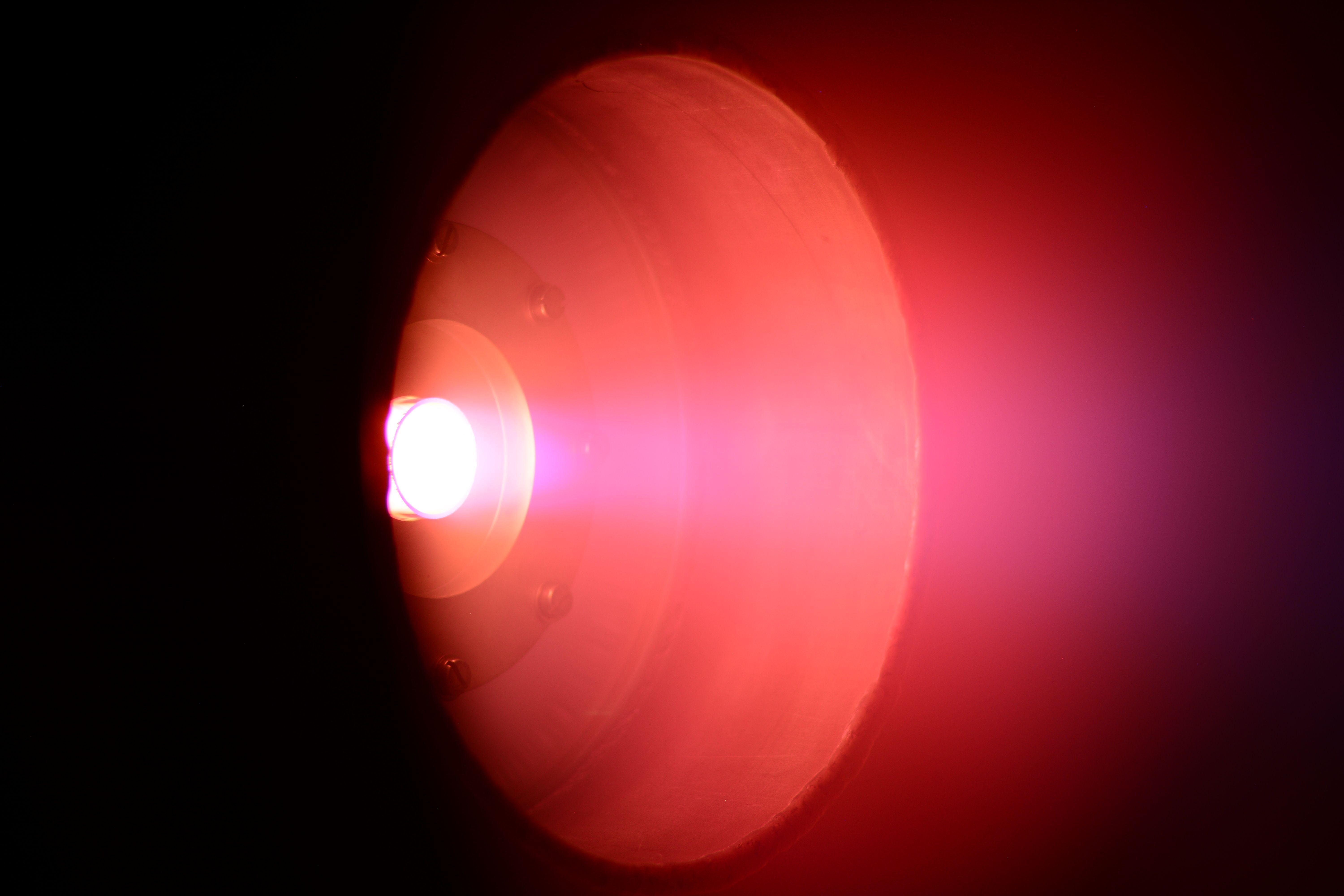}
            \caption{$I_S=\SI{6.5}{\ampere}$, $P_r=0-\SI{10}{\watt}$}
      \end{subfigure}%
      ~ 
      \begin{subfigure}[b]{0.5\textwidth}
                \centering
                \includegraphics[width=.7\textwidth, trim={12cm 0cm 2cm 0cm},clip]{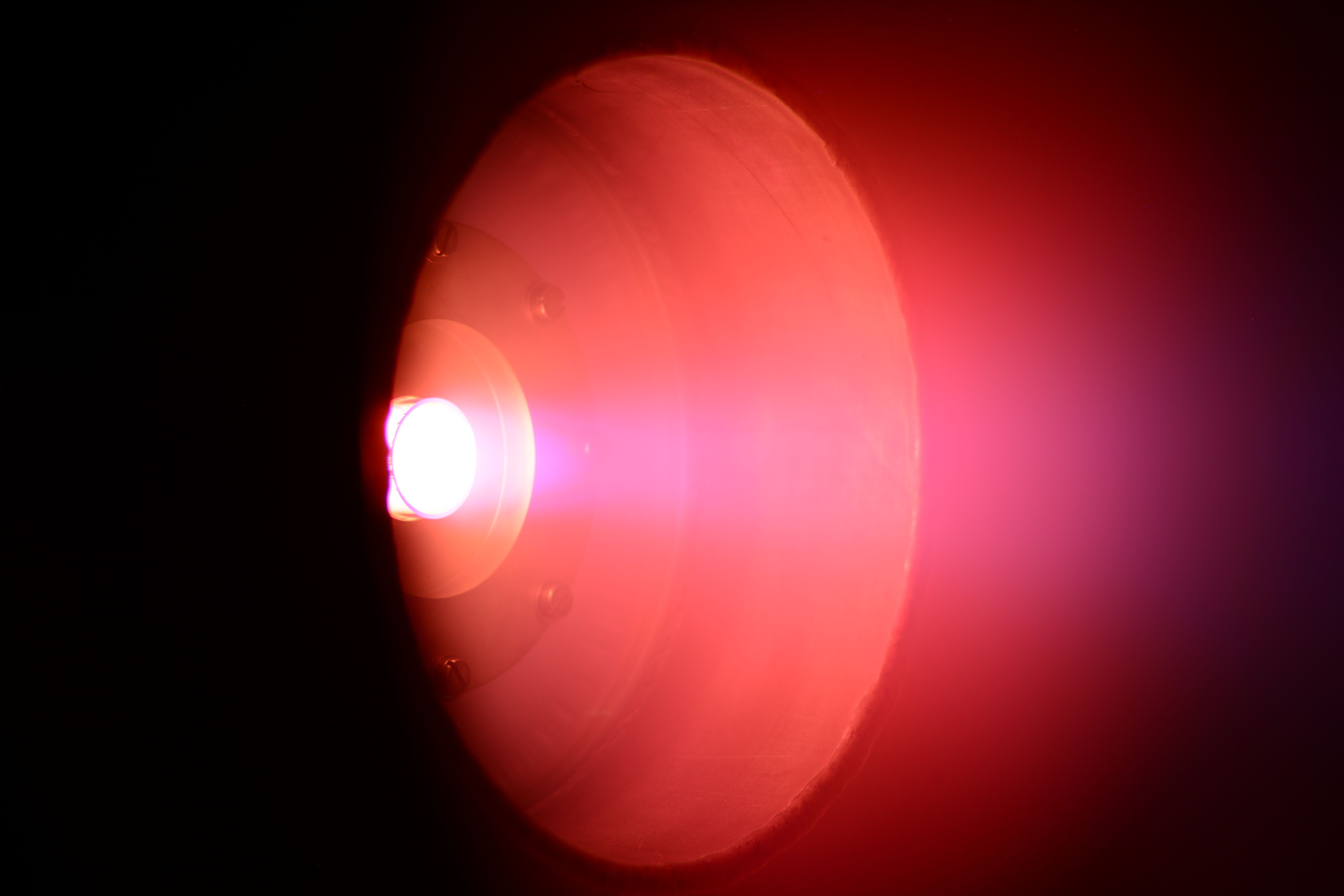}
                \caption{$I_S=\SI{6.3}{\ampere}$, $P_r=1-\SI{13}{\watt}$}
      \end{subfigure}%
      ~\\
      \begin{subfigure}[b]{0.5\textwidth}
                    \centering
                    \includegraphics[width=.7\textwidth, trim={12cm 0cm 2cm 0cm},clip]{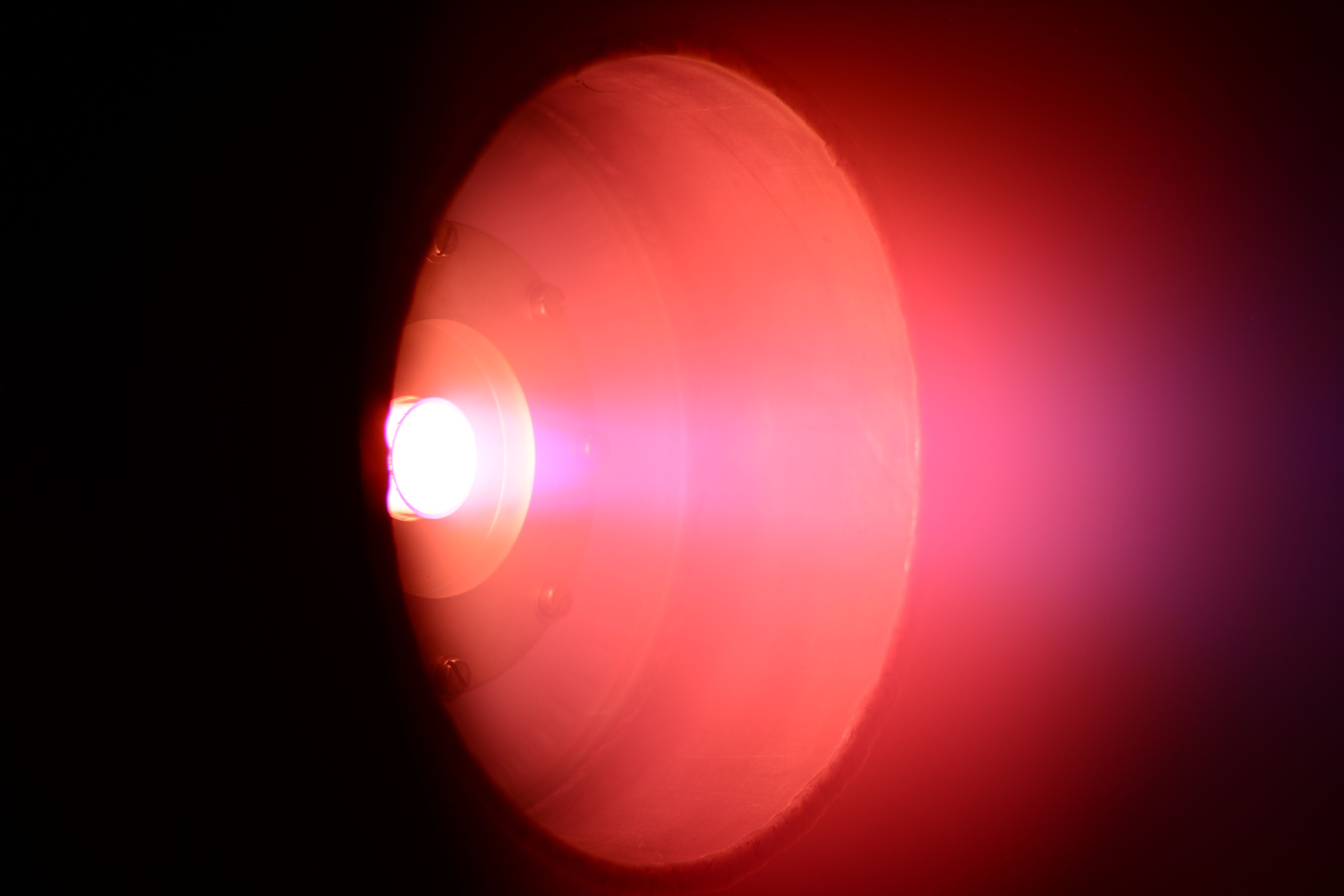}
                    \caption{$I_S=\SI{6.0}{\ampere}$, $P_r=2-\SI{11}{\watt}$}
      \end{subfigure}%
                ~ 
      \begin{subfigure}[b]{0.5\textwidth}
                        \centering
                        \includegraphics[width=.7\textwidth, trim={12cm 0cm 2cm 0cm},clip]{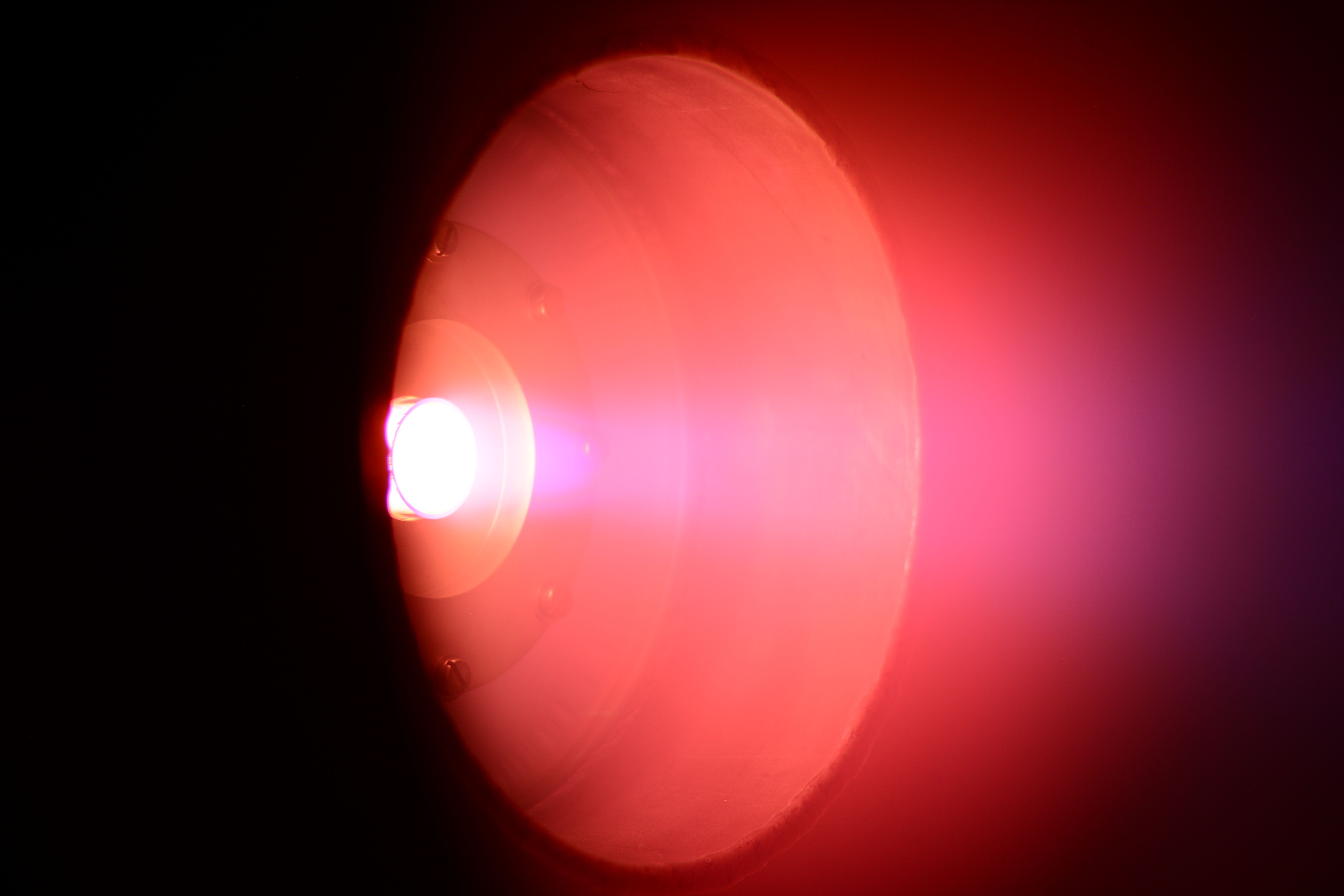}
                        \caption{$I_S=\SI{5.7}{\ampere}$, $P_r=2-\SI{13}{\watt}$}
      \end{subfigure}%
                    ~ \\
      \begin{subfigure}[b]{0.5\textwidth}
                       \centering
                       \includegraphics[width=.7\textwidth, trim={12cm 0cm 2cm 0cm},clip]{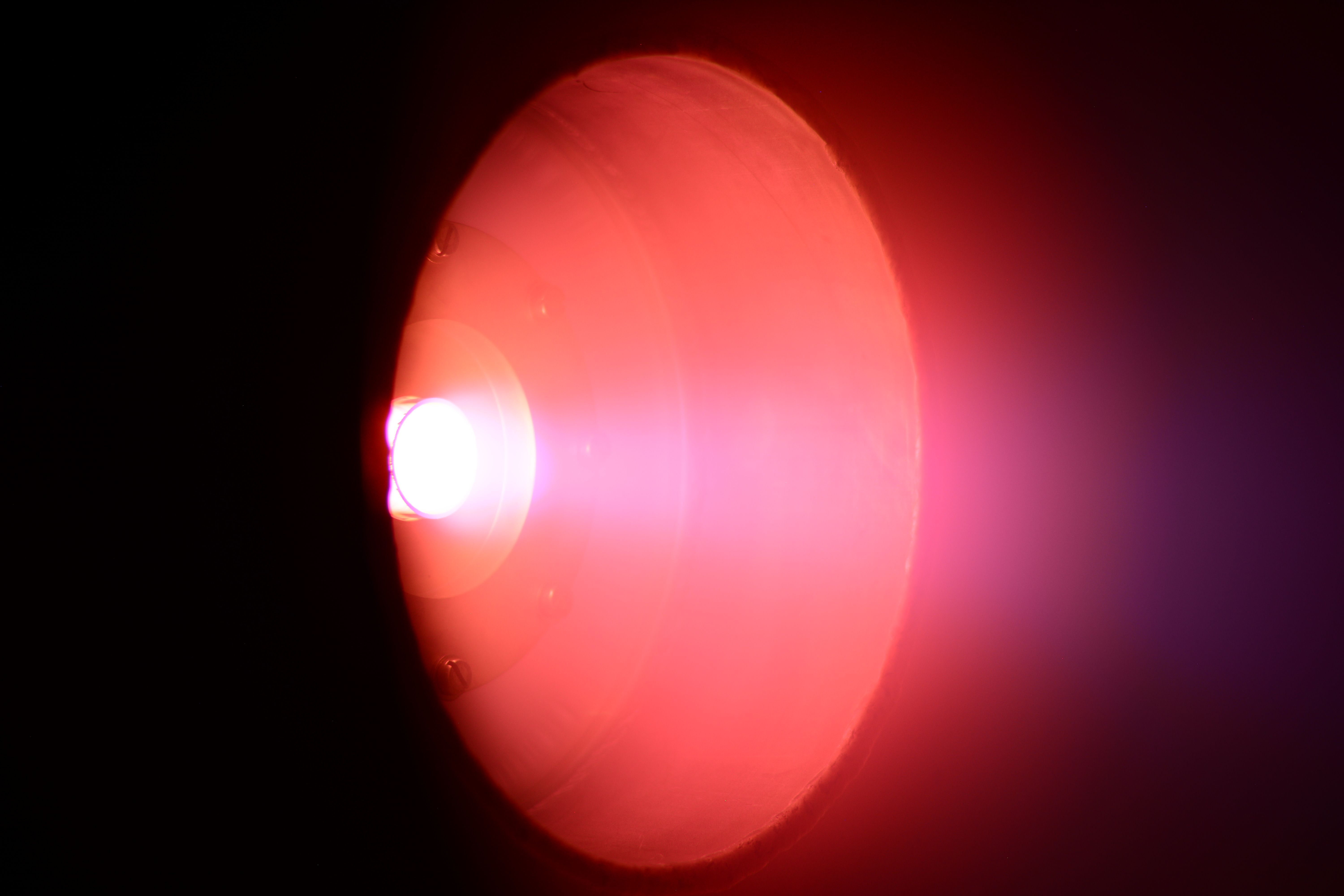}
                       \caption{$I_S=\SI{3.5}{\ampere}$, $P_r=2-\SI{14}{\watt}$}
      \end{subfigure}%
                        ~
\caption{IPT \ce{N2}, $P_f=\SI{60}{\watt}$,~$\dot{m}=\SI{0.426}{\milli\gram\per{\second}}$, Focal length \SI{50}{\milli\meter}, Aperture $f/4$, Exposure time $1/5\SI{}{\second}$, Solenoid at \SI{35}{\milli\meter}.}
\label{fig:N2_Bfield}
\end{figure}

\begin{figure}[H]
\centering
\includegraphics[width=\textwidth]{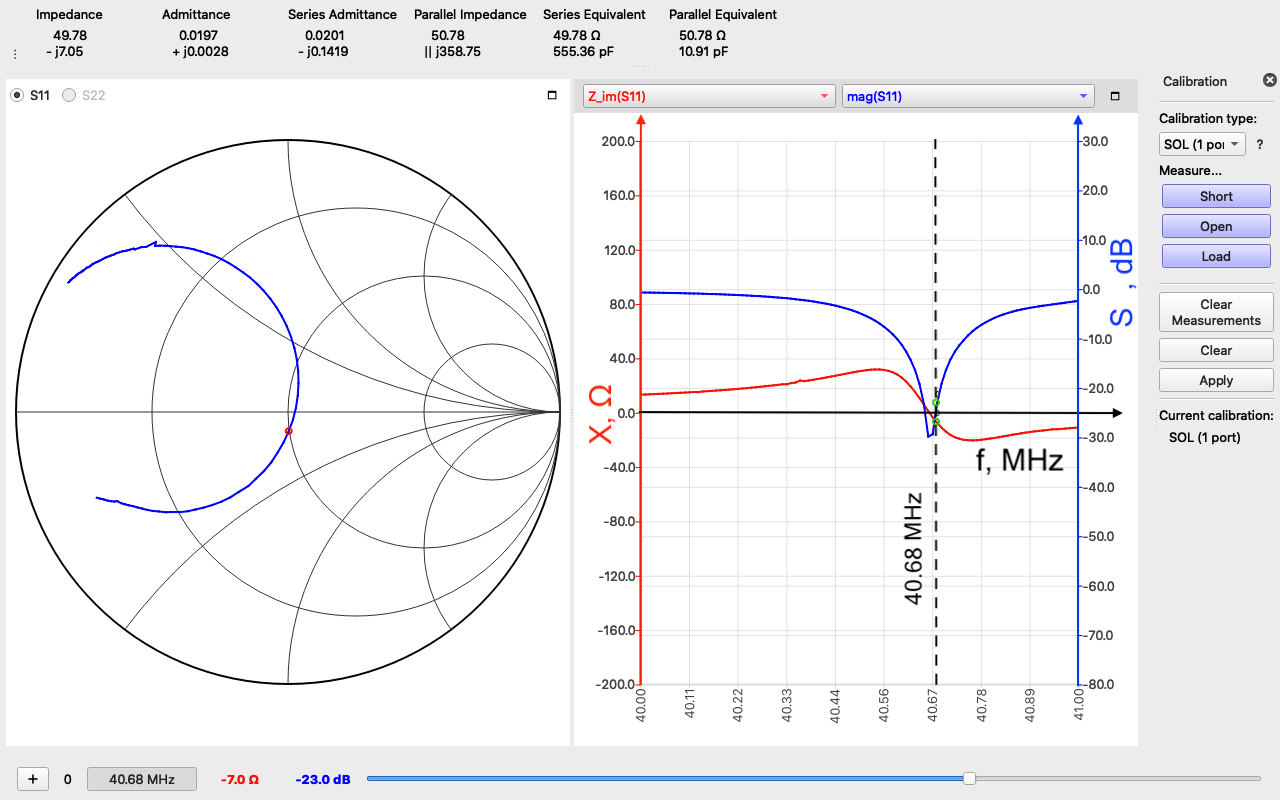}
\caption{IPT $Z_{IPT}$ and $S_{11}$ after \ce{N2} test.}
\label{fig:resonance_N2}
\end{figure}

\section{IPT operating on \ce{O2}} 
The discharge analysis test with \ce{O2} has been performed by injecting the same particle flux as that of the corresponding \ce{Ar} test, see Tab.~\ref{table:ArN2O2}.
With no externally applied magnetic field $B_0$, $P_f$ is injected, after ignition the power is fixed to $P_f=\SI{60}{\watt}$, and the solenoid is switched on. Ignition is achieved between $\dot{m}=0.487-\SI{0.650}{\milli\gram\per{\second}}$ for $P_f\sim50-\SI{70}{\watt}$, while between $\dot{m}=0.081-\SI{0.323}{\milli\gram\per{\second}}$, $P_f=\SI{110}{\watt}$ is required, while the operation can be still maintained, after ignition, at $P_f=\SI{60}{\watt}$. Furthermore, $P_f$ can be reduced to $P_f=\SI{43}{\watt}$ at $\dot{m}=\SI{0.323}{\milli\gram\per{\second}}$ while having the plasma still ignited. However $P_r=\SI{37}{\watt}$ and the plasma became unstable.
The input powers have been tested from an initial condition of $P_f=\SI{60}{\watt}$ with an applied solenoid current of $I_S=\SI{10.07}{\ampere}$ based on $P_r=\SI{0}{\watt}$ as in tests with \ce{Ar} and \ce{N2}. Furthermore, by applying  $I_S=6.5-\SI{6.7}{\ampere}$ a visually more collimated jet and an overall brighter plasma can be observed, and $P_r$ increases, compared to $I_S=\SI{10.07}{\ampere}$.
 The most representative cases are shown in Fig.~\ref{fig:O2_1} and Fig.~\ref{fig:O2_2}, highlighting the difference between a $B_0$ with $I_S=\SI{10.07}{\ampere}$ and $I_S=\SI{6.5}{\ampere}$.
  The $P_r$ can be reduced by tuning the applied magnetic field $B_0$ as well. Different plume configurations can be achieved by varying the applied magnetic field $B_0$ intensity similarly to the \ce{Ar} and \ce{N2} cases. The test to visualize the effect of the decreasing applied magnetic field $B_0$ has been performed by applying the same decreasing $I_S$ of the \ce{Ar} and \ce{N2} tests, see Fig.~\ref{fig:O2_Bfield}, and the behaviour is visually similar to the \ce{Ar} and \ce{N2} cases.
 
 \begin{figure}[H]
 \centering
 \includegraphics[width=.65\textwidth, trim={11cm 0cm 2cm 0cm},clip]{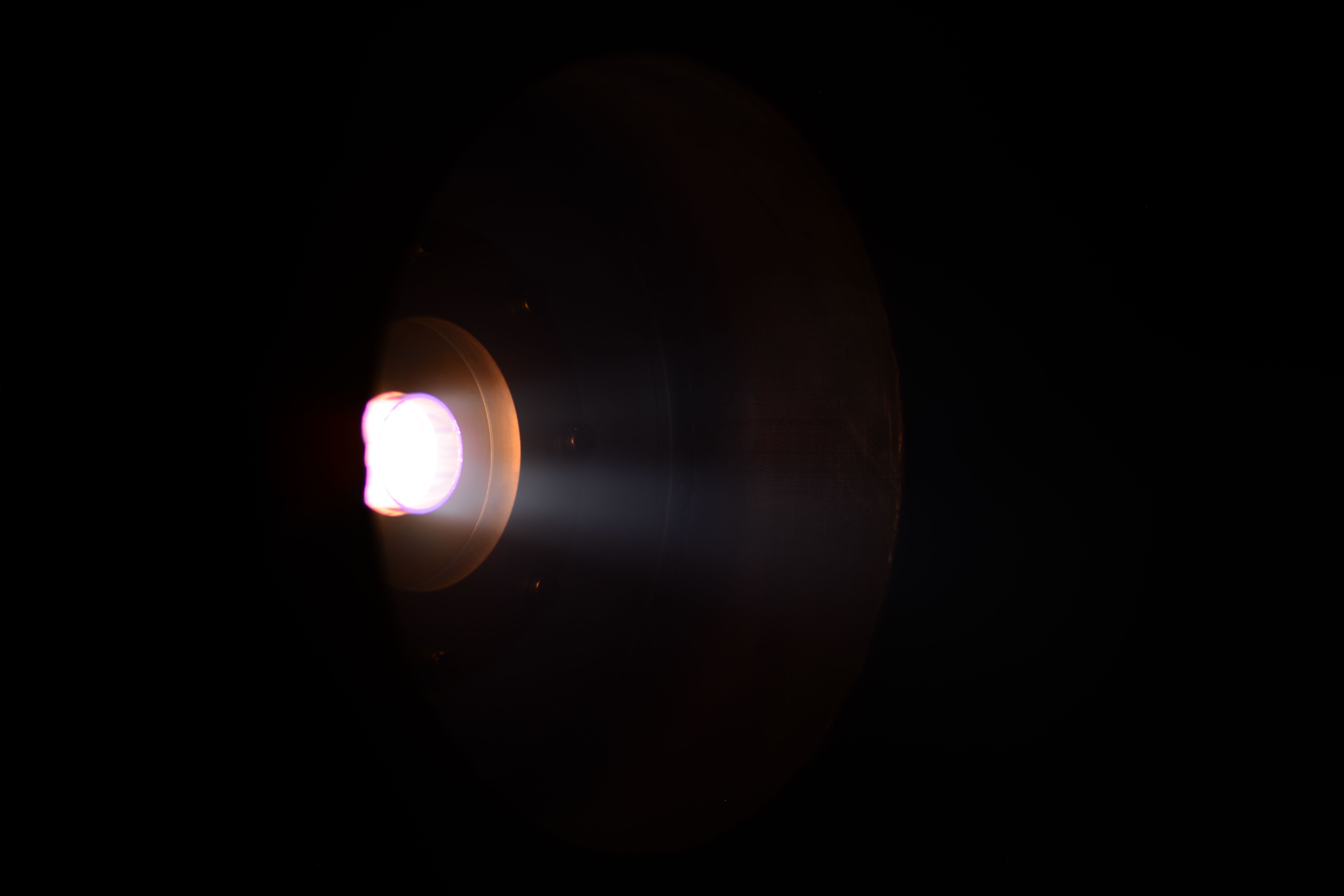}
 \caption{IPT \ce{O2} at $\dot{m}=\SI{0.650}{\milli\gram\per{\second}}, P_f=\SI{60}{\watt}$,  $P_r=\SI{0}{\watt}$,  $I_{S}=\SI{10.07}{\ampere}$, Focal length \SI{50}{\milli\meter}, Aperture $f/4$, Exposure time $1/5\SI{}{\second}$.}
 \label{fig:O2_1}
 \end{figure}
 \begin{figure}[H]
 \centering
 \includegraphics[width=.65\textwidth, trim={11cm 0cm 2cm 0cm},clip]{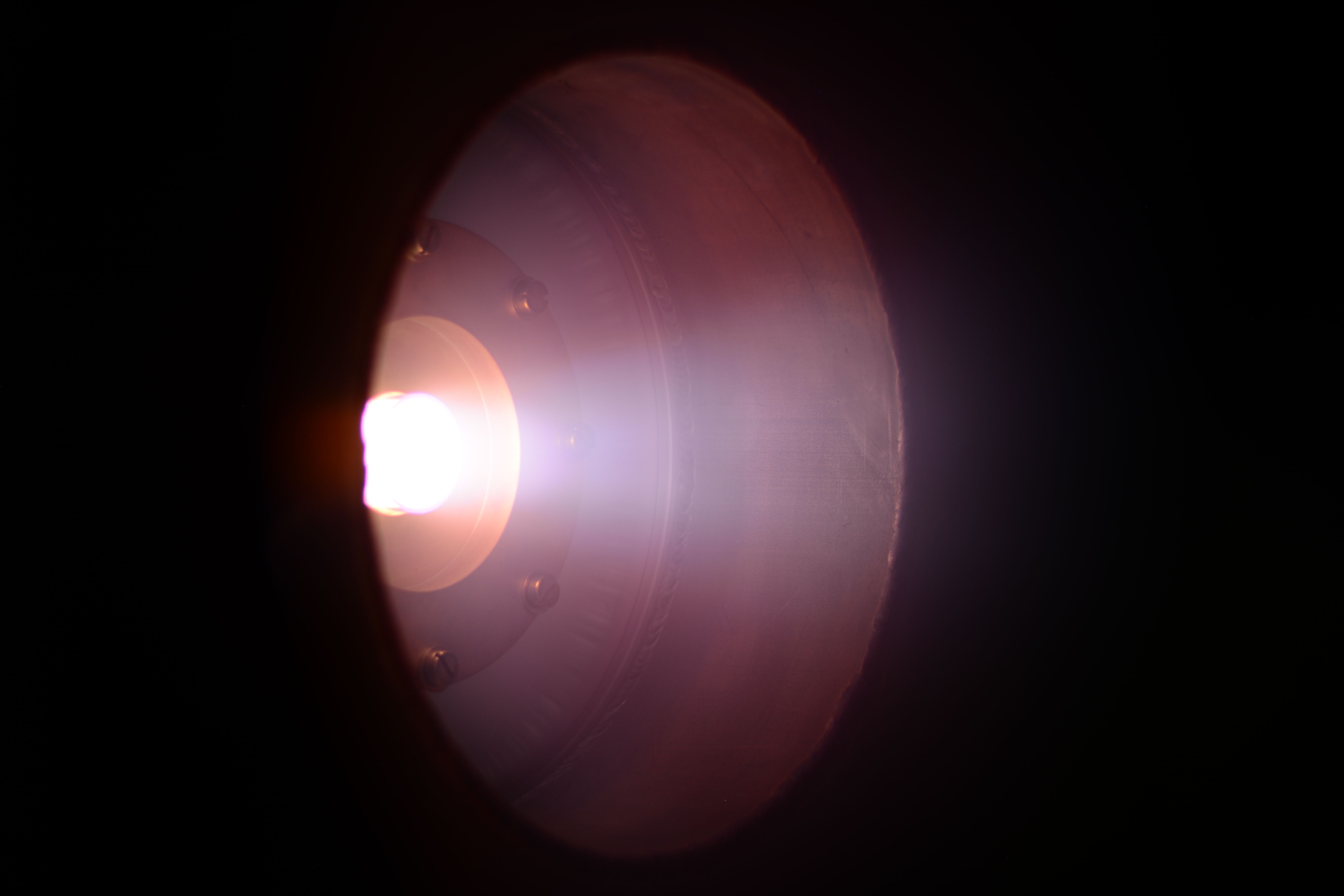}
 \caption{IPT \ce{O2} at $\dot{m}=\SI{0.650}{\milli\gram\per{\second}}, P_f=\SI{60}{\watt}$,  $P_r=7-\SI{14}{\watt}$,  $I_{S}=\SI{6.5}{\ampere}$, Focal length \SI{50}{\milli\meter}, Aperture $f/4$, Exposure time $1/5\SI{}{\second}$.}
 \label{fig:O2_2}
 \end{figure}

\begin{figure}[H]
    \centering
    \begin{subfigure}[b]{0.5\textwidth}
        \centering
        \includegraphics[width=.7\textwidth, trim={11cm 0cm 2cm 0cm},clip]{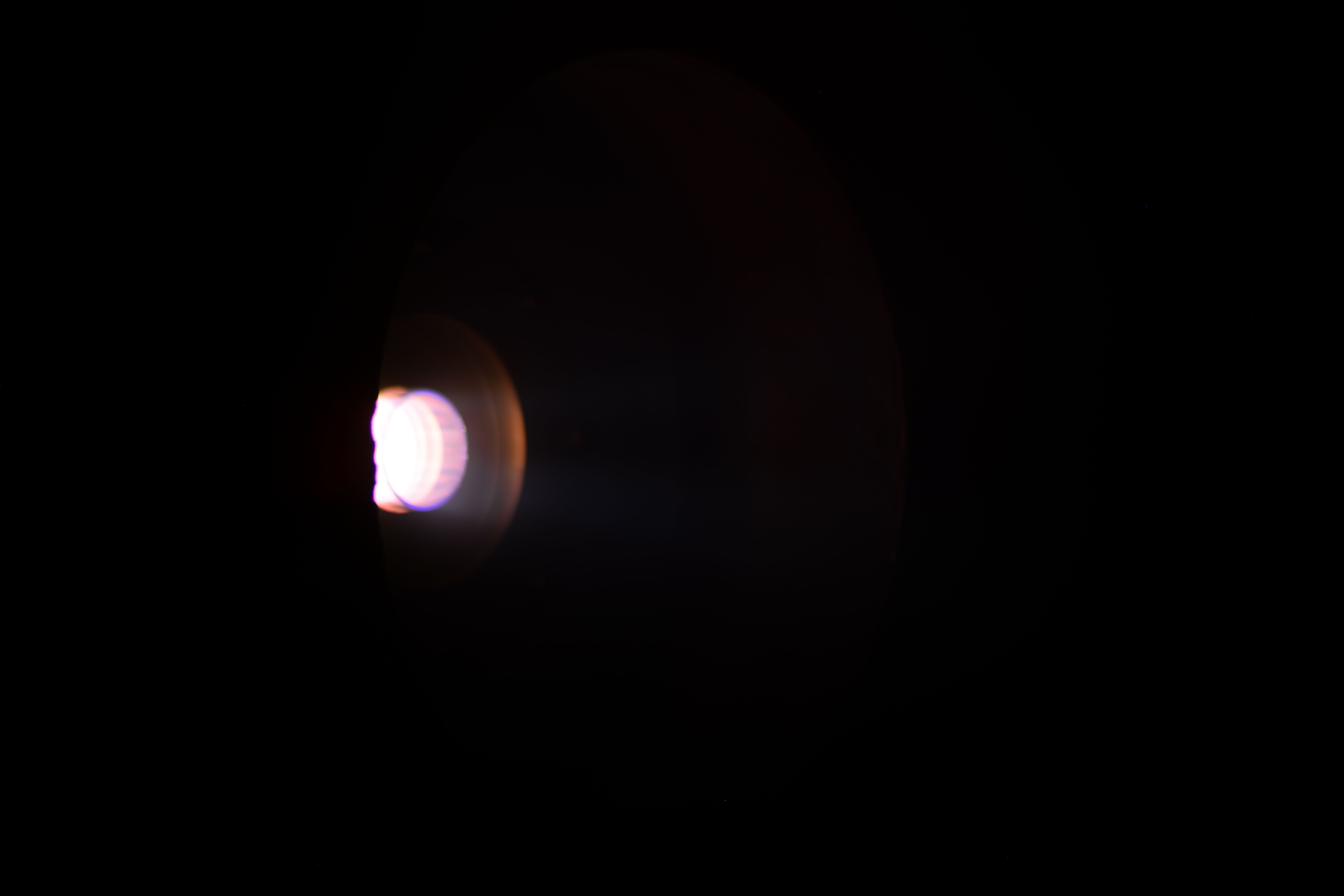}
        \caption{$I_S=\SI{10.07}{\ampere}$, $P_r=8-\SI{0}{\watt}$}
    \end{subfigure}%
    ~ 
      \begin{subfigure}[b]{0.5\textwidth}
            \centering
            \includegraphics[width=.7\textwidth, trim={11cm 0cm 2cm 0cm},clip]{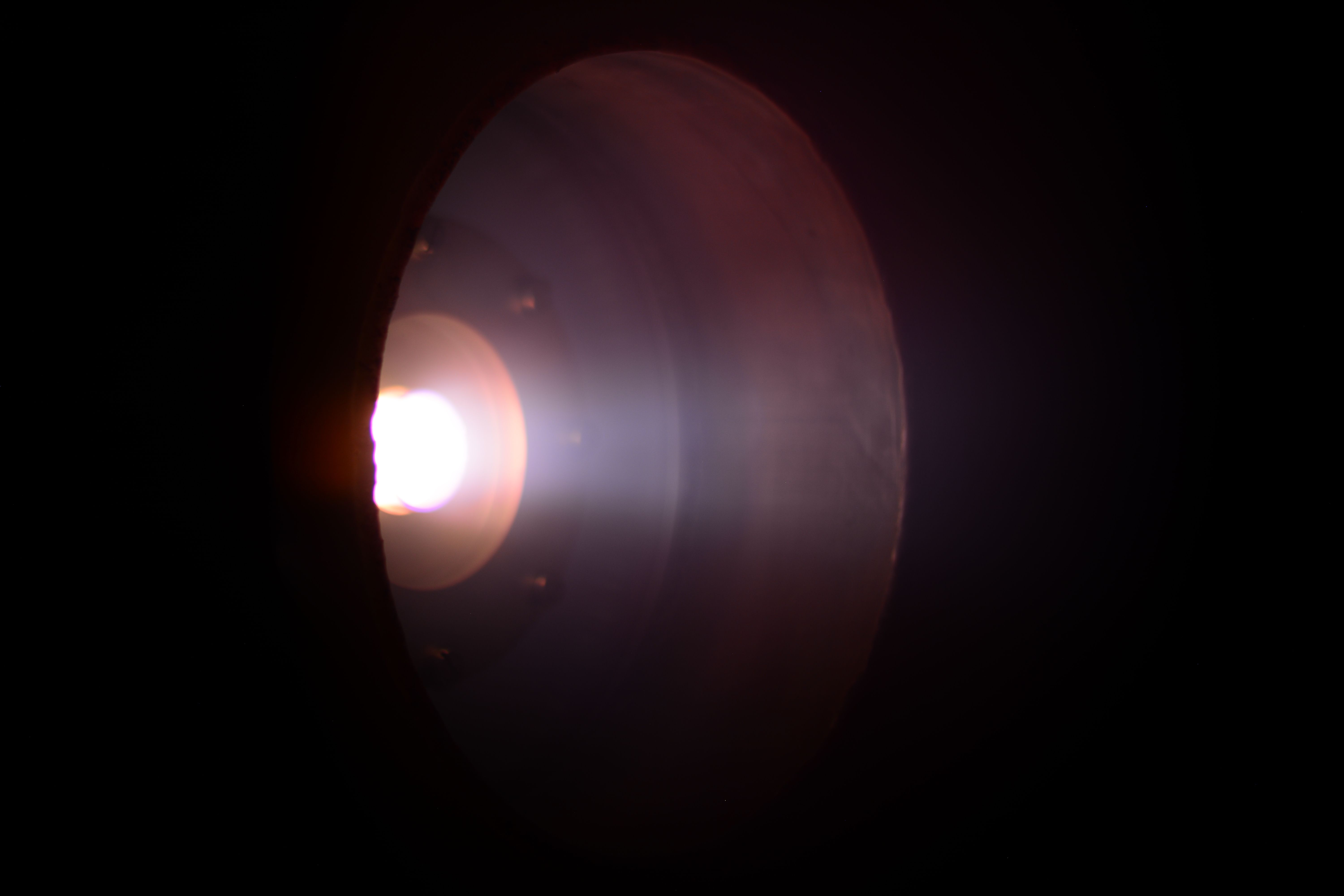}
            \caption{$I_S=\SI{6.7}{\ampere}$, $P_r=5-\SI{15}{\watt}$}
      \end{subfigure}%
      ~ \\
      \begin{subfigure}[b]{0.5\textwidth}
                \centering
                \includegraphics[width=.7\textwidth, trim={11cm 0cm 2cm 0cm},clip]{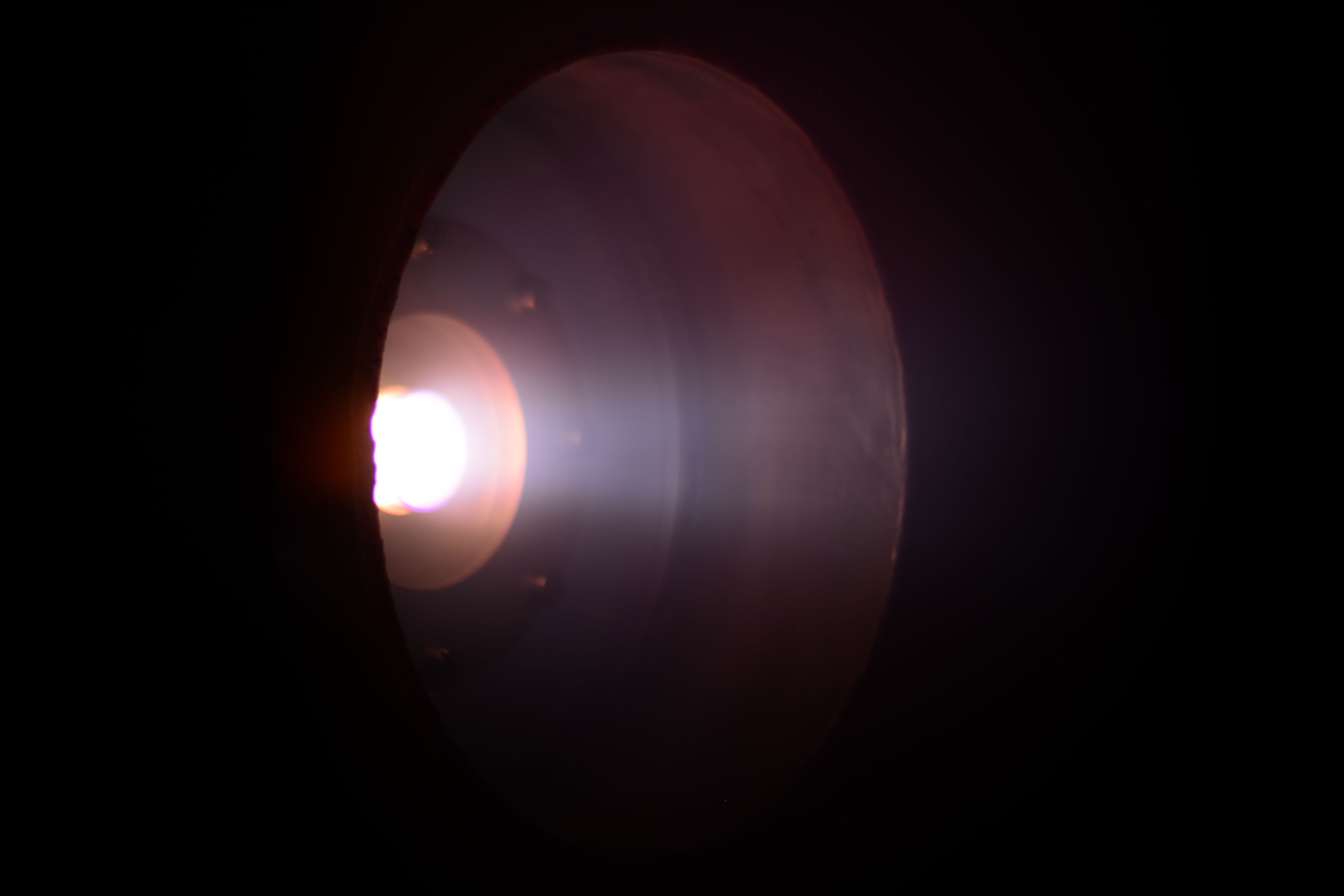}
                \caption{$I_S=\SI{6.5}{\ampere}$, $P_r=5-\SI{16}{\watt}$}
      \end{subfigure}%
      ~
         \begin{subfigure}[b]{0.5\textwidth}
                     \centering
                     \includegraphics[width=.7\textwidth, trim={11cm 0cm 2cm 0cm},clip]{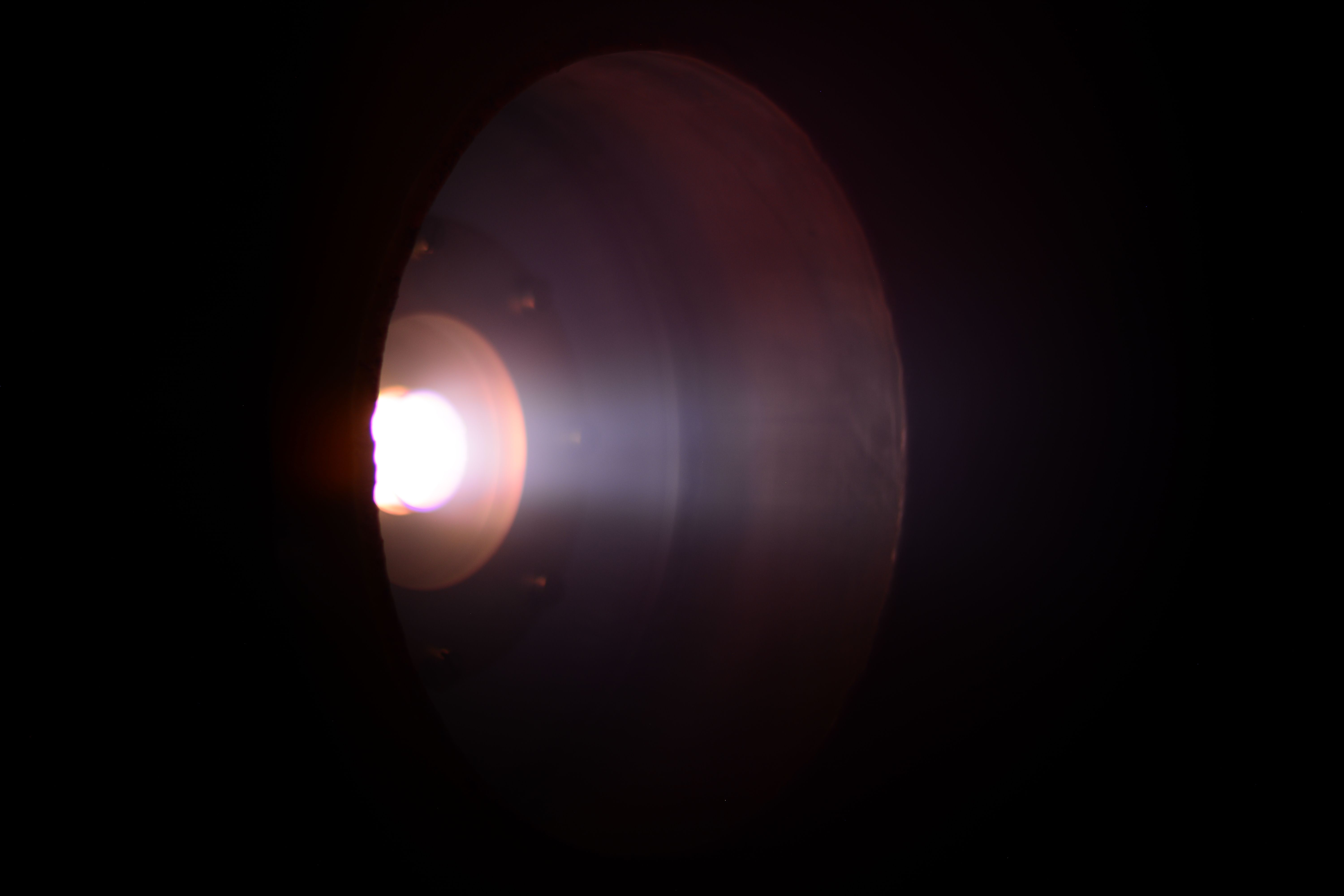}
                     \caption{$I_S=\SI{6.3}{\ampere}$, $P_r=5-\SI{16}{\watt}$}
           \end{subfigure}%
           ~\\
           \begin{subfigure}[b]{0.5\textwidth}
                    \centering
                    \includegraphics[width=.7\textwidth, trim={11cm 0cm 2cm 0cm},clip]{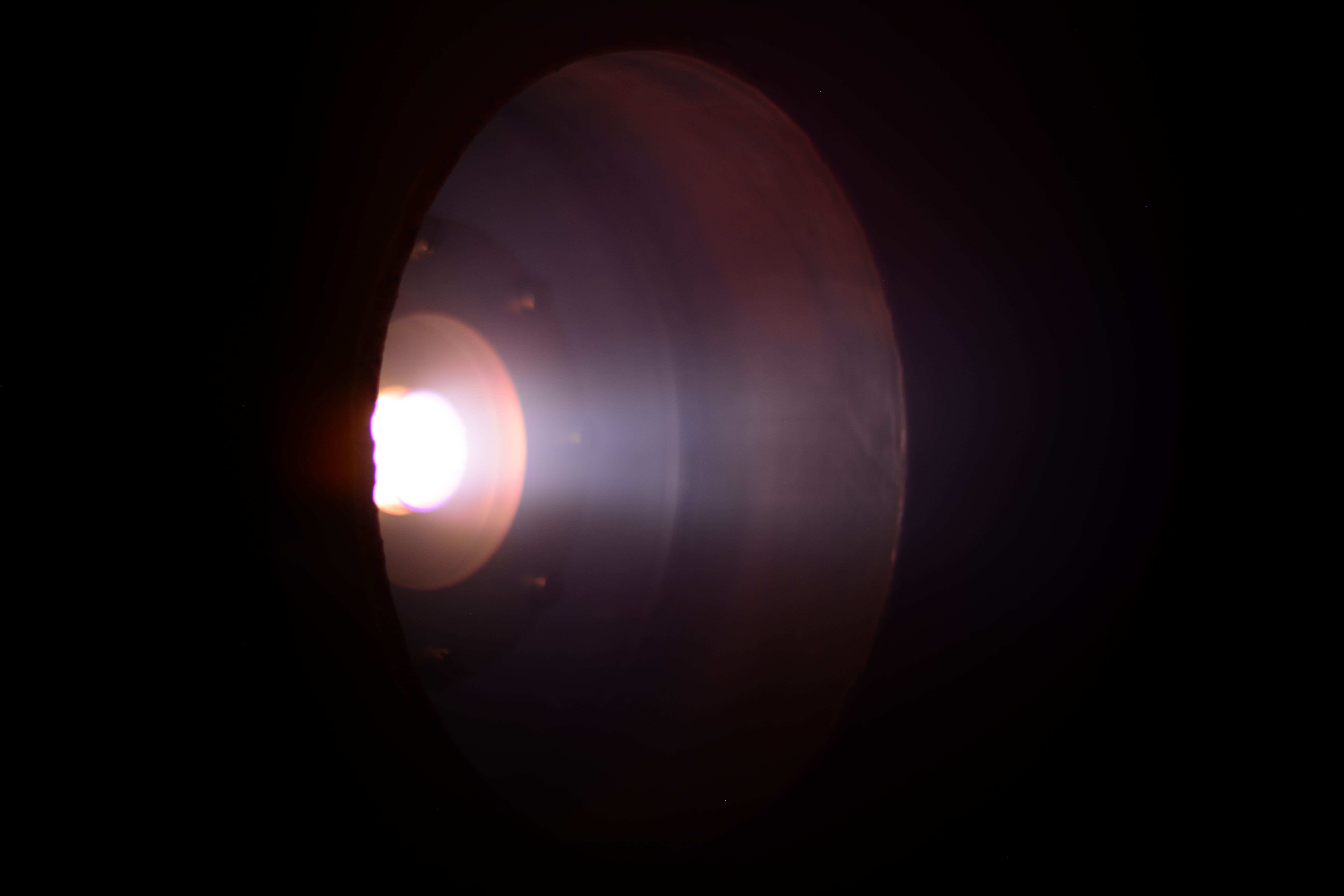}
                    \caption{$I_S=\SI{6.0}{\ampere}$, $P_r=5-\SI{16}{\watt}$}
      \end{subfigure}%
                ~ 
      \begin{subfigure}[b]{0.5\textwidth}
                        \centering
                        \includegraphics[width=.7\textwidth, trim={11cm 0cm 2cm 0cm},clip]{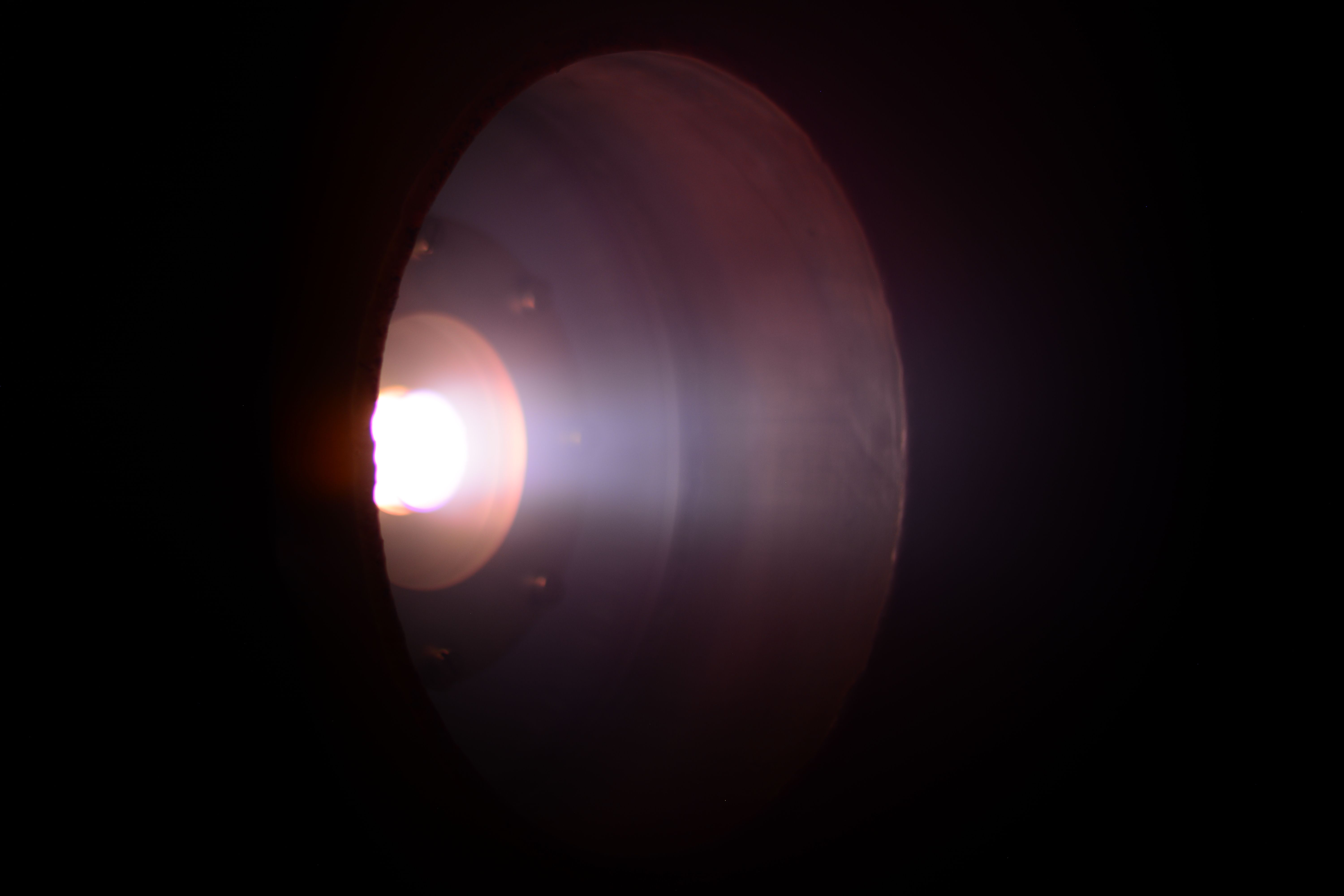}
                        \caption{$I_S=\SI{5.7}{\ampere}$, $P_r=5-\SI{17}{\watt}$}
      \end{subfigure}%
                    ~\\
      \begin{subfigure}[b]{0.5\textwidth}
                        \centering
                        \includegraphics[width=.7\textwidth, trim={11cm 0cm 2cm 0cm},clip]{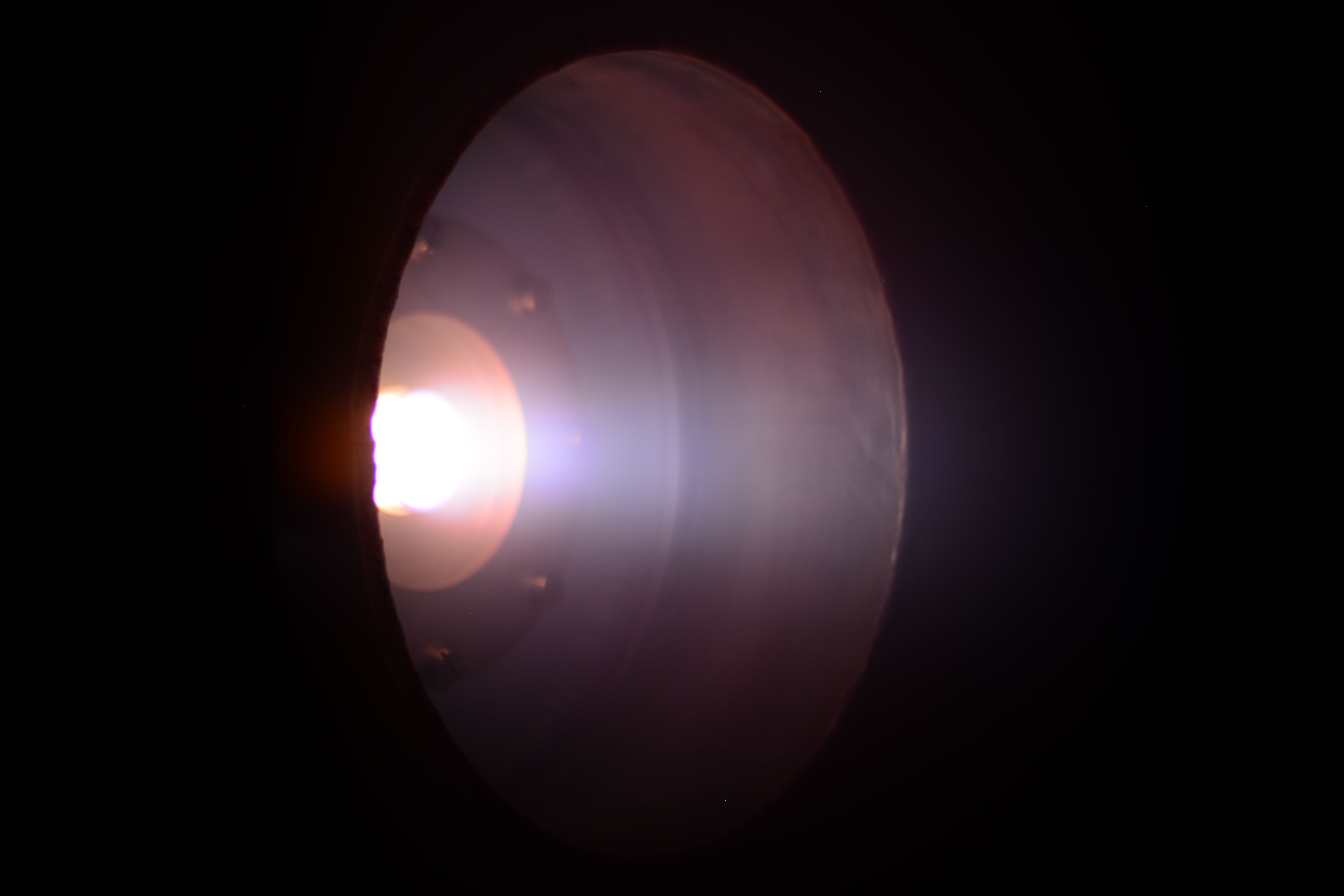}
                        \caption{$I_S=\SI{3.5}{\ampere}$, $P_r=4-\SI{16}{\watt}$}
      \end{subfigure}%
                    ~\\
\caption{$P_f=\SI{60}{\watt}$,~$\dot{m}=\SI{0.487}{\milli\gram\per{\second}}$, Focal length \SI{50}{\milli\meter}, Aperture $f/4$, Exposure time $1/5\SI{}{\second}$, Solenoid at \SI{35}{\milli\meter}.}
\label{fig:O2_Bfield}
\end{figure}

By comparing photographs in Fig.~\ref{fig:O2_Bfield} with the same exposure time, it can be observed how the brightness at the IPT outlet section and the respective plasma jet increases by decreasing $B_0$. The behaviour is similar to that of the \ce{Ar} and \ce{N2} cases. After testing, the IPT performance slightly changed with a decrease of the resonance frequency and an increase in $S_{11}$, while the condition at $f=\SI{40.68}{\mega\hertz}$ improved to $S_{11}=-\SI{43.4}{\decibel}$ and $Z_{IPT}=49.84-j0.66\SI{}{\ohm}$, see Fig.~\ref{fig:resonance_O2}. The absolute increase of the $S_{11}$ parameter is ought to be due to the improvement of the respective soldering spots of the capacitors while operating at relatively higher temperatures. 

\begin{figure}[H]
\centering
\includegraphics[width=.9\textwidth]{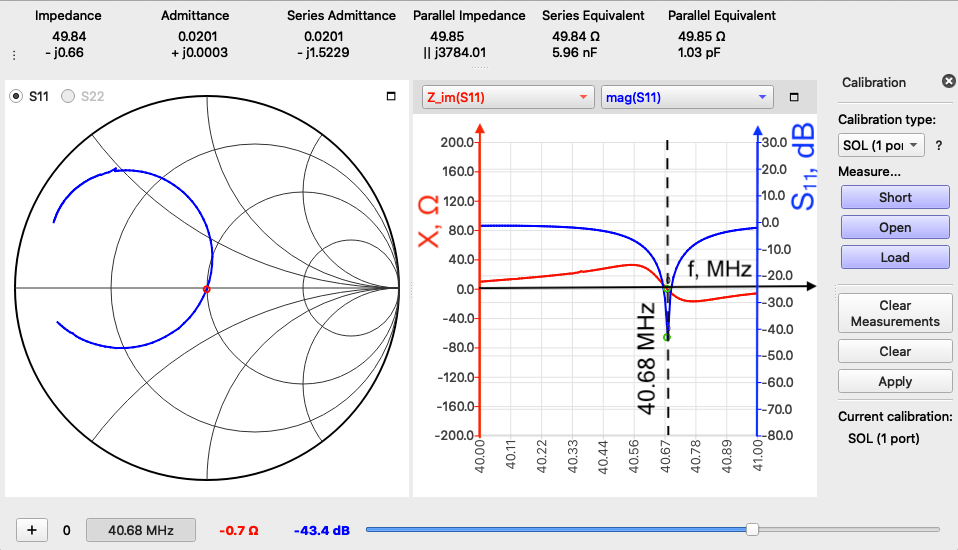}
\caption{IPT $Z_{IPT}$ and $S_{11}$ after test.}
\label{fig:resonance_O2}
\end{figure}
%
\section{Test Campaign Discussion}
Within the IPT test campaign, the operation of the IPT is performed on \ce{Ar}, \ce{N2}, and \ce{O2}. The thruster operates at relatively low input powers $P_f\sim\SI{60}{\watt}$ and could be ignited and operated with all three propellants within the complete given envelope of $\dot{N}$. Fine tuning with the injector head has been successful, leading to a matched load, and the electrical properties of the thruster have been maintained throughout the test campaign. In all cases, the ignition without applied $B_0$, showed a plasma contained within the discharge channel of the IPT. By switching on the solenoid, therefore applying $B_0$, the formation of a plasma jet is observed and, for all the three propellants the configuration, lead to a visibly focused plasma jet for $I_S=6.5-\SI{6.7}{\ampere}$, corresponding to $B_0\sim 28.7-\SI{29.6}{\milli\tesla}$ at the centre of the solenoid, with $P_r<\SI{10}{\watt}$. A zeroed $P_r\sim \SI{0}{\watt}$, instead, is achieved for $I_S=\SI{10.07}{\ampere}$, corresponding to $B_0\sim \SI{44.5}{\milli\tesla}$ at the centre of the solenoid, highlighting a less focused plasma jet with a conical bluish structure that could present higher $T_e$ and/or be a source of further ionization in the plasma jet, according to the similar phenomena observed in~\cite{taka2018,takahashi2019helicon}. Moreover, according to~\cite{detach}, a lower $B_0$ corresponds to lower ion magnetization that would allow them to unfollow magnetic field lines and, instead, leave the discharge channel axially, improving plasma extraction and, therefore, thrust. The tests for the decreasing applied $B_0$, corresponding to the range $B_0=44.5-\SI{14.7}{\milli\tesla}$ at the centre of the solenoid, highlighted for \ce{Ar}, \ce{N2}, and \ce{O2}, the increase of $P_r$ by decreasing $B_0$ and a slightly brighter exhaust section of the discharge channel for $I_S=\SI{3.5}{\ampere}$ corresponding to $B_0\sim \SI{14.7}{\milli\tesla}$, therefore, it is suggested to evaluate the beam divergence of the plasma jet by means of a Faraday probe.

Concerning the Rogowski coil measurements, the current flowing to the IPT for the analysed conditions at $P_f\sim\SI{60}{\watt}$ is of $I_{IPT}\sim\SI{1}{\ampere}$. However, the resolution of the reading does not allow for a preciser measurement, therefore, a Rogowski coil with higher sensitivity is needed for the low power measurements. 

In Tab.~\ref{tab:discharge} a preliminary required $P_f$ for ignition for each condition. Higher $P_f$ is required for the \ce{O2} test case as well as for the lowest $\dot{N}$ in the \ce{N2} case. This can be due to the different ionization energies for \ce{N2} and \ce{O2} compared to \ce{Ar}~\cite{heliconABEP}. Also, the available RF generator provides a reliable $P_f$ only for $P_f>\SI{50}{\watt}$, therefore it is suggested to perform the test with a smaller RF generator that provides reliable $P_f<\SI{50}{\watt}$, and compare with measurement of  i.e. $n_e$ and $T_e$. Finally, the measured $P_r$ is at the generator, and differences of $P_r$ between thruster and matching network for different gases cannot be seen, therefore, a measurement of the thruster impedance during operation, see Chapter~\ref{ch:impedance}, would aid in evaluating such differences.
\begin{table}[H]
\centering
\caption{IPT Estimated $P_f$ Required for Ignition for \ce{Ar}, \ce{N2}, and \ce{O2}.}
\label{tab:discharge}
\begin{tabular}{cccc}
$P_f$, \ce{Ar} &  $P_f$, \ce{N2} &  $P_f$, \ce{O2} & $\dot{N}$\\ 
\toprule
 \SI{}{\watt} & \SI{}{\watt} & \SI{}{\watt} & $\SI{}{1\per{\second}}$\\ 
\midrule
$50-60$ & $50-70$ & $50-70$ & 20.305\\ 
$50-60$ & $50-70$ & $50-70$ & 15.228\\ 
$50-60$ & $50-70$ & $110$ & 10.152\\ 
$50-60$ & $50-70$ & $110$ & 5.076\\ 
$50-60$ & $100$ & $110$ & 2.538\\ 
\bottomrule
\end{tabular}
\end{table}



\chapter{Discussion and Conclusion}
\label{ch:discussion}
Within this dissertation, the main input parameters for the design and evaluation of an ABEP system, namely the intake area $A_{in}$ and its efficiency $\eta_c$, and the thruster efficiency $\eta_T$, are investigated. Those data have to be integrated with an atmospheric model, in this case the NRLMSISE-00 for Earth orbit, required for determining density $\rho(h)$ and composition $n_i(h)$, mainly \ce{N2} and AO, over altitude $h$, location $Lat.,~Long.$, and time (solar activity). Furthermore, the spacecraft frontal area $A_f$ is utilized to estimate the aerodynamic drag $D(h)$. By estimating the collectible mass flow that can be delivered to the thruster $\dot{m}_{thr}$, the required power for ABEP operation $P_{ABEP}$ can be estimated, see Eq~\ref{eq:reqpower}. The calculations have shown that, for example, the $P_{ABEP}$ can variate, over one orbit due to the natural variation of $\rho(h,Lat.,Long.)$, up to $30\%$. It is important to stress out the strong link existing between ABEP and mission requirements leading to the fact that they are finally depending on each other. Ideally, the ABEP system has to operate within a certain range of conditions, especially in terms of input power, propellant density and composition, requiring a certain degree of flexibility due to the variability of the VLEO environment. This is true also for the application to atmospheres of celestial bodies different than Earth. 

The \textbf{intake design} is based on the free molecular flow condition, in which particles do not collide with each other, and GSI models. 

The use of fully diffuse reflecting materials leads to intake designs in which particles are collected and thermalized. Those then proceed to the thruster discharge channel via thermal diffusion. In the diffuse reflection based intake design, a honeycomb structure of small ducts in the front of the intake provides a barrier for the collected particles to escape back into space, basically operating as molecular trap. Fundamental is the area ratio $A_{in}/A_{out}$ that directly influences $\eta_c$, and cannot be changed arbitrarily. Those intakes can be partially evaluated by means of the balance model (BM), that is based on fully diffuse reflecting materials and respective transmission probabilities for each section. The latter, in particular, can only be partially calculated via analytical methods, therefore, for the final design, full DSMC of the intake are required. The intakes based on diffuse reflective materials that are hereby designed, are for a fixed $A_{out}$ given by the thruster discharge channel $d_{out}=\SI{37}{\milli\meter}$ and provide $\eta_c<0.50$ with a relatively compact shape. Furthermore, the presence of the structure of small ducts at the front generally lowers delivers $A_{in,eff}<A_{in}$ and also worsen the collection with an angle $\beta$ to the flow, making the intake $\eta_c$ highly sensitive to flow misalignments, requiring a fast-responsive and precise ADCS to avoid ABEP performance loss. Furthermore, flow misalignment can also arise due to strong winds at high altitudes as seen by GOCE~\cite{GOCE}.

 Instead, if fully specular reflecting materials are utilized, the intake design can be based on optics and directly drive the atmospheric particles inside the discharge channel. The proposed design is based on a parabolic shape, and $\eta_c$ directly depends on the position of the parabola focus. The intake hereby designed provides $\eta_c<0.95$ and is much less sensitive to flow misalignments, therefore relaxing the ADCS requirements compared to the diffuse intake. To achieve the final design, full DSMC are required. 
 
 Moreover, the use of both specular and diffuse reflecting materials leads to hybrid intake designs that can use specular reflective materials to collect and direct the particles in the desired location, and diffuse reflective materials to thermalize them, fro example by using a two stage scheme. While $\eta_c$ is lower compared to a purely specular design, the resulting $\dot{m}_{thr}$ and $p_{ch}$ can be just enough for the thruster requirements.
 
 Finally, the GSI material properties of fully diffuse and fully specular reflections are based on theoretical assumptions. The proposed materials, titanium alloy \ce{Ti-6Al-4V} oxide, nitride, or gold coated for the fully diffuse case, and highly oriented pyrolytic graphite (HOPG) coating or \ce{SiO2} for the fully specular case do not provide the ideal value of accommodation coefficient $\alpha=1$ and $\alpha=0$ respectively, but, based on the current knowledge, are some of the closest $\alpha$ possible while providing a certain degree of resistance to AO. The intake designs shall be re-iterated via real material properties as well as validated in laboratory. This last point is being investigated within the H2020 DISCOVERER project in which sub-scaled intake designs are to be tested in an atomic oxygen wind tunnel (ATOX) facility named ROAR~\cite{oikoiac4}, with the aim to measure, at  first, pressure and particle flux at the outlet. To date, April 2021, the sub-scaled intake are moving into the manufacturing phase. Last but not least, the intake design geometry is to be adapted, based on an iteration loop with the thruster performance, and design. 
 
 The \textbf{thruster} is based on RF frequency and is contactless and neutralizer-less. It does not have any component in direct contact with the plasma, therefore minimizing erosion and the respective performance degradation over time that is typical of conventional EP, especially if operating on chemically reactive propellants such as AO. The thruster produces a quasi-neutral plasma exhaust that removes the need of a neutralizer that would required additional propellant storage and/or a design capable to operate on atmospheric propellant as well, finally increasing the design complexity. The formation of helicon waves within the plasma provides a more efficient discharge but requires an applied magnetic field $B_0$. In particular, for a given plasma density $n_p$ and $B_0$, a peak exists for which the plasma resistance $R_P$ is higher, therefore maximizing the coupling. The effect of input frequency $f$ is to increase $R_P$ for a given $n_p$ but also increases the required $B_0$. In particular $f>\SI{27.12}{\mega\hertz}$ is desirable. Finally a higher $f$ also means for lower $n_p$, $R_P>\SI{1}{\ohm}$ meaning that ignition and operation at lower densities, therefore pressures is enabled, a key requirement for ABEP operation. 
 
 Crucial part of the plasma thruster is the antenna and it is the main responsible for the ionization and (at least partially) acceleration of the propellant gas for thrust generation. From the electrical circuit point of view, the antenna is connected to an RF generator through a matching network. The latter, has the function to adapt the load impedance $Z_L$ to that required by the source $Z_S$. It must be pointed out that the matching network does not improve the load, but only operates as protection for the RF generator by partially absorbing the reflected power within its circuitry and, combined with the load, creates an impedance that is $Z=Z_{Match}+Z_L\sim Z_S$. Moreover, to precisely know power, voltage, and current effectively reaching the thruster, those have to be measured at the output of the matching network. Finally, by designing the antenna for a specified impedance $Z$, the power losses in the electrical circuit can be \textit{a priori} minimized. In this particular case, the birdcage antenna operates at resonance, and it is tuned to resonate at the frequency of the RF generator, meaning that its reactance $X=\SI{0}{\ohm}$. The real part of the impedance is tuned by modifying the relative position of the input and ground port finally reaching to a matched impedance (without plasma) of $Z_{IPT}\sim50+j0\SI{}{\ohm}$. Once the plasma is ignited, $Z_{IPT}$ changes, and the resonance frequency shifts, causing $P_r$ to increase. In this case, due to the relatively small discharge channel and based on the experimental results, it is effectively compensated by the matching network, and $P_r$ is limited. However, to maintain the condition of maximum power coupling, the thruster shall be tuned further, requiring an active method that adjusts the impedance at the thruster itself. 
 
 As RF plasma thrusters are based on emitting EM waves, those need to be shielded. By using applying a Faraday shield around the antenna region, two advantages are simultaneously achieved: the external EM fields do not perturb the antenna EM fields, and the outer environment (e.g. the spacecraft electronics) is not perturbed by the antenna's EM fields.
 From the plasma point of view, the addition of an applied magnetic field aligned with the symmetry axis of the discharge channel leads to many advantages: it provides one of the boundary conditions necessary for the formation of helicon waves within the plasma discharge that increases its efficiency, it tunes $P_r$ by directly changing $Z_{plasma}$, it provides confinement of the plasma from the discharge channel, and its divergence at the outlet section of the discharge channel provides a magnetic nozzle effect for efficient quasi-neutral plasma acceleration.
 From the point of view of EM fields, antennae can provide a variety of configurations that can more or less effectively ionize and accelerate the plasma. It is therefore important to analyse the EM fields generated by the antenna and evaluate their effects on the plasma. It is of particular interest those that create $\vec{E}\times\vec{B}$ acceleration as they act on both ions and electrons at the same time. 
 The testing of the thruster has shown that plasma could be ignited relatively easy with \ce{Ar}, \ce{N2}, and \ce{O2} within all the tested $\dot{m}_{thr}$ range requiring $P_f=50-\SI{100}{\watt}$ with $P_r=0-\SI{15}{\watt}$. For all propellants, the amount of $P_r$ can be tuned by changing the amplitude of the applied $B$-field $B_0$, also (visually) affecting the plasma plume. At the highest magnetic field condition $I_S=\SI{10.07}{\ampere}$, corresponding to $B_0\sim \SI{44.5}{\milli\tesla}$ at the centre of the solenoid, a cone of bluish colour is observed in the plasma exhaust, for all gases, suggesting further ionization and also being visually similar to what observed in~\cite{taka2018,takahashi2019helicon}. By reducing $B_0$ the plasma plume is more focused suggesting a less wide beam divergence. Moreover, according to~\cite{detach}, a lower $B_0$ corresponds to lower ion magnetization that would allow them to unfollow magnetic field lines and, instead, leave the discharge channel axially, improving plasma extraction and, therefore, thrust. In particular, the case of $I_S=\SI{6.7}{\ampere}$, corresponding to $B_0\sim \SI{29.6}{\milli\tesla}$ at the centre of the solenoid, it suggested as the condition at which the beam is less divergent and also minimizes $P_r$. If the $B_0$ is not applied, the plasma is contained within the discharge channel and no visible plasma plume is observed. Instead, once the $B_0$ start to be applied from lower $I_S$, the visible plasma plume is immediately triggered. A jump of brightness has not been observed, even up to $P_f=\SI{300}{\watt}$ suggesting that, due to the efficient antenna design, the discharge already starts in the inductive or helicon regime. Some condition also show hysteresis effect in which, after ignition, $P_f$ could be reduced while still having stable plasma and $P_r\sim\SI{0}{\watt}$. All in all the applied $B_0$ fits within the estimated required $B_0$ based on HELIC calculations.
 
 To be noted, that the current RF generator provides reliable power output only for $P_f>\SI{50}{\watt}$. Therefore, it is possible that the ignition requires even lower powers. Similarly does the matching network. Moreover, it is expected that in the current laboratory setup, between RF generator and thruster some losses can arise, especially within the Rogowski coil adapter which impedance could not be measured due to technical reasons. All in all, $P_c=P_f-P_r=P_{IPT}+P_{loss}$, and it is the power flowing between matching network and the thruster, therefore making the power effectively reaching the thruster itself $P_{IPT}<P_f$. This, finally highlights the fact the power that is effectively required for thruster operation is less than the forward power $P_{IPT}<P_f$. The solenoid shall also be taken into account, however for the future final version an optimum solenoid design is to be designed to minimize the power requirement while providing $B_0$ in the required range. The use of stronger $B_0$ up to $B_0=\SI{100}{\milli\tesla}$ is also desired. For the current thruster set-up an impedance model is presented that can be used to estimate the thruster's impedance given the real time position of the capacitor within the matching network. This can enhance the knowledge of the thruster behaviour for different mass flows $\dot{m}_{thr}$, applied magnetic field $B_0$, and forward power $P_f$ leading to the effectively power that is coupled to the plasma as well as the respective $Z_{plasma}$ variation. 
 
 There is of course the necessity of performing plasma diagnostics measurements, especially with RF compensated probes. In particular, it is of scientific interest to verify the presence of helicon waves within the plasma. This can be done by employing a 3-axes $B$-dot magnetic inductive probe in the plasma plume to measure (if present) the time-rotation of the alternating $B$-field of the helicon wave. The common sets of (RF compensated) Langmuir and  Faraday probes, and Retarding Potential Analyzer (RPA) can be used especially to extract fundamental plasma parameters such as $n_e$, $T_e$, and only then reliably evaluate the thrust based on the model presented in Sec.~\ref{sec:thrustmodel}, as well as measuring the beam divergence and finally extract the ion velocity by the using the RPA. To directly measure thrust a baffle plate can be applied to the current thruster as it is mounted directly on the vacuum chamber.

\label{ch:conclusion}
\subsubsection{Conclusion}

The development of an ABEP system is investigated within the dissertation. 
 At first, a literature review of state-of-the-art ABEP studies is presented and resumed, outlining the main relevant ABEP-based parameters and their range of values. 
 
 Second, the analysis of an \textbf{ABEP} system implementation in VLEO is presented along with a detailed investigation of the main parameters variations, in particular the atmospheric condition vs time, altitude, and location around the globe, as well as intake and thruster efficiency $\eta_c,~\eta_T$ to derive required power $P_{ABEP}$ based on spacecraft's frontal area $A_f$. An example of a GOCE-like ABEP mission in Earth orbit using ABEP is presented, finally requiring $P_{ABEP}<\SI{1.6}{\kilo\watt}$ for $h=190-\SI{250}{\kilo\meter}$, $A_f=\SI{1.1}{\meter^2}$, and $C_D=3.7$. Additionally, the case a Mars orbiting ABEP-based mission is also investigated, requiring $P_{ABEP}<\SI{1.6}{\kilo\watt}$ for $h=120-\SI{160}{\kilo\meter}$, $A_f=\SI{1.1}{\meter^2}$, and $C_D=3$. An ABEP system can be theoretically applied to any celestial object with atmosphere given that enough electric power is provided.
  
 Three \textbf{intake} designs have been developed, two based on fully diffusive reflecting materials, the "EFD Intake" and the "Diffuse Intake", and one based on specular reflecting materials, the "Specular Intake". The "EFD Intake" and the "Diffuse Intake" deliver $\eta_c=0.430$, and $\eta_c=0.458$ respectively. They are optimized for the thruster discharge channel's diameter of $d_{out}=\SI{37}{\milli\meter}$ being the outlet section of the intake. The "EFD Intake" has $A_{in}=\SI{0.005}{\meter^2}$, while the "Diffuse Intake" has $A_{in}=\SI{0.008}{\meter^2}$. The "Diffuse Intake" is $82\%$ shorter than the "EFD Intake". The "Diffuse Intake" has been also analysed in terms of $\eta_c$ sensitivity to the inflow angle $\beta$, showing that $\eta_c$ drop significantly, $-17\%$, already with only $\beta=\SI{5}{\degree}$.
 The "Specular Intake", instead, delivers $\eta_c=0.943$ for an $A_{in}=\SI{0.019}{\meter^2}$ and is less sensitive to $\beta$, showing a $-8\%$ drop in $\eta_c$ only for $\beta=\SI{15}{\degree}$. The "Specular Intake" is selected as the best candidate for being more efficient and less sensitive to $\beta$ compared to the EFD and Diffuse intakes. However, the latter designs can also be applied, given that the requirement for the thruster in terms of $\dot{m}_{thr}$ and $p_{ch}$ is matched. Moreover, an hybrid intake design concept is also proposed. 
 
 The laboratory model of the RF plasma \textbf{thruster} is designed, built, and tested for operation at $f=\SI{40.68}{\mega\hertz}$ with a respective optimized RF power system. The RF plasma thruster is contactless, minimizing components in direct contact with the plasma, and produces a quasi-neutral plasma plume that removes the need of a neutralizer.
 
 A new concept of antenna is applied to a plasma thruster: the cylindrical birdcage antenna to maximize the electrical efficiency and providing a matched load to the PPU,  It operates at resonance, and allows for customized matching and tuning by applying capacitors on the antenna itself as well as including a mechanism of further fine tuning. The thruster design includes a Faraday shield to shield out the thruster's antenna from the environment EM fields, as well as to shield out the environment from the thruster's generated EM fields, therefore minimizing possible interferences with spacecraft electronics. An applied $B$-field $B_0$ is generated by a solenoid, this theoretically enables the boundary conditions for triggering helicon waves in the plasma discharge for higher efficiency, as well as a magnetic nozzle effect at the exhaust due to its divergence, providing thrust. Furthermore, the birdcage antenna has a linearly polarized EM field configuration that also theoretically provides drift velocity to both ions and electrons at the same time and along the same direction, therefore adding additional thrust.  
 The thruster based on a cylindrical birdcage antenna operating at resonance lead to the final impedance (without plasma) of $Z_{IPT}\sim50+j0\SI{}{\ohm}$, being a matched load to the RF generator. The design has been verified in accordance with XFdtd\textsuperscript{\textregistered}-based simulations. As well, the mechanism of fine tuning with a movable conductive injector head demonstrated its operation and has been used to tune the thruster's resonance frequency. The testing has been successful on the tested gases of \ce{Ar},~\ce{N2}, and~\ce{O2}, and has shown easy ignition and low power consumption for steady operation, requiring $P_f = \SI{60}{\watt}$ with $P_r\sim\SI{0}{\watt}$ at the RF generator for any of the injected propellants, providing stable plasma and a respective jet in the vacuum chamber. The applied magnetic field $B_0$ has shown to tune $P_r$, as well as visually change the shape of the plasma jet. In particular, for $I_S=\SI{6.7}{\ampere}$, corresponding to $B_0\sim\SI{29.6}{\milli\tesla}$ at the centre of the solenoid, seem to provide a visually collimated plasma jet with the minimum $P_r$ at the generator, that fits within the range of required $B_0$ predicted by HELIC. 
 
 An impedance model for the current set-up has been provided with respective equations that can be used to estimate the thruster impedance during operation based on the matching network's capacitors position.
 
 To verify the presence of helicon waves by measuring the rotating magnetic field within the plasma plume of the thruster, a 3-axis magnetic inductive B-dot probe has been designed and assembled. 
 
An ABEP unit can be derived based on the analysis carried within this dissertation. Hereby, three conceptual schematics of an ABEP unit are shown based on the developed thruster, and the three intakes: EFD, Diffuse, and Specular Intake, see Fig.~\ref{fig:EFDABEP},~Fig.~\ref{fig:DIFFUSEABEP}, and~Fig.~\ref{fig:SPECULARABEP}. Such units can be clustered, included into the spacecraft main frame, or mounted on pods within nacelles therefore also providing capability of steering by having different power applied to each unit, as well as scaled up, or down, in terms of geometry and input power.

\begin{figure}[h]
	\centering
	\includegraphics[width=\textwidth]{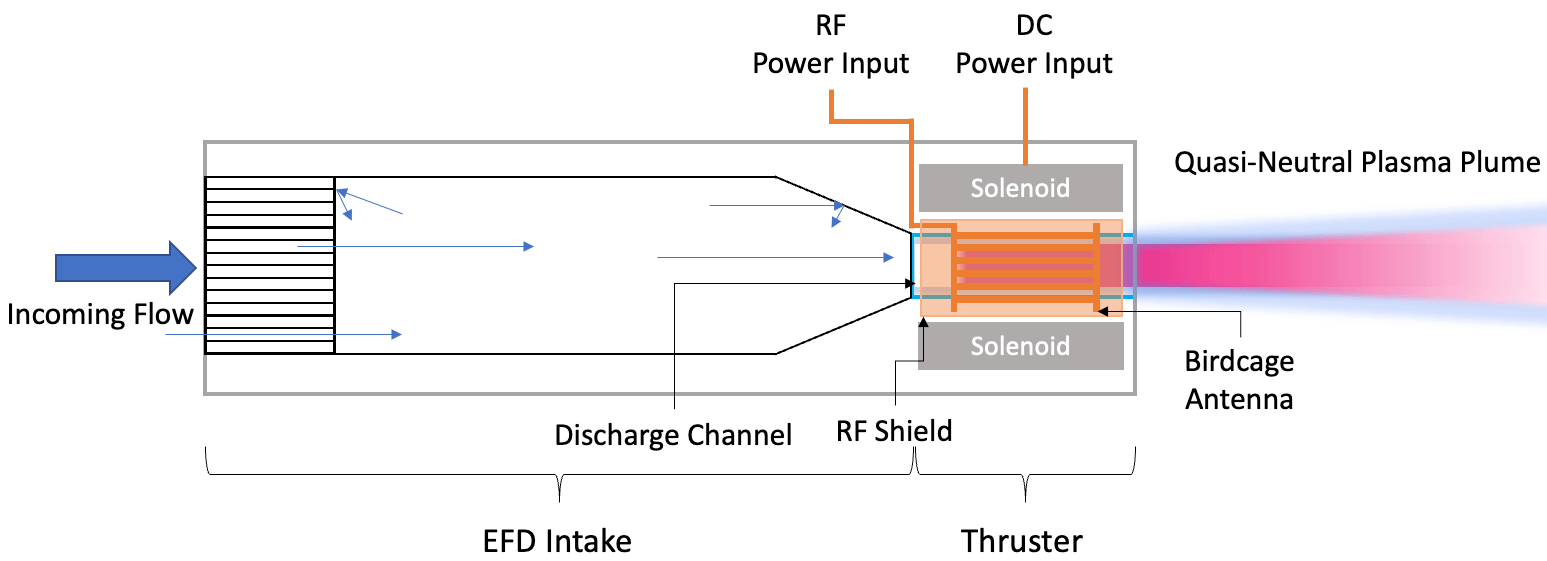}
	\caption{ABEP Unit Concept based on EFD Intake.}
	\label{fig:EFDABEP}
\end{figure}

\begin{figure}[h]
	\centering
	\includegraphics[width=.78\textwidth]{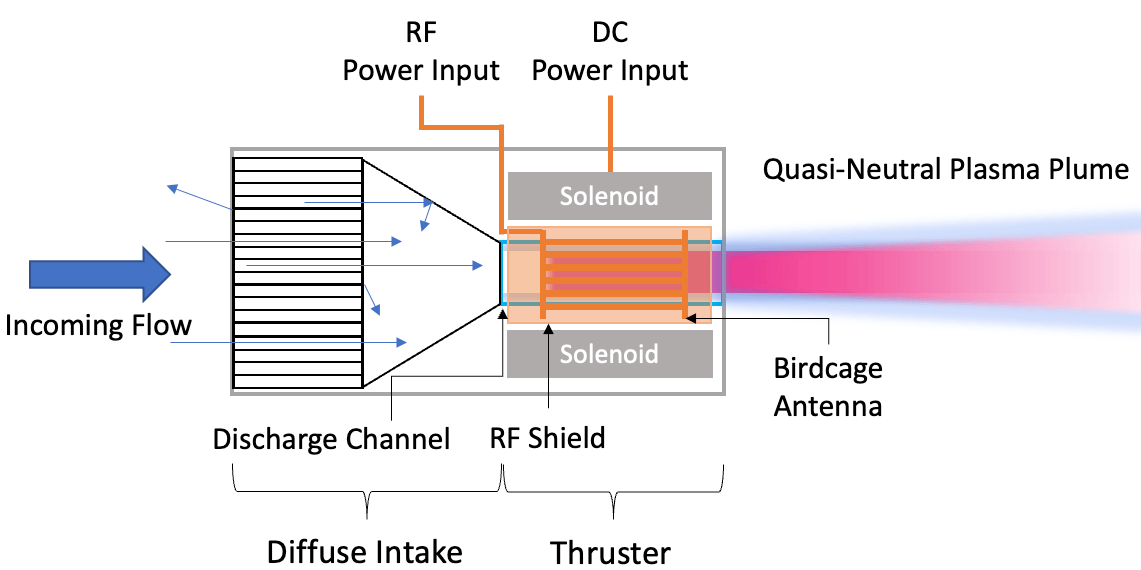}
	\caption{ABEP Unit Concept based on Diffuse Intake.}
	\label{fig:DIFFUSEABEP}
\end{figure}

\begin{figure}[h]
	\centering
	\includegraphics[width=1\textwidth]{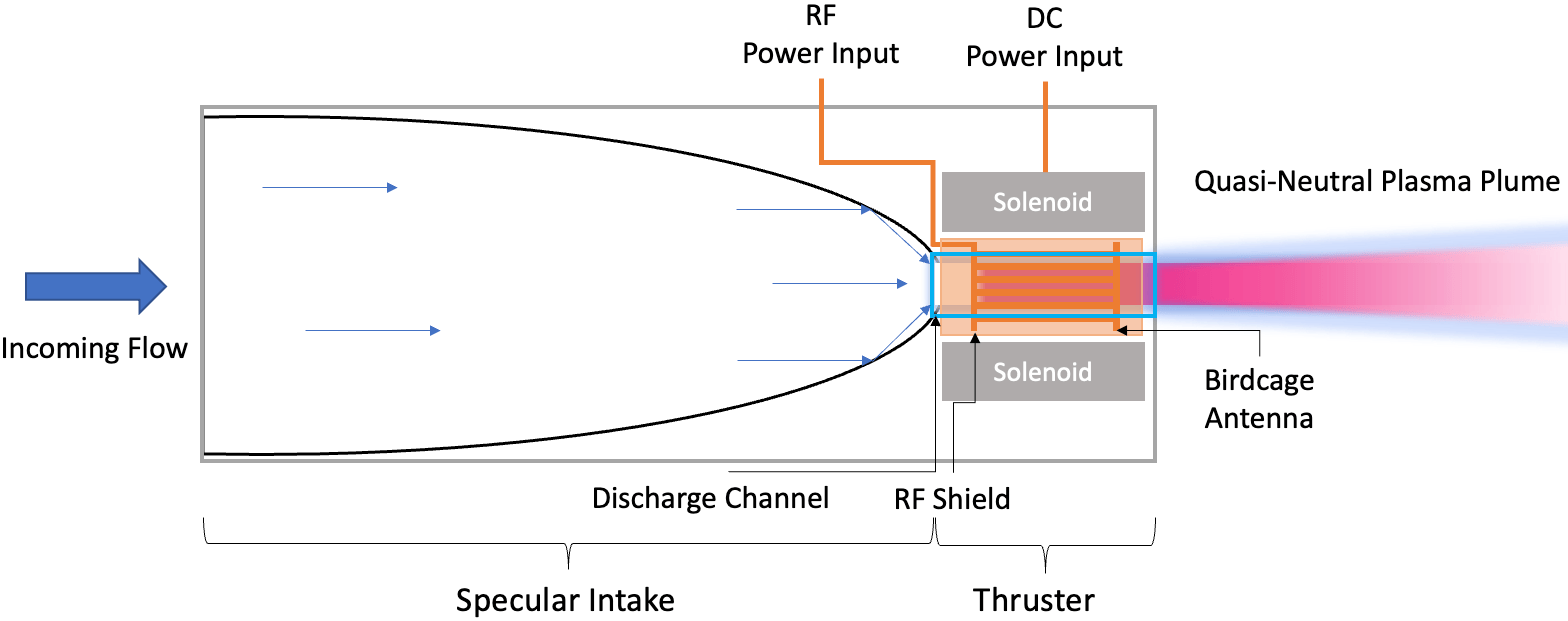}
	\caption{ABEP Unit Concept based on Specular Intake.}
	\label{fig:SPECULARABEP}
\end{figure}
\vspace{-15pt}
The configuration shown in Fig.~\ref{fig:ABEP_Config}(c) displays four ABEP unit mounted outside the spacecraft core in pods, this also has steering capability, but with a shorter arm, moreover, the symmetric solar arrays can provide stability and by rolling, the solar array angle to the Sun can be controlled. The configuration shown in Fig.~\ref{fig:ABEP_Config}(d) is the clustered one. It can have multiple ABEP units, or intakes converging onto one thruster with the spacecraft core around it. In Fig.~\ref{fig:ABEP_Config}(e), the ABEP units are contained within the spacecraft main frame, similar to the concept presented shown in~\cite{BUSEK2}. Furthermore. the solar array fins can be steerable to provide a certain degree of aerodynamic control, this is currently under investigation within DISCOVERER~\cite{CRISP202185}. Finally, for large required powers or for spacecraft orbiting further away than the asteroid belt, the use of solar array can become counter productive or not applicable at all, therefore. other power sources have to be taken into account, for example nuclear based power.
\begin{figure}[h] 
	\centering
	\includegraphics[width=.85\textwidth]{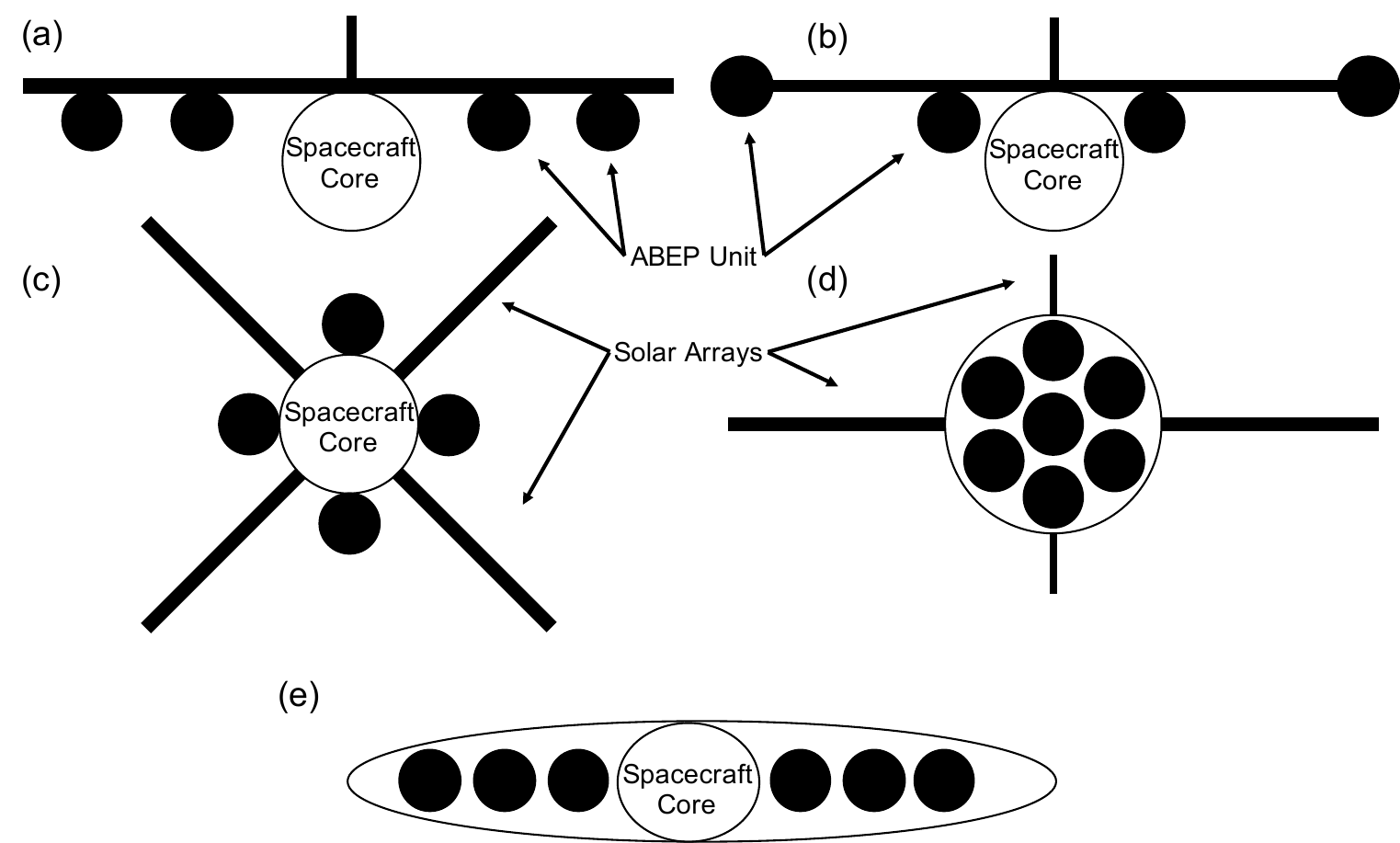}
	\caption{ABEP Unit Configurations on a Spacecraft.}
	\label{fig:ABEP_Config}
\end{figure}

\chapter{Outlook}
To further advance the development of an ABEP system, there are many steps that can be taken.

Concerning the intake, it is needed account for real materials and their properties in terms of accommodation coefficient and GSI into the intake design that needs to be evaluated based on simulation at first and secondly be tested on ground. It is also very important to evaluate their erosion behaviour given the expected fluxes of particle species in the target orbit, to ensure the performance is maintained over the mission lifetime. 

In terms of thruster development, it is needed to estimate the thrust both numerically, with the model provided within this dissertation and based on the measured plasma properties, as well via direct thrust measurements that can be performed either by using a baffle plate in case of the currently developed thruster, or on a thrust balance once the vacuum version of the thruster is built. The latter needs to ensure operation within the required range of temperatures, that can become critical for the capacitors soldering spot, therefore a more robust way of integration can be investigated. Its shielded design, however, can ensure an interference-free operation on a satellite. 

From the electrical point of view, an impedance operation envelope of the thruster is needed to design an optimized matching network that can be either a separate unit, or a mechanism on the thruster itself that can perform the matching, for example movable electrically conductive surfaces or variable capacitors on the antenna. Those can be controlled by an active thruster control system, much like an automatic gearbox of a car, that continuously measures the thruster impedance and acts on the moving surfaces or changes the capacitors capacitances to ensure the thruster to be operating at maximum coupling condition, with zeroed reflected power to finally minimize not only power requirements but also heating of the thruster's antenna. 

From the physics point of view, the thruster shall be analysed by means of a full set of plasma diagnostics equipped with respective RF compensation circuit. The results can be used to study the power coupling mechanism of helicon discharge-based thrusters that, with an impedance-matched design as the one shown in this dissertation, can drastically reduced uncertainties. 
The developed B-dot probe needs to be calibrated and inserted into the thruster's plasma plume to measure the rotating magnetic field and confirm its operation in the helicon regime. Such results can further define the range of required magnetic field strength for thruster operation. This will feed directly into an optimized magnetic field generation system that can be controlled by the thruster control system mentioned above. 

Finally, joining the thruster together with an intake enables the operation of the system as one unit. However, to test the system on ground, a way to generate flows that resemble VLEO condition is a great challenge. A solution would be to use a second thruster unit for it, as it generates a quasi-neutral plasma, at a distance such that the ionization level of the flow reaching the intake could be negligible, but the particle velocity still high enough.

For the flight version, a hybrid system could be equipped on a first mission, in which the thruster is to be operated either by feeding it with a propellant from a tank, or by using the intake. The use of multiple ABEP-unit in a clustered version, as well the upscaling of the system would enable even relatively large spacecraft to orbit at very low altitude for an unlimited period of time. 

%

\cleardoublepage
\addcontentsline{toc}{chapter}{A. Balance Model Equations}
\chapter*{A. Balance Model Equations}
The following main equations are used in the BM to evaluate the intake performance (full accommodation), especially the intake efficiency, $\eta_c$, and the density ratio $\frac{n_{ch}}{n_{in}}$ that are then used to calculated the mass flow to the thruster $\dot{m}_{thr}$.
\begin{equation}
\dot{N}_{inl.{1}}=\dot{N}_{in}\Theta_{inl.{1}}
\label{eq:BM1a}
\end{equation}

The thermal mass flux $\Gamma$ is defined as in Eq.~\ref{eq:effusion}~\cite{romanoiepc}.
\begin{equation}
\Gamma(n,T)_{x_i}=n\sqrt{\frac{m_{p}k_{B}T_{ch}}{2\pi}}=m_p n \bar{v}_{x_i}
\label{eq:effusion}
\end{equation}

Therefore, it is possible to apply $\Gamma$ to determine the backflow, Eq.~\ref{eq:back}, and the outflow in the chamber of the intake as in Eq. \ref{eq:out} due to the assumption of full accommodation.
\begin{equation}
\dot{N}_{inl.2}=\frac{\Gamma(n_{ch},T_{ch})}{m_p}A_{in}\Theta_{inl.2}
\label{eq:back}
\end{equation}
\begin{equation}
\dot{N}_{out}=\frac{\Gamma(n_{ch},T_{ch})}{m_p}A_{out}\Theta_{out}+\dot{N}_{accel.}
\label{eq:out}
\end{equation}
The flows are balanced as in Eq.~\ref{eq:cont} and Eq.~\ref{eq:cont1}.
\begin{equation}
\dot{N}_{inl.1}=\dot{N}_{inl.2}+\dot{N}_{out}
\label{eq:cont} 
\end{equation}

\begin{equation}
\dot{N}_{in}\Theta_{inl.1}=\frac{\Gamma(n_{ch}, T_{ch})}{m_p}(A_{in}\Theta_{inl.2}+A_{out}\Theta_{out})
\label{eq:cont1}
\end{equation}

Therefore $\Gamma$ can be extracted and, thus, the density $n_{ch}$ inside the chamber from Eq.~\ref{eq:effusion} results in Eq.~\ref{eq:n}.
\begin{equation}
n_{ch}=\Gamma(n_{ch},T_{ch})\sqrt{\frac{2\pi}{m_p k_B T_{ch}}}
\label{eq:n}
\end{equation}

The pressure can be calculated by applying the ideal gas condition as: in Eq.~\ref{eq:pch}.
\begin{equation}
p_{ch}=n_{ch}k_B T_{ch}
\label{eq:pch}
\end{equation}

The intake efficiency is in Eq.~\ref{eq:effcap}, and the respective pressure ratio and number density ratio in Eq.~\ref{eq:pratio} and Eq.~\ref{eq:nratio}.
\begin{equation}
\eta_{c}=\frac{\dot{N}_{out}}{\dot{N}_{in}}=\frac{\Theta_{inl.1}}{\frac{A_{in}}{A_{out}}\frac{\Theta_{inl.2}}{\Theta_{out}}+1}
\label{eq:effcap}
\end{equation}

\begin{equation}
\frac{p_{ch}}{p_{in}}=\frac{m_p\dot{N}_{in}\Theta_{inl.1}}{A_{in}\Theta_{inl.2}+A_{out}\Theta_{out}} \sqrt{\frac{2\pi}{m_p k_B T_{ch}}}\frac{ T_{ch}}{T_{in}n_{in} }
\label{eq:pratio}
\end{equation}
\begin{equation}
\frac{n_{ch}}{n_{in}}=\frac{m_p v_{in}\Theta_{inl.1}}{\Theta_{inl.2}+\frac{A_{out}}{A_{in}}\Theta_{out}}\sqrt{\frac{2\pi}{m_p k_B T_{ch}}}
\label{eq:nratio}
\end{equation}
\begin{table}[h]
 \caption{Inflow Conditions for Clausing $\Theta$}
 \label{tab:conditions}
 \centering
\begin{tabular}{ccc}
 \toprule
& $v_{in}$ & $m_p$\\
& \SI{}{\meter\per{\second}}& $\SI{E-26}{\kilo\gram}$\\
\midrule
1 & 7800 & $2.66/\ce{O}$\\
2 & 3500 & $2.66/\ce{O}$\\
3 & 7800 & $7.31/\ce{CO2}$\\
\bottomrule
\end{tabular}
\end{table}

\begin{table}[h]
 \caption{Squared EFD with Squared Ducts Geometrical Design}
 \label{tab:EFDVLMO}
 \centering
 \begin{tabular}{cccccccc}
 \toprule
$A_{in}$ & $F_f$ & $N_{ducts}$ & $L/R_{ducts}$ & $R_{ducts}$ & $L/R_{tube}$ & $R_{tube}=R_{in}$ & $\eta_c$\\
$\SI{}{\meter^2}$ & - & - & - & $\SI{}{\centi\meter}$ & - & $\SI{}{\centi\meter}$ & -\\
\midrule
0.01075 & 0.940 & 467 & 9 & 0.25 & 10.667 & 5.85 & 0.43\\
 \bottomrule
 \end{tabular}
 \end{table} 
\label{ch:appendixB}
\chaptermark{Appendix A}



\cleardoublepage

\addcontentsline{toc}{chapter}{Bibliography}
\phantomsection
\bibliography{bibliography}
\bibliographystyle{elsarticle-num}

\end{document}